\documentclass[a4paper,11pt]{article}
\pdfoutput=1 

\usepackage{jheppub} 

\usepackage[T1]{fontenc} 

\usepackage{slashed}
\usepackage{cancel}

\usepackage{mathtools}

\usepackage{tikz}
\RequirePackage{xspace}

\usepackage[acronym]{glossaries}
\glsdisablehyper
\glsaddkey
    {hyphenated}            
    {\relax}                
    {\glsentryhyphtext}     
    {\Glsentryhyphtext}     
    {\glshyphtext}          
    {\Glshyphtext}          
    {\GLShyphtext}          
\newcommand{\glshyph}[1]{%
  \ifglsused{#1}{\glsentryshort{#1}}{\glshyphtext{#1} (\glsentryshort{#1})\glsunset{#1}}%
}
\newacronym[hyphenated={initial-state}]{is}{IS}{initial state}
\newacronym[hyphenated={final-state}]{fs}{FS}{final state}
\newacronym{ym}{YM}{Yang--Mills}
\newacronym{ewsm}{EWSM}{Electroweak Standard Model}
\newacronym{ckm}{CKM}{Cabibbo--Kobayashi--Maskawa}
\newacronym{brst}{BRST}{Becchi--Rouet--Stora--Tyutin}
\newacronym{qed}{QED}{Quantum Electrodynamics}
\newacronym{qcd}{QCD}{Quantum Chromodynamics}
\newacronym{ir}{IR}{infrared}
\newacronym[hyphenated={leading-order}]{lo}{LO}{leading order}
\newacronym[hyphenated={next-to-leading-order}]{nlo}{NLO}{next-to-leading order}
\newacronym[hyphenated={next-to-next-to-leading-order}]{nnlo}{NNLO}{next-to-next-to-leading order}
\newacronym{ew}{EW}{electroweak}
\newacronym{ewa}{EWA}{effective W approximation}
\newacronym{gbet}{GBET}{Goldstone-boson equivalence theorem}
\newacronym{geg}{GEG}{Goldstone Equivalence Gauge}
\newacronym[longplural={Ward identities}]{wi}{WI}{Ward identity}
\newacronym{st}{ST}{Slavnov--Taylor}
\newacronym{pi}{PI}{path integral}
\newacronym{ssb}{SSB}{spontaneous symmetry breaking}
\newacronym[longplural={equations of motion}]{eom}{EOM}{equation of motion}
\newacronym{vev}{vev}{vacuum expectation value}
\newacronym{gf}{GF}{Green function}
\newacronym{lhc}{LHC}{Large Hadron Collider}

\newcommand{\genfield}[1]{\varphi_{#1}}

\newcommand{\genantifield}[1]{\bar{\varphi}_{#1}}
\newcommand{\nonfermfield}[1]{\varrho_{#1}}
\newcommand{\nonfermantifield}[1]{\bar{\varrho}_{#1}}
\newcommand{\nonfermtildefield}[1]{\tilde{\varrho}_{#1}}
\newcommand{\vbfield}[1]{V_{#1}}
\newcommand{\vbantifield}[1]{\bar{V}_{#1}}
\newcommand{\vbtildefield}[1]{\tilde{V}_{#1}}
\newcommand{\gbfield}[1]{G_{#1}}
\newcommand{\gbantifield}[1]{\bar{G}_{#1}}
\newcommand{\gbtildefield}[1]{\tilde{G}_{#1}}
\newcommand{\higgsfield}[1]{H_{#1}}
\newcommand{\higgsantifield}[1]{\bar{H}_{#1}}
\newcommand{\fermfield}[1]{f_{#1}}
\newcommand{\antifermfield}[1]{\bar{f}_{#1}}
\newcommand{\restpol}{r}
\newcommand{\chir}{\sigma}
\newcommand{\pol}{\kappa}
\newcommand{\polferm}{\kappa}
\newcommand{\polvb}{\lambda}
\newcommand{\doubcptone}{\zeta}
\newcommand{\doubcpttwo}{\rho}
\newcommand{\doubcptthree}{\gamma}
\newcommand{\chgWindone}{a}
\newcommand{\chgWindtwo}{b}

\newcommand{\TeV}{\ensuremath{\text{TeV}}}
\newcommand{\scalcpt}{\ensuremath{\phi}} 
\newcommand{\scalmult}{\ensuremath{\phi}}
\newcommand{\cplxhiggscpt}{\ensuremath{\phi}}
\newcommand{\cplxhiggsmult}{\ensuremath{\phi}}

\newcommand{\gaugegroup}{\ensuremath{G}} 
\newcommand{\subgroup}{\ensuremath{G'}} 
\newcommand{\epsepsdiff}{\ensuremath{%
    \bar{\varepsilon}_+^{\mu}
    \bar{\varepsilon}_+^{\nu \ast}
    -
    \bar{\varepsilon}_-^{\mu}
    \bar{\varepsilon}_-^{\nu \ast}
  }}%
\newcommand{\epsepssum}{\ensuremath{%
    \bar{\varepsilon}_+^{\mu}
    \bar{\varepsilon}_+^{\nu \ast}
    +
    \bar{\varepsilon}_-^{\mu}
    \bar{\varepsilon}_-^{\nu \ast}
  }}%
\newcommand{\imagi}{\ensuremath{\mathrm{i}}}

\newcommand{\expe}{\ensuremath{\mathrm{e}}}
\newcommand{\D}[1]{\ensuremath{\mathop{}\!\mathrm{d}#1}}
\newcommand{\orderof}[1]{\ensuremath{\mathcal{O}\left(#1\right)}}
\newcommand{\RM}[1]{\MakeUppercase{\romannumeral #1}}
\newcommand{\diagtype}[1]{\RM{#1}}%
\newcommand\mathinheadbold[2]{\texorpdfstring{$\boldsymbol{#1}$}{#2}}
\newcommand\mathinhead[2]{\texorpdfstring{$#1$}{#2}}
\makeatletter
\newcommand{\smallerstyle}[1]{%
  \ifx#1\displaystyle\scriptstyle\else
  \ifx#1\textstyle\scriptstyle\else
  \scriptscriptstyle\fi\fi
}
\newcommand*{\xsim}[1]{\mathrel{\mathpalette\x@sim{#1}}}
\newcommand*{\x@sim}[2]{%
   \sbox0{\m@th$\smallerstyle#1\,#2\,$}%
   \sbox1{\m@th$#1\sim$}%
   #1\mathop{%
      \ifdim\wd0>\wd1
         \resizebox{\wd0}{\height}{\box1}%
      \else
         \box1
      \fi
   }\limits_{#2}%
}
\makeatother
\newcommand{\ftntmath}[1]{\mbox{\footnotesize$\displaystyle#1$}}
\newcommand{\MContrNoFields}[1]{\ensuremath{
  \mathcal{M}^{(n+1)}_{(#1)}
  }}
\newcommand{\MContrExplFields}[3]{\ensuremath{
    \Big[\MContrNoFields{#1}\Big]_{#2 #3 X}
  }}
\newcommand{\MContr}[3]{
  \MContrExplFields{\genfield{#1}}{\genfield{#2}}{\genfield{#3}}
  }

\newcommand{\revision}[1]{{#1}}%
\newcommand{\revisiontwo}[1]{{#1}}%

\DeclareRobustCommand{\ensuremathrm}[1]{\ensuremath{\mathrm{#1}}}
\DeclareRobustCommand{\Pp}{{\ensuremathrm{p}}\xspace}
\DeclareRobustCommand{\PH}{{\ensuremathrm{H}}\xspace}
\DeclareRobustCommand{\Pphot}{{\ensuremathrm{\gamma}}\xspace}
\DeclareRobustCommand{\PZ}{{\ensuremathrm{Z}}\xspace}
\DeclareRobustCommand{\PW}{{\ensuremathrm{W}}\xspace}

\usetikzlibrary{
  calc,
  arrows.meta,
  arrows,
  decorations,
  decorations.pathmorphing,
  decorations.markings,
  decorations.pathreplacing,
  shapes.misc, 
  positioning,
  backgrounds, 
}

\def\stealtharrow{{\arrow[xshift=1.1pt+2.25\pgflinewidth,scale=1.4]{stealth}}}
\newcommand\blob{} 
\def\blob (#1)#2#3#4{%
  \draw [fill=#4] (#1) circle (#2); 
  \draw (#1) node {#3}; 
}
\tikzset{photon/.style={
  draw=none,
  decoration={name=none},
  postaction={
    draw,
    decoration={snake,amplitude=.05cm,segment length=.2cm},
    decorate=true
  }}}
\tikzset{gluon/.style={
  draw=none,
  decoration={name=none},
  postaction={
    draw,
    decoration={coil,amplitude=.1cm, segment length=.13cm,aspect=.8},
    decorate=true
  }}}
\tikzset{scalar/.style={
  draw=none,
  decoration={name=none},
  postaction={
    draw,
    dashed, 
  }}}
\tikzset{warrow/.style={
  decoration={
    markings,
    mark=at position .6 with {
      \arrow[scale=1.4]{latex}
    },
  },
  postaction={decorate=true},
}}
\tikzset{warrowrev/.style={
  decoration={
    markings,
    mark=at position .55 with {
      \arrow[scale=1.4]{latex reversed},
    },
  },
  postaction={decorate=true},
}}
\tikzset{fermion/.style={
  draw=none,
  decoration={name=none},
  postaction={
    draw,
    warrow
}}}
\tikzset{antifermion/.style={
  draw=none,
  decoration={name=none},
  postaction={
    draw,
    warrowrev
}}}
\tikzset{
  momentum/momlabdist/.store in=\momlabdist,
  momentum/momlabdist=0pt,
  momentum/arrdist/.store in=\momarrdist,
  momentum/arrdist=2mm,
  momentum/startpos/.store in=\startpos,
  momentum/startpos=0.2,
  momentum/endpos/.store in=\endpos,
  momentum/endpos=0.8,
  }
\tikzset{momentum/.style args={[#1]#2}{
  decoration={show path construction,
    lineto code={
      \tikzset{momentum/.cd,#1}
      \path
        (\tikzinputsegmentfirst) -- (\tikzinputsegmentlast)
        coordinate [pos=\startpos] (m1)
        coordinate [pos=\endpos] (m2);
      \draw [-Stealth]
      ($(m1)!\momarrdist!90:(m2)$) -- ($(m2)!\momarrdist!-90:(m1)$)
      node [pos=0.5, auto,
        inner sep=0.333pt,
        outer sep=\momlabdist,
        ] {#2};
    }
  }, postaction=decorate
}}
\tikzset{momentum'/.style args={[#1]#2}{
  decoration={show path construction,
    lineto code={
      \tikzset{momentum/.cd,#1}
      \path
        (\tikzinputsegmentfirst) -- (\tikzinputsegmentlast)
        coordinate [pos=\startpos] (m1)
        coordinate [pos=\endpos] (m2);
      \draw [-Stealth]
      ($(m1)!\momarrdist!-90:(m2)$) -- ($(m2)!\momarrdist!90:(m1)$)
      node [pos=0.5, auto, swap,
        inner sep=0.333pt,
        outer sep=\momlabdist,
        ] {#2};
    }
  }, postaction=decorate
}}

\newcommand{\halfline}{\the\dimexpr\fontdimen22\textfont2\relax}%
\newif\ifdrawmom\drawmomtrue%
\newif\ifdrawblob\drawblobtrue%
\newif\ifdrawvertdot\drawvertdotfalse%
\newcommand{\sclfac}{}%
\newcommand{\extsclfac}{}%
\newcommand{\intsclfac}{}%
\newcommand{\len}{}%
\newcommand{\rad}{}%
\newcommand{\dotrad}{}%
\newcommand{\diagang}{}%
\newcommand{\resetdiagargs}{
  \renewcommand{\sclfac}{1}%
  \renewcommand{\extsclfac}{1}%
  \renewcommand{\intsclfac}{.85}%
  \renewcommand{\len}{1.2cm}%
  \renewcommand{\rad}{6mm}%
  \renewcommand{\dotrad}{.7mm}%
  \renewcommand{\diagang}{60}%
}%
\resetdiagargs%
\newcommand{\mytemp}{inc}%
\newcommand{\feynmanfieldline}[2]{%
  \begin{tikzpicture}[
    momentum/momlabdist=1.2mm,
    momentum/arrdist=2.5mm,
    baseline=-\halfline]
    \node (l) {};
    \node at ($(l)+(0:\sclfac*\len)$) (r)  {};
    \draw [#1] (l.center) -- (r.center);
    \newcommand{\inc}{inc};
    \newcommand{\outg}{outg};
    \renewcommand{\mytemp}{#2}\ifx\mytemp\inc
      \blob (r) {\sclfac*\dotrad}{}{black};
    \fi
    \renewcommand{\mytemp}{#2}\ifx\mytemp\outg
      \blob (l) {\sclfac*\dotrad}{}{black};
    \fi
  \end{tikzpicture}
}%
\newcommand{\feynmanprop}[4]{%
  \begin{tikzpicture}[
    momentum/momlabdist=1.2mm,
    momentum/arrdist=2.5mm,
    baseline=-\halfline]
    \node (l) {};
    \node at ($(l)+(0:\sclfac*\len)$) (r)  {};
    \node [left = 0mm of l] (labl) {\(#2\)};
    \node [right = 0mm of r] (labr) {\(#3\)};
    \draw [#1, momentum={[]\(#4\)}] (r.center) -- (l.center);
    \blob (l) {\sclfac*\dotrad}{}{black};
    \blob (r) {\sclfac*\dotrad}{}{black};
  \end{tikzpicture}
}%
\newcommand{\feynvertex}[9]{%
  \begin{tikzpicture}[
    momentum/momlabdist=1.2mm,
    momentum/arrdist=2.5mm,
    baseline=-\halfline]
    \node (e1) {\(#4\)};
    \node at ($(e1)+(0        :\sclfac*\len)$) (v)  {};
    \node at ($(v) +(-\diagang:\sclfac*\len)$) (e2) {\(#5\)};
    \node at ($(v) +( \diagang:\sclfac*\len)$) (e3) {\(#6\)};
    \renewcommand{\mytemp}{#7}\ifx\mytemp\empty
      \draw [#1] (e1) -- (v.center);
    \else
      \draw [#1,momentum={[]\(#7\)}] (e1) -- (v.center);
    \fi
    \renewcommand{\mytemp}{#8}\ifx\mytemp\empty
      \draw [#2] (e2) -- (v.center);
    \else
      \draw [#2,momentum={[]\(#8\)}] (e2) -- (v.center);
    \fi
    \renewcommand{\mytemp}{#9}\ifx\mytemp\empty
      \draw [#3] (e3) -- (v.center);
    \else
      \draw [#3,momentum={[]\(#9\)}] (e3) -- (v.center);
    \fi
    \ifdrawvertdot
      \blob (v) {\sclfac*\dotrad}{}{black};
    \fi
  \end{tikzpicture}
}%
\newcommand{\splittingdiagIndivFS}[7]{%
  \begin{tikzpicture}[baseline=-\halfline]%
    \node (t) {};%
    \ifdrawblob%
      \node at ($(t)+(\sclfac*\rad,0)$) (ij) {};%
    \else%
      \node at ($(t)$) (ij) {};%
    \fi%
    \ifdrawmom%
      \tikzset{thismomij/.style={%
        momentum={[momlabdist = 2pt]\(p_{ij}\)}%
      }};%
      \tikzset{thismomi/.style={%
        momentum= {[]\(p_{i}\)}%
      }};%
      \tikzset{thismomj/.style={%
        momentum'={[]\(p_{j}\)}%
      }};%
    \else%
      \tikzset{thismomij/.style={}};%
      \tikzset{thismomi/.style={}};%
      \tikzset{thismomj/.style={}};%
    \fi%
    \draw [#2,thismomij]%
      (ij.center) -- ++(0       :\intsclfac*\sclfac*\len)%
      node (v) {}%
      node [midway,below] {\(#5\)};%
    \draw [#3,thismomi]%
      (v.center)  -- ++(\diagang:\extsclfac*\sclfac*\len)%
      node [anchor=west] (i) {\(#6\)};%
    \draw [#4,thismomj]%
      (v.center) -- ++(-\diagang:\extsclfac*\sclfac*\len)%
      node [anchor=west] (j) {\(#7\)};%
    \ifdrawblob%
      \blob (t) {\sclfac*\rad}{\(#1\)}{white};%
    \fi%
    \ifdrawvertdot%
      \blob (v) {\sclfac*\dotrad}{}{black};%
    \fi%
  \end{tikzpicture}%
}%
\newcommand{\splittingdiagIndivIS}[7]{%
  \begin{tikzpicture}[baseline=-\halfline]%
    \node (a) {};%
    \node at ($(a)-(\rad,0)$) (al) {};%
    \draw [opacity=0] (al) circle (\rad);%
    \node [anchor=east%
      ,inner sep=0%
      ] at (a) {\(#5\)\hspace*{2pt}};%
    \ifdrawmom%
      \tikzset{thismoma/.style={%
        momentum={[momlabdist = 2pt]\(p_{a}\)}%
      }};%
      \tikzset{thismomi/.style={%
        momentum= {[]\(p_{i}\)}%
      }};%
      \tikzset{thismomai/.style={%
        momentum'={[]\(p_{ai}\)}%
      }};%
    \else%
      \tikzset{thismoma/.style={}};%
      \tikzset{thismomi/.style={}};%
      \tikzset{thismomai/.style={}};%
    \fi%
    \draw [#2,thismoma]%
      (a.center) -- ++(0:       \intsclfac*\sclfac*\len)%
      node (v) {};%
    \draw [#3,thismomi]%
      (v.center) -- ++(\diagang:\extsclfac*\sclfac*\len)%
      node [anchor=west] (i) {\(#6\)};%
    \draw [#4,thismomai]%
      (v.center) -- ++(-\diagang:\intsclfac*\sclfac*\len)%
      node (ai) {}%
      node [midway,anchor=south west,inner sep=1.5pt%
      ] {\(#7\)};%
    \node at ($(ai)+(-\diagang:\rad)$) (b) {};%
    \blob (b) {\sclfac*\rad}{\(#1\)}{white};%
    \ifdrawvertdot%
      \blob (v) {\sclfac*\dotrad}{}{black};%
    \fi%
  \end{tikzpicture}%
}%
\newcommand{\spdtypeail}{}%
\newcommand{\spdtypeair}{}%
\newcommand{\spdtypeijl}{}%
\newcommand{\spdtypeijr}{}%
\newcommand{\spdtypea}{}%
\newcommand{\spdtypei}{}%
\newcommand{\spdtypej}{}%
\newcommand{\splittingdiagSqFS}[6]{%
  \begin{tikzpicture}[inner sep=0mm,baseline=-\halfline]
    \newcommand{\dashlen}{1.44cm};
    \newcommand{\totlen}{6.42cm};
    \node (b1) {};
    \node at ($(b1)+(\sclfac*\totlen,0)$) (b2) {};
    \node at ($(b1)+(\sclfac*\rad,0)$) (i1) {};
    \node at ($(b2)-(\sclfac*\rad,0)$) (i2) {};
    \node at ($(i1)+(\sclfac*\len,0)$) (v1) {};
    \node at ($(i2)-(\sclfac*\len,0)$) (v2) {};
    \node at ($(v1)+( \diagang: \sclfac*\extsclfac*\len)$) (o1) {};
    \node at ($(v2)+(-\diagang:-\sclfac*\extsclfac*\len)$) (o2) {};
    \node at ($(v1)+(-\diagang: \sclfac*\extsclfac*\len)$) (u1) {};
    \node at ($(v2)+( \diagang:-\sclfac*\extsclfac*\len)$) (u2) {};
    \draw [style=\spdtypeijl,
      momentum={[
      momlabdist = 2pt
      ]\(p_{ij}\)}]
      (i1.center) -- (v1.center)
      node[midway,below,
      inner sep = .8mm
      ] {\(#1\)};
    \draw [style=\spdtypei,
      momentum={[
      momlabdist = 1pt
      ]\(p_{i}\)}]
      (v1.center) -- (o1.center);
    \draw [style=\spdtypej,
      momentum'={[
      momlabdist = 1pt
      ]\(p_{j}\)}]
      (v1.center) -- (u1.center);
    \draw [\spdtypeijr,
      momentum'={[
      momlabdist = 2pt
      ]\(p_{ij}\)}]
      (i2.center) -- (v2.center)
      node[midway,below,
      inner sep = .8mm
      ] {\(#2\)};
    \draw [\spdtypei,
        momentum'={[
        momlabdist = 1pt
        ]\(p_{i}\)}]
        (v2.center) -- (o2.center);
    \draw [\spdtypej,
        momentum={[
        momlabdist = 1pt
        ]\(p_{j}\)}]
        (v2.center) -- (u2.center);
    \blob (b1) {\sclfac*\rad}{#5}{white};
    \blob (b2) {\sclfac*\rad}{#6}{white};
    \ifdrawvertdot%
      \blob (v1) {\sclfac*\dotrad}{}{black};%
      \blob (v2) {\sclfac*\dotrad}{}{black};%
    \fi%
    \node at ($(b1)!0.5!(b2)+(0,\sclfac*\dashlen)$) (mu) {};
    \node at ($(b1)!0.5!(b2)-(0,\sclfac*\dashlen)$) (ml) {};
    \draw [dashed,semithick] (mu) -- (ml);
    \node [fill=white,inner sep=1.3]
      at ($(mu)!0.8!(ml)+(0,\sclfac*\dashlen)$) (texti) {\(#3\)};
    \node [fill=white,inner sep=1.3]
      at ($(mu)!0.2!(ml)-(0,\sclfac*\dashlen)$) (textj) {\(#4\)};
  \end{tikzpicture}%
}%
\newcommand{\splittingdiagSqIS}[6]{%
  \begin{tikzpicture}[inner sep=0mm,baseline=-\halfline]
    \newcommand{\dashlen}{1.44cm};
    \newcommand{\totlen}{7.35cm};
    \newcommand{\buf}{1.35mm};
    \node (i1) {};
    \node at ($(i1)+(\sclfac*\totlen,0)$) (i2) {};
    \node at ($(i1)+(\sclfac*\extsclfac*\len,0)$) (v1) {};
    \node at ($(i2)-(\sclfac*\extsclfac*\len,0)$) (v2) {};
    \node at ($(v1)+( \diagang: \sclfac*\extsclfac*\len)$) (o1) {};
    \node at ($(v2)+(-\diagang:-\sclfac*\extsclfac*\len)$) (o2) {};
    \node at ($(v1)+(-\diagang: \sclfac*\len)$) (u1) {};
    \node at ($(v2)+( \diagang:-\sclfac*\len)$) (u2) {};
    \node at ($(i1)-(\buf,0)$) (ibuf1) {};
    \node at ($(i2)+(\buf,0)$) (ibuf2) {};
    \draw [style=\spdtypea,
      momentum={[
      momlabdist = 2pt
      ]\(p_{a}\)}]
      (i1.center) -- (v1.center)
      node[midway,below,
      inner sep = .8mm
      ] {\(#3\)};
    \draw [style=\spdtypei,
      momentum={[
      momlabdist = 1pt
      ]\(p_{i}\)}]
      (v1.center) -- (o1.center);
    \draw [style=\spdtypeail,
      momentum'={[
      momlabdist = 1pt
      ]\(p_{ai}\)}]
      (v1.center) -- (u1.center)
      node[midway,above right,
      ] {\(#1\)};
    \draw [style=\spdtypea,
      momentum'={[
      momlabdist = 2pt
      ]\(p_{a}\)}]
      (i2.center) -- (v2.center)
      node[midway,below,
      inner sep = .8mm
      ] {\(#3\)};
    \draw [style=\spdtypei,
      momentum'={[
      momlabdist = 1pt
      ]\(p_{i}\)}]
      (v2.center) -- (o2.center);
    \draw [style=\spdtypeair,
      momentum={[
      momlabdist = 1pt
      ]\(p_{ai}\)}]
      (v2.center) -- (u2.center)
      node[midway,above left,
      ] {\(#2\)};
    \node at ($(u1)+(-\diagang: \rad)$) (b1) {};
    \node at ($(u2)+( \diagang:-\rad)$) (b2) {};
    \blob (b1) {\sclfac*\rad}{#5}{white};
    \blob (b2) {\sclfac*\rad}{#6}{white};
    \ifdrawvertdot%
      \blob (v1) {\sclfac*\dotrad}{}{black};%
      \blob (v2) {\sclfac*\dotrad}{}{black};%
    \fi%
    \node at ($(i1)!0.5!(i2)+(0,\sclfac*\dashlen)$) (mu) {};
    \node at ($(i1)!0.5!(i2)-(0,\sclfac*\dashlen)$) (ml) {};
    \draw [dashed,semithick] (mu) -- (ml);
    \node[fill=white,inner sep=1.3]
      at ($(mu)!0.8!(ml)+(0,\sclfac*\dashlen)$) (texti) {\(#4\)};
  \end{tikzpicture}
}%

\allowdisplaybreaks

\makeatletter
\divide\@tempdima\baselineskip
\@tempcnta=\@tempdima
\makeatother

\def\draftdate{\relax}
\def\mda{\relax}
\def\mua{\relax}
\def\mla{\relax}
\def\draft{
\def\thtystars{******************************}
\def\sixtystars{\thtystars\thtystars}
\typeout{}
\typeout{\sixtystars**}
\typeout{* Draft mode!
         For final version remove \protect\draft\space in source file *}
\typeout{\sixtystars**}
\typeout{}
\def\draftdate{\today}
\def\mua{\marginpar[\boldmath\hfil$\uparrow$]%
                   {\boldmath$\uparrow$\hfil}%
                    \typeout{marginpar: $\uparrow$}\ignorespaces}
\def\mda{\marginpar[\boldmath\hfil$\downarrow$]%
                   {\boldmath$\downarrow$\hfil}%
                    \typeout{marginpar: $\downarrow$}\ignorespaces}
\def\mla{\marginpar[\boldmath\hfil$\rightarrow$]%
                   {\boldmath$\leftarrow $\hfil}%
                    \typeout{marginpar: $\leftrightarrow$}\ignorespaces}
\def\Mua{\marginpar[\boldmath\hfil$\Uparrow$]%
                   {\boldmath$\Uparrow$\hfil}%
                    \typeout{marginpar: $\uparrow$}\ignorespaces}
\def\Mda{\marginpar[\boldmath\hfil$\Downarrow$]%
                   {\boldmath$\Downarrow$\hfil}%
                    \typeout{marginpar: $\downarrow$}\ignorespaces}
\def\Mla{\marginpar[\boldmath\hfil{$\Rightarrow$}]%
                   {\boldmath{$\Leftarrow $}\hfil}%
                    \typeout{marginpar:$\leftrightarrow$}\ignorespaces}
\overfullrule 5pt
\oddsidemargin 15mm
\marginparwidth 29mm
}

\preprint{{\small FR-PHENO-2025-003}}

\title{Electroweak splitting functions \\
in the Standard Model and beyond}

\author{Stefan Dittmaier}
\author{and Max Reyer}
\affiliation{%
  Albert-Ludwigs-Universität Freiburg, Physikalisches Institut,\\%
  Hermann-Herder-Str. 3, 79104 Freiburg, Germany
  }%

\emailAdd{stefan.dittmaier@physik.uni-freiburg.de}
\emailAdd{physics@mxreyer.slmail.me}

\abstract{We derive quasi-collinear factorization formulas in generic
spontaneously broken 
gauge theories with scalars, fermions, and vector bosons.
Specifically, we obtain polarized leading-order splitting functions for
all possible final-state and initial-state $1 \to 2$ processes in the
considered gauge theory.
The main complication lies in the presence of mass-singular terms in
longitudinal polarization vectors, prohibiting the direct application
of the usual factorization procedure known from Quantum Electrodynamics
and Quantum Chromodynamics.
We overcome this issue with two different strategies, using gauge
invariance and Ward identities as guiding principle.
Our derivations do not use any explicit component-wise parametrizations
of momenta and wave functions and bear no reference to a particular
Lorentz frame.
Furthermore, our results are valid for completely general definitions of
the spin reference axes of the individual external particles
and are formulated in the standard Dirac formalism to facilitate their
application.
The various massless limits, the special case of the Electroweak
Standard Model, the reproduction of existing literature results, and
symmetry relations among our splitting functions are 
discussed in detail.
}

\begin{document} 
\maketitle

\flushbottom

\section{Introduction}
\label{sect:introduction}
The 2012 discovery~\cite{ATLAS:2012yve,CMS:2012qbp} of the Higgs boson
and the ensuing confirmation of its properties provided the last missing
piece of experimental confirmation of the \gls{ewsm}.
Since then, no clear signs of physics 
beyond the Standard Model have
been observed~\cite{Workman:2022ynf}, despite the plethora of data taken
at the \gls{lhc} at different center-of-mass energies of up to $\sqrt{s}
= 13.6\, \TeV$.
Such new physics is needed to explain, for example, the observation of
dark matter on cosmological scales in the framework of particle physics.
In recent years, much of the experimental effort has therefore shifted
to the determination of observables and cross sections
with increasing precision.
For a meaningful comparison with the expectations from the \gls{ewsm},
the uncertainties from the corresponding theoretical predictions should
ideally be lower than those from the measurements.
For many key processes
at the \gls{lhc}, this requires computations at \gls{nnlo}
in \gls{qcd} and \gls{nlo} in \gls{ew} theory.
On top of that, resummation techniques such as \emph{parton-shower
simulations} (see~\cite{Ellis1991qcd}, for example) have proven to be of
great importance for the proper prediction of kinematic distributions.


Parton-shower simulations account for the well-known fact that
scattering amplitudes of gauge theories become large when two external
particles are approximately \emph{collinear}, 
if the typical energy
scale of the considered process is much greater than the masses of the
involved particles.
This strongly enhances the production of additional, collimated
particles.
Specifically, parton-shower simulations re-sum the leading corrections
associated to the production of arbitrarily many additional collinear
particles.
Therein, the involved matrix elements are approximated via
\emph{quasi-collinear factorization formulas}, which express squared
matrix elements as products of underlying matrix elements 
for a hard mother process and universal radiation factors.
These quasi-collinear \glshyph{lo} \emph{splitting
functions}~\cite{Altarelli:1977zs} capture the collinear enhancements.
For massless particles, splitting functions are textbook knowledge;
for massive fermions,
they are known for
\gls{qed}~\cite{Baier:1973ms,Kleiss:1986ct,dittmaier1999photonrad,dittmaier2008polarized}
and \gls{qcd}~\cite{Keller:1998tf,Catani:2000ef,catani2002massivedipole}.
Apart from parton-shower algorithms,
splitting functions also are essential ingredients to \emph{subtraction
methods} for \gls{nlo}
computations~\cite{Ellis:1980wv,Frixione:1995ms,catani1996general,dittmaier1999photonrad,Phaf:2001gc,catani2002massivedipole,dittmaier2008polarized,Frederix:2009yq}.

For scattering energies in the TeV and multi-TeV range, \gls{ew} corrections have been
found to be substantial, often reaching 10--30\% already for a few TeV with a
tendency of logarithmic increase, which renders this effect already relevant in LHC
analyses. This observation triggered many studies of the
logarithmic structure of virtual \gls{ew} corrections at 
NLO~\cite{denner2001oneloop1,denner2001oneloop2}, 
NNLO~\cite{Hori:2000tm,Beenakker:2001kf,Denner:2003wi,Jantzen:2005xi}, and 
beyond~\cite{Ciafaloni:1999ub,Fadin:1999bq,Kuhn:1999nn,Melles:2001ye,Chiu:2007yn,Chiu:2008vv}
(see, e.g., \cite{Denner:2019vbn} for further references).
Real-emission effects, on the other hand, were mostly left for process-specific LO studies with
multi-purpose Monte Carlo generators
(see, e.g., \cite{Baur:2006sn}).
A potential future hadron-collider experiment
is expected to reach center-of-mass energies of
about~$100\,\TeV$~\cite{FCC:2018vvp}, much larger even than the masses
of \gls{ew} bosons.
This renders \gls{ew} multi-particle emission an important effect and calls for the extension 
of parton-shower and subtraction algorithms
to \gls{ew} theory.
At its core, this requires the derivation of quasi-collinear
factorization formulas in the full \gls{ewsm}, which is the main focus
of this paper.
In fact, we derive splitting functions in the broader setting of generic
spontaneously broken gauge theories with scalars, fermions, and vector bosons.
This is supported by a discussion of gauge couplings, quantization,
\gls{ssb}, and \glspl{wi} in this general context.
We pay particular attention to symmetry relations among 
the derived splitting functions,
their various massless limits, and the comparison to results 
existing in the literature.


First results on \glshyph{lo} \gls{ew} splitting functions in the
\emph{unbroken phase} of the \gls{ewsm} have been reported
in~\cite{Hebenstreit:2006,Christiansen:2014kba,Bauer:2018xag}.
These derivations strictly assume the high-energy limit and neglect all
effects of the \gls{ew} symmetry-breaking scale~$v$.
In particular, all fields are assumed massless and usually taken in the
interaction basis.
Amplitudes for longitudinally polarized vector bosons are replaced by
amplitudes with would-be Goldstone bosons via the
\gls{gbet}~\cite{Gounaris:1986cr,Cornwall:1974km,Chanowitz:1985hj,Lee:1977eg,Vayonakis:1976vz,Denner:1996gb}.
While constituting an important first step, this approach is neither
capable of capturing the evolution of \gls{ew} showers all the way down
to the \gls{ew} scale~$v$, nor of extracting all collinearly enhanced
terms needed, e.g., for the construction of \gls{ew} subtraction methods.


More recently, \glshyph{lo} 
double-collinear splitting functions in the \emph{broken
phase} of the \gls{ewsm} have been
obtained~\cite{Chen:2016wkt,Cuomo:2019siu,Kleiss:2020rcg,Brooks:2021kji,%
Masouminia:2021kne,Nardi:2024tce}.\footnotemark\
\footnotetext{%
First steps towards LO triple-collinear splitting functions involving
massive fermions, which are relevant for NNLO QCD and QED calculations,
have been taken in \cite{Dhani:2023uxu,Craft:2023aew}.}
All of these derivations are based on the evaluation of \emph{splitting
amplitudes} for $1 \to 2$ splitting processes with off-shell incoming
momentum and on-shell polarization vectors and spinors.
This relies on non-trivial, well-defined decompositions of the
numerators of off-shell propagators into on-shell polarization sums.
The consistency of this approach, however, is explicitly proven 
only in~\cite{Chen:2016wkt,Cuomo:2019siu,Nardi:2024tce}.
By nature of the derivations
presented in these references, the frame of reference of the
polarization vectors and spinors for the mother particles is assumed to
be identical to those of the daughter particles and identified with the lab frame.
The decomposition into transverse and longitudinal polarization vectors
for massive vector bosons is, however, frame dependent.
This is problematic in view of subtraction formalisms.\footnotemark\
\footnotetext{%
  In the dipole subtraction formalism~\cite{catani1996general,dittmaier1999photonrad,%
Phaf:2001gc,catani2002massivedipole,dittmaier2008polarized}, for example, the individual
  dipole terms need to be integrated in auxiliary frames determined by the dipole
  kinematics.
  Performing the azimuthal-angle integral around the collinear axis requires
  the separation of longitudinal and transverse components in the
  specific frame of integration.
  Because this frame differs from the lab frame, it is not clear whether
  the existing splitting functions from the literature are readily
  suited for the construction of dipole functions.
}%
Additionally, for practical applications, the splitting amplitudes first
have to be combined into factorization formulas for the \emph{squared}
real-emission amplitudes.
Where this has been done explicitly in the literature, flavor and spin
interference effects have at least been partially neglected.
Finally,~\cite{Chen:2016wkt,Cuomo:2019siu,Masouminia:2021kne} rely on
explicit, component-wise, and frame-dependent parametrizations of
momenta, spinors, and polarization vectors.
The derivations given in~\cite{Kleiss:2020rcg,Brooks:2021kji} are
conducted in a way that bears no explicit reference to a specific frame.
 In~\cite{Nardi:2024tce}, ``amplitude tensors'' (in spin space) for $1
 \to 2$ splitting processes are expressed in terms of generic bi-spinors.
 While their explicit evaluation is performed using a component-wise
 parametrization of momenta and a representation of bi-spinors that
 singles out a specific frame, the Lorentz invariance of the
obtained amplitude objects is emphasized.


In order to overcome the mentioned drawbacks,
we aim at a factorization directly at
the level of squared matrix elements and employ 
general \emph{Sudakov parametrizations}
to describe the quasi-collinear kinematics.\footnotemark\
\footnotetext{%
We do not explicitly consider the {\it soft} emission limits which are ruled
by the (massive) {\it eikonal approximation}. The are, for instance, 
investigated in \cite{Platzer:2022nfu,Nardi:2024tce}.}
The latter express the splitting momenta in terms of generic
four-vectors in a Lorentz-covariant way.
This allows us to write collinear factorization formulas via
contractions of polarization-vector-stripped hard matrix elements and
tensor-valued splitting functions, 
encoding potential spin
correlations.
In this approach, the decomposition of internal propagator numerators
into frame-dependent polarization vectors is circumvented.
We avoid the precise definition of polarization altogether by
not making any assumption on the spin reference axes for the individual
external particles.
This provides a further reduction of assumptions compared to the
existing literature.\footnotemark\
\footnotetext{%
  The generic amplitude tensors from Appendix~D.2 of~\cite{Nardi:2024tce} do not
    rely on the specific choice of spin reference axes as well.
    In~\cite{Nardi:2024tce}, however, final results are obtained
    (and given in Appendix~D.3 therein) by using explicit
    representations of bi-spinors, which fix these reference axes.
}%
We note that our approach does not involve any modification of the
standard Dirac formalism for computing amplitudes in
Quantum Field Theory, unlike the
framework chosen in~\cite{Cuomo:2019siu,Nardi:2024tce}.
Moreover, we use standard $R_\xi$ gauge (with generic gauge parameters
$\xi$), in contrast to the computations described in~\cite{Chen:2016wkt}.
Finally, working in a generic spontaneously broken gauge theory is a feature shared
only by~\cite{Cuomo:2019siu}.


The derivation of splitting functions in gauge theories with massive
vector bosons provides a serious challenge.
Specifically, in the high-energy limit, the polarization vector of a
longitudinally polarized vector boson with momentum~$p^\mu$ and mass
$M_V$ diverges like $\varepsilon_\mathrm{L}^\mu (p) \sim p^\mu/M_V$.
Gauge invariance, however, dictates that this mass-singular factor
cancels in the computation of full amplitudes.
These gauge cancellations usually involve all kinds of diagrams for a
given process.
When deriving approximate factorization formulas, however,
only certain diagram types are considered.
In such a procedure, great care must be taken to not destroy the
sensitive gauge cancellations.
In previous publications, this issue has been dealt with in multiple
ways.
In~\cite{Chen:2016wkt}, a customary axial gauge is chosen for which this
problem is absent.
An axial gauge has also been used in~\cite{Borel:2012by} for proving the
{\it\gls{ewa}}~\cite{Dawson:1984gx}.
In~\cite{Cuomo:2019siu}, \glspl{wi} are employed to remove the
problematic terms~$\propto p^\mu/M_V$ from~$\varepsilon_\mathrm{L}^\mu$
and to replace them with well-behaved would-be Goldstone-boson amplitudes.
This is achieved by a redefinition of the
four-vector~$\varepsilon_\mathrm{L}^\mu$, introducing an extra fifth
component associated to the corresponding would-be Goldstone boson.
A similar strategy, although in much less generality, has first been
used by~\cite{Dawson:1984gx} for the derivation of the \gls{ewa}.
In this specific situation, which concerns the emission of a vector
boson off a massless fermion, the contributions~$\propto p^\mu/M_V$
become zero.
Without modification,the authors of \cite{Masouminia:2021kne} 
coin this assumption
``Dawson's strategy'' and apply it to \emph{all} kinds of \gls{ew} splitting
processes, despite not being generally valid.
The results of~\cite{Masouminia:2021kne} are, therefore, in conflict
both with the existing
literature~\cite{Chen:2016wkt,Cuomo:2019siu,Kleiss:2020rcg,Brooks:2021kji}
and our findings.
Lastly, the derivation in~\cite{Kleiss:2020rcg,Brooks:2021kji} relies on
carefully identifying 
and eliminating all ill-behaved terms that must cancel against
contributions from neglected diagrams.
\looseness-1


In this paper, we address the mass-singular terms in longitudinal
polarization vectors with two different strategies.
Our default approach is similar to~\cite{Cuomo:2019siu}.
However, we work in the standard formalism, avoiding the redefinition of
polarization vectors.
As an alternative, we outline a procedure close
to~\cite{Kleiss:2020rcg,Brooks:2021kji}, where we identify and manually
cancel ill-behaved terms.
We show that the two
methods produce equivalent results, 
providing a check on 
the obtained splitting functions.


Collinear enhancements also occur in one-loop virtual corrections.
The corresponding leading double and single logarithmic contributions
have been derived in~\cite{denner2001oneloop1,denner2001oneloop2}, where
a generic notation for gauge-theory couplings is used.
We adopt and extend this notation in the following, facilitating the
combination of these results with our factorization formulas. 
In particular, this is of interest for the assessment of partial
cancellations between real and virtual corrections for specific
processes and observables, as for instance observed in~\cite{Baur:2006sn}.
This is, however, left for future work.
Moreover, while we take special care that our results are suitable for
the use in subtraction algorithms, we leave this application for a
future publication as well.


This paper is structured as follows.
We start with a detailed discussion of generic 
spontaneously broken gauge theories
in Section~\ref{sect:generic-ym-theory}.
Particular emphasis is put on 
gauge-invariance relations like \glspl{wi} and on the resulting
relations between gauge and would-be Goldstone-boson couplings.
In Section~\ref{sect:quasi-coll-fact}, we describe our quasi-collinear
factorization procedure and detail the issues that arise with
longitudinally polarized vector bosons.
This is spelled out explicitly in Section~\ref{sect:splfunc-derivation},
where we derive our \glshyph{fs} splitting functions, discuss their
properties, and compute the azimuthal averages.
We also describe symmetry relations and use them to derive
\glshyph{is} splitting functions.
\revision{Moreover, we give a brief outlook to further steps that are
needed for the construction of an EW parton shower or of a subtraction formalism
for soft/collinear singularities arising from EW splittings at high energies.}
In Section~\ref{sect:lit-review-splittings}, we compare our methodology
and results to the existing
literature~\cite{Chen:2016wkt,Cuomo:2019siu,Kleiss:2020rcg,Brooks:2021kji,Masouminia:2021kne},
before we conclude in Section~\ref{sect:summary}.
We close this paper with supplementary material.
First, Appendix~\ref{sect:ewsm} specializes the discussion from
Section~\ref{sect:generic-ym-theory} to the \gls{ewsm} and provides
explicit expressions for the corresponding coupling factors.
In Appendix~\ref{sect:generic-gb-cpl-relat-derivations}, we derive the
coupling relations presented in Section~\ref{sect:generic-gb-cpl-relat}.
Our alternative strategy for the treatment of the problematic,
mass-singular terms in longitudinal polarization vectors is explained in
Appendix~\ref{sect:alt-approach}.
Appendix~\ref{sect:qed-qcd-litres} details the recovery of \gls{qed} and
\gls{qcd} literature results from our splitting functions.
Finally, Appendix~\ref{sect:details-fermion-splittings} provides further
details on \glshyph{fs} splittings in fermionic processes.

\section{Generic treatment of spontaneously broken gauge theories}
\label{sect:generic-ym-theory}

\subsection{Generic gauge couplings}
\label{sect:generic-gauge-cpl}
We consider \gls{ym} theories built from matter
fields that transform as multiplets~$\varphi$ in unitary
representations of a gauge group~\gaugegroup.
We consider fermion and scalar multiplets~$\varphi=f$
and~$\varphi=\scalmult$, respectively.
Denoting by~$\varphi_k$ the~$k$-th component of the multiplet,
the variation of $\varphi_{k}$ under global gauge
transformations with infinitesimal group parameter~$\delta \theta_{V_a}$ reads
\begin{align}
  \delta \varphi_k (x)
  =
  \imagi e I^{V_a}_{\varphi_k \varphi_l} \varphi_l (x) \delta \theta_{V_a}
  ,
  \label{eq:inf-trf}
\end{align}
where~$e$ is the coupling strength, playing the role of the electric unit
charge in the \gls{ewsm}.
Following the notation of~\cite{denner2001oneloop1,denner2001oneloop2},
the objects~$I^{V_a}$ correspond to the
generators of the
gauge group~\gaugegroup, and the matrices~$I^{V_a}_{\varphi_k \varphi_l}$ to their
representation on the space of~$\varphi_l$.
The fields $V_a$ denote the respective gauge fields and transform under
global gauge transformations via~\eqref{eq:inf-trf} themselves,
with~$I^{V_a}_{V_b V_c}$ corresponding to the generators in the adjoint
representation.

The antifield of $\varphi_k$ is denoted by $\bar \varphi_k$.
Up to manipulations in the internal space of the fields that commute
with the gauge group generators, 
$\bar \varphi_k$ is obtained by complex
conjugation.
Specifically, we have $\bar \scalmult_k = \scalmult_k^\ast$ for scalars, while
$\bar f_k=f_k^\dagger \gamma_0$ for fermions, and $\bar V_a (x) =
V_a^\ast (x)$ for gauge bosons.
The transformation behavior of the multiplets~$\bar \varphi$ is
therefore obtained by conjugation of~\eqref{eq:inf-trf}.

In \gls{ym} theories, the components~$I^{V_a}_{\varphi_j \varphi_k}$
of the generators determine the couplings between the gauge fields~$V_a$
and multiplet components of~$\varphi$.
For fermion and scalar multiplets, the couplings stem from the covariant
derivative
\begin{align}
  D^\mu_{\varphi_k \varphi_l}
  =
  \delta_{\varphi_k \varphi_l} \partial^\mu
  - \imagi e I^{V_a}_{\varphi_k \varphi_l} V_a^{\mu} (x)
  ,
  \label{eq:cov-der}
\end{align}
which appear in the Lagrangian as terms
\begin{align}
  \imagi \bar f_k (x) \slashed D_{f_k f_l} f_l (x)
  \quad
  \mathrm{and}
  \quad
  &
  \big(D_{\scalmult_k \scalmult_l, \mu} \scalmult_l (x)\big)^\dagger
  \big(D^\mu_{\scalmult_k \scalmult_m} \scalmult_m (x)\big)
  \label{eq:lag-cpl-ferm-scal}
\end{align}
that are invariant under local gauge transformations.
For matter fields, the local transformations correspond to~\eqref{eq:inf-trf}
with parameters~$\delta \theta_{V_a} \equiv \delta \theta_{V_a}(x)$ depending on
the space--time point~$x$, while
\begin{align}
  \delta V^\mu_a (x)
  =
  \imagi e I^{V_b}_{V_a V_c} V^\mu_c (x) \delta \theta_{V_b} (x)
  +
  \partial^\mu \delta \theta_{V_a} (x)
  .
  \label{eq:inf-trf-local-gauge-boson}
\end{align}
The pure gauge-field couplings originate from the field-strength tensor
\begin{align}
  F_{\mu\nu}^{V_a} =
  \partial_\mu V^a_\nu(x) - \partial_\nu V^a_\mu(x)
  - \imagi e I^{V_b}_{V_a V_c} V^b_\mu (x) V^c_\nu (x)
  ,
  \label{eq:field-strength-tensor}
\end{align}
which appears in the Lagrangian as locally gauge-invariant term
\begin{align}
  - \frac{1}{4} F^{\bar V_a}_{\mu \nu} F^{V_a \mu \nu}
  \label{eq:kin-gauge-boson-lagr}
  .
\end{align}

The coupling factor corresponding to the~$V_a \bar \varphi_k
\varphi_l$ vertex reads~$\imagi e I^{V_a}_{\varphi_k \varphi_l}$, where
the fields~$V_a$, $\bar \varphi_k$, and~$\varphi_l$ are defined as
incoming.
The coupling~$\bar V_a \varphi_k \bar \varphi_l$ is determined by
\begin{align}
  I^{\bar V_a}_{\bar \varphi_k \bar \varphi_l}
  =
  -\left(I^{V_a}_{\varphi_k \varphi_l}\right)^\ast
  ,
  \label{eq:cpl-mtr-symmetry-1}
\end{align}
which follows from conjugation of~\eqref{eq:inf-trf}.
Unitarity of global gauge transformations implies
\begin{align}
  I^{V_a}_{\bar \varphi_l \bar \varphi_k} = -I^{V_a}_{\varphi_k \varphi_l}
  \label{eq:cpl-mtr-symmetry-2}
\end{align}
and provides a relation to the~$V_a \varphi_l \bar \varphi_k$ vertex.
Relations~\eqref{eq:cpl-mtr-symmetry-1}
and~\eqref{eq:cpl-mtr-symmetry-2} combine to
\begin{align}
  I^{\bar V_a}_{\varphi_l \varphi_k} = \left(I^{V_a}_{\varphi_k \varphi_l}\right)^\ast,
  \label{eq:cpl-mtr-hermiticity}
\end{align}
meaning that generators for real $V_a$, i.e.\ generators from the real
Lie algebra of the gauge transformation, are hermitian.
On top of the symmetries~\eqref{eq:cpl-mtr-symmetry-1}
and~\eqref{eq:cpl-mtr-symmetry-2}, the gauge-boson coupling
matrices~$I^{V_a}_{\bar V_b V_c}$ fulfill the relations
\begin{align}
  I^{V_a}_{\bar V_b V_c}
  = I^{V_b}_{\bar V_c V_a}
  = I^{V_c}_{\bar V_a V_b}
  .
  \label{eq:cpl-mtr-symmetry-3}
\end{align}

Couplings that do not involve gauge bosons cannot
immediately be read off from the gauge structure alone.
They depend on the specifics of the considered model, i.e.\
on terms of the Lagrangian that are not dictated solely by
the requirement of local gauge invariance.
In this paper, we assume \gls{ym} theories that experience
\gls{ssb} via a potential~$V(\scalmult)$ in the
scalar sector.
Such theories contain a set of scalar would-be Goldstone bosons~$G_a$,
which are intimately linked to the longitudinal degrees of freedom of
the massive gauge bosons~$V_a$ that correspond to broken symmetries.
While general scalar couplings of type~$\scalmult_m \varphi_k \varphi_l$
depend on the specific structure of the theory, the couplings~$G_a
\varphi_k \varphi_l$ of would-be Goldstone bosons are related to the
corresponding gauge-boson couplings~$V_a \varphi_k \varphi_l$ 
by gauge invariance expressed in terms of \glspl{wi}.
These coupling relations are discussed in Section~\ref{sect:generic-gb-cpl-relat}.
Expressing gauge invariance of the theory,
these relations are crucial to guarantee the 
absence of unphysical behavior in physical amplitudes~$\mathcal{M}$ in general and
the independence of $\mathcal{M}$ of gauge parameters of the gauge-fixing procedure
in particular.
In the derivation of EW splitting functions these relations lead to
a meaningful combination of Feynman diagrams related
by the interchange of would-be Goldstone bosons and gauge bosons on
both \emph{internal and external} lines.

The specialization of the generic couplings to the case of the \gls{ewsm} is 
described in App.~\ref{sect:ewsm}.

\subsection{Quantization and gauge fixing}
\label{sect:quantization}
Employing the standard Faddeev--Popov prodecure to define the functional
integral of the quantized gauge theory, we extend the classical
Lagrangian $\mathcal{L}_\mathrm{class}$ to
the effective Lagrangian
\begin{align} 
  \mathcal{L}_{\mathrm{eff}} [\varphi,u,\bar{u};x] &=
  \mathcal{L}_\mathrm{class} [\varphi;x] + \mathcal{L}_{\mathrm{fix}} [V,G;x]
  + \mathcal{L}_{\mathrm{ghost}} [V,G,H,u,\bar{u};x] ,
\end{align}
which does not only involve the matter and gauge fields, generically denoted 
by~$\varphi$, but also
auxiliary ghost fields $u^{V_a}$ and antighost fields $\bar{u}^{V_a}$ corresponding 
to the gauge field $V^a$.
The gauge-fixing Lagrangian
\begin{align}
  \mathcal{L}_{\mathrm{fix}} [V,G;x] &= -
  \frac{1}{2 \xi_{V_a}} C^{\bar{V}_a}[V,G;x] C^{V_a}[V,G;x]
  \label{eq:gauge-fixing-lagrangian}
\end{align}
breaks gauge invariance by the gauge-non-invariant functionals 
$C^{V_a}$ for the gauge degree of freedom associated with $V^a$.
Here, $\xi_{V_a}=\xi_{\bar V_a}$ is a (real)
gauge-fixing parameter which is widely arbitrary.
For spontaneously broken symmetries, $C^{V_a}$ also involves the corresponding 
would-be Goldstone-boson field~$G^a$.
The ghost Lagrangian
\begin{align}
  \mathcal{L}_{\mathrm{ghost}} [V,G,H,u,\bar{u};x] &=
  - \int \D^4 y \, \bar{u}^{V_a}(x) \,
  \frac{ \delta C^{V_a}[V,G;x] }{ \delta \theta_{V_b}(y) } \,
  u^{V_b}(y) ,
\end{align}
is derived from the gauge variations of the functionals $C^{V_a}$,
which in general depend
on gauge fields~$V$, would-be Goldstone fields~$G$, and Higgs fields~$H$
that combine with the would-be Goldstone fields to complete gauge multiplets.

Unlike $\mathcal{L}_\mathrm{class}$, the effective Lagrangian $\mathcal{L}_{\mathrm{eff}}$
can be directly used to evaluate Green functions and amplitudes via perturbation theory.
It is neither invariant under global nor local gauge transformations, but
possesses the global \gls{brst} symmetry by extending gauge transformations to the
ghost sector.
For the original fields~$\varphi$ of the theory, the \gls{brst} transformations
just take the form of local gauge transformations~\eqref{eq:inf-trf}
and~\eqref{eq:inf-trf-local-gauge-boson} with gauge group
parameters~$\delta \theta^{V_a} (x) = \delta \lambda \, u^{V_a}(x)$,
\begin{align}
  \delta_\mathrm{BRST} \, \varphi_k (x) &= 
  \imagi e \, \delta \lambda \, I^{V_a}_{\varphi_k \varphi_l} u^{V_a} (x) \varphi_l (x) ,
  \nonumber\\
  \delta_\mathrm{BRST} \, V_a^\mu (x) &= \imagi e \, \delta \lambda \,
  I^{V_b}_{V_a V_c} u^{V_b} (x) V_c^\mu (x) + \delta \lambda \, \partial^\mu u^{V_a} (x) ,
  \label{eq:def-brst-trf-matter-flds}
\end{align}
where ~$\delta \lambda$ is an anticommuting infinitesimal parameter, so that
the \gls{brst} transformations leave the classical
Lagrangian invariant, $\delta_\mathrm{BRST} \, \mathcal{L}_\mathrm{class}=0$.
The \gls{brst} invariance of $\mathcal{L}_{\mathrm{eff}}$ is guaranteed
upon defining the \gls{brst} transformations of the ghost and antighost fields
according to
\begin{align}
  \delta_\mathrm{BRST} \, u^{V_a} (x) = 
   \frac{\imagi e}{2} \delta \lambda I^{V_b}_{V_a V_c} u^{V_b} (x) u^{V_c} (x) ,\qquad
  \delta_\mathrm{BRST} \, \bar{u}^{V_a} (x) =
  - \frac{\delta \lambda}{\xi_{V_a}} C^{\bar{V}_a}\big[V,G;x\big],
  \label{eq:def-brst-trf-ghosts}
\end{align}
from which 
$\delta_\mathrm{BRST}\,\big(\mathcal{L}_\mathrm{fix}+\mathcal{L}_\mathrm{ghost}\big)=0$
can be seen easily.
Defining the \gls{brst} operator~$s$ as
\begin{align}
  \delta_\mathrm{BRST} \, \varphi_k(x) = \delta \lambda \, (s \varphi_k(x)) ,
  \label{eq:brst-op}
\end{align}
the ghost Lagrangian can be rewritten as
\begin{align}
  \mathcal{L}_\mathrm{ghost} \big[V,G,H,u,\bar{u};x\big] =
  - \bar{u}^{V_a}(x) s C^{V_a} [V,G;x] .
  \label{eq:ghost-lagr-brst-op}
\end{align}
The explicit form of $s$ for the \gls{ewsm} can be found, for example,
in \cite{bohm2001gauge,Denner:2019vbn}.

\subsection{Spontaneous symmetry breaking}
\label{sect:ssb}
We consider \gls{ym} theories whose gauge symmetry is spontaneously
broken via a non-zero \gls{vev}~$\langle \scalmult \rangle$ of the
scalar fields, representing the location of a minimum of the gauge-invariant scalar
potential $V(\scalmult)$.
Inserting the parametrization
\begin{align}
  \scalcpt_i(x) = \scalcpt_i^\prime(x) + \langle \scalcpt_i \rangle ,
  \label{eq:scalar-param-vev}
\end{align}
of a complex scalar field $\phi$ into its kinetic Lagrangian 
(second term in~\eqref{eq:lag-cpl-ferm-scal}), the \gls{vev} produces
a bilinear gauge-boson term
\begin{align}
  \mathcal{L}_{M_V} =
  - e^2 \left[ I^{V_a}_{\scalcpt_i \scalcpt_j}
    I^{\bar V_b}_{\scalcpt_i^\ast \scalcpt_k^\ast}
    \langle \scalcpt_j \rangle \langle \scalcpt_k^\ast \rangle \right]
  V_{a,\mu}(x) \bar V_b^\mu(x)
  \label{eq:vb-mass-term}
\end{align}
as well as a scalar--gauge-boson mixing term
\begin{align}
  \mathcal{L}_{GV} =
  - \imagi e \left[
    \left(\partial_\mu \scalcpt_i^{\prime \ast}\right)
    I^{V_a}_{\scalcpt_i \scalcpt_j}
    \langle \scalcpt_j \rangle +
    \left(\partial_\mu \scalcpt_i^\prime\right)
    I^{V_a}_{\scalcpt_i^\ast \scalcpt_j^\ast}
    \langle \scalcpt_j^\ast \rangle
  \right]
  V_a^\mu .
  \label{eq:vbgb-mix-term}
\end{align}
The \gls{vev}~$\langle \scalmult \rangle$ is invariant under the action
of a subgroup \subgroup\ of the gauge group \gaugegroup.
The generators $I^{V'_a}$
of $\subgroup$ correspond to \emph{unbroken} symmetries, and their
representations on the space of~$\scalmult$ fulfill
\begin{align}
  I^{V'_a}_{\scalcpt_{i} \scalcpt_{j}} \langle \scalcpt_{j} \rangle =
  I^{V'_a}_{\scalcpt_{i}^\ast \scalcpt_{j}^\ast} \langle \scalcpt_{j}^\ast \rangle = 0 ,
  \label{eq:unbroken-gen}
\end{align}
while the generators of \emph{broken} symmetries do not produce zero in all these
equations.
Diagonalizing the mass terms in~\eqref{eq:vb-mass-term} yields
$\dim \subgroup$ gauge fields of massless gauge bosons and 
$\dim \gaugegroup - \dim \subgroup$ fields of massive gauge bosons.

For each of the massive fields $V_a$,~\eqref{eq:vbgb-mix-term}
describes mixing with the associated scalar would-be Goldstone-boson
field $G_a$,
\begin{align}
  G_a(x) &=
  - \frac{e \, \eta_{V_a}}{M_{V_a}} \left(
    I^{\bar V_a}_{\scalcpt_{i}^\ast \scalcpt_{j}^\ast}
    \langle \scalcpt_{j}^\ast \rangle
    \scalcpt_{i}(x) + I^{\bar V_a}_{\scalcpt_{i} \scalcpt_{j}}
    \langle \scalcpt_{j} \rangle
    \scalcpt^\ast_{i}(x)
  \right) ,
\end{align}
with $\langle G_a \rangle = 0$ and a phase $\eta_{V_a}$ which is fixed by
some convention.
Specifically, denoting the remaining scalar fields in a mass-diagonal field
basis as $H_m$, we get
\begin{align}
  \mathcal{L}_{GV} =
  - \imagi e \left(\partial_\mu G_b^\ast\right)
  I^{V_a}_{G_b H_m} \langle H_m \rangle V_a^\mu ,
  \qquad
  I^{V_a}_{G_b H_m} \langle H_m \rangle =
  \frac{\eta_{V_a} M_{V_a}}{e} \delta_{V_a V_b} .
  \label{eq:VGH-cpl-vb-mass}
\end{align}
Requiring that the would-be Goldstone-boson field for~$\bar V_a$ is given
by $\bar G_a = G_a^\ast$ yields
\begin{align}
  \eta_{\bar V_a}=-\eta_{V_a}^\ast .
\end{align}

In practice it is very convenient to eliminate 
the mixing terms upon choosing so-called
$R_\xi$ gauge-fixing functionals
in $\mathcal{L}_\mathrm{fix}$,
\begin{align}
  C^{\bar V_a} =
  \partial_\mu \bar V_{a}^{\mu} + \imagi \xi_{V_a} \eta_{V_a} M_{V_a} \bar G_a 
  \label{eq:gauge-fixing-functionals}
  .
\end{align}
Massless gauge-boson fields~$V_a$ do not possess would-be
Goldstone-boson fields, i.e.\ the second term on the r.h.s.\
in~\eqref{eq:gauge-fixing-functionals} vanishes.
For massive gauge bosons $V_a$, $\mathcal{L}_\mathrm{fix}$
introduces mass terms for the corresponding would-be Goldstone bosons $G^a$.
The squared masses of $V_a$ and $G_a$ are related by
$M_{G_a}^2 = \xi_{V_a} M_{V_a}^2$.

From the gauge transformations
of~\eqref{eq:gauge-fixing-functionals}, the ghost Lagrangian in~$R_\xi$ gauge
is readily obtained as
\begin{align}
  \mathcal{L}_{\mathrm{ghost}} &=
  - \bar u^{V_a} \left( \partial_\mu D^\mu_{V_a V_b} - e \xi_{V_a} \eta_{\bar{V}_a} M_{V_a}
    \left[ I^{V_b}_{G_a G_c} G_c + I^{V_b}_{G_a H_m} H_m \right] \right) u^{V_b} ,
  \label{eq:gen-ghostlagr}
\end{align}
which contains mass terms for the ghost fields corresponding to massive gauge bosons with
$M^2_{u^{V_a}} = \xi_{V_a} M^2_{V_a}$.
Moreover, \eqref{eq:gen-ghostlagr} yields
gauge-boson--ghost couplings proportional to gauge group generators.

Both the ghost and would-be Goldstone-boson fields are unphysical, i.e.\ they do not
create and annihilate particle states. 
The ghost fields are mere auxiliary fields
introduced to define a proper separation of physical and unphysical (gauge) modes
of the gauge fields, and the would-be Goldstone fields deliver the necessary
degrees of freedom to open
physical longitudinal polarizations of massive gauge bosons.
The remaining scalar fields $H_m$ are physical, i.e.\ Higgs fields if they
are contained in scalar multiplets together with would-be Goldstone fields.

Finally, \gls{ssb} may account for Dirac mass terms via
Yukawa-type terms
\begin{align}
\mathcal{L}_{\mathrm{Yuk}} =
  \bar f^\mathrm{L}_i(x) \Upsilon^{\scalcpt_{m}}_{f^\mathrm{L}_i f^\mathrm{R}_j}
  f^\mathrm{R}_j(x) \scalcpt_{m}(x) + \bar f^\mathrm{L}_i(x)
  \Upsilon^{\scalcpt_{m}^\ast}_{f^\mathrm{L}_i f^\mathrm{R}_j}
  f^\mathrm{R}_j(x) \scalcpt_{m}^\ast(x) + \mathrm{h.c.} ,
  \label{eq:general-ferm-yukawa-cpl}
\end{align}
see, for example,~\cite{Weinberg:1996kr,bohm2001gauge}.\footnotemark\
\footnotetext{%
  Note that both~\cite{bohm2001gauge} and~\cite{Weinberg:1996kr} assume
  a theory with a real scalar multiplet.
  Furthermore, the sign in front of the second term in (21.2.13)
  of~\cite{Weinberg:1996kr} should be $-1$, and (4.1.77)
  in~\cite{bohm2001gauge} is also incorrect as printed.
  A corrected version of the latter is available in the official
  errata~\cite{bohm2001gauge-errata}.
}%
These respect gauge invariance if
\begin{align}
  - I^{V_a}_{f^\mathrm{L}_i f^\mathrm{L}_k}
  \Upsilon^{\scalcpt_{m}}_{f^\mathrm{L}_k f^\mathrm{R}_j}
  + I^{V_a}_{f^\mathrm{R}_k f^\mathrm{R}_j}
  \Upsilon^{\scalcpt_{m}}_{f^\mathrm{L}_i f^\mathrm{R}_k}
  + I^{V_a}_{\scalcpt_{n} \scalcpt_{m}}
  \Upsilon^{\scalcpt_{n}}_{f^\mathrm{L}_i f^\mathrm{R}_j}
  &= 0,
  \nonumber \\
  - I^{V_a}_{f^\mathrm{L}_i f^\mathrm{L}_k}
  \Upsilon^{\scalcpt_{m}^\ast}_{f^\mathrm{L}_k f^\mathrm{R}_j}
  + I^{V_a}_{f^\mathrm{R}_k f^\mathrm{R}_j}
  \Upsilon^{\scalcpt_{m}^\ast}_{f^\mathrm{L}_i f^\mathrm{R}_k}
  + I^{V_a}_{\scalcpt_{n}^\ast \scalcpt_{m}^\ast}
  \Upsilon^{\scalcpt_{n}^\ast}_{f^\mathrm{L}_i f^\mathrm{R}_j}
  &= 0 .
  \label{eq:general-ferm-yukawa-cpl-cond}
\end{align}
Upon inserting~\eqref{eq:scalar-param-vev}, we obtain Dirac mass terms
with the mass matrix
\begin{align}
  M_{f^\mathrm{L}_i f^\mathrm{R}_j} =
  - \Upsilon^{\scalcpt_m}_{f^\mathrm{L}_i f^\mathrm{R}_j} \langle \scalcpt_m \rangle
  - \Upsilon^{\scalcpt_m^\ast}_{f^\mathrm{L}_i f^\mathrm{R}_j}
  \langle \scalcpt_m^\ast \rangle ,
  \label{eq:ferm-mass-matrix-from-yuk}
\end{align}
which can be diagonalized with real, non-negative eigenvalues via a
bi-unitary field transformation.

\subsection{Feynman rules}
\label{sect:feynmanrules-gengaugetheo}
\resetdiagargs%
Following the preceding discussion, we obtain generic Feynman rules
for a spontaneously broken gauge theory with
scalars, fermions, and vector bosons, in~$R_\xi$ gauge.
We adapt them from~\cite{bohm2001gauge,Denner:2019vbn}.


%
Incoming scalars~$S$, fermions~$f$, antifermions~$\bar f$, and vector
bosons~$V$ with on-shell momentum~$p$
are assigned wave function factors according to
\begin{align}
    \feynmanfieldline{scalar}{inc}
    =1
    ,\quad
    \feynmanfieldline{fermion}{inc}
    =u_\polferm (p)
    ,\quad
    \feynmanfieldline{antifermion}{inc}
    =\bar{v}_\polferm (p)
    ,\quad
    \feynmanfieldline{photon}{inc}
    =\varepsilon^{\mu}_\polvb (p)
    ,
\end{align}
respectively, where $u_\polferm$ and $v_\polferm$ are the particle and
antiparticle solutions of the free Dirac
equation, and $\varepsilon^{\mu}_\polvb$ is the polarization vector for on-shell
vector bosons,
\begin{align}
  (\slashed p - m_f) u_\polferm(p) = 0
  ,\qquad
  (\slashed p + m_f) v_\polferm(p) = 0
  ,\qquad
  p_\mu\varepsilon^\mu_\polvb(p) = 0
  ,
  \label{eq:dirac-equation}
\end{align}
with polarization labels~$\polferm,\polvb$.
More detail on the definition of polarization is provided below.
Dots denote a vertex inside the attached diagram.
For outgoing lines, we have
\begin{align}
    \feynmanfieldline{scalar}{outg}
    =1
    ,\quad
    \feynmanfieldline{fermion}{outg}
    =\bar{u}_\polferm (p)
    ,\quad
    \feynmanfieldline{antifermion}{outg}
    =v_\polferm (p)
    ,\quad
    \feynmanfieldline{photon}{outg}
    =\varepsilon^{\mu}_\polvb (p)^\ast
    .
\end{align}
%
%
The propagators read
\begin{alignat}{2}
  G^{f\bar f}(-p,p) ={}
  &
  \feynmanprop{fermion}{f}{\bar f}{p}
  &&
  ={} \dfrac{\imagi (\slashed{p} + m_f)}{p^2 - m_f^2}
  ,
  \nonumber\\
  G^{S \bar S}(-p,p) ={}
  &
  \feynmanprop{scalar}{S}{\bar S}{p}
  &&
  ={} \dfrac{\imagi}{p^2 - m_S^2}
  ,
  \nonumber\\
  G^{V_{\mu_1} \bar V_{\mu_2}}(-p,p) ={}
  &
  \feynmanprop{photon}{V_{\mu_1}}{\bar V_{\mu_2}}{p}
  &&
  ={} \dfrac{- \imagi g_{\mu_1 \mu_2}}{p^2 - M_V^2}
    + \dfrac{\imagi (1-\xi_{V}) p_{\mu_1} p_{\mu_2}}{(p^2 - M_V^2)(p^2-\xi_{V} M_{V}^2)}
  ,
  \label{eq:propagators}
\end{alignat}
where the mass of a would-be Goldstone boson~$S=G_a$ depends
on~$\xi_{V_a}$ and reads~$M_{G_a}=\sqrt{\xi_{V_a}} M_{V_a}$.
\par
%
%
In the following, we list the 3-particle vertices.
All fields and momenta are defined as incoming.
Momentum assignments are explicitly provided only where the
momenta appear in
the kinematic structure of the vertices.
We have
\drawmomtrue%
\drawvertdottrue%
\begin{flalign}
  \feynvertex{photon}{photon}{photon}{V_{a,\mu_1}}{V_{b,\mu_2}}{V_{c,\mu_3}}{p_1}{p_2}{p_3}
  &= \imagi e I^{V_a}_{\bar V_b V_c}
  [
    g_{{\mu_1}{\mu_2}}(p_1 -p_2)_{\mu_3}
    +g_{{\mu_2}{\mu_3}}(p_2 -p_3)_{\mu_1}
    +g_{{\mu_3}{\mu_1}}(p_3 -p_1)_{\mu_2}
  ]
  ,
  \label{eq:app-VVV-vertex}
  \\
  \feynvertex{scalar}{photon}{photon}{S}{V_{a,\mu_1}}{V_{b,\mu_2}}{}{}{}
  &= \imagi e C_{S V_a V_b} g_{{\mu_1}{\mu_2}}
  ,
  \label{eq:app-SVV-vertex}
  \\
  \feynvertex{photon}{scalar}{scalar}{V_\mu}{S_a}{S_b}{}{p_1}{p_2}
  &= - \imagi e I^{V}_{\bar S_a S_b} (p_1-p_2)_{\mu}
  ,
  \label{eq:app-VSS-vertex}
  \\
  \feynvertex{scalar}{scalar}{scalar}{S_a}{S_b}{S_c}{}{}{}
  &= \imagi e C_{S_a S_b S_c}
  ,
  \label{eq:app-SSS-vertex}
  \\
  \feynvertex{photon}{antifermion}{fermion}{V_{\mu}}{\bar{f}_k}{f_l}{}{}{}
  &=  \imagi e I^{V}_{ f^\chir_k f^\chir_l}
      \gamma_\mu \omega_\chir
  ,
  \label{eq:app-Vff-vertex}
  \\
  \feynvertex{scalar}{antifermion}{fermion}{S}{\bar{f}_k}{f_l}{}{}{}
  &=
  \imagi e
  C^{\chir}_{S \bar f_k f_l}
  \omega_\chir
  .
  \label{eq:app-Sff-vertex}
\end{flalign}
In~\eqref{eq:app-Vff-vertex} and~\eqref{eq:app-Sff-vertex}, we sum over
the right- and left-handed fermion
chiralities~$\chir=\mathrm{R}/\mathrm{L}=\pm$ and employ the chirality
projection matrices
\begin{align}
  \omega_\pm = \omega_{\mathrm{R}/\mathrm{L}} = \frac{1 \pm \gamma^5}{2}
  .
  \label{eq:chir-proj-op}
\end{align}
As described in Section~\ref{sect:generic-gauge-cpl}, the various
factors~$I^{V_a}_{\varphi_{k} \varphi_{l}}$ correspond to the generators
of the considered gauge theory.
For the \gls{ewsm}, they are provided explicitly in
Appendix~\ref{sect:explicit-cpl-matrices}.
For would-be Goldstone bosons, $C_{G_a \varphi_{k} \varphi_{l}}$ is
related to $I^{V_a}_{\varphi_{k} \varphi_{l}}$ via \glspl{wi}, as
discussed in Sect.~\ref{sect:generic-gb-cpl-relat}.
The remaining couplings depend on further details of the theory.
In the \gls{ewsm}, for example, they are given in~\eqref{eq:app-HHH-cpl}
and~\eqref{eq:app-HFF-cpl}.


In order to unify the above definitions, we generically denote couplings
between three incoming fields~$\varphi_k$, $\varphi_l$, and~$\varphi_m$ by
$\mathcal{C}^{(\chir)}_{\varphi_k \varphi_l \varphi_m}$,
including a chirality label~$\chir$ if fermions are among these fields.
Specifically, we write
\begin{align}
  \mathcal{C}_{V_a V_b V_c}&=I^{V_a}_{\bar V_b V_c},
  &\mathcal{C}_{S V_a V_b}&=C_{S V_a V_b},
  &\mathcal{C}_{V_c S_a S_b}&=I^{V_c}_{\bar S_a S_b},
  \nonumber\\
  \mathcal{C}_{S_a S_b S_c}&=C_{S_a S_b S_c},
  &\mathcal{C}^\chir_{V_a \bar f_k f_l}&=I^{V_a}_{f^\chir_k f^\chir_l},
  &\mathcal{C}^\chir_{S_a \bar f_k f_l}&=C^\chir_{S_a \bar f_k f_l}
  .
  \label{eq:generic-couplings-expl-def}
\end{align}

\subsection{Ward and Slavnov--Taylor identities}
\label{sect:wardids}
In quantized field theories, the invariance of the action functional under some
field transformation gives rise to relations between \glspl{gf}
(see e.g.~\cite{Weinberg:1996kr,bohm2001gauge,Denner:2019vbn}).
For example, field shifts $\varphi_k (x) \to \varphi_k (x) + \epsilon
f_k(x)$ with infinitesimal~$\epsilon$ and arbitrary functions~$f_k(x)$
result in the \glspl{eom}~\cite{bohm2001gauge}.
Specifically, for the antighost field,
\begin{align}
  \Big \langle T \frac{\delta}{\delta \bar u^{V_a}} \prod_{j} \varphi_{l_j} \Big \rangle
  = \imagi \Big \langle T s C^{V_a} \prod_{j} \varphi_{l_j} \Big \rangle,
  \label{eq:eom-antighost}
\end{align}
where $\langle T\dots\rangle$ stands for the vacuum expectation value of the
time-ordered product of fields~\cite{bohm2001gauge}.
If the field transformations are a symmetry of the action, we obtain the
well-known \glspl{wi} (see \cite{bohm2001gauge,Denner:2019vbn}, and
references therein).
In case of the \gls{brst} symmetry of a quantized gauge theory, these are represented by
the famous \gls{st} identities.
Following the discussion in~\cite{bohm2001gauge,Denner:2019vbn},
the general form of the \gls{st} identities reads
\begin{align}
  s \Big \langle T \prod_{k=1}^{n} \varphi_{l_k} \Big \rangle = 0.
  \label{eq:general-ST-id}
\end{align}
An important class of identities is obtained by taking $\varphi_{l_1}$ as an
antighost field $\bar u^{\bar V_{a_1}}$, and $\varphi_{l_i}$ as
gauge-fixing functionals $C^{V_{a_i}}$ for $i=2, \dots, m$.
The remaining fields are chosen as ``physical'', i.e.\ the propagators pointing to the
corresponding field points are
truncated and replaced by on-shell wave functions.
In the notation, these physical fields are
indicated by~$\varphi_{l_j}^\mathrm{phys}$.
Note that they are characterized by 
$\langle T \dots s \varphi_{l_j}^\mathrm{phys} \dots \rangle=0$,
where the dots stand for any physical fields,
cf.~\cite{bohm2001gauge}.
Considering connected \glspl{gf} only,
using both~\eqref{eq:def-brst-trf-ghosts} and the
\gls{eom}~\eqref{eq:eom-antighost} for the antighost field, 
inserting~\eqref{eq:gauge-fixing-functionals}, 
and finally Fourier transforming to
momentum space, we obtain
\begin{align}
  \Big \langle T \prod_{i=1}^{m}
    \Big[ p_{i, \mu_i} \bar V_{a_i}^{\mu_i}(p_i) 
    + \xi_{V_{a_i}} \eta_{V_{a_i}} M_{V_{a_i}} \bar G_{a_i}(p_i) \Big]
    \prod_{j=m+1}^{n} \varphi_{l_j}^\mathrm{phys}
  \Big \rangle_\mathrm{c} = 0,
  \label{eq:multi-Ca-ST-id-connected}
\end{align}
where the ``c'' indicates the restriction to the connected part of the GF.
Equation~\eqref{eq:multi-Ca-ST-id-connected} represents a very convenient form
of the \gls{st} identities for reducible, connected \glspl{gf}.
Note that for massless vector-boson fields~$V_a$, no associated would-be
Goldstone-boson fields~$G_a$ exist, i.e.\ the corresponding terms with ~$G_a$
are absent in this case.

In order to obtain statements about $S$-matrix elements, we
now reformulate these identities in terms of connected, truncated \glspl{gf}
$G_\mathrm{c}^{\underline{ \varphi_{l_1}}, \dots}(p,\dots)$.
Following the conventions of~\cite{bohm2001gauge,Denner:2019vbn} for
\glspl{gf} throughout, we indicate truncated fields by underlining
and take these fields and their momenta as incoming.
By contrast, fields that are not truncated are taken as outgoing 
(with incoming momenta).
Since we are only interested in the relations among couplings resulting
from the \gls{st} identities, it is sufficient to restrict the following
discussion to the lowest 
perturbative order, i.e.\ to tree level.
Due to the choice of $R_\xi$ gauge-fixing functionals, the lowest-order
propagator~$G_0^{\varphi_k \varphi_l}$ neither mixes vector-boson
fields~$V_a$ with
scalar fields~$\phi_k$ nor would-be Goldstone-boson
fields~$G_a$ with
the remaining scalar fields~$H_m$.
With~\eqref{eq:propagators}, we obtain the \gls{st} identity for
\emph{truncated}, connected \glspl{gf} in lowest order (indicated by subscript ``0''),
\begin{align}
  & \Big \langle T \prod_{i=1}^{m} \Big[ p_{i, \mu_i} \underline{V_{a_i}^{\mu_i}}(p_i)
      - \eta_{V_{a_i}} M_{V_{a_i}} \underline{G_{a_i}}(p_i) \Big]
    \prod_{j=m+1}^{n} \varphi_{l_j}^\mathrm{phys}
  \Big \rangle_{\mathrm{c},0} = 0 .
  \label{eq:multi-ward-id}
\end{align}

Due to its relevance for the subsequent sections, we conclude by
explicitly spelling this identity out for $m=1,2,3$.
For a single, possibly off-shell external vector boson, we obtain
\begin{align}
  p^{\mu} G_{0, \mathrm{c}, \mu}^{ \underline{V^{a}} \varphi^\mathrm{phys}_{l_1} \dots }(p,\dots)
  = \eta_{V_{a}} M_{V_{a}} G_{0,\mathrm{c}}^{ \underline{G^{a}} \varphi^\mathrm{phys}_{l_1} \dots
  }(p,\dots) ,
  \label{eq:single-ward-id}
\end{align}
where we have trivially renumbered the $\varphi^\mathrm{phys}_{l_j}$.
For two vector bosons,~\eqref{eq:multi-ward-id} involves cross terms, leading to
\begin{align}
  0 ={}&
  p_1^{\mu_1} p_2^{\mu_2} G_{0, \mathrm{c}, \mu_1 \mu_2}^{
    \underline{V^{a_1}} \underline{V^{a_2}} \varphi^\mathrm{phys}_{l_1} \dots }(p_1,p_2,\dots)
  + \eta_{V_{a_1}} M_{V_{a_1}} \eta_{V_{a_2}} M_{V_{a_2}}
  G_{0,\mathrm{c}}^{ \underline{G^{a_1}} \underline{G^{a_2}} \varphi^\mathrm{phys}_{l_1} \dots
  }(p_1,p_2,\dots)
  \nonumber\\
  &- p_1^{\mu_1} \eta_{V_{a_2}} M_{V_{a_2}}
  G_{0, \mathrm{c}, \mu_1}^{
    \underline{V^{a_1}} \underline{G^{a_2}} \varphi^\mathrm{phys}_{l_1} \dots }(p_1,p_2,\dots) 
  - p_2^{\mu_2} \eta_{V_{a_1}} M_{V_{a_1}}
  G_{0, \mathrm{c}, \mu_2}^{
    \underline{G^{a_1}} \underline{V^{a_2}} \varphi^\mathrm{phys}_{l_1} \dots }(p_1,p_2,\dots) ,
  \label{eq:double-ward-id}
\end{align}
Finally, for three vector bosons we get
\begin{align}
  0 ={}&
  %
  %
  p_1^{\mu_1} p_2^{\mu_2} p_3^{\mu_3}
  G_{0, \mathrm{c}, \mu_1 \mu_2 \mu_3}^{
    \underline{V^{a_1}} \underline{V^{a_2}} \underline{V^{a_3}} \varphi^\mathrm{phys}_{l_1} \dots
  }(p_1,p_2,p_3,\dots)
  \nonumber\\
  %
  %
  &-
  p_1^{\mu_1} p_2^{\mu_2} \eta_{V_{a_3}} M_{V_{a_3}}
  G_{0, \mathrm{c}, \mu_1 \mu_2}^{
    \underline{V^{a_1}} \underline{V^{a_2}} \underline{G^{a_3}} \varphi^\mathrm{phys}_{l_1} \dots
  }(p_1,p_2,p_3,\dots) 
  \nonumber\\
  &- p_2^{\mu_2} p_3^{\mu_3} \eta_{V_{a_1}} M_{V_{a_1}}
  G_{0, \mathrm{c}, \mu_2 \mu_3}^{
    \underline{G^{a_1}} \underline{V^{a_2}} \underline{V^{a_3}} \varphi^\mathrm{phys}_{l_1} \dots
  }(p_1,p_2,p_3,\dots) 
  \nonumber\\
  &- p_1^{\mu_1} p_3^{\mu_3} \eta_{V_{a_2}} M_{V_{a_2}}
  G_{0, \mathrm{c}, \mu_1 \mu_3}^{
    \underline{V^{a_1}} \underline{G^{a_2}} \underline{V^{a_3}} \varphi^\mathrm{phys}_{l_1} \dots
  }(p_1,p_2,p_3,\dots) 
  \nonumber\\
  %
  %
  &+ p_1^{\mu_1} \eta_{V_{a_2}} M_{V_{a_2}} \eta_{V_{a_3}} M_{V_{a_3}}
  G_{0, \mathrm{c}, \mu_1}^{
    \underline{V^{a_1}} \underline{G^{a_2}} \underline{G^{a_3}} \varphi^\mathrm{phys}_{l_1} \dots
  }(p_1,p_2,p_3,\dots)
  \nonumber\\
  &+ p_2^{\mu_2} \eta_{V_{a_1}} M_{V_{a_1}} \eta_{V_{a_3}} M_{V_{a_3}}
  G_{0, \mathrm{c}, \mu_2}^{
    \underline{G^{a_1}} \underline{V^{a_2}} \underline{G^{a_3}} \varphi^\mathrm{phys}_{l_1} \dots
  }(p_1,p_2,p_3,\dots)
  \nonumber\\
  &+ p_3^{\mu_3} \eta_{V_{a_1}} M_{V_{a_1}} \eta_{V_{a_2}} M_{V_{a_2}}
  G_{0, \mathrm{c}, \mu_3}^{
    \underline{G^{a_1}} \underline{G^{a_2}} \underline{V^{a_3}} \varphi^\mathrm{phys}_{l_1} \dots
  }(p_1,p_2,p_3,\dots)
  \nonumber\\
  %
  %
  &- \eta_{V_{a_1}} M_{V_{a_1}} \eta_{V_{a_2}} M_{V_{a_2}} \eta_{V_{a_3}} M_{V_{a_3}}
  G_{0, \mathrm{c}, \mu_2}^{
    \underline{G^{a_1}} \underline{G^{a_2}} \underline{G^{a_3}} \varphi^\mathrm{phys}_{l_1} \dots
  }(p_1,p_2,p_3,\dots) .
  \label{eq:triple-ward-id}
\end{align}

\subsection{Generic would-be Goldstone couplings}
\label{sect:generic-gb-cpl-relat}
The lowest-order \glspl{wi}~\eqref{eq:single-ward-id}, \eqref{eq:double-ward-id},
and~\eqref{eq:triple-ward-id}
imply identities between couplings
that are obtained by replacing individual vector bosons by would-be
Goldstone bosons.
Here, we explicitly spell out these relations, thereby extending the
notation of generic couplings
from~\cite{denner2001oneloop1,denner2001oneloop2}.
The corresponding derivations are presented in
Appendix~\ref{sect:generic-gb-cpl-relat-derivations}.


\paragraph{\boldmath{$VGH$} couplings.}%
Couplings of type $HVV$ and $VGH$ are connected via
\begin{align}
  C_{H_m V_a V_b}
  =
  - 2 \eta_{V_a} M_{V_a} I^{V_b}_{\bar G_a H_m}
  =
  - 2 \eta_{V_b} M_{V_b} I^{V_a}_{\bar G_b H_m}
  ,
  \label{eq:coupling-relat-SVV-VSG}
\end{align}
which also implies the following identity between $VGH$ couplings,
\begin{align}
  I^{V_a}_{\bar G_b H_m}
  =
  \frac{
    \eta_{V_a} M_{V_a}
  }{
    \eta_{V_b} M_{V_b}
  }
  I^{V_b}_{\bar G_a H_m}
  .
  \label{eq:coupling-relat-VSG-VSG}
\end{align}
By virtue of~\eqref{eq:VGH-cpl-vb-mass}, $HVV$ couplings are related to
vector-boson masses,
\begin{align}
  C_{H_m V_a V_b} \langle H_m \rangle
  =
  \frac{
    2 M_{V_a}^2
  }{
    e
  }
  \delta_{V_a \bar V_b}
  ,
  \label{eq:coupling-relat-SVV-VSG-insertmass}
\end{align}
which involves a sum over the fields~$H_m$.
In the \gls{ewsm}, this becomes
\begin{align}
  C_{H V_a V_b}
  =
  \frac{2 M_{V_a}^2}{e v} \delta_{V_a \bar V_b}
  .
  \label{eq:coupling-relat-SVV-VSG-insertmass-EWSM}
\end{align}
\paragraph{\boldmath{$GVV$} couplings.}%
$GVV$ couplings are related to pure-gauge $VVV$ couplings through
\begin{align}
  C_{G_a V_b V_c}
  = I^{V_a}_{\bar V_b V_c}\frac{M_{V_c}^2 - M_{V_b}^2}{\eta_{V_a} M_{V_a}}
  .
  \label{eq:coupling-relat-VVV-GVV}
\end{align}

\paragraph{\boldmath{$VGG$} couplings.}%
The couplings of type~$VGG$ can be obtained from the corresponding gauge
group generators.
However, they are also related to~$VVV$ couplings via
\begin{align}
  I^{V_c}_{\bar G_a G_b}
  &= 
  I^{V_c}_{\bar V_a V_b}
  \frac{M_{V_c}^2-M_{V_a}^2-M_{V_b}^2}{2 \eta_{V_a} \eta_{V_b} M_{V_a} M_{V_b}}
  .
  \label{eq:coupling-relat-VVV-VGG}
\end{align}
Through~\eqref{eq:coupling-relat-VVV-GVV}, this connects $GGH$ and
$VGG$ couplings, as well.

\paragraph{\boldmath{$GGH$} couplings.}%
$GGH$ couplings are obtained from the $VGH$ counterparts according to
\begin{align}
  C_{G_a G_b H_m}
  =
  - \frac{M_{H_m}^2}{\eta_{V_b} M_{V_b}} I^{V_b}_{\bar G_a H_m}
  =
  - \frac{M_{H_m}^2}{\eta_{V_a} M_{V_a}} I^{V_a}_{\bar G_b H_m}
  ,
  \label{eq:coupling-relat-GGH-VGH}
\end{align}
which combines with~\eqref{eq:coupling-relat-SVV-VSG} into a relation
between $GGH$ and $HVV$ couplings.
Furthermore,~\eqref{eq:VGH-cpl-vb-mass} implies
\begin{align}
  C_{G_a G_b H_m} \langle H_m \rangle
  =
  - \frac{M_{H_m}^2}{e} 
  \delta_{V_a \bar V_b}
  ,
  \label{eq:coupling-relat-GGH-VGH-insertmass}
\end{align}
with \gls{ewsm} form
\begin{align}
  C_{G_a G_b H}
  =
  - \frac{M_\mathrm{H}^2}{e v} \delta_{V_a \bar V_b}
  .
\end{align}
\paragraph{\boldmath{$GHH$} couplings.}%
The relation between $GHH$ and $VHH$ couplings reads
\begin{align}
  C_{G_a H_m H_n}
  =
  \frac{
    M_{H_m}^2 - M_{H_n}^2
  }{\eta_{V_a} M_{V_a}}
  I^{V_a}_{\bar H_m H_n}
  .
  \label{eq:coupling-relat-VHH-GHH}
\end{align}
Both types of couplings vanish in the \gls{ewsm}.
\paragraph{\boldmath{$GGG$} couplings.}%
All $GGG$ couplings vanish,
\begin{align}
  C_{G_a G_b G_c}=0
  .
  \label{eq:coupling-GGG-is-zero}
\end{align}

\paragraph{\boldmath{$G\bar ff$} couplings.}%
The couplings of type $G\bar ff$ and $V\bar ff$ satisfy the identity
\begin{align}
  C^{\chir}_{G_a \bar f_k f_l}
  =
  \frac{
    m_{f_k} I^{V_a}_{f^\chir_k f^\chir_l}
    - m_{f_l} I^{V_a}_{f^{-\chir}_k f^{-\chir}_l}
  }{\eta_{V_a} M_{V_a}}.
  \label{eq:coupling-relat-Vff-Gff}
\end{align}

\paragraph{\boldmath{$HHH$ and $H\bar ff$} couplings.}%
The remaining couplings of type~$HHH$ and~$H\bar ff$ depend on the
scalar potential and the Yukawa sector of the theory, respectively.
Thus, they are not related to other couplings via gauge symmetries.
For the example of the \gls{ewsm}, see~\eqref{eq:app-HHH-cpl}
and~\eqref{eq:app-HFF-cpl}.


\paragraph{Massless limits.}%
The coupling relations~%
\eqref{eq:coupling-relat-VSG-VSG},
\eqref{eq:coupling-relat-VVV-GVV},
\eqref{eq:coupling-relat-VVV-VGG},
\eqref{eq:coupling-relat-GGH-VGH},
\eqref{eq:coupling-relat-VHH-GHH},
and~%
\eqref{eq:coupling-relat-Vff-Gff}
involve factors that feature the masses of at least one gauge
boson~$V_a$ in their denominator.
Any meaningful way of taking the limit~$M_{V_a} \to 0$ in some gauge
theory should keep couplings well defined.
Consequently, the mass-singular factors should cancel in the massless limit.
This is, however, not directly obvious, as the corresponding
numerators involve combinations of masses of various other fields.
Yet, because the derivation of the above relations is based on
\glspl{wi}, this fact is a consequence of the gauge structure of the
theory.


Consider, for example,~\eqref{eq:coupling-relat-VVV-GVV} in the
\gls{ewsm}, where~$V_a=A$ is the photon field.\footnotemark\
This relation is only consistent if~$I^A_{\bar V_b V_c}=0$,
unless~$M_{V_b}=M_{V_c}$.
This is indeed the case:
The only non-vanishing coupling of the photon to other gauge bosons is
that to two W bosons.
Similarly,~\eqref{eq:coupling-relat-Vff-Gff} requires the photon to
couple to fermions of the same mass in a chirality-independent
way.
Again, this is the case.
\footnotetext{%
  In theories that involve massless gauge bosons~$V_a$ from the outset,
  the formulation of the above coupling relations is not sensible,
  since~$V_a$ is not associated to a would-be Goldstone boson~$G_a$.
  However, in the \glspl{wi} used for the derivation of the
  coupling relations, all terms that involve would-be Goldstone bosons
  vanish exactly due to the proportionality to~$M_{V_a}$.
  Without ever dividing by $M_{V_a}$, we instead end up with equations
  stating that the numerators on the r.h.s.\ of~%
  \eqref{eq:coupling-relat-VSG-VSG},
  \eqref{eq:coupling-relat-VVV-GVV},
  \eqref{eq:coupling-relat-VVV-VGG},
  \eqref{eq:coupling-relat-GGH-VGH},
  \eqref{eq:coupling-relat-VHH-GHH},
  and~%
  \eqref{eq:coupling-relat-Vff-Gff}
  vanish, and this is what we refer to in this \gls{ewsm} example.
  Alternatively, we could implement the \gls{ewsm} as the limit~$m_\gamma \to 0$ of 
  a more general theory with a massive photon~$\gamma$.
  Specifically, this requires embedding the \gls{ewsm} into a
  larger theory with additional scalars.
  These need to decouple from \gls{ewsm} fields in the massless limit.
  This means that the l.h.s.\ of the referred relations must vanish
  (i.e.\ the couplings of the would-be Goldstone boson associated to the
  massive photon).
  The \gls{ewsm} limit $m_\gamma\to0$ is hence taken by replacing the
  whole factor on the r.h.s.\ of these equations by zero.
}


The limit~$M_\mathrm{W} \to 0$, $M_\mathrm{Z} \to 0$ from the \gls{ewsm}
to the corresponding unbroken theory is taken by~$v \to 0$, which can be
realized by~$\mu^2 \to 0$ in~\eqref{eq:higgs-pot}.
The would-be Goldstone-boson fields for W and Z bosons become
physical scalar degrees of freedom and replace longitudinal components of
W- and Z-boson fields.
For example, the massless limit of the r.h.s.\
in~\eqref{eq:coupling-relat-VVV-VGG} is finite.
Accordingly, the l.h.s.\ describes the coupling of these physical
scalars to vector bosons, which are now massless.
On the other hand, the r.h.s.\ of~\eqref{eq:coupling-relat-VVV-GVV}
vanishes because of the quadratic masses in the numerator, implying
zero coupling of a single scalar to two vector bosons.
The limit~$v\to0$ does not only cause the masses of all vector boson
fields, but also those of the fermion fields to vanish.
Hence, the r.h.s.\ of~\eqref{eq:coupling-relat-Vff-Gff} remains well
defined for~$M_\mathrm{W}\to0$, $M_\mathrm{Z}\to0$ despite the
parity-violating nature of the weak interaction.
Again, the limit is finite and results in non-vanishing,
chiral couplings of physical scalars to massless fermions,
for~$v=0$.


As we discuss in Sections~\ref{sect:expl-splfuncs-FS-bosonic}, 
\ref{sect:expl-splfuncs-FS-fermionic},
and~\ref{sect:expl-splfuncs-IS}, these considerations are important for
the various massless limits of our splitting functions, involving
similar, seemingly mass-singular terms.

\subsection{Goldstone-boson equivalence theorem}
\label{sect:gbet}
For vector bosons with mass $M_V\neq0$, there are three independent
polarization vectors.
In any frame of reference, they can be subdivided into two transverse
vectors $\varepsilon^\mu_{\mathrm{T},a}$, $a=1,2$, with $\boldsymbol{p}
\boldsymbol{\varepsilon}_{\mathrm{T},a} = 0$
for the 3-dimensional space-like parts $\boldsymbol{p},
\boldsymbol{\varepsilon}_{\mathrm{T},a}$ 
and one longitudinal vector
$\varepsilon^\mu_\mathrm{L}$.
We may choose~$\varepsilon^\mu_{\mathrm{T},a}$
as helicity eigenvectors~$\varepsilon^\mu_\pm$ with eigenvalues $\pm 1$,
while the helicity eigenvalue of $\varepsilon^\mu_\mathrm{L}$ is 0.
The separation into $\varepsilon^\mu_{\mathrm{T},a}$ and
$\varepsilon^\mu_\mathrm{L}$, however, is frame dependent.
Generally, we can parametrize
\begin{align}
  \varepsilon^\mu_{\mathrm{L}}(p)
  =
  \frac{p^\mu}{M_V}
  +
  \restpol^\mu(p)
  ,
  \label{eq:long-pol-vec}
\end{align}
with a gauge vector $\restpol^\mu$ that fulfills $\restpol^2=0$, $p \restpol = - M_V$ and
is of $\orderof{M_V}$ for $p^0\gg M_V$.
It constrains the remaining transverse
polarization vectors via
\begin{align}
  \restpol \varepsilon_{\mathrm{T},a} = 0 .
  \label{eq:trv-pol-vec-cond}
\end{align}
Specifically, the four-vector $\restpol^\mu$ defines the frame in which
$\varepsilon^\mu_{\mathrm{T},a}$ and $\varepsilon^\mu_\mathrm{L}$ are
transverse and longitudinal w.r.t.~$p^\mu$ as three-vectors,
respectively, via the requirement~$\boldsymbol{p} \parallel \boldsymbol{\restpol}$.


The \gls{wi}~\eqref{eq:multi-ward-id} makes an important statement about
the high-energy behavior of amplitudes involving external longitudinal
gauge bosons.
The first term in~\eqref{eq:long-pol-vec} grows indefinitely in the
limit where the energy~$p^0$ of the vector boson becomes large w.r.t.\
its mass~$M_V$.
This suggests problematic behavior of the associated matrix elements.
However, it turns out that these mass-singular terms cancel
systematically when computing full amplitudes, as a consequence of the
\gls{wi}~\eqref{eq:multi-ward-id}.
In fact, we have~\cite{Chanowitz:1985hj,bohm2001gauge,Denner:2019vbn}
\begin{align}
  &
  \Big \langle T
    \prod_{i=1}^{l}
    \Big[
      \varepsilon^{\mu_i}_{\mathrm{L}}(p_i)
      \underline{V^{a_i}_{\mu_i}}(p_i)
    \Big]
    \prod_{j=1}^{n} \varphi_{l_j}^\mathrm{phys}
  \Big \rangle_{0, \mathrm{c}}
  =
  \Big \langle T
    \prod_{i=1}^{l}
    \Big[
      \restpol^{\mu_i}_i
      \underline{V^{a_i}_{\mu_i}}(p_i)
      +
      \eta_{V_{a_i}}
      \underline{G_{a_i}}(p_i)
    \Big]
    \prod_{j=1}^{n} \varphi_{l_j}^\mathrm{phys}
  \Big \rangle_{0, \mathrm{c}}
  ,
  \label{eq:multi-gbet-exact}
\end{align}
where the r.h.s.\ is manifestly well-behaved in the high-energy limit.
In~\cite{Chanowitz:1985hj}, this statement is
shown by summing all possible versions of~\eqref{eq:multi-ward-id} that
involve the insertion of~$m$ gauge-fixing functionals ($1\leq m\leq l$) 
and~$l-m$ additional longitudinal vector bosons
inside~$\varphi^\mathrm{phys}_{l_j}$.\footnotemark\
\footnotetext{%
  For $l>1$,
  this proof is much more involved than the naive approach of simply
  applying~\eqref{eq:single-ward-id} to the contributions proportional
  to $p^{\mu_i}_i/M_{V_{a_i}}$, which is invalid due to unphysical
  external fields. A less cumbersome, but fundamentally analogous version of this proof is
provided by~\cite{Cuomo:2019siu}.
}%


\begin{sloppypar}
In the following, we use~\eqref{eq:multi-gbet-exact} for the treatment
of longitudinally polarized vector bosons in the derivation of collinear
factorization formulas, as it allows for the \emph{exact} rearrangement
of terms in the corresponding matrix elements.
For completeness, we note that the tree-level
\gls{gbet}~\cite{Gounaris:1986cr,Cornwall:1974km,Chanowitz:1985hj,Lee:1977eg,Vayonakis:1976vz,Denner:1996gb}
\begin{align}
  &
  \Big \langle T
    \prod_{i=1}^{l}
\left[
\varepsilon^{\mu_i}_{\mathrm{L}}(p_i)
    \underline{V^{a_i}_{\mu_i}}(p_i)
\right]
    \prod_{j=1}^{n} \varphi_{l_j}^\mathrm{phys}
  \Big \rangle_{0, \mathrm{c}}
  ={}
  \Big \langle T
    \prod_{i=1}^{l}
\left[
\eta_{V_{a_i}}
    \underline{G_{a_i}}(p_i)
\right]
    \prod_{j=1}^{n} \varphi_{l_j}^\mathrm{phys}
  \Big \rangle_{0, \mathrm{c}}
  +
  \orderof{\max_{i=1,\dots,l}
  \left\{
    \frac{M_{V_{a_i}}}{p_i^0}
  \right\}} 
  \label{eq:multi-gbet-leadmassorder}
\end{align}
is obtained as an \emph{approximation} of~\eqref{eq:multi-gbet-exact},
because all terms involving a factor~$\restpol^{\mu_i}_i$ scale
like~$\mathcal{O}\big(M_{V_{a_i}}/p^0_i\big)$.
\end{sloppypar}


Finally, we comment on the limit $M_{V_a} \to 0$
of~\eqref{eq:multi-gbet-exact}, for which $\varepsilon^\mu_\mathrm{L}$
is ill-defined due to its mass-singular component.
Naively, this also suggests singular behavior of the corresponding
amplitudes, i.e.\ the l.h.s.\ of~\eqref{eq:multi-gbet-exact}.
However, because $\restpol^\mu \to 0$, the r.h.s.\ becomes equal to the
would-be Goldstone-boson amplitude and thereby remains finite;
it even turns into an amplitude for a physical scalar boson, since
the Goldstone boson turns into a physical degree of freedom.
On the other hand, massless vector bosons do not possess longitudinal
polarization states, and there are no associated physical matrix
elements.
Instead, the finiteness of \eqref{eq:multi-gbet-leadmassorder} 
in the massless limit implies the transversality
condition for truncated amplitudes for massless gauge bosons, 
i.e.\ the l.h.s.\ of \eqref{eq:multi-gbet-leadmassorder} becomes zero
after replacing the polarization vector
$\varepsilon^{\mu_i}_{\mathrm{L}}(p_i)$ of the gauge boson $V_{a_i}$
by its momentum $p_i^{\mu_i}$ if $M_{V_{a_i}} \to 0$.

\section{Quasi-collinear factorization}
\label{sect:quasi-coll-fact}

\subsection{Collinear and quasi-collinear factorization in QED and QCD}
\label{sect:quasi-coll-fact-qed-qcd}
We are interested in the behavior of matrix elements for arbitrary~$2
\to n+1$ processes
\begin{align}
  \genfield{l_{\mathrm{A}}}(p_{\mathrm{A}}) \genfield{l_{\mathrm{B}}}(p_{\mathrm{B}})
  \to \genfield{l_1}(p_1) \dots \genfield{l_{n+1}}(p_{n+1}) ,
  \label{eq:generic-real-process}
\end{align}
in the limit where any two momenta become collinear.
If both of them correspond to particles in the \gls{fs}, we denote the
latter by~$i$ and~$j$.
This situation is called \gls{fs} case in the following.
If one of them lies in the \gls{is}, while the other one is part of the
\gls{fs}, we denote them by~$a$ and~$i$, respectively.
This is referred to as \gls{is} case.
In the following, we simplify our notation by referring
to~$\genfield{l_a}$, $\genfield{l_i}$, and~$\genfield{l_j}$
as~$\genfield{a}$, $\genfield{i}$, and~$\genfield{j}$, respectively.
Essentially, this corresponds to absorbing the \emph{multiplet} label
$l$ into~$\genfield{}$ and thereby to the reinterpretation of the latter as
\emph{field}.
The matrix elements for the FS and IS cases are written as
\begin{align}
  \mathcal{M}^{(n+1)\, \pol_i \pol_j}_{\genfield{i} \genfield{j} X}
  \quad
  \text{(\gls{fs} case),}
\qquad
\qquad
  \mathcal{M}^{(n+1)\, \pol_a \pol_i}_{\genfield{a} \genfield{i} X}
  \quad
  \text{(\gls{is} case),}
  \label{eq:notation-realme}
\end{align}
where all remaining fields are collectively denoted by $X$.
Polarization is denoted by the labels~$\pol$.


It is well known that the matrix elements~\eqref{eq:notation-realme} of
massless \gls{qed} and \gls{qcd}
exhibit singularities in the collinear limit.
At leading order, in appropriate gauges, these originate from diagram
topologies with an internal line that splits into two external lines,
\drawmomtrue%
\drawvertdottrue%
\renewcommand{\diagang}{45}%
\begin{align}
  \splittingdiagIndivFS{}
    {}{}{}
    {\genfield{ij}}{\genfield{i}}{\genfield{j}}
  \quad
  \text{(\gls{fs} case),}
\qquad\qquad
  \splittingdiagIndivIS{}
    {}{}{}
    {\genfield{a}}{\genfield{i}}{\genfield{ai}}
  \quad
  \text{(\gls{is} case).}
  \label{eq:enhanced-diag-topologies}
\drawmomfalse%
\resetdiagargs%
\end{align}
The intermediate line is denoted by~$ij$ in the \gls{fs} case and~$ai$
in the \gls{is} case
\revision{and carries the momentum $p_{ij} = p_i+p_j$ and $p_{ai}=p_a-p_i$, respectively.}
In these graphs, we have suppressed the remaining \gls{fs} and \gls{is}
particles~$X$.
We use white blobs to denote the sum over all possible subdiagrams that
connect the internal line~$\genfield{ij}$ or~$\genfield{ai}$
with~$X$.
For now, we consider 't~Hooft--Feynman gauge, i.e.\ $R_\xi$ gauge with $\xi_{\vbfield{}}=1$,
for which spurious propagator poles are absent.
The collinear singularities stem from the denominators of the propagators
associated with~$\genfield{ij}$ or~$\genfield{ai}$.
More specifically, evaluating the diagrams
from~\eqref{eq:enhanced-diag-topologies} in the collinear limit, one
observes \emph{factorization} of the matrix
elements~\eqref{eq:notation-realme} into underlying $n$-particle
matrix elements
\begin{align}
  \mathcal{M}^{(n)\, \pol_{ij}}_{\genfield{ij} X}
  \quad
  \text{(\gls{fs} case),}
\qquad\qquad
  \mathcal{M}^{(n)\, \pol_{ai}}_{\genfield{ai} X}
  \quad
  \text{(\gls{is} case),}
  \label{eq:notation-underlyingme}
\end{align}
with external~$\genfield{ij}$ or~$\genfield{ai}$ lines, and universal factors.
The latter are singular in the collinear limit and depend on the species
of the collinear particles and their polarization, but not on the
remaining details of the full process~\eqref{eq:generic-real-process}.
Collinear factorization has first been formulated
in~\cite{Altarelli:1977zs} for massless \gls{qcd} 
and is treated by many textbooks, such
as~\cite{Ellis1991qcd,Peskin:1995ev,bohm2001gauge,Dissertori:2009,Schwartz:2014sze}.


In \gls{qed} and \gls{qcd} with massive fermions, the non-zero masses
regularize the collinear singularities.
The propagators get enhanced, but remain finite, when
$p_{ij}^2 \sim \orderof{m^2}$ or $p_{ai}^2 \sim \orderof{m^2}$ and the
remaining energy scales~$Q^2$ of the
process~\eqref{eq:generic-real-process} are much larger
than~$\orderof{m^2}$.
Here, $m$ stands symbolically for any of the participating masses.
In this situation, which is called 
{\it quasi-collinear
limit}~\cite{Catani:2000ef,catani2002massivedipole}, factorization
remains
valid~\cite{dittmaier1999photonrad,dittmaier2008polarized,Baier:1973ms,Kleiss:1986ct,Catani:2000ef,catani2002massivedipole,Keller:1998tf}.


Collinear singularities and quasi-collinear enhancements are
conveniently investigated in the \emph{Sudakov parametrization} of the
external momenta.
For two \gls{fs} particles~$i$ and~$j$ it reads~\cite{catani2002massivedipole}
\begin{align}
  p_i^{\mu}
  &=z \bar p^{\mu}
  +k_{\perp}^\mu 
  -\underbrace{
    \frac{k_{\perp}^2 + z^2 m_{ij}^2 -m_i^2}{2 z (\bar p n)}
    }_{\equiv A_i}n^{ \mu} ,
  \qquad
  p_i^2=m_i^2 ,
  \nonumber\\
  p_j^{\mu}
  &=(1-z) \bar p^{ \mu}
  -k_{\perp}^\mu 
  -\underbrace{
    \frac{k_{\perp}^2+(1-z)^2m_{ij}^2-m_{j}^2}{2 (1-z) (\bar p n)}
    }_{\equiv A_j} n^{\mu} ,
  \qquad
  p_j^2=m_j^2 ,
  \nonumber\\
  0&=
  \bar p k_\perp
  =n k_\perp
  =n^2 ,
  \qquad \bar p^2=m_{ij}^2 .
  \label{eq:sudakov-param-fs}
\end{align}
In this decomposition, the four-vector $\bar p^\mu$ fulfills $\bar
p^2=m_{ij}^2$ and defines the collinear direction.
The variable $z$ determines how the momentum $\bar p^\mu$ is shared by particles~$i$ and $j$.
The mass-shell conditions~$p_i^2=m_i^2$ and~$p_j^2=m_j^2$ are ensured
via the auxiliary light-like vector $n^\mu$.
The space-like vector $k_\perp^\mu$ denotes the transverse
momentum component of the splitting and cancels in the sum
of $p_i^\mu$ and $p_j^\mu$.
The frame in which its spatial components are perpendicular to those of
$\bar{p}^\mu$ depends 
on the precise choice of $n^\mu$.
For the \gls{is} case, we generalize the parametrization
from~\cite{catani2002massivedipole} to allow for
a massive incoming particle $a$ and a
massive collinear momentum~$\bar p^\mu$.
We get
\begin{align}
  p_a^\mu
  &=\frac{1}{x} \bar p^\mu
  - \underbrace{
    \frac{m_{ai}^2-x^2 m_a^2}{2 x (\bar p n)}
    }_{\equiv B_a} n^\mu ,
  \qquad
  p_a^2=m_a^2 ,
  \nonumber\\
  p_i^\mu
  &=\frac{1-x}{x} \bar p^\mu
  + k_\perp^\mu
  - \underbrace{
    \frac{(k_\perp^2 -m_i^2) x^2 + m_{ai}^2 (1-x)^2}{2 x (1-x) (\bar p n)}
    }_{\equiv B_i} n^\mu ,
  \qquad
  p_i^2=m_i^2 ,
  \nonumber\\
  0&=
  \bar p k_\perp
  =n k_\perp
  =n^2 ,
  \qquad
  \bar p^2=m_{ai}^2 ,
  \label{eq:sudakov-param-is}
\end{align}
where the variable~$x$ determines how the initial momentum
\revisiontwo{$p_a^\mu$}
is shared by the
particles~$i$ and $ai$.
\revisiontwo{Note that in this case $p_a^\mu$ is defining the collinear
direction $\bar p^\mu$, and not the off-shell momentum $p_{ai}^\mu$; this is in contrast to 
the case of final-state emission where $\bar p^\mu$ is defined by the off-shell momentum
$p_{ij}^\mu$.}
Again, the precise definition of~$n^\mu$ determines how exactly this
limit is taken.
In the fully massless case, the propagator denominators
of~$\genfield{ij}$ and~$\genfield{ai}$ read
\begin{align}
    2 p_{i} p_{j} = - \frac{k_\perp^2}{z (1-z)} ,
    \qquad
    2 p_{a} p_{j} = - \frac{k_\perp^2}{1-x} ,
\end{align}
and vanish in the collinear limit $k_\perp^\mu \to 0$.
In the massive case, we have
\begin{align}
  p_{ij}^2 - m_{ij}^2
  &= - \frac{k_\perp^2}{z (1-z)}
  + \frac{m_i^2}{z}
  + \frac{m_j^2}{1-z}
  - m_{ij}^2
  ,
  \nonumber\\
  p_{ai}^2 - m_{ai}^2
  &= \frac{k_\perp^2}{1-x} + m_a^2 x - \frac{x m_i^2}{1-x} - m_{ai}^2
  .
  \label{eq:prop-denom-sudakov-massive}
\end{align}
The quasi-collinear enhancement of propagators thus corresponds to the
kinematic regime where $k_\perp^2 \sim \orderof{m^2}$.
More precisely, the quasi-collinear limit may be
defined by $k_\perp^\mu,m \to 0$, while $k_\perp^\mu/m$ is kept fixed.
In practice, this can be realized by uniform rescaling of~$k_\perp^\mu$
and all masses~$m$ with a parameter~$\lambda$,
\begin{alignat}{5}
  &k_\perp^\mu \to \lambda k_\perp^\mu, \quad
  &&m_i \to \lambda m_i,  \quad
  &&m_j \to \lambda m_j,  \quad
  &&m_{ij} \to \lambda m_{ij},  \quad
  &&\text{(FS case)}
  ,
  \nonumber\\
  &k_\perp^\mu \to \lambda k_\perp^\mu,  \quad
  && m_a \to \lambda m_a, \quad
  &&m_i \to \lambda m_i,  \quad
  &&m_{ai} \to \lambda m_{ai},  \quad
  &&\text{(IS case)}
  ,
\label{eq:quasi-coll-rescaling-lambda}
\end{alignat}
followed by sending $\lambda \rightarrow 0$~\cite{Catani:2000ef,catani2002massivedipole}.
Thus, quasi-collinear propagators scale with $\lambda^{-2}$.
This scheme reproduces the collinear limit if all masses vanish.
We conventionally write $\orderof{\lambda} \equiv \orderof{m}$ and call
it the \emph{mass order}.


In this framework, the derivation of quasi-collinear factorization
formulas in \gls{qed} and \gls{qcd} proceeds as follows.
First, select and compute the
diagrams~\eqref{eq:enhanced-diag-topologies}, which we refer to as
type~\diagtype{1}.
Avoiding gauges that provide further sources of enhanced factors, other
diagrams may be considered as subleading and neglected.\footnotemark%
\footnotetext{%
  Vector-boson propagators in axial gauges as well as external polarization sums for
  massless vector bosons contain terms~$\propto \frac{1}{p g}$, where $p$ is
  the momentum of the vector boson and $g$ is some gauge vector.
  Depending on the choice of~$g$, such contributions can be enhanced in the
  quasi-collinear limit. This happens, for example, if~$\genfield{i}$ represents
  a vector boson, and the momentum $p_j$ is used as gauge vector for the
  polarization vectors of~$\genfield{i}$.
}%
\textsuperscript{,}\footnotemark\
\footnotetext{%
  Of course, we also require the absence of effects that suppress the
  diagram topologies~\eqref{eq:enhanced-diag-topologies} by factors
  of~$\orderof{m}$ w.r.t.\ the remaining diagrams contributing
  to~\eqref{eq:notation-realme}.
  This will always be assumed in the following.
}%
We call them type~\diagtype{2} in the following.
Second, evaluate the squared sum of
diagrams of type~\diagtype{1} in the
parametrizations~\eqref{eq:sudakov-param-fs}
and~\eqref{eq:sudakov-param-is}.
Lastly, take \eqref{eq:quasi-coll-rescaling-lambda} and keep the terms
of leading~\orderof{\lambda}.

\subsection{Factorization in gauge theories with massive vector bosons}
\label{sect:fact-gen-gaugetheo}
The outlined procedure requires more care in theories with massive
vector bosons.
This is due to the mass-singular contributions~$\propto 1/M_V$ in the
longitudinal polarization vectors~\eqref{eq:long-pol-vec} for external
particles~$i$, $j$, or~$a$.
Specifically, these factors may also render 
contributions involving diagrams of type~\diagtype{2}
formally enhanced in the quasi-collinear limit.
At first sight, this mandates their inclusion, which is
not readily possible in a process-independent manner.
As discussed in Section~\ref{sect:gbet} in the context of the
\gls{gbet}, the mass-singular factors
must systematically cancel when computing full amplitudes.
These \emph{gauge cancellations}, however, involve \emph{all
contributing diagrams} and cannot be expected to take place within the
subsets of diagrams \diagtype{1} and \diagtype{2} separately.
Explicitly, in some contributions from the diagrams of
type~\diagtype{1}, the propagator denominator 
that is responsible for the collinear enhancement
may fully cancel against
corresponding factors in the numerator.\footnotemark\
\footnotetext{%
  This mechanism is illustrated in Appendix~\ref{sect:alt-approach}.
}%
This yields mass-singular terms without collinear propagator
enhancement.
These terms exhibit the same structure as mass-singular terms from
diagrams of type~\diagtype{2}, allowing for compensations between the
two types of diagrams.
Simply dropping those of type~\diagtype{2} may, thus, disturb gauge
cancellations.

\revision{
Figure~\ref{fig:ud2www} illustrates type~I and type~II graphs for the
partonic process $u\bar d\to \PW^+ \PW^- \PW^+$ that contributes to the
process of triple-W production at pp~colliders, 
$\Pp\Pp\to \PW^+ \PW^- \PW^+ +X$, where we concentrate on the limit
in which $\PW^+_{(i)}$ and $\PW^-_{(j)}$ become collinear.
\begin{figure}
\begin{tikzpicture}
[line width=0.5pt,
phot/.style={decorate,decoration={snake, segment length=#1, amplitude=2pt}},
phot/.default=6.2pt,
ferm/.style={postaction={decorate},decoration={markings, mark=at position #1 with {\stealtharrow}}},
ferm/.default=0.5]
\small
\coordinate (v1) at (0.8,1.5);
\coordinate (v2) at (0.8,0.75);
\coordinate (v3) at (0.8,0);
\coordinate (lab) at (0.8,0);
\coordinate (i1) at (0,1.5);
\coordinate (i2) at (0,0);
\coordinate (o1) at (1.6,1.5);
\coordinate (o2) at (1.6,0.75);
\coordinate (o3) at (1.6,0);
\draw[phot=6.1pt] (v1) -- (o1);
\draw[phot=6.1pt] (v2) -- (o2);
\draw[phot=6.1pt] (v3) -- (o3);
\draw[ferm] (i1) -- (v1);
\draw[ferm] (v1) -- (v2);
\draw[ferm] (v2) -- (v3);
\draw[ferm] (v3) -- (i2);
\draw[fill=black] (v1) circle (1pt);
\draw[fill=black] (v2) circle (1pt);
\draw[fill=black] (v3) circle (1pt);
\node[xshift=-6pt] at (i1) {$u$};
\node[xshift=-6pt] at (i2) {$\bar{d}$};
\node[xshift=+11pt] at (o1) {$\PW^+_{(i)}$};
\node[xshift=+11pt] at (o2) {$\PW^-_{(j)}$};
\node[xshift=+10pt] at (o3) {$\PW^+$};
\node[xshift=0pt,yshift=-15pt] at (lab) {(a)};
\end{tikzpicture}
\enskip
\begin{tikzpicture}
[line width=0.5pt,
phot/.style={decorate,decoration={snake, segment length=#1, amplitude=2pt}},
phot/.default=6.2pt,
ferm/.style={postaction={decorate},decoration={markings, mark=at position #1 with {\stealtharrow}}},
ferm/.default=0.5]
\small
\coordinate (v1) at (0.4,1.125);
\coordinate (v2) at (1.0,1.125);
\coordinate (v3) at (0.4,0);
\coordinate (lab) at (0.8,0);
\coordinate (i1) at (0,1.125);
\coordinate (i2) at (0,0);
\coordinate (o1) at (1.6,1.5);
\coordinate (o2) at (1.6,0.75);
\coordinate (o3) at (1.6,0);
\draw[phot=6.3pt] (v2) -- (o1);
\draw[phot=6.3pt] (v2) -- (o2);
\draw[phot=6pt] (v3) -- (o3);
\draw[ferm] (i1) -- (v1);
\draw[phot] (v1) -- node[above]{\Pphot/\PZ} (v2);
\draw[ferm] (v1) -- (v3);
\draw[ferm] (v3) -- (i2);
\draw[fill=black] (v1) circle (1pt);
\draw[fill=black] (v2) circle (1pt);
\draw[fill=black] (v3) circle (1pt);
\node[xshift=-6pt] at (i1) {$u$};
\node[xshift=-6pt] at (i2) {$\bar{d}$};
\node[xshift=+11pt] at (o1) {$\PW^+_{(i)}$};
\node[xshift=+11pt] at (o2) {$\PW^-_{(j)}$};
\node[xshift=+10pt] at (o3) {$\PW^+$};
\node[xshift=0pt,yshift=-15pt] at (lab) {(b)};
\end{tikzpicture}
\enskip
\begin{tikzpicture}
[line width=0.5pt,
phot/.style={decorate,decoration={snake, segment length=#1, amplitude=2pt}},
phot/.default=6.2pt,
ferm/.style={postaction={decorate},decoration={markings, mark=at position #1 with {\stealtharrow}}},
ferm/.default=0.5]
\small
\coordinate (v1) at (0.3,0.75);
\coordinate (v2) at (1.0,0.75);
\coordinate (v3) at (1.3,1.125);
\coordinate (lab) at (0.8,0);
\coordinate (i1) at (0,1.5);
\coordinate (i2) at (0,0);
\coordinate (o1) at (1.6,1.5);
\coordinate (o2) at (1.6,0.75);
\coordinate (o3) at (1.6,0);
\draw[phot=6.4pt] (v3) -- (o1);
\draw[phot=6.4pt] (v3) -- (o2);
\draw[phot=6.6pt] (v2) -- (o3);
\draw[ferm] (i1) -- (v1);
\draw[phot=6.1pt] (v1) -- node[below] {\PW} (v2);
\draw[phot] (v2) -- node[above left=-1pt] {\Pphot/\PZ} (v3);
\draw[ferm] (v1) -- (i2);
\draw[fill=black] (v1) circle (1pt);
\draw[fill=black] (v2) circle (1pt);
\draw[fill=black] (v3) circle (1pt);
\node[xshift=-6pt] at (i1) {$u$};
\node[xshift=-6pt] at (i2) {$\bar{d}$};
\node[xshift=+11pt] at (o1) {$\PW^+_{(i)}$};
\node[xshift=+11pt] at (o2) {$\PW^-_{(j)}$};
\node[xshift=+10pt] at (o3) {$\PW^+$};
\node[xshift=0pt,yshift=-15pt] at (lab) {(c)};
\end{tikzpicture}
\enskip
\begin{tikzpicture}
[line width=0.5pt,
phot/.style={decorate,decoration={snake, segment length=#1, amplitude=2pt}},
phot/.default=6.2pt,
ferm/.style={postaction={decorate},decoration={markings, mark=at position #1 with {\stealtharrow}}},
ferm/.default=0.5,
higgs/.style=dashed]
\small
\coordinate (v1) at (0.3,0.75);
\coordinate (v2) at (1.0,0.75);
\coordinate (v3) at (1.3,1.125);
\coordinate (lab) at (0.8,0);
\coordinate (i1) at (0,1.5);
\coordinate (i2) at (0,0);
\coordinate (o1) at (1.6,1.5);
\coordinate (o2) at (1.6,0.75);
\coordinate (o3) at (1.6,0);
\draw[phot=6.4pt] (v3) -- (o1);
\draw[phot=6.4pt] (v3) -- (o2);
\draw[phot=6.6pt] (v2) -- (o3);
\draw[ferm] (i1) -- (v1);
\draw[phot=6.1pt] (v1) -- node[below] {\PW}(v2);
\draw[higgs] (v2) -- node[above left=-1pt] {\PH} (v3);
\draw[ferm] (v1) -- (i2);
\draw[fill=black] (v1) circle (1pt);
\draw[fill=black] (v2) circle (1pt);
\draw[fill=black] (v3) circle (1pt);
\node[xshift=-6pt] at (i1) {$u$};
\node[xshift=-6pt] at (i2) {$\bar{d}$};
\node[xshift=+11pt] at (o1) {$\PW^+_{(i)}$};
\node[xshift=+11pt] at (o2) {$\PW^-_{(j)}$};
\node[xshift=+10pt] at (o3) {$\PW^+$};
\node[xshift=0pt,yshift=-15pt] at (lab) {(d)};
\end{tikzpicture}
\enskip
\begin{tikzpicture}
[line width=0.5pt,
phot/.style={decorate,decoration={snake, segment length=#1, amplitude=2pt}},
phot/.default=6.2pt,
ferm/.style={postaction={decorate},decoration={markings, mark=at position #1 with {\stealtharrow}}},
ferm/.default=0.5]
\small
\coordinate (v1) at (0.4,0.75);
\coordinate (v2) at (1.2,0.75);
\coordinate (i1) at (0,1.5);
\coordinate (i2) at (0,0);
\coordinate (o1) at (1.6,1.5);
\coordinate (o2) at (1.6,0.75);
\coordinate (o3) at (1.6,0);
\draw[phot=6.6pt] (v2) -- (o1);
\draw[phot=7.0pt] (v2) -- (o2);
\draw[phot=6.6pt] (v2) -- (o3);
\draw[ferm] (i1) -- (v1);
\draw[phot=6.1pt] (v1) -- node[above] {\PW} (v2);
\draw[ferm] (v1) -- (i2);
\draw[fill=black] (v1) circle (1pt);
\draw[fill=black] (v2) circle (1pt);
\node[xshift=-6pt] at (i1) {$u$};
\node[xshift=-6pt] at (i2) {$\bar{d}$};
\node[xshift=+11pt] at (o1) {$\PW^+_{(i)}$};
\node[xshift=+11pt] at (o2) {$\PW^-_{(j)}$};
\node[xshift=+10pt] at (o3) {$\PW^+$};
\node[xshift=0pt,yshift=-15pt] at (lab) {(e)};
\end{tikzpicture}
\caption{\revision{Some representative tree-level diagrams contributing to the 
partonic process $u\bar d\to \PW^+ \PW^- \PW^+$ (adapted from Ref.~\cite{Dittmaier:2017bnh}).
In the limit in which $\PW^+_{(i)}$ and $\PW^-_{(j)}$ become collinear,
graphs (b), (c), (d) are of type~I, while graphs (a), (e) are of type~II.}}
\label{fig:ud2www}
\end{figure}
Graphs~(a) and (e) are of type~II without collinear enhancement if 
$(p_i+p_j)^2=\orderof{m^2}=\orderof{M_{\PW}^2}$;
graphs (b), (c), and (d) are of type~I featuring collinearly enhanced splittings
$\genfield{ij}\to \PW^+_{(i)} \PW^-_{(j)}$ with $\genfield{ij}=\gamma/\PZ$
for graphs~(b) and (c) and $\genfield{ij}=\PH$ for graph~(d).
}

In this paper, we devise two different strategies for the solution of
these issues.
Both approaches modify the expressions for the diagrams of
type~\diagtype{1} in a universal manner, completing gauge cancellations
within this set of diagrams.
Henceforth, dropping all remaining 
contributions from diagrams of type~\diagtype{2} becomes
possible, and the manifest process independence of the factorization
procedure is retained.
In both approaches, the \gls{gbet} and its implications for the cancellation of
mass-singular factors are employed as guiding principle.
In summary, our default strategy rewrites 
amplitudes with longitudinally polarized vector bosons
from the outset via~\eqref{eq:multi-gbet-exact}, immediately
replacing the associated mass-singular factors.
We discuss this procedure in Section~\ref{sect:VVV}.
Our alternative procedure, on the other hand, manually identifies and
performs the gauge cancellations implied by the \gls{gbet}, closely
following the above discussion.
This approach is detailed in Appendix~\ref{sect:alt-approach}.


In the rest of this section, we describe the remaining details
regarding our derivation of quasi-collinear factorization formulas,
assuming longitudinal polarization vectors are treated properly.
The actual computation of factorization formulas for individual
splitting processes is provided in Sections~\ref{sect:expl-splfuncs-FS-bosonic}, \ref{sect:expl-splfuncs-FS-fermionic},
and~\ref{sect:expl-splfuncs-IS}, based on the approach outlined here.

We aim to factorize \emph{squared} amplitudes from the
outset, i.e.\ we start from 
\drawvertdottrue%
\renewcommand{\sclfac}{1}%
\renewcommand{\extsclfac}{1.2}%
\renewcommand{\intsclfac}{.85}%
\renewcommand{\len}{1.2cm}%
\renewcommand{\rad}{6mm}%
\renewcommand{\dotrad}{.7mm}%
\renewcommand{\diagang}{45}%
\begin{align}
  \renewcommand{\spdtypeijl}{}%
  \renewcommand{\spdtypeijr}{}%
  \renewcommand{\spdtypei}{}%
  \renewcommand{\spdtypej}{}%
  \Big|
  \mathcal{M}
    ^{(n+1)\, \pol_i \pol_j}
    _{\genfield{i} \genfield{j} X}
  \Big|^2
  \xsim{\lambda \to 0}
  \sum_{\genfield{ij^{\vphantom{\prime}}}, \, \genfield{ij^\prime}}
  \splittingdiagSqFS
    {\genfield{ij^{\vphantom{\prime}}}}{\genfield{ij^\prime}}
    {\genfield{i} , \pol_i}{\genfield{j} , \pol_j}
    {$T^{(n)}_{\genfield{ij} X}$}{$T^{(n)}_{\genfield{ij^\prime} X}$}%
  ,
  \nonumber
  \\
  \renewcommand{\spdtypeail}{}%
  \renewcommand{\spdtypeair}{}%
  \renewcommand{\spdtypea}{}%
  \renewcommand{\spdtypei}{}%
  \Big|
  \mathcal{M}
    ^{(n+1)\, \pol_a \pol_i}
    _{\genfield{a} \genfield{i} X}
  \Big|^2
  \xsim{\lambda \to 0}
  \sum_{\genfield{ai^{\vphantom{\prime}}}, \, \genfield{ai^\prime}}
  \splittingdiagSqIS
    {\genfield{ai^{\vphantom{\prime}}}}{\genfield{ai^\prime}}
    {\genfield{a} , \pol_a}{\genfield{i} , \pol_i}
    {$T^{(n)}_{\genfield{ai} X}$}{$T^{(n)}_{\genfield{ai^\prime} X}$}%
  ,
  \label{eq:decomp-sq-realme-by-mother}
\end{align}
\resetdiagargs%
where the dashed line denotes the product of interfering diagrams.
Here, we account for the fact that various mother particles~$ij$ or~$ai$
may occur in the diagrams~\eqref{eq:enhanced-diag-topologies} by
labeling them with~$ij/ij^\prime$ in the \gls{fs} case
and~$ai/ai^\prime$ in the \gls{is} case.
The corresponding multiplet indices and masses are written
as~$l_{ij}/l_{ij}^\prime$, $l_{ai}/l_{ai}^\prime$
and~$m_{ij}/m_{ij}^\prime$, $m_{ai}/m_{ai}^\prime$, respectively.
For example, in the \gls{ewsm}, a $\gamma \mathrm{W}^\pm$-pair can
originate from the splitting of a W or its associated would-be Goldstone
boson, while a $\mathrm{W}^\pm \mathrm{W}^\mp$-pair can be produced via
the splitting of a photon, Higgs boson, Z, or the corresponding would-be
Goldstone boson. 
In~\eqref{eq:decomp-sq-realme-by-mother}, all possible interferences may
occur.
Accordingly, we denote the contributions from the individual diagrams of
type~\diagtype{1} with \emph{fixed} intermediate $\genfield{ij}$
or~$\genfield{ai}$ as \emph{partial matrix elements}
\begin{align}
  \MContr{ij}{i}{j}^{\pol_i \pol_j} ,
  \qquad
  \MContr{ai}{a}{i}^{\pol_a \pol_i}
  .
  \label{eq:def-partialM}
\end{align}


While being more involved than the consideration of individual
amplitudes, this approach has several advantages.
First, the obtained results may be directly applied to shower
formulations or \gls{nlo} subtraction algorithms, avoiding the necessity
of non-trivial combinations of formulas for individual amplitudes.
Second, external wave functions $\mathcal{W}$ always occur in pairs
in~\eqref{eq:decomp-sq-realme-by-mother}, which allows for their
replacement via (spin-projected) completeness relations of the form
\begin{align}
  \mathcal{W}^\pol_{\genfield{}}(p)
  \overline{\mathcal{W}}^\pol_{\genfield{}}(p)
  =
  C_{\genfield{}}(p^\mu,\restpol^\mu,\pol)
  ,
  \label{eq:general-completeness-relation}
\end{align}
where $C_{\genfield{}}$ is a tensor-valued function (in Lorentz or Dirac
space) and the bar denotes conjugation, i.e.~$\bar \psi = \psi^\dagger
\gamma^0$ for spinors $\psi$ and complex conjugation otherwise.
The light-like gauge vector $\restpol^\mu$ is defined as
in~\eqref{eq:long-pol-vec} and provides the precise definition of the
spin reference axis for the considered particle.
For vector bosons, the role of $\restpol^\mu$ has been discussed in Section~\ref{sect:gbet},
and the
\emph{transverse} polarization sum for an on-shell vector boson of 
momentum~$p$ and mass $M_V$ ($p^2=M_V^2$) reads
\begin{align}
  \sum_{\polvb=\pm}
  \varepsilon^{\mu \ast}_{\polvb}(p)
  \varepsilon^{\nu}_{\polvb}(p)
  &=
  - g^{\mu\nu}
  - \frac
    {p^\mu \restpol^\nu + p^\nu \restpol^\mu}
    {M_V}
  - \restpol^\mu \restpol^\nu
  ,
  \qquad
  p \restpol = - M_V ,
\qquad r^2=0.
  \label{eq:trv-poln-sum}
\end{align}
The spinors for on-shell fermions and antifermions of 
momentum~$p$ and mass $m_f$ ($p^2=m_f^2$) fulfill
\begin{align}
    u_{\polferm}(p) \bar{u}_{\polferm}(p)
    &=
    \Sigma_{\polferm}
    (\slashed p + m_f)
    ,\qquad
    v_{\polferm}(p) \bar{v}_{\polferm}(p)
    =
    \Sigma_{\polferm}
    (\slashed p - m_f)
    ,
    \label{eq:spinproj-complrelat-ferm}
\end{align}
respectively, with the spin projection matrix
\begin{align}
  \Sigma_{\polferm}
  &=
  \frac{
    1
    +
    \polferm \gamma^5
    \slashed s
  }{2}
  ,
  \qquad
  s^\mu
  =
  \frac{p^\mu}{m_f}
  +
  \restpol^\mu
  ,
  \qquad
  p \restpol = - m_f,
\qquad r^2=0.
  \label{eq:spin-projection-op}
\end{align}
We recall that $\Sigma_\pm$ projects onto
the fermion and antifermion spinors $u_\pm$ and $v_\pm$, which both correspond to
states with spin projection $\pm1/2$
in the reference direction.
In the rest frame of $p^\mu$, this direction is
provided by the unit vector
$\hat{\boldsymbol{r}}$.
In any frame with
$\boldsymbol{p} \parallel \boldsymbol{r}$,
these spinors correspond to states with helicity $\pm1/2$.
In summary, the use of~\eqref{eq:general-completeness-relation} bypasses
the need for explicit parametrizations of spinors and polarization
vectors and thereby enables us to work in a way that is both manifestly
independent of the Lorentz frame and the precise definition of
polarization.
In our approach, the latter 
is reflected by the fact that the
vectors $\restpol^\mu$ are found to drop out of our results, without making any
further assumptions.


Plugging propagators, vertices, and wave functions from
Section~\ref{sect:feynmanrules-gengaugetheo}
into~\eqref{eq:decomp-sq-realme-by-mother}, and employing completeness
relations like~\eqref{eq:general-completeness-relation}, we obtain
asymptotic expressions
\begin{align}
  \left|
  \mathcal{M}
    ^{(n+1)\, \pol_i \pol_j}
    _{\genfield{i} \genfield{j} X}
  \right|^2
  \xsim{\lambda \to 0}
  e^2
  \sum_{\genfield{ij^{\vphantom{\prime}}}, \genfield{ij^\prime}}
  &
  \frac{1}{p_{ij}^2  - m_{ij}^2} \frac{1}{p_{ij}^2 - m_{ij}^{\prime\, 2}}
  \nonumber\\
  &
  \cdot T^{(n)}_{\genfield{ij^{\vphantom{\prime}}} X}(p_{ij}) \otimes
  \tilde{\mathcal{P}}
    {}^{\genfield{ij^{\vphantom{\prime}}}}_{\genfield{i}}
    {}^{\genfield{ij^\prime}}_{\genfield{j}}
  (p_i,p_j,\restpol_i,\restpol_j,\pol_i,\pol_j)
  \otimes \bar{T}^{(n)}_{\genfield{ij^\prime} X}(p_{ij})
  ,
  \nonumber\\
  \left|
  \mathcal{M}
    ^{(n+1)\, \pol_a \pol_i}
    _{\genfield{a} \genfield{i} X}
  \right|^2
  \xsim{\lambda \to 0}
  e^2 
  \sum_{\genfield{ai^{\vphantom{\prime}}}, \genfield{ai^\prime}}
  &
  \frac{1}{p_{ai}^2  - m_{ai}^2} \frac{1}{p_{ai}^2 - m_{ai}^{\prime\, 2}}
  \nonumber\\
  &
  \cdot T^{(n)}_{\genfield{ai^{\vphantom{\prime}}} X}(p_{ai}) \otimes
  \tilde{\bar{\mathcal{P}}}
    {}^{\genfield{ai^{\vphantom{\prime}}}}_{\genfield{a}}
    {}^{\genfield{ai^\prime}}_{\genfield{i}}
  (p_a,p_i,\restpol_a,\restpol_i,\pol_a,\pol_i)
  \otimes \bar{T}^{(n)}_{\genfield{ai^\prime} X}(p_{ai})  
  \label{eq:realme-sq-fact-nonsudakov}
\end{align}
in terms of some functions~$\tilde{\mathcal{P}}$
and~$\tilde{\bar{\mathcal{P}}}$,
which are precursors of the splitting function we are after.
These may carry Lorentz or Dirac indices, are completely independent of
the underlying $n$-particle process, and only depend on the splitting
type and kinematics.
All process dependence is contained in the \emph{stripped matrix
elements}~$T$, defined by factoring wave functions off full matrix
elements $\mathcal{M}$,
\begin{align}
  \mathcal{M}
    ^{\pol_{ij}}
    _{\genfield{ij} X}
  &=
  T_{\genfield{ij} X}(p_{ij})
  \otimes
  \mathcal{W}
    _{\genfield{ij}}
    ^{\pol_{ij}}
    (p_{ij})   
  ,
\qquad&
  \mathcal{M}
    ^{\pol_{ai}}
    _{\genfield{ai} X}
  &=
  T_{\genfield{ai}X}(p_{ai})
  \otimes
  \mathcal{W}
    _{\genfield{ai}}
    ^{\pol_{ai}}
    (p_{ai})
  .
  \label{eq:notation-ubornme}
\end{align}
The symbolized product~$\otimes$ represents multiplication, including
contraction of Lorentz and Dirac indices, depending on the types of the
particles~$ij$ and~$ai$.
Whenever~$\otimes$ appears, the order of factors does not necessarily
represent the position of the various (Lorentz and Dirac) indices.
This must be inferred from the specific case.
For example, if~$ij$ and~$ai$ are vector
bosons,~\eqref{eq:notation-ubornme} reads
\begin{align}
  \mathcal{M}^{(n)\, \polvb_{ij}}_{\vbfield{ij} X}
  & =
  T^{(n) \mu}_{\vbfield{ij} X}(p_{ij})
  \,
  \varepsilon_{\polvb_{ij},\mu}^\ast(p_{ij})
  ,
\qquad&
  \mathcal{M}^{(n)\, \polvb_{ai}}_{\vbfield{ai} X}
  & =
  T^{(n) \mu}_{\vbfield{ai} X}(p_{ai})
  \,
  \varepsilon_{\polvb_{ai},\mu}(p_{ai})
  .
  \label{eq:notation-stripped-ubornme-vector}
\end{align}
If they are fermions, we instead have 
\begin{align}
  \mathcal{M}^{(n)\, \polferm_{ij}}_{\fermfield{ij} X}
  & =
  \bar u_{\polferm_{ij}}(p_{ij})
  \,
  T^{(n)}_{\fermfield{ij} X}(p_{ij})
  ,
\qquad&
  \mathcal{M}^{(n)\, \polferm_{ai}}_{\fermfield{ai} X}
  & =
  \bar T^{(n)}_{\fermfield{ai} X}(p_{ai})
  \,
  u_{\polferm_{ai}}(p_{ai})
  ,
  \label{eq:notation-stripped-ubornme-fermion}
\end{align}
and in the case of antifermions
\begin{align}
  \mathcal{M}^{(n)\, \polferm_{ij}}_{\antifermfield{ij} X}
  & =
  \bar T^{(n)}_{\antifermfield{ij} X}(p_{ij})
  \,
  v_{\polferm_{ij}}(p_{ij})
  ,
\qquad&
  \mathcal{M}^{(n)\, \polferm_{ai}}_{\antifermfield{ai} X}
  & =
  \bar v_{\polferm_{ai}}(p_{ai})
  \,
  T^{(n)}_{\antifermfield{ai} X}(p_{ai})
  .
  \label{eq:notation-stripped-ubornme-antifermion}
\end{align}
Here and in the following, we include a bar in the definition of~$T$ for
outgoing antifermions~$\fermfield{ij}$ and incoming fermions~$\fermfield{ai}$ 
to indicate the Dirac structure.


In order to evaluate~\eqref{eq:realme-sq-fact-nonsudakov} in the quasi-collinear
limits, we employ the Sudakov parametrizations given
in~\eqref{eq:sudakov-param-fs} and~\eqref{eq:sudakov-param-is}.
We assign auxiliary invariant masses $\tilde{m}_{ij}$ and $\tilde{m}_{ai}$
to the collinear four-momentum~$\bar p$ to properly account for the cases
$m_{ij} \neq m^\prime_{ij}$ and $m_{ai} \neq m^\prime_{ai}$.
The auxiliary masses are consistently counted as~$\orderof{m}$ by adding
$\tilde{m}_{ij} \to \lambda \tilde{m}_{ij}$ and
$\tilde{m}_{ai} \to \lambda \tilde{m}_{ai}$
to the rescaling procedure~\eqref{eq:quasi-coll-rescaling-lambda}.
\revision{We note that we do not expand the momenta
$p_{ij}$ and $p_{ai}$ in the hard matrix elements
$T^{(n)}_{\genfield{ij} X}\left(p_{ij}\right)$
and
$T^{(n)}_{\genfield{ai} X}\left(p_{ai}\right)$
explicitly, or replace them by on-shell counterparts.
Introducing a corresponding procedure
(potentially also shifting other external momenta to respect momentum conservation),
however, would not modify our
results as long as only subleading $\orderof{m}$-terms are introduced.}
Special care has to be taken when any of the intermediate particles is a
vector boson~$\vbfield{}$.
In this case,~\eqref{eq:realme-sq-fact-nonsudakov}
involves contractions of $\bar p^\mu$ with
$T^{(n) \mu}_{\vbfield{} X}(p)$.
Via the \gls{wi}~\eqref{eq:single-ward-id}, this becomes proportional to
$M_{\vbfield{}} T^{(n)}_{\gbfield{} X}(p)$, which is of $\orderof{m}$.\footnotemark\
\footnotetext{%
  The \gls{wi} requires the contraction with $p = \bar p + \orderof{m}$,
  but that does not change this argument.
}%
Therefore,~$\bar p^\mu$ must be counted as~$\orderof{m}$ when contracted
with the stripped underlying amplitude.
We stress that no assumption on~$\restpol^\mu$ other than~$\restpol^\mu = \orderof{m}$,
and thereby no constraints on the precise definition of polarizations,
are made when evaluating the quasi-collinear limits.
Keeping only terms of lowest order in $\lambda$
from~\eqref{eq:realme-sq-fact-nonsudakov} finally provides expressions
of~$\orderof{m^{-2}}$.
Similar to~$\tilde{\mathcal{P}}$ and~$\tilde{\bar{\mathcal{P}}}$, we can
read off the \gls{fs} and \gls{is} splitting functions
\begin{align}
  \mathcal{P}
    {}^{\genfield{ij^{\vphantom{\prime}}}}_{\genfield{i}}
    {}^{\genfield{ij^\prime}}_{\genfield{j}}
    \big(z,k_\perp^\mu,\bar p^\mu,\pol_i,\pol_j\big)
  ,\qquad
  \bar{\mathcal{P}}
      {}^{\genfield{ai^{\vphantom{\prime}}}}_{\genfield{a}}
      {}^{\genfield{ai^\prime}}_{\genfield{i}}
    \big(x,k_\perp^\mu,\bar p^\mu,\pol_a,\pol_i\big)
  ,
  \label{eq:spl-func}
\end{align}
respectively, which are of $\orderof{m^{2}}$.
These correspond to the splitting processes
\begin{align}
  [\genfield{ij}^\ast/\genfield{ij^\prime}^\ast]
  \to
  \genfield{i} \, \genfield{j}
  ,\qquad
  \genfield{a} \to \genfield{i} \,
  [\genfield{ai}^\ast/\genfield{ai^\prime}^\ast]
  ,
  \label{eq:spl-proc}
\end{align}
according to the corresponding subdiagrams
in~\eqref{eq:decomp-sq-realme-by-mother}.
The square brackets denote interfering mother particles, while the
asterisk indicates off-shellness.
The procedure of taking the quasi-collinear limits is demonstrated more
explicitly in Section~\ref{sect:VVV}.


%
As indicated by the arguments of the splitting functions~$\mathcal{P}$
and~$\bar{\mathcal{P}}$ in~\eqref{eq:spl-func}, they are found to be
independent of auxiliary quantities such as the vectors~$n^\mu$,
$\restpol^\mu$, and the masses~$\tilde{m}_{ij}$, $\tilde{m}_{ai}$.
As mentioned above, this implies that our results are
independent of the precise definition of spin reference axes for the
individual external particles.
Moreover, working in general $R_\xi$ gauge, we observe that all
dependence on the gauge parameters~$\xi_{\vbfield{}}$ drops out.


Our computations are performed within \textsc{Mathematica} 13, employing
\textsc{FeynArts} 3.11~\cite{Hahn:2000kx} for the generation of diagrams of
type~\eqref{eq:def-partialM} and their translation into analytical
expressions.
For the handling of Lorentz and Dirac algebra, we
use~\textsc{FeynCalc} 9.3.1~\cite{Shtabovenko:2016sxi,Shtabovenko:2020gxv,Kublbeck:1990xc}.
Most of the results have been checked in a second, independent calculation
with inhouse \textsc{Mathematica} routines.

\section{Derivation of splitting functions}
\label{sect:splfunc-derivation}

\subsection{Symmetries of splitting functions}
\label{sect:splfunc-symm}
The splitting functions are subject to certain symmetry and crossing
relations, which are detailed in this section.
Moreover, there are additional relations under charge conjugation for
splitting functions involving fermions.
This is detailed in Section~\ref{sect:expl-splfuncs-FS-fermionic}.

\subsubsection[%
  Symmetries under
  \mathinhead{i \leftrightarrow j}{i <-> j}
  and
  \mathinhead{a \leftrightarrow i}{a <-> i}
  ]{%
  Symmetries under
  \mathinheadbold{i \leftrightarrow j}{i <-> j}
  and
  \mathinheadbold{a \leftrightarrow i}{a <-> i}
  }
\label{sect:splfunc-symm-i-j-and-a-i}
The diagrams that determine
the \gls{fs} splitting
functions~$\mathcal{P}$ are invariant under simultaneous exchange of the
fields $\genfield{i}$, $\genfield{j}$, polarization labels~$\pol_i$,
$\pol_j$, and momenta~$p_i$, $p_j$.
Via~\eqref{eq:sudakov-param-fs}, this means that $\mathcal{P}$ is
symmetric under
\begin{align}
  i\leftrightarrow j
  \equiv
  \big\{
    \genfield{i} \leftrightarrow \genfield{j}
    ,\
    m_i \leftrightarrow m_j
    ,\
    \pol_i \leftrightarrow \pol_j
    ,\
    z \to (1-z)
    ,\
    k_\perp^\mu \to - k_\perp^\mu , \dots
  \big\},
  \label{eq:i-j-symmru}
\end{align}
i.e.\
\begin{align}
  \mathcal{P}
    {}^{\genfield{ij^{\vphantom{\prime}}}}_{\genfield{i}}
    {}^{\genfield{ij^\prime}}_{\genfield{j}}
  \big(
    z,k_\perp^\mu,\bar p^\mu,
    \pol_i,\pol_j
  \big)
  &=
  \mathcal{P}
    {}^{\genfield{ij^{\vphantom{\prime}}}}_{\genfield{j}}
    {}^{\genfield{ij^\prime}}_{\genfield{i}}
  \big(
    1-z,-k_\perp^\mu,\bar p^\mu,
    \pol_j,\pol_i
  \big)
  \Big|_{m_i \leftrightarrow m_j}
  .
  \label{eq:i-j-symmru-splfunc}
\end{align}
The dots in \eqref{eq:i-j-symmru} indicate that there might be more properties
of $i$ and $j$ relevant to characterize the splitting functions, such as
the phases of would-be Goldstone-boson
fields, which are suppressed in the notation here (see Section~\ref{sect:VHHV}).
To avoid confusion, we stress that we understand the fields and
polarization labels in the definition~\eqref{eq:spl-func} of
$\mathcal{P}$ and $\bar{\mathcal{P}}$ to be assigned to the particles of
the splitting processes~\eqref{eq:spl-proc} by their \emph{position} in
the respective lists of arguments, regardless of what \emph{indices} the
fields and polarization labels may carry themselves.
In contrast, the masses $m_a$, $m_i$, and $m_j$ (and all
remaining parameters such as phases of would-be Goldstone-boson fields)
are assigned \emph{implicitly by their indices}.
Specifically, in the functions
\begin{align}
  \mathcal{P}
    {}^{\alpha}_{\gamma}
    {}^{\beta}_{\delta}
  (
    z,k_\perp^\mu,\bar p^\mu,
    \pol_1,\pol_2
  ),
  \qquad
  \bar{\mathcal{P}}
    {}^{\alpha}_{\gamma}
    {}^{\beta}_{\delta}
  (
    x,k_\perp^\mu,\bar p^\mu,
    \pol_1,\pol_2
  ),
\end{align}
the fields $\gamma$ and $\delta$, the polarizations $\kappa_1$ and
$\kappa_2$, and the masses $m_{i/a}$ and $m_{j/i}$ correspond to the
first and second external particle in the \gls{fs} and \gls{is}
splitting process of~\eqref{eq:spl-proc}, respectively.
Therefore, $m_{i/a}$ and $m_{j/i}$ \emph{always} belong
to the fields $\gamma$ and $\delta$, with polarizations $\pol_1$ and
$\pol_2$.
Thus, before the interchange $m_i\leftrightarrow m_j$,
the mass $m_i$ in the splitting function on the r.h.s.\
of~\eqref{eq:i-j-symmru-splfunc} corresponds to the
field~$\genfield{j}$, \emph{not} to~$\genfield{i}$.\footnotemark
\footnotetext{%
  To spell this out even more explicitly
  in a specific example, in the \gls{fs} splitting function
    $\mathcal{P}
      {}^{\fermfield{ij^{\vphantom{\prime}}}}_{\vbfield{j}}
      {}^{\fermfield{ij^\prime}}_{\fermfield{i}}
      \big(z,k_\perp^\mu,\bar p^\mu,\polvb,\polferm\big)$
  the vector boson~$\vbfield{j}$ carries the mass $m_i$,
  polarization $\polvb$, and momentum fraction~$z$, while the fermion
  $\fermfield{i}$ carries the mass $m_j$, polarization $\polferm$, and
  momentum fraction~$1-z$.
}%
\textsuperscript{,}\footnotemark\
\footnotetext{%
  In the following, similar issues do not arise for the polarization
  labels $\kappa_{a/i/j}$.
  Their position in the argument lists of splitting functions is always
  consistent with the position of the fields with the respective index.
}%
The splitting functions on the r.h.s.\
of~\eqref{eq:i-j-symmru-splfunc} may, finally, be related to splitting functions
for fields with indices in conventional order by modification of
couplings, see~\eqref{eq:VVV-TL-split-func}
and~\eqref{eq:VFF-swap-cpl-wo-chir} below.


Analogously, the diagrams corresponding to the \gls{is} splitting
functions~$\bar{\mathcal{P}}$ are symmetric under~\emph{crossing} of the
fields~$\genfield{a}$ and~$\genfield{i}$, reassigning as well as
flipping the polarizations~$\pol_a \leftrightarrow -\pol_i$, and
swapping the momenta~$p_a$ and $-p_i$.
Crossing involves charge conjugation of the fields,
i.e.\ fermions turn 
into antifermions and vice versa.
Via~\eqref{eq:sudakov-param-is}, this corresponds to the symmetry
of~$\bar{\mathcal{P}}$ under
\begin{align}
  a\leftrightarrow i
  \equiv
  \bigg\{
    &
    m_a \leftrightarrow m_i
    ,\
    \genfield{a} \leftrightarrow \genantifield{i}
    ,\
    \pol_a \leftrightarrow -\pol_i
    ,
    \nonumber\\
    &
    x \to \frac{x}{x-1}
    ,\
    k^\mu_\perp \to \frac{-1}{x-1} k^\mu_\perp
    ,\
    \bar p^\mu \to \bar p^\mu - \frac{x}{x-1} k^\mu_\perp , \dots
  \bigg\}
  .
  \label{eq:a-i-symmru}
\end{align}
The kinematic part of these substitutions results from the interchange 
$p_a \leftrightarrow -p_i$ neglecting some 
mass-suppressed terms that are not relevant in the following
symmetry relation of the splitting function expressing crossing.\footnotemark\
\footnotetext{%
  Denoting the invariant mass of $\bar{p}^\mu$
  in~\eqref{eq:sudakov-param-is} with $\tilde{m}_{ai}$, according to
  Section~\ref{sect:fact-gen-gaugetheo}, the mapping $p_a
  \leftrightarrow -p_i$ also requires the replacement $\tilde{m}_{ai}^2
  \rightarrow \tilde{m}_{ai}^2 + \frac{k_\perp^2 x^2}{(x-1)^2}$.
  This can be neglected in~\eqref{eq:a-i-symmru} as well because splitting
  functions are independent of the auxiliary mass parameter
  $\tilde{m}_{ai}$.
  The physical mass $m_{ai}$ corresponding to $\genfield{ai}$, on the
  other hand, is obviously not affected by the crossing $a \leftrightarrow i$.
}%
Note that the transformation of~$\bar p^\mu$ and~$k^\mu_\perp$ 
maps $\bar p^\mu - k^\mu_\perp$ onto itself, so that 
$p_{ai}^\mu = \bar p^\mu - k^\mu_\perp + \orderof{m^2}$ is unchanged.
For the splitting functions, we get
\begin{align}
  \bar{\mathcal{P}}
      {}^{\genfield{ai^{\vphantom{\prime}}}}_{\genfield{a}}
      {}^{\genfield{ai^\prime}}_{\genfield{i}}
  \big(
    x,k_\perp^\mu,\bar p^\mu,
    \pol_a,\pol_i
  \big)
  &=
  \bar{\mathcal{P}}
      {}^{\genfield{ai^{\vphantom{\prime}}}}_{\genantifield{i}}
      {}^{\genfield{ai^\prime}}_{\genantifield{a}}
  \Big(
    \frac{x}{x-1},
    \frac{-1}{x-1} k_\perp^\mu,
    \bar p^\mu - \frac{x}{x-1} k^\mu_\perp
    ,-\pol_i,-\pol_a
  \Big)
  \Big|_{
    m_a \leftrightarrow m_i
  }
  .
  \label{eq:a-i-symmru-splfunc}
\end{align}

\subsubsection[%
  Symmetry under
  \mathinhead{ij \leftrightarrow ij^\prime}{ij <-> ijprime}
  and
  \mathinhead{ai \leftrightarrow ai^\prime}{ai <-> aiprime}
  ]{%
  Symmetry under
  \mathinheadbold{ij \leftrightarrow ij^\prime}{ij <-> ijprime}
  and
  \mathinheadbold{ai \leftrightarrow ai^\prime}{ai <-> aiprime}
  }
\label{sect:splfunc-symm-ij-ijpr-and-ai-aipr}
The \gls{fs} splitting functions are invariant under complex conjugation
combined with exchange of the intermediate particles and the associated
parameters.
Together, the symmetry of~$\mathcal{P}$ reads
\begin{align}
  \mathcal{P}
    {}^{\genfield{ij^{\vphantom{\prime}}}}_{\genfield{i}}
    {}^{\genfield{ij^\prime}}_{\genfield{j}}
  \big(z,k_\perp^\mu,\bar p^\mu,\pol_i,\pol_j\big)
  &=
  \bigg[
  \mathcal{P}
    {}^{\genfield{ij^{\prime}}}_{\genfield{i}}
    {}^{\genfield{ij^{\vphantom{\prime}}}}_{\genfield{j}}
  \big(z,k_\perp^\mu,\bar p^\mu,\pol_i,\pol_j\big)
  \bigg]
  ^\ast
  _{m_{ij} \leftrightarrow m_{ij}^\prime}
  .
  \label{eq:ij-ijpr-symmru-splfunc}
\end{align}
In the same way, the \gls{is} splitting functions possess the symmetry
relation
\begin{align}
  \bar{\mathcal{P}}
      {}^{\genfield{ai^{\vphantom{\prime}}}}_{\genfield{a}}
      {}^{\genfield{ai^\prime}}_{\genfield{i}}
  \big(x,k_\perp^\mu,\bar p^\mu,\pol_a,\pol_i\big)
  &=
  \bigg[
  \bar{\mathcal{P}}
      {}^{\genfield{ai^{\prime}}}_{\genfield{a}}
      {}^{\genfield{ai^{\vphantom{\prime}}}}_{\genfield{i}}
  \big(x,k_\perp^\mu,\bar p^\mu,\pol_a,\pol_i\big)
  \bigg]
  ^\ast
  _{m_{ai} \leftrightarrow m_{ai}^\prime}
  .
  \label{eq:ai-aipr-symmru-splfunc}
\end{align}

\subsubsection{Crossing symmetry between FS and IS splitting functions}
\label{sect:splfunc-symm-crossing}
The \gls{fs} and \gls{is} splitting functions can be related to each
other via the crossing symmetry of the associated matrix elements,
combined with a mapping between the kinematic
parametrizations~\eqref{eq:sudakov-param-fs}
and~\eqref{eq:sudakov-param-is}.
In other words, crossing relates the 
\gls{fs} splitting functions~$\mathcal{P}$ for
the process
$\genantifield{ai}^\ast/\genantifield{ai^\prime}^\ast \to \genantifield{a} \genfield{i}$
to the 
\gls{is} splitting functions~$\bar{\mathcal{P}}$ for
$\genfield{a} \to \genfield{i} [\genfield{ai}^\ast/\genfield{ai^\prime}^\ast]$.
This expressed by
\begin{align}
  \mathrm{FS} \to \mathrm{IS}
  \equiv
  \bigg\{ 
    &
    \genfield{ij} \to \genantifield{ai}
    ,\
    \genfield{ij^\prime} \to \genantifield{ai^\prime}
    ,\
    \genfield{i} \to \genantifield{a}
    ,\
    \genfield{j} \to \genfield{i}
    ,
    \nonumber\\
    &
    m_i \to m_a
    ,\
    m_j \to m_i
    ,\
    m_{ij} \to m_{ai}
    ,\
    m_{ij}^{\prime} \to m_{ai}^{\prime}
    ,\
    \pol_i \to -\pol_a
    ,\
    \pol_j \to \pol_i
    ,
    \nonumber\\
    &
    z \to \frac{1}{x}
    ,\
    k_\perp^\mu \to \frac{- k_\perp^\mu}{x}
    ,\
    \bar{p}^\mu \to - \bar{p}^\mu + k_\perp^\mu , \dots
  \bigg\}
  ,
  \label{eq:fs-to-is-rules}
\end{align}
where the kinematic substitutions derive from 
$p_i\to -p_a$ and $p_j\to p_i$ up to mass-suppressed terms that are not
relevant in the following crossing relation of the splitting functions.
Explicitly, we have 
\begin{align}
  \bar{\mathcal{P}}
      {}^{\genfield{ai^{\vphantom{\prime}}}}_{\genfield{a}}
      {}^{\genfield{ai^\prime}}_{\genfield{i}}
    \big(x,k_\perp^\mu,\bar{p}^\mu,\pol_a,\pol_i\big)
  &=
  s_{\genfield{a}}
  \mathcal{P}
      {}^{\genfield{ij^{\vphantom{\prime}}}}_{\genfield{i}}
      {}^{\genfield{ij^\prime}}_{\genfield{j}}
  \bigg(
    z,k_\perp^\mu,\bar p^\mu,\pol_i,\pol_j
  \bigg)
  \bigg|_{\eqref{eq:fs-to-is-rules}}
  \nonumber\\
  &=
  s_{\genfield{a}}
  \mathcal{P}
      {}^{\genfield{ij^{\vphantom{\prime}}}}_{\genfield{i}}
      {}^{\genfield{ij^\prime}}_{\genfield{j}}
  \bigg(
    \frac{1}{x},-\frac{k_\perp^\mu}{x},-\bar p^\mu+k_\perp^\mu,-\pol_a,\pol_i
  \bigg)
  \bigg|_{\genfield{\mathrm{FS}} \to \genfield{\mathrm{IS}},m_\mathrm{FS} \to m_\mathrm{IS}}
  ,%
  \label{eq:fs-to-is-splfunc-mapping}
\end{align}
where 
the substitutions $\varphi_\mathrm{FS} \to \varphi_\mathrm{IS}$ and
$m_\mathrm{FS} \to m_\mathrm{IS}$ are the ones 
for fields and masses in \eqref{eq:fs-to-is-rules},
respectively, and 
the sign~$s_{\genfield{a}}$ is due to the crossing of the
field~$\genfield{a}$ between the \gls{fs} and \gls{is} of the 
process~\eqref{eq:enhanced-diag-topologies}.
It takes the value~$s_{\genfield{a}}=-1$ if~$\genfield{a}$ is a
fermion or antifermion and~$s_{\genfield{a}}=+1$ otherwise.%

\subsection{Bosonic FS splitting functions}
\label{sect:expl-splfuncs-FS-bosonic}
We now turn to the explicit derivation of \gls{fs} splitting functions
for specific splitting processes~\eqref{eq:spl-proc}, applying
the procedure described in Section~\ref{sect:fact-gen-gaugetheo}.
We work in a generic gauge theory with scalars, fermions, and vector
bosons, in general $R_\xi$ gauge.
First, we focus on splitting processes that involve \emph{bosonic fields} only.
For selected
situations, we demonstrate the application of the symmetries
from Section~\ref{sect:splfunc-symm}.
Similarly, symmetry relations may be used to obtain all splitting
functions that are not explicitly detailed in the following.
We work in $D=4$ space--time dimensions throughout, assuming that 
collinear singularities are regularized by the masses of the particles 
taking part in the splitting process.

%
\subsubsection[%
  \mathinhead{V^\ast \to VV}{V* to V V}
  splitting functions%
  ]{%
  \mathinheadbold{V^\ast \to VV}{V* to V V}
  splitting functions}
\label{sect:VVV}
First, we consider the splitting functions for the splitting process
\begin{align}
  [\vbfield{ij}^\ast/\vbfield{ij^\prime}^\ast]
  \to
  \vbfield{i} \, \vbfield{j}
  .
\end{align}
The diagrams~\eqref{eq:decomp-sq-realme-by-mother} involving such
a subprocess are always accompanied by corresponding diagrams with
internal would-be Goldstone-boson lines, i.e.\ with the splitting
processes
\begin{align}
  [\vbfield{ij}^\ast/\gbfield{ij^\prime}^\ast]
  \to
  \vbfield{i} \, \vbfield{j}
  ,\qquad
  [\gbfield{ij}^\ast/\vbfield{ij^\prime}^\ast]
  \to
  \vbfield{i} \, \vbfield{j}
  ,\qquad
  [\gbfield{ij}^\ast/\gbfield{ij^\prime}^\ast]
  \to
  \vbfield{i} \, \vbfield{j}
  .
\end{align}
We combine all of these contributions into a common splitting
function.
As detailed below, this proceeds by relating the
amplitudes~$T^{(n), \mu}_{\vbfield{} X}$ and~$T^{(n)}_{\gbfield{} X}$
via \glspl{wi}
and simplifying the sum 
of contributions 
upon collecting terms according to their propagator structure.
In this way we merge the contributions from
intermediate vector-bosons and would-be Goldstone-bosons
to~\eqref{eq:realme-sq-fact-nonsudakov} into a single term
\begin{align}
  e^2
  \frac{1}{p_{ij}^2  - m_{ij}^2} \frac{1}{p_{ij}^2 - m_{ij}^{\prime\, 2}}
  \
  T^{(n)}_{\vbfield{ij^{\vphantom{\prime}}} X, \mu}(p_{ij})
  \left[
    \tilde{\mathcal{P}}
      {}^{\vbfield{ij^{\vphantom{\prime}}}}_{\vbfield{i}}
      {}^{\vbfield{ij^\prime}}_{\vbfield{j}}
    (p_i,p_j,\polvb_i,\polvb_j)
  \right]^{\mu \nu}
  T^{(n) \ast}_{\vbfield{ij^\prime} X, \nu}(p_{ij})
  .
  \label{eq:realme-sq-fact-nonsudakov-intvec}
\end{align}
In this procedure, a meaningful combination of the vector-boson and
would-be Goldstone-boson parts is facilitated by the coupling relations
from Section~\ref{sect:generic-gb-cpl-relat}.


As already mentioned in Section~\ref{sect:fact-gen-gaugetheo}, the
presence of mass-singular factors in longitudinal polarization vectors
poses a severe issue for the diagram selection procedure.
Care must be taken to consistently capture all contributions that are
relevant at the lowest order in~$\lambda$ and to restore the gauge
cancellations destroyed by naively dropping diagrams of
type~\diagtype{2}.
The easiest way to deal with this problem is to turn the argument from
Section~\ref{sect:gbet} around and employ the \emph{exact}
version~\eqref{eq:multi-gbet-exact} of the \gls{gbet} to \emph{enforce}
all gauge cancellations of mass-singular terms from the
outset.
Specifically, we employ this formula to rewrite the 
real-emission matrix elements~\eqref{eq:notation-realme}
\begin{align}
  \mathcal{M}
    ^{(n+1)\, \polvb_i \polvb_j}
    _{\vbfield{i} \vbfield{j} X}
  &=
  \left[
    T^{(n+1)}_{\vbfield{i} \vbfield{j} X}
  \right]_{\mu_i \mu_j}
  \varepsilon_{\polvb_i}^{\mu_i \ast}(p_i)
  \varepsilon_{\polvb_j}^{\mu_j \ast}(p_j)
  \label{eq:LL-pol-ampl}
\end{align}
with~$\polvb_i=\mathrm{L}$ and/or~$\polvb_j=\mathrm{L}$ in terms of
\emph{scalar} amplitudes.
It is then obvious that diagrams of type~\diagtype{2} do not contribute at
leading power in~$\lambda$.
This is very similar to the approach taken in~\cite{Cuomo:2019siu}.


For $\polvb_i=\mathrm{L}$ and $\polvb_j=\mathrm{T}$, 
employing \eqref{eq:multi-gbet-exact} for the outgoing field $V_i$
amounts to
\begin{align}
  \mathcal{M}^{(n+1)\, \mathrm{L}\mathrm{T}}_{\vbfield{i} \vbfield{j} X}
  &=
  \eta^\ast_{\vbfield{i}}
  \mathcal{M}^{(n+1)\, \mathrm{G}\mathrm{T}}_{\gbfield{i} \vbfield{j} X}
  +
  \mathcal{M}^{(n+1)\, \mathrm{R}\mathrm{T}}_{\vbfield{i} \vbfield{j} X}
  ,
  \label{eq:LT-goldstone-strat}
\end{align}
where we defined the partial polarization vector
$\varepsilon_{\mathrm{R}}^\mu(p)=\restpol^\mu(p)$ and labeled it with
$\polvb=\mathrm{R}$.
Then, we \emph{separately} carry out the procedure from
Section~\ref{sect:fact-gen-gaugetheo} for the two amplitudes on the
r.h.s.\ of \eqref{eq:LT-goldstone-strat}.
This means that we evaluate the sum of the partial matrix elements
\begin{align}
  \eta^\ast_{\vbfield{i}}
  \MContrExplFields{\genfield{ij}}{\gbfield{i}}{\vbfield{j}}^{\mathrm{G} \mathrm{T}}
  +
  \MContrExplFields{\genfield{ij}}{\vbfield{i}}{\vbfield{j}}^{\mathrm{R} \mathrm{T}}
  \label{eq:VVV-LT-partial-amp-redef}
\end{align}
instead of the corresponding longitudinally polarized partial matrix element.\footnotemark\
\footnotetext{%
  Note that this explicitly retains the R-polarized parts, which are
  considered as subleading in the
  \gls{gbet}~\eqref{eq:multi-gbet-leadmassorder}.
  However, as we explicitly observe below, both parts contribute at the
  same order in the quasi-collinear limit.
}%
While~\eqref{eq:LT-goldstone-strat} is an identity,
switching to its r.h.s.\
effectively \emph{modifies} the diagram selection procedure for
the terms~$\propto p^{\mu_i}_{i}/M_{\vbfield{i}}$ from the
amplitude with longitudinally polarized $V_i$, because \emph{scalar} diagrams of type
\diagtype{1} instead of the corresponding \emph{vector-boson} diagrams
are kept.


For two longitudinally polarized external vector bosons
($\polvb_i=\polvb_j=\mathrm{L}$), we have
\begin{align}
  \mathcal{M}^{(n+1)\, \mathrm{L}\mathrm{L}}_{\vbfield{i} \vbfield{j} X}
  =
  \mathcal{M}^{(n+1)\, \mathrm{R}\mathrm{R}}_{\vbfield{i} \vbfield{j} X}
  +
  \eta^\ast_{\vbfield{i}}
  \mathcal{M}^{(n+1)\, \mathrm{G}\mathrm{R}}_{\gbfield{i} \vbfield{j} X}
  +
  \eta^\ast_{\vbfield{j}}
  \mathcal{M}^{(n+1)\, \mathrm{R}\mathrm{G}}_{\vbfield{i} \gbfield{j} X}
  +
  \eta^\ast_{\vbfield{i}}
  \eta^\ast_{\vbfield{j}}
  \mathcal{M}^{(n+1)\, \mathrm{G}\mathrm{G}}_{\gbfield{i} \gbfield{j} X}
  ,
  \label{eq:LL-goldstone-strat}
\end{align}
as starting point that is manifestly free of mass-singular contributions.
Again, the diagram selection is now performed for each term separately,
which amounts to the evaluation of
\begin{align}
  \MContrExplFields{\genfield{ij}}{\vbfield{i}}{\vbfield{j}}^{\mathrm{R} \mathrm{R}}
  +
  \eta^\ast_{\vbfield{i}}
  \MContrExplFields{\genfield{ij}}{\gbfield{i}}{\vbfield{j}}^{\mathrm{G} \mathrm{R}}
  +
  \eta^\ast_{\vbfield{j}}
  \MContrExplFields{\genfield{ij}}{\vbfield{i}}{\gbfield{j}}^{\mathrm{R} \mathrm{G}}
  +
  \eta^\ast_{\vbfield{i}}
  \eta^\ast_{\vbfield{j}}
  \MContrExplFields{\genfield{ij}}{\gbfield{i}}{\gbfield{j}}^{\mathrm{G} \mathrm{G}}
  .
  \label{eq:VVV-LL-partial-amp-redef}
\end{align}


In order to follow this procedure, we have to consider the following types of diagrams,
\newcommand{\FSDiagGeo}{%
  \renewcommand{\sclfac}{1}%
  \renewcommand{\intsclfac}{1}%
  \renewcommand{\extsclfac}{1}%
  \renewcommand{\len}{1.1cm}%
  \renewcommand{\rad}{.6cm}%
  \renewcommand{\dotrad}{1mm}%
  \renewcommand{\diagang}{45}%
}%
\drawmomtrue%
\drawvertdottrue%
\begin{alignat}{2}%
  \ftntmath{%
    \MContrExplFields{\vbfield{ij}}{\vbfield{i}}{\vbfield{j}}^{\polvb_i \polvb_j}%
  }%
  &\ftntmath{{}={}}%
  \FSDiagGeo%
  \ftntmath{%
    \splittingdiagIndivFS{T^{(n)}_{\vbfield{ij} X}}%
      {photon}{photon}{photon}%
      {\vbfield{ij}}{\vbfield{i}, \polvb_i}{\vbfield{j}, \polvb_j}%
    ,%
  }%
  \resetdiagargs%
  &\quad
  \ftntmath{%
    \MContrExplFields{\gbfield{ij}}{\vbfield{i}}{\vbfield{j}}^{\polvb_i \polvb_j}%
  }%
  &\ftntmath{{}={}}%
  \FSDiagGeo%
  \ftntmath{%
    \splittingdiagIndivFS{T^{(n)}_{\gbfield{ij} X}}%
      {scalar}{photon}{photon}%
      {\gbfield{ij}}{\vbfield{i}, \polvb_i}{\vbfield{j}, \polvb_j}%
    ,%
  }%
  \resetdiagargs%
  \nonumber\\%
  \ftntmath{%
    \MContrExplFields{\vbfield{ij}}{\gbfield{i}}{\vbfield{j}}^{\mathrm{G} \polvb_j}%
  }%
  &\ftntmath{{}={}}%
  \FSDiagGeo%
  \ftntmath{%
    \splittingdiagIndivFS{T^{(n)}_{\vbfield{ij} X}}%
      {photon}{scalar}{photon}%
      {\vbfield{ij}}{\gbfield{i}}{\vbfield{j}, \polvb_j}%
    ,%
  }%
  \resetdiagargs%
  &%
  \ftntmath{%
    \MContrExplFields{\gbfield{ij}}{\gbfield{i}}{\vbfield{j}}^{\mathrm{G} \polvb_j}%
  }%
  &\ftntmath{{}={}}%
  \FSDiagGeo%
  \ftntmath{%
    \splittingdiagIndivFS{T^{(n)}_{\gbfield{ij} X}}%
      {scalar}{scalar}{photon}%
      {\gbfield{ij}}{\gbfield{i}}{\vbfield{j}, \polvb_j}%
    ,%
  }%
  \resetdiagargs%
  \nonumber\\%
  \ftntmath{%
    \MContrExplFields{\vbfield{ij}}{\vbfield{i}}{\gbfield{j}}^{\polvb_i \mathrm{G}}%
  }%
  &\ftntmath{{}={}}%
  \FSDiagGeo%
  \ftntmath{%
    \splittingdiagIndivFS{T^{(n)}_{\vbfield{ij} X}}%
      {photon}{photon}{scalar}%
      {\vbfield{ij}}{\vbfield{i}, \polvb_i}{\gbfield{j}}%
    ,%
  }%
  \resetdiagargs%
  &%
  \ftntmath{%
    \MContrExplFields{\gbfield{ij}}{\vbfield{i}}{\gbfield{j}}^{\polvb_i \mathrm{G}}%
  }%
  &\ftntmath{{}={}}%
  \FSDiagGeo%
  \ftntmath{%
    \splittingdiagIndivFS{T^{(n)}_{\gbfield{ij} X}}%
      {scalar}{photon}{scalar}%
      {\gbfield{ij}}{\vbfield{i}, \polvb_i}{\gbfield{j}}%
    ,%
  }%
  \resetdiagargs%
  \nonumber\\%
  \ftntmath{%
    \MContrExplFields{\vbfield{ij}}{\gbfield{i}}{\gbfield{j}}^{\mathrm{G}\mathrm{G}}%
  }%
  &\ftntmath{{}={}}%
  \FSDiagGeo%
  \ftntmath{%
    \splittingdiagIndivFS{T^{(n)}_{\vbfield{ij} X}}%
      {photon}{scalar}{scalar}%
      {\vbfield{ij}}{\gbfield{i}}{\gbfield{j}}%
    ,%
  }%
  \resetdiagargs%
  &%
  \ftntmath{%
    \MContrExplFields{\gbfield{ij}}{\gbfield{i}}{\gbfield{j}}^{\mathrm{G}\mathrm{G}}%
  }%
  &\ftntmath{{}={}}%
  \FSDiagGeo%
  \ftntmath{%
    \splittingdiagIndivFS{T^{(n)}_{\gbfield{ij} X}}%
      {scalar}{scalar}{scalar}%
      {\gbfield{ij}}{\gbfield{i}}{\gbfield{j}}%
    .%
  }%
  \resetdiagargs%
  \label{eq:VVV-diags-FS}
\end{alignat}%
After employing momentum conservation, the \gls{wi} \eqref{eq:single-ward-id} for the
underlying would-be Goldstone amplitude (with outgoing $V_{ij}, G_{ij}$),
\begin{align}
  T^{(n)}_{\gbfield{ij} X}(p_{ij})
  =
  \eta_{\vbfield{ij}}
  \frac{p_{ij}^\mu}{m_{ij}}
  T^{(n)}_{\vbfield{ij} X, \mu}(p_{ij})
  ,
  \label{eq:VVV-ward-id-ul-me}
\end{align}
and inserting the coupling relations~\eqref{eq:coupling-relat-VVV-GVV}
and~\eqref{eq:coupling-relat-VVV-VGG},
they translate into the expressions
\begin{align}
  %
  %
  \MContrExplFields{\vbfield{ij}}{\vbfield{i}}{\vbfield{j}}^{\polvb_i \polvb_j}
  ={}&
  - \imagi e
  I_{\vbfield{i} \vbantifield{j}}^{\vbfield{ij}}
  T^{(n)\, \sigma}_{\vbfield{ij} X}\left(p_{ij}\right)
  G^{\vbantifield{ij} \vbfield{ij}}_{\sigma \delta}(p_{ij},-p_{ij})
  g^{\delta \alpha} 
  \varepsilon_{\polvb_i}^{\beta  \ast}(p_i)
  \varepsilon_{\polvb_j}^{\gamma  \ast}(p_j)
  \nonumber\\*
  &
  \cdot
  \left[
    g_{\beta \gamma } \left(p_i-p_j\right)_{\alpha }+
    g_{\alpha \gamma } \left(p_i+2 p_j\right)_{\beta }+
    g_{\alpha \beta } \left(-2 p_i-p_j\right)_{\gamma }
  \right]
  ,
  \nonumber\\
  %
  %
  \MContrExplFields{\gbfield{ij}}{\vbfield{i}}{\vbfield{j}}^{\polvb_i \polvb_j}
  ={}&
  \imagi e
  I_{\vbfield{i} \vbantifield{j}}^{\vbfield{ij}}
  \frac{m_j^2-m_i^2}{m_{ij}^2}
  G^{\gbantifield{ij} \gbfield{ij}} (p_{ij},-p_{ij})
  T_{\vbfield{ij} X}^{(n)\, \alpha }\left(p_{ij}\right)
  p_{ij,\alpha }
  g_{\beta \gamma }
  \varepsilon_{\polvb_i}^{\beta  \ast}(p_i)
  \varepsilon_{\polvb_j}^{\gamma  \ast}(p_j)
  ,
  \nonumber\\
  %
  %
  \MContrExplFields{\vbfield{ij}}{\gbfield{i}}{\vbfield{j}}^{\mathrm{G} \polvb_j}
  ={}&
  \imagi e
  I_{\vbfield{i} \vbantifield{j}}^{\vbfield{ij}}
  \eta_{\vbfield{i}}
  \frac{m_{j}^2-m_{ij}^2}{m_{i}}
  T^{(n)\, \sigma}_{\vbfield{ij} X}\left(p_{ij}\right)
  G^{\vbantifield{ij} \vbfield{ij}}_{\sigma \gamma}(p_{ij},-p_{ij})
  \varepsilon_{\polvb_j}^{\gamma  \ast}(p_j)
  ,
  \nonumber\\
  %
  %
  \MContrExplFields{\gbfield{ij}}{\gbfield{i}}{\vbfield{j}}^{\mathrm{G} \polvb_j}
  ={}&
  \imagi e
  I_{\vbfield{i} \vbantifield{j}}^{\vbfield{ij}}
  \eta_{\vbfield{i}}
  \frac{m_{j}^2-m_{ij}^2-m_i^2}{2 m_{ij}^2 m_{i}}
  \nonumber\\*
  &
  \cdot
  G^{\gbantifield{ij} \gbfield{ij}} (p_{ij},-p_{ij})
  T_{\vbfield{ij} X}^{(n)\, \alpha }\left(p_{ij}\right)
  p_{ij,\alpha }
  \varepsilon_{\polvb_j}^{\gamma  \ast}(p_j)
  \left(2p_i+p_j\right)_{\gamma }
  ,
  \nonumber\\
  %
  %
  \MContrExplFields{\vbfield{ij}}{\vbfield{i}}{\gbfield{j}}^{\polvb_i \mathrm{G}}
  ={}&
  \imagi e
  I_{\vbfield{i} \vbantifield{j}}^{\vbfield{ij}}
  \eta_{\vbfield{j}}
  \frac{m_{ij}^2-m_i^2}{m_{j}}
  T^{(n)\, \sigma}_{\vbfield{ij} X}\left(p_{ij}\right)
  G^{\vbantifield{ij} \vbfield{ij}}_{\sigma \beta}(p_{ij},-p_{ij})
  \varepsilon_{\polvb_i}^{\beta  \ast}(p_i)
  ,
  \nonumber\\
  %
  %
  \MContrExplFields{\gbfield{ij}}{\vbfield{i}}{\gbfield{j}}^{\polvb_i \mathrm{G}}
  ={}&
  - \imagi e
  I_{\vbfield{i} \vbantifield{j}}^{\vbfield{ij}}
  \eta_{\vbfield{j}}
  \frac{m_{i}^2-m_{ij}^2-m_j^2}{2 m_{ij}^2 m_{j}}
  \nonumber\\*
  &
  \cdot
  G^{\gbantifield{ij} \gbfield{ij}} (p_{ij},-p_{ij})
  T_{\vbfield{ij} X}^{(n)\, \alpha }\left(p_{ij}\right)
  p_{ij,\alpha }
  \varepsilon_{\polvb_i}^{\beta  \ast}(p_i)
  \left(p_i+2p_j\right)_{\beta }
  ,
  \nonumber\\
  %
  %
  \MContrExplFields{\vbfield{ij}}{\gbfield{i}}{\gbfield{j}}^{\mathrm{G} \mathrm{G}}
  ={}&
  - \imagi e
  I_{\vbfield{i} \vbantifield{j}}^{\vbfield{ij}}
  \eta_{\vbfield{i}}
  \eta_{\vbfield{j}}
  \frac{m_{ij}^2-m_i^2-m_j^2}{2 m_i m_{j}}
  T^{(n)\, \sigma}_{\vbfield{ij} X}\left(p_{ij}\right)
  G^{\vbantifield{ij} \vbfield{ij}}_{\sigma \alpha}(p_{ij},-p_{ij})
  \left(p_j-p_i\right)_{\alpha }
  ,
  \nonumber\\
  %
  %
  \MContrExplFields{\gbfield{ij}}{\gbfield{i}}{\gbfield{j}}^{\mathrm{G} \mathrm{G}}
  ={}&
  0
  .
  \label{eq:VVV-EnhAmp}
\end{align}
The last equality is due to~\eqref{eq:coupling-GGG-is-zero}.
Here, we left the propagators symbolic, which are the only places where
the gauge parameters $\xi_{\vbfield{}}$ appear.
The independence of the 
splitting functions from gauge parameters serves
as an important check on our derivation.
Therefore, we split
\begin{align}
  G^{\vbantifield{}, \vbfield{}}_{\mu \nu}(p,-p)
  &=
  \frac
    {\imagi}
    {p^2 - M_{\vbfield{}}^2}
  \left(
    - g_{\mu \nu}+
    \frac
      {(1-\xi_{\vbfield{}})}
      {p^2-\xi_{\vbfield{}} M_{\vbfield{}}^2}
      p_{\mu} p_{\nu}
  \right)
  =
  G^{\vbantifield{} \vbfield{}}_{\mathrm{tHF}, \, \mu \nu}(p,-p)
  +
  G^{\vbantifield{} \vbfield{}}_{\xi, \, \mu \nu}(p,-p)
  ,
  \nonumber\\
  G^{\gbantifield{} \gbfield{}}(p,-p)
  &
  =
  \frac{\imagi}{p^2 - \xi_{\vbfield{}} M_{\vbfield{}}^2}
  =
  \frac{\imagi}{p^2 - M_{\vbfield{}}^2}
  \left(
    1+
    \frac
    {(\xi_{\vbfield{}}-1) M_{\vbfield{}}^2}
    {p^2-\xi_{\vbfield{}} M_{\vbfield{}}^2}
  \right)
  =
  G^{\gbantifield{} \gbfield{}}_{\mathrm{tHF}}(p,-p)
  +
  G^{\gbantifield{} \gbfield{}}_{\xi}(p,-p)
  \label{eq:decomp-prop-in-tHf-and-xi}
\end{align}
into their expressions in 't~Hooft--Feynman gauge 
and remainder terms that explicitly depend on~$\xi_{\vbfield{}}$,
\begin{align}
  G^{\vbantifield{} \vbfield{}}_{\mathrm{tHF}, \, \mu \nu}(p,-p)
  &=
  \frac
    {- \imagi g_{\mu \nu}}
    {p^2 - M_{\vbfield{}}^2}
  ,
  &
  G^{\vbantifield{} \vbfield{}}_{\xi, \, \mu \nu}(p,-p)
  &=
  \frac
    {\imagi}
    {p^2 - M_{\vbfield{}}^2}
  \frac
    {(1-\xi_{\vbfield{}}) p_{\mu} p_{\nu}}
    {p^2-\xi_{\vbfield{}} M_{\vbfield{}}^2}
  ,
  \nonumber\\
  G^{\gbantifield{} \gbfield{}}_{\mathrm{tHF}}(p,-p)
  &=
  \frac{\imagi}{p^2 - M_{\vbfield{}}^2}
  &
  G^{\gbantifield{} \gbfield{}}_{\xi}(p,-p)
  &=
  \frac{-\imagi}{p^2 - M_{\vbfield{}}^2}
  \frac
    {(1-\xi_{\vbfield{}}) M_{\vbfield{}}^2}
    {p^2-\xi_{\vbfield{}} M_{\vbfield{}}^2}
  .
  \label{eq:decomp-prop-in-tHf-and-xi-definitions}
\end{align}
According to this, we decompose the partial matrix elements
from~\eqref{eq:VVV-EnhAmp} into
\begin{align}
  \MContrNoFields{\vbfield{ij}}
  =
  \MContrNoFields{\vbfield{ij}, \mathrm{tHF}}
  +
  \MContrNoFields{\vbfield{ij}, \xi}
  ,\qquad
  \MContrNoFields{\gbfield{ij}}
  =
  \MContrNoFields{\gbfield{ij}, \mathrm{tHF}}
  +
  \MContrNoFields{\gbfield{ij}, \xi}
  .
  \label{eq:decomp-partial-amp-in-tHf-and-xi}
\end{align}
The second, gauge-parameter dependent terms do not vanish.
They do, however, cancel in the sum of the two objects
from~\eqref{eq:decomp-partial-amp-in-tHf-and-xi}.
This implies that the final results of our splitting functions are free
from gauge parameters.


For two transversely polarized external vector bosons with
$\polvb_i=\polvb_j=\mathrm{T}$, this follows straightforwardly from
\begin{align}
  G^{\vbantifield{} \vbfield{}}_{\xi, \, \mu \nu}(p,-p)
  =
  -
  \frac{
    p_{\mu} p_{\nu}
  }{
    M_{\vbfield{}}^2
  }
  G^{\gbantifield{} \gbfield{}}_{\xi}(p,-p)
  \label{eq:xi-part-prop-VV-GG-relat}
\end{align}
and the application of the \gls{wi}~\eqref{eq:single-ward-id}
to both~$T^{(n)}_{\vbfield{},\mu}$ and the splitting vertex from the first
diagram in~\eqref{eq:VVV-diags-FS}.
If at least one of the external vector bosons is longitudinal, the
cancellation of gauge parameters is more intricate.
This is because the partial matrix elements in
\eqref{eq:VVV-LT-partial-amp-redef}
and~\eqref{eq:VVV-LL-partial-amp-redef} have \emph{unphysical} external
would-be Goldstone bosons and R-polarized gauge bosons.
The corresponding splitting vertices do not fulfill the
requirements of the \gls{wi}~\eqref{eq:single-ward-id}.
However, the gauge-parameter dependent parts cancel within the sum of
\emph{all} unphysical partial amplitudes
from~\eqref{eq:VVV-LT-partial-amp-redef}
and~\eqref{eq:VVV-LL-partial-amp-redef}, due to the
\glspl{wi}~\eqref{eq:single-ward-id}, \eqref{eq:double-ward-id},
and~\eqref{eq:triple-ward-id}.
In our explicit calculations, this cancellation is made possible by the
fact that these \glspl{wi} are comprised by the coupling relations from
Section~\ref{sect:generic-gb-cpl-relat}.
This provides self-consistency checks for the implementation of these
relations and the decompositions
\eqref{eq:LT-goldstone-strat},
\eqref{eq:LL-goldstone-strat}, in particular.


Naively, the presence of masses in the denominators
of~\eqref{eq:VVV-EnhAmp} suggests bad behavior in the various massless
limits, despite the treatment of longitudinal polarization vectors.
These factors, however, stem either from the application of the
\gls{wi}~\eqref{eq:VVV-ward-id-ul-me}
or the insertion of the coupling
relations~\eqref{eq:coupling-relat-VVV-GVV}
and~\eqref{eq:coupling-relat-VVV-VGG}.
In both cases, the gauge structure of the considered theory implies that
the 
various massless limits~$m\to0$ are well defined, as discussed in
Sections~\ref{sect:generic-gb-cpl-relat} and~\ref{sect:gbet}.
We provide further comments on this when presenting our final results
for~$V^\ast \to V V$ splitting functions.


The following computations are supported by
Appendix~\ref{sect:VVV-alt-approach}, where our alternative strategy for
the treatment of mass-singular factors in longitudinal polarization
vectors is discussed.
We find full agreement between the results from the two different methods.

\paragraph{Polarization \boldmath{$\polvb_i=\mathrm{T}$}, \boldmath{$\polvb_j=\mathrm{T}$}.}
The TT configuration is free of problems regarding longitudinal
polarization vectors, allowing for a straightforward evaluation of the
quasi-collinear limit.
As mentioned, the fact that the
sum of the
$\xi$-dependent parts from the
decompositions~\eqref{eq:decomp-partial-amp-in-tHf-and-xi} vanishes in
our explicit computation,
\begin{align}
  \MContrExplFields
    {\vbfield{ij},\xi}
    {\vbfield{i}}
    {\vbfield{j}}^{\mathrm{T}\mathrm{T}}
  +
  \MContrExplFields
    {\gbfield{ij},\xi}
    {\vbfield{i}}
    {\vbfield{j}}^{\mathrm{T}\mathrm{T}}
  =
  0
  ,
\end{align}
provides an important check.
For the $\xi$-independent contributions in 't~Hooft--Feynman gauge, we get 
\begin{alignat}{2} 
  &
  \lefteqn{
    \MContrExplFields
      {\vbfield{ij},\mathrm{tHF}}
      {\vbfield{i}}
      {\vbfield{j}}^{\mathrm{T}\mathrm{T}}
    +
    \MContrExplFields
      {\gbfield{ij},\mathrm{tHF}}
      {\vbfield{i}}
      {\vbfield{j}}^{\mathrm{T}\mathrm{T}}
  }
  \nonumber\\*
  &=
  \frac{
    e I_{\vbfield{i} \vbantifield{j}}^{\vbfield{ij}}
  }{
    p_{ij}^2 - m_{ij}^2
  }
  \bigg\{
    &&
    2 \big(p_i  \varepsilon^\ast _{\mathrm{T}} (p_j)\big)
    \Big(\varepsilon^\ast _{\mathrm{T}}(p_i)  T^{(n)}_{\vbfield{ij} X}(p_{ij})\Big)
    -
    2 \big(p_j  \varepsilon^\ast _{\mathrm{T}}(p_i)\big)
    \Big(\varepsilon^\ast _{\mathrm{T}}(p_j)  T^{(n)}_{\vbfield{ij} X}(p_{ij})\Big)
    \nonumber\\
    &&&+
    \big(\varepsilon^\ast _{\mathrm{T}}(p_i)  \varepsilon^\ast _{\mathrm{T}} (p_j)\big)
    \rho (m_{ij},m_i,m_j)
    \Big(p_j  T^{(n)}_{\vbfield{ij} X}(p_{ij})\Big)
    \nonumber\\
    &&&-
    \big(\varepsilon^\ast _{\mathrm{T}}(p_i)  \varepsilon^\ast _{\mathrm{T}} (p_j)\big)
    \rho (m_{ij},m_j,m_i)
    \Big(p_i  T^{(n)}_{\vbfield{ij} X}(p_{ij})\Big)
  \bigg\}
  ,
  \label{eq:VVV-TT-indiv-amp-fullL}
\end{alignat}
where terms of higher orders
in~$\lambda$ have not been neglected yet and
we defined the shorthand
\begin{align}
  \rho(m_1,m_2,m_3)
  =
  \frac{m_1^2+m_2^2-m_3^2}{m_1^2}
  ,
  \label{eq:def-rho-abb}
\end{align}
which fulfills the relations
\begin{align}
  m_1^2 \, \rho(m_1,m_2,m_3)
  =
  m_2^2 \, \rho(m_2,m_1,m_3)
  ,
  \nonumber
  \\
  \rho(m_1,m_3,m_2)
  =
  2-\rho(m_1,m_2,m_3)
  .
\end{align}
The form of~$\rho$ is linked to the factors in the coupling
relations~\eqref{eq:coupling-relat-VVV-GVV}
and~\eqref{eq:coupling-relat-VVV-VGG}.
In particular, the discussion at the end of Section~\ref{sect:generic-gb-cpl-relat}
implies that
\begin{align}
  I_{\bar{V}_{b} V_{c}}^{V_{a}} m_1 \rho(m_1,m_2,m_3)
  ,\qquad
  I_{\bar{V}_{b} V_{c}}^{V_{a}} \frac{m_1}{m_2} \rho(m_1,m_2,m_3)
  \label{eq:rho-ml-limit}
\end{align}
are well behaved in the limits~$m_1 \to 0$ and~$m_2 \to 0$, by virtue of
the gauge symmetry of the theory.
Here, $m_1,m_2,m_3$ represent any permutation
of~$M_{V_a},M_{V_b},M_{V_c}$.


We proceed by multiplying~\eqref{eq:VVV-TT-indiv-amp-fullL} with its
conjugate, with
$\vbfield{ij}$ and $\gbfield{ij}$
replaced by
$\vbfield{ij^\prime}$ and $\gbfield{ij^\prime}$, and insert
the transverse polarization sum~\eqref{eq:trv-poln-sum} for
$\varepsilon^\mu_\mathrm{T}(p_i)$ and $\varepsilon^\mu_\mathrm{T}(p_j)$,
with separately defined 
$\restpol^\mu_i$ and $\restpol^\mu_j$.\footnotemark\
\footnotetext{%
  The factors~$\propto p^\mu/M_V$ in~\eqref{eq:trv-poln-sum}
  do not pose a problem for the naive factorization procedure.
  While this is not straightforward to see at the level of squared amplitudes,
  it can be inferred by considering individual
  amplitudes, as transverse polarization vectors are free of spurious
  factors~$\orderof{\lambda^{-1}}$.
}%
This is followed by applying the Sudakov
parametrization~\eqref{eq:sudakov-param-fs} and inserting explicit
expressions for known scalar products, such as on-shell relations and
transversality conditions.
Finally, in order to identify contributions beyond the leading order
in~$\lambda$, we evaluate the mass order of all remaining scalar
products, entirely using
\begin{align}
  k_\perp^\mu
    = \orderof{\lambda},
  \quad
  p_i p_j
    = \orderof{\lambda^2},
  \quad
  \bar p^\mu \, T^{(n)}_{\vbfield{ij} X, \mu}(p_{ij})
    = \orderof{\lambda},
  \quad
  \restpol^\mu_{i/j}
  = \orderof{\lambda}.
  \label{eq:oc-summary}
\end{align}
In particular, no other assumptions on the scaling of scalar products
involving the gauge vectors~$\restpol^\mu$ and thereby on their precise
definition are made.
Finally, we employ
\begin{align}
  0
  &=
  p_i \varepsilon_{\mathrm{L}}(p_i)
  =
  m_i
  +
  \big(z \bar p - A_i n + k_\perp\big) \restpol_i
  =
  m_i + z \, \big(\bar p \restpol_i\big) + \orderof{\lambda^2}
  ,
  \nonumber\\
  0
  &=
  p_j \varepsilon_{\mathrm{L}}(p_j)
  =
  m_j
  +
  \big((1-z) \bar p - A_j n - k_\perp\big) \restpol_j
  =
  m_j + (1-z) \, \big(\bar p \restpol_j\big) + \orderof{\lambda^2}
  ,
\end{align}
with $A_i$ and $A_j$ defined in \eqref{eq:sudakov-param-fs},
to replace $\bar p \restpol_i$ and $\bar p \restpol_j$ at leading mass order.
We arrive at the transverse splitting function
\begin{align}
  \bigg[
    \mathcal{P}
      {}^{\vbfield{ij^{\vphantom{\prime}}}}_{\vbfield{i}}
      {}^{\vbfield{ij^\prime}}_{\vbfield{j}}
      \big(z,k_\perp^\mu,\bar p^\mu,\mathrm{T},\mathrm{T}\big)
  \bigg]^{\mu \nu}
  ={} &
  I_{\vbfield{i} \vbantifield{j}}^{\vbfield{ij}} 
  I_{\vbfield{i} \vbantifield{j}}^{\vbfield{ij^\prime} \ast}
  \bigg\{
    2 \bigg(\frac{1}{(z-1) z}+2\bigg)
    \bigg(2 z-\rho (m_{ij}^{\prime },m_i,m_j)\bigg)
    \bar{p}^{\nu } k_{\perp }^{\mu }
    \nonumber\\
    &\quad {}+
    2 \bigg(\frac{1}{(z-1) z}+2\bigg)
    \bigg(2 z-\rho (m_{ij},m_i,m_j)\bigg)
    \bar{p}^{\mu } k_{\perp }^{\nu }
    \nonumber\\
    &\quad {}+
    2
    \bigg(2 z-\rho (m_{ij},m_i,m_j)\bigg)
    \bigg(2 z-\rho (m_{ij}^{\prime },m_i,m_j)\bigg)
    \bar{p}^{\mu } \bar{p}^{\nu }
    \nonumber\\
    &\quad {}+
    8
    k_{\perp }^{\mu } k_{\perp }^{\nu }
    +
    4
    \bigg(\frac{1}{z^2}+\frac{1}{(z-1)^2}\bigg)
    k_\perp^2
    g^{\mu \nu }
  \bigg\}
  .
  \label{eq:VVV-TT-split-func}
\end{align}
The assignment of the Lorentz indices is clarified
by~\eqref{eq:realme-sq-fact-nonsudakov-intvec}.
This result is well behaved for~$m_{ij}\to0$
and~$m_{ij}^\prime\to0$, due to the properties of~$\rho$ discussed
with~\eqref{eq:rho-ml-limit} and the
\gls{wi}~\eqref{eq:VVV-ward-id-ul-me}.
%
%
The disappearance of 
$\restpol^\mu_i$ and~$\restpol^\mu_j$ from this result, together with
the fact that no constraints on these gauge vectors were placed, implies
independence from the frames in which the transverse polarization
vectors are defined.

\paragraph{Polarization \boldmath{$\polvb_i=\mathrm{L}$}, \boldmath{$\polvb_j=\mathrm{T}$}.}
Next, we turn to the LT case.
Again, the gauge-parameter-dependent parts cancel,
\begin{align}
  0
  =
  \eta_{\vbfield{i}}^\ast
  \MContrExplFields
    {\vbfield{ij},\xi}
    {\gbfield{i}}
    {\vbfield{j}}^{\mathrm{G}\mathrm{T}}
  +
  \MContrExplFields
    {\vbfield{ij},\xi}
    {\vbfield{i}}
    {\vbfield{j}}^{\mathrm{R}\mathrm{T}}
  +
  (\vbfield{ij} \rightarrow \gbfield{ij})
  .
\end{align}
For the tHF contributions to~\eqref{eq:VVV-LT-partial-amp-redef}, we get
\begin{alignat}{2}
  &
  \lefteqn{
    \eta_{\vbfield{i}}^\ast
    \MContrExplFields
      {\vbfield{ij},\mathrm{tHF}}
      {\gbfield{i}}
      {\vbfield{j}}^{\mathrm{G}\mathrm{T}}
    +
    (\vbfield{ij} \rightarrow \gbfield{ij})
  }
  \nonumber\\* 
  &=
  \frac{
    e I_{\vbfield{i} \vbantifield{j}}^{\vbfield{ij}}
  }{
    p_{ij}^2 - m_{ij}^2
  }
  \Bigg\{
    &&
    \frac{m_j^2 -m_{ij}^2}{m_i}
    \Big(\varepsilon^\ast _{\mathrm{T}}(p_j)  T^{(n)}_{\vbfield{ij} X}(p_{ij})\Big)
    \nonumber\\
    &&&{}+
    \frac{\rho (m_{ij},m_i,m_j)}{m_i}
    \big(p_i \varepsilon^\ast _{\mathrm{T}}(p_j)\big)
    \Big(
      p_i  T^{(n)}_{\vbfield{ij} X}(p_{ij})
      +
      p_j  T^{(n)}_{\vbfield{ij} X}(p_{ij})
    \Big)
  \Bigg\}
  ,
  \nonumber\\
  &
  \lefteqn{
    \MContrExplFields
      {\vbfield{ij},\mathrm{tHF}}
      {\vbfield{i}}
      {\vbfield{j}}^{\mathrm{R}\mathrm{T}}
    +
    (\vbfield{ij} \rightarrow \gbfield{ij})
  }
  \nonumber\\* 
  &=
  \frac{
    e I_{\vbfield{i} \vbantifield{j}}^{\vbfield{ij}}
  }{
    p_{ij}^2 - m_{ij}^2
  }
  \Bigg\{
    &&
    \Big(m_i-2 \big(p_j  \restpol_i\big)\Big)
    \Big(\varepsilon^\ast _{\mathrm{T}}(p_j)  T^{(n)}_{\vbfield{ij} X}(p_{ij})\Big)
    +
    2 \big(p_i  \varepsilon^\ast _{\mathrm{T}}(p_j)\big)
    \Big(\restpol_i  T^{(n)}_{\vbfield{ij} X}(p_{ij})\Big)
    \nonumber\\
    &&&{}-
    \rho (m_{ij},m_j,m_i)
    \Big(\restpol_i  \varepsilon^\ast _{\mathrm{T}}(p_j)\Big)
    \Big(p_i  T^{(n)}_{\vbfield{ij} X}(p_{ij})\Big)
    \nonumber\\
    &&&{}+
    \rho (m_{ij},m_i,m_j)
    \Big(\restpol_i  \varepsilon^\ast _{\mathrm{T}}(p_j)\Big)
    \Big(p_j  T^{(n)}_{\vbfield{ij} X}(p_{ij})\Big)
  \Bigg\}
  .
  \label{eq:VVV-LT-single-amp-GandR}
\end{alignat}
As before, terms of higher order in~$\lambda$ have not been neglected at
this point.
The convention-dependent phase factors $\eta_{\vbfield{}}$ have dropped out.
Summing these two contributions and proceeding as for the TT case by
forming the interference contributions with 
their counterparts for intermediate fields $\vbfield{ij^\prime}$
and~$\gbfield{ij^\prime}$, employing the transverse polarization
sum for~$\vbfield{j}$, and dropping terms of subleading mass order
produces the LT splitting function
\begin{align}
  \bigg[
    \mathcal{P}
      {}^{\vbfield{ij^{\vphantom{\prime}}}}_{\vbfield{i}}
      {}^{\vbfield{ij^\prime}}_{\vbfield{j}}
      \big(z,k_\perp^\mu,\bar p^\mu,\mathrm{L},\mathrm{T}\big)
  \bigg]^{\mu \nu}
  ={}&
  I_{\vbfield{i} \vbantifield{j}}^{\vbfield{ij}} 
  I_{\vbfield{i} \vbantifield{j}}^{\vbfield{ij^\prime} \ast}
  \bigg\{
    \frac{\rho (m_{ij},m_i,m_j)}{(z-1) z}
    \left( 2 - z\rho (m_i,m_{ij}^{\prime },m_j) \right)
    \bar{p}^{\mu } k_{\perp }^{\nu }
    \nonumber\\
    &\quad{}+
    \frac{\rho (m_{ij}^{\prime },m_i,m_j)}{(z-1) z}
    \left( 2 - z\rho (m_i,m_{ij},m_j) \right)
    \bar{p}^{\nu } k_{\perp }^{\mu }
    \nonumber\\
    &\quad{}-
    k_\perp^2
    \frac{
      \rho (m_{ij},m_i,m_j)
      \rho (m_{ij}^{\prime },m_i,m_j)
    }{
      (z-1)^2 m_i^2
    }
    \bar{p}^{\mu } \bar{p}^{\nu }
    \nonumber\\
    &\quad{}-
    \frac{m_i^2}{z^2}
    \left(2 - z\rho (m_i,m_{ij},m_j)\right)
    \left(2 - z\rho (m_i,m_{ij}^{\prime },m_j)\right)
    g^{\mu \nu }
  \bigg\}
  .
  \label{eq:VVV-LT-split-func}
\end{align}
Once more, the properties of~$\rho$ and~\eqref{eq:VVV-ward-id-ul-me}
imply that this result is well behaved for~$m_{ij} \to 0$,
$m_{ij}^\prime \to 0$, and~$m_{i} \to 0$, which becomes apparent by
rewriting
\begin{alignat}{2}
  &
  \lefteqn{
    \bigg[
      \mathcal{P}
        {}^{\vbfield{ij^{\vphantom{\prime}}}}_{\vbfield{i}}
        {}^{\vbfield{ij^\prime}}_{\vbfield{j}}
        \big(z,k_\perp^\mu,\bar p^\mu,\mathrm{L},\mathrm{T}\big)
    \bigg]^{\mu \nu}
  }
  \nonumber\\*
  &=
  I_{\vbfield{i} \vbantifield{j}}^{\vbfield{ij}} 
  I_{\vbfield{i} \vbantifield{j}}^{\vbfield{ij^\prime} \ast}
  \bigg\{
    &&
    \frac{m_{i}}{m_{ij}}
    \rho (m_{i},m_{ij},m_j)
    \bigg(
      \frac{2 m_i}{(z-1) z}
      -
      \frac{m_i}{z-1}
      \rho (m_i,m_{ij}^{\prime },m_j)
    \bigg)
    \frac{\bar{p}^{\mu }}{m_{ij}}
    k_{\perp }^{\nu }
    \nonumber\\
    &&&{}+
    \frac{m_i}{m_{ij}^{\prime }}
    \rho (m_i,m_{ij}^{\prime },m_j)
    \bigg(
      \frac{2 m_i}{(z-1) z}
      -
      \frac{m_i}{z-1}
      \rho (m_i,m_{ij},m_j)
    \bigg)
    \frac{\bar{p}^{\nu }}{m_{ij}^\prime}
    k_{\perp }^{\mu }
    \nonumber\\
    &&&{}-
    \frac{
      k_\perp^2
    }{
      (z-1)^2
    }
    \bigg(
      \frac{m_i}{m_{ij}}
      \rho (m_i,m_{ij},m_j)
    \bigg)
    \bigg(
      \frac{m_i}{m_{ij}^\prime}
      \rho (m_i,m_{ij}^{\prime },m_j)
    \bigg)
    \frac{\bar{p}^{\mu }}{m_{ij}}
    \frac{\bar{p}^{\nu }}{m_{ij}^\prime}
    \nonumber\\
    &&&{}-
    \bigg(\frac{2 m_i}{z}-m_i \rho (m_i,m_{ij},m_j)\bigg)
    \bigg(\frac{2 m_i}{z}-m_i \rho (m_i,m_{ij}^{\prime },m_j)\bigg)
    g^{\mu \nu }
  \bigg\}
  .
  \label{eq:VVV-LT-split-func-for-ml-limit}
\end{alignat}
In particular, consider the limit~$m_i\to0$.
As discussed at the end of Section~\ref{sect:gbet}, the longitudinal
polarization state of the vector boson~$\vbfield{i}$ ceases to exist and is
replaced by the associated scalar~$\gbfield{i}$, which becomes physical.
Accordingly, we relabel $\gbfield{i} \to \higgsfield{i}$.
With
\begin{align}
  \frac{m_i}{m_{ij}}
  \rho (m_i,m_{ij},m_j)
  I_{\vbfield{i} \vbantifield{j}}^{\vbfield{ij}} 
  =
  -2
  \eta_{\vbantifield{i}}
  \eta_{\vbfield{ij}}
  I_{\gbantifield{ij} \gbantifield{i}}^{\vbantifield{j}} 
  &\;\xrightarrow[m_i \to 0]{\gbfield{i} \to \higgsfield{i}}\;
  2
  \eta_{\vbantifield{i}}
  \eta_{\vbantifield{j}}
  \frac{m_j}{m_{ij}}
  I_{\higgsfield{i} \gbantifield{j}}^{\vbfield{ij}} 
  ,
  \nonumber\\
  m_i
  \rho (m_i,m_{ij},m_j)
  I_{\vbfield{i} \vbantifield{j}}^{\vbfield{ij}} 
  =
  m_i
  I_{\vbfield{i} \vbantifield{j}}^{\vbfield{ij}} 
  +
  \eta_{\vbantifield{i}}
  C_{\gbantifield{i} \vbantifield{j} \vbfield{ij}} 
  &\;\xrightarrow[m_i \to 0]{\gbfield{i} \to \higgsfield{i}}\;
  2
  \eta_{\vbantifield{i}}
  \eta_{\vbantifield{j}}
  m_j
  I_{\higgsfield{i} \gbantifield{j}}^{\vbfield{ij}} 
  ,
  \label{eq:VVV-LT-FS-ml-limit}
\end{align}
we see that the splitting function~\eqref{eq:VVV-LT-split-func} does not
necessarily vanish for~$m_i\to0$, but that it instead reproduces the
splitting
function~\eqref{eq:VHV-T-split-func} for~$V^\ast \to H V$,
\begin{align}
  &
  \bigg[
    \mathcal{P}
      {}^{\vbfield{ij^{\vphantom{\prime}}}}_{\vbfield{i}}
      {}^{\vbfield{ij^\prime}}_{\vbfield{j}}
      \big(z,k_\perp^\mu,\bar p^\mu,\mathrm{L},\mathrm{T}\big)
  \bigg]^{\mu \nu}
  \xrightarrow[m_i \to 0]{\gbfield{i} \to \higgsfield{i}}
  \bigg[
    \mathcal{P}
      {}^{\vbfield{ij^{\vphantom{\prime}}}}_{\higgsfield{i}}
      {}^{\vbfield{ij^\prime}}_{\vbfield{j}}
      \big(z,k_\perp^\mu,\bar p^\mu,\mathrm{T}\big)
    \bigg]^{\mu \nu}_{m_i=0}
  ,
  \label{eq:VVV-LT-to-VHV-ml-limit}
\end{align}
which is given in Section~\ref{sect:VHV} below.

\paragraph{Polarization \boldmath{$\polvb_i=\mathrm{T}$}, \boldmath{$\polvb_j=\mathrm{L}$}.}
To obtain the TL result, we proceed in the same way.
Alternatively, it can be obtained by application of the
symmetry~\eqref{eq:i-j-symmru-splfunc},
\begin{align}
  \bigg[
    \mathcal{P}
      {}^{\vbfield{ij^{\vphantom{\prime}}}}_{\vbfield{i}}
      {}^{\vbfield{ij^\prime}}_{\vbfield{j}}
      \big(z,k_\perp^\mu,\bar p^\mu,\mathrm{T},\mathrm{L}\big)
  \bigg]^{\mu \nu}
  &=
  \bigg[
    \mathcal{P}
      {}^{\vbfield{ij^{\vphantom{\prime}}}}_{\vbfield{j}}
      {}^{\vbfield{ij^\prime}}_{\vbfield{i}}
      \big(1-z,-k_\perp^\mu,\bar p^\mu,\mathrm{L},\mathrm{T}\big)
  \bigg]^{\mu \nu}
  _{m_i \leftrightarrow m_j}
  \nonumber\\
  &=
  \bigg[
    \mathcal{P}
      {}^{\vbfield{ij^{\vphantom{\prime}}}}_{\vbfield{i}}
      {}^{\vbfield{ij^\prime}}_{\vbfield{j}}
      \big(1-z,-k_\perp^\mu,\bar p^\mu,\mathrm{L},\mathrm{T}\big)
  \bigg]^{\mu \nu}
  _{m_i \leftrightarrow m_j}
  ,
  \label{eq:VVV-TL-split-func}
\end{align}
where the last line is due to
$
I_{\vbfield{j} \vbantifield{i}}^{\vbfield{ij\smash{{}^{(\prime)}}}} 
=
-I_{\vbfield{i} \vbantifield{j}}^{\vbfield{ij\smash{{}^{(\prime)}}}}
$.

\paragraph{Polarization \boldmath{$\polvb_i=\mathrm{L}$}, \boldmath{$\polvb_j=\mathrm{L}$}.}
Once more, the gauge-dependent contributions
drop out in the combination of
all contributions from~\eqref{eq:VVV-LL-partial-amp-redef},
\begin{align}
  0
  ={}&
  \eta_{\vbfield{i}}^\ast \eta_{\vbfield{j}}^\ast
  \MContrExplFields
    {\vbfield{ij},\xi}
    {\gbfield{i}}
    {\gbfield{j}}^{\mathrm{G}\mathrm{G}}
  +
  \MContrExplFields
    {\vbfield{ij},\xi}
    {\vbfield{i}}
    {\vbfield{j}}^{\mathrm{R}\mathrm{R}}
  \nonumber\\
  &{}+
  \eta_{\vbfield{j}}^\ast
  \MContrExplFields
    {\vbfield{ij},\xi}
    {\vbfield{i}}
    {\gbfield{j}}^{\mathrm{R}\mathrm{G}}
  +
  \eta_{\vbfield{i}}^\ast
  \MContrExplFields
    {\vbfield{ij},\xi}
    {\gbfield{i}}
    {\vbfield{j}}^{\mathrm{G}\mathrm{R}}
  +
  (\vbfield{ij} \rightarrow \gbfield{ij})
  .
\end{align}
For the gauge-parameter-independent tHF parts, we get
\begin{alignat}{3}
  &
  \lefteqn{
    \eta_{\vbfield{i}}^\ast \eta_{\vbfield{j}}^\ast
    \MContrExplFields
      {\vbfield{ij},\mathrm{tHF}}
      {\gbfield{i}}
      {\gbfield{j}}^{\mathrm{G}\mathrm{G}}
    +
    \MContrExplFields
      {\vbfield{ij},\mathrm{tHF}}
      {\vbfield{i}}
      {\vbfield{j}}^{\mathrm{R}\mathrm{R}}
  }
  \nonumber\\
  &
  \lefteqn{
    \phantom{{}={}}
    +
    \eta_{\vbfield{j}}^\ast
    \MContrExplFields
      {\vbfield{ij},\mathrm{tHF}}
      {\vbfield{i}}
      {\gbfield{j}}^{\mathrm{R}\mathrm{G}}
    +
    \eta_{\vbfield{i}}^\ast
    \MContrExplFields
      {\vbfield{ij},\mathrm{tHF}}
      {\gbfield{i}}
      {\vbfield{j}}^{\mathrm{G}\mathrm{R}}
    +
    (\vbfield{ij} \rightarrow \gbfield{ij})
  }
  \nonumber\\
  &=
  \frac
    {e I_{\vbfield{i} \vbantifield{j}}^{\vbfield{ij}}}
    {p_{ij}^2 - m_{ij}^2}
  \bigg\{
    &&
    \lefteqn{
      \Big(m_i \rho (m_i,m_j,m_{ij})-2 \big(p_j  \restpol_i\big)\Big)
      \Big(\restpol_j  T^{(n)}_{\vbfield{ij} X}(p_{ij})\Big)
    }
    \nonumber\\
    &&&
    \lefteqn{
      {} -
      \Big(m_j \rho (m_j,m_i,m_{ij})-2 \big(p_i  \restpol_j\big)\Big)
      \Big(\restpol_i  T^{(n)}_{\vbfield{ij} X}(p_{ij})\Big)
    }
    \nonumber\\
    &&&
    {} +
    \bigg[
      &&
      \lefteqn{
        \rho (m_{ij},m_j,m_i)
        \left(
          \frac{1}{2}\frac{m_i}{m_j} \rho (m_i,m_j,m_{ij})- \frac{p_j  \restpol_i}{m_j}
        \right)
      }
    \nonumber\\
    &&&&&
        {} +
        \rho (m_{ij},m_i,m_j)
        \left(
          \restpol_i  \restpol_j+\frac{p_i  \restpol_j}{m_i}
        \right)
      \bigg]
      \Big(p_j  T^{(n)}_{\vbfield{ij} X}(p_{ij})\Big)
    \nonumber\\
    &&&
    {} -
    \bigg[
      &&
      \lefteqn{
        \rho (m_{ij},m_i,m_j)
        \left(
          \frac{1}{2}\frac{m_j}{m_i} \rho (m_j,m_i,m_{ij})- \frac{p_i  \restpol_j}{m_i}
        \right)
      }
    \nonumber\\
    &&&&&
        {} +
        \rho (m_{ij},m_j,m_i)
        \left(
          \restpol_i  \restpol_j+\frac{p_j  \restpol_i}{m_j}
        \right)
      \bigg]
    \Big(p_i  T^{(n)}_{\vbfield{ij} X}(p_{ij})\Big)
  \bigg\}
  ,
  \label{eq:VVV-LL-indiv-amp-fullL}
\end{alignat}
which does not depend on $\eta_{\vbfield{}}$.
Proceeding as before, with the exception that no transverse polarization
sums have to be employed, we obtain the LL splitting function
\begin{align}
  \bigg[
    \mathcal{P}
      {}^{\vbfield{ij^{\vphantom{\prime}}}}_{\vbfield{i}}
      {}^{\vbfield{ij^\prime}}_{\vbfield{j}}
      \big(z,k_\perp^\mu,\bar p^\mu,\mathrm{L},\mathrm{L}\big)
  \bigg]^{\mu \nu}
  ={}&
  I_{\vbfield{i} \vbantifield{j}}^{\vbfield{ij}} 
  I_{\vbfield{i} \vbantifield{j}}^{\vbfield{ij^\prime} \ast}
  \bigg\{
    \frac{
      m_i
      \rho (m_i,m_j,m_{ij})
    }{
      m_j
    }
    k_{\perp }^{\mu }
    +
    \frac{
      \chi (z,m_i,m_j,m_{ij})
    }{
      2 (z-1) z m_i m_j m_{ij}^2
    }
    \bar{p}^{\mu }
  \bigg\}
  \nonumber\\*
  & \cdot
  \bigg\{
    \frac{
      m_i
      \rho (m_i,m_j,m_{ij}^{\prime })
    }{
      m_j
    }
    k_{\perp }^{\nu }
    +
    \frac{
      \chi (z,m_i,m_j,m_{ij}^{\prime })
    }{
      2 (z-1) z m_i m_j m_{ij}^{\prime\, 2}
    }
    \bar{p}^{\nu }
  \bigg\}
  ,
  \label{eq:VVV-LL-split-func}
\end{align}
with
\begin{align}
  \chi (z,m_1,m_2,m_3)
  =
  m_3^2 \big[
    &
    (z-1) z m_1^2 \rho (m_1,m_2,m_3)
    \big(2 z-\rho (m_3,m_1,m_2)\big)
    \nonumber \\
    &
    - 2 (z-1)  m_1^2 \rho (m_3,m_2,m_1)
    - 2 z      m_2^2 \rho (m_3,m_1,m_2)
  \big]
  ,
  \label{eq:def-chi-abb}
\end{align}
fulfilling the symmetry relations
\begin{align}
  \chi (z,m_1,m_2,m_3)
  &=
  -\chi (1-z,m_2,m_1,m_3)
  ,
  \nonumber\\
  \chi (z,m_1,m_2,m_3)
  &=
  z^3 \chi (1/z,m_3,m_2,m_1)
  . 
  \label{eq:def-chi-relat}
\end{align}
Similar to the previous results,~\eqref{eq:VVV-LL-split-func} is well
behaved in all massless limits.
For the second terms in curly brackets, this can easily be seen by writing
\begin{align}
  \frac{\chi (z,m_i,m_j,m_{ij})}{m_i m_j m_{ij}^2}
  \bar p^\mu
  =
  \bigg[
    &
    (z-1) z \frac{m_i}{m_j} \rho (m_i,m_j,m_{ij})
    \big(2 z m_{ij} - m_{ij} \rho (m_{ij},m_i,m_j)\big)
    \nonumber \\
    &
    - 2 (z-1)  m_i \frac{m_{ij}}{m_j} \rho (m_{ij},m_j,m_i)
    - 2 z      m_j \frac{m_{ij}}{m_i} \rho (m_{ij},m_i,m_j)
  \bigg]
  \frac{\bar p^\mu}{m_{ij}}
  ,
  \label{eq:chi-ml-limit}
\end{align}
recalling~\eqref{eq:VVV-ward-id-ul-me}, and
identifying the factors from~\eqref{eq:rho-ml-limit}.
Focusing on the limits~$m_i \to 0$ and~$m_j \to 0$, 
in analogy to~\eqref{eq:VVV-LT-to-VHV-ml-limit},
we find
\begin{align}
  \bigg[
    \mathcal{P}
      {}^{\vbfield{ij^{\vphantom{\prime}}}}_{\vbfield{i}}
      {}^{\vbfield{ij^\prime}}_{\vbfield{j}}
      \big(z,k_\perp^\mu,\bar p^\mu,\mathrm{L},\mathrm{L}\big)
  \bigg]^{\mu \nu}
  &\xrightarrow[m_i \to 0]{\gbfield{i} \to \higgsfield{i}}
  \bigg[
    \mathcal{P}
      {}^{\vbfield{ij^{\vphantom{\prime}}}}_{\higgsfield{i}}
      {}^{\vbfield{ij^\prime}}_{\vbfield{j}}
      \big(z,k_\perp^\mu,\bar p^\mu,\mathrm{L}\big)
    \bigg]^{\mu \nu}_{m_i=0}
  ,
  \nonumber\\
  \bigg[
    \mathcal{P}
      {}^{\vbfield{ij^{\vphantom{\prime}}}}_{\vbfield{i}}
      {}^{\vbfield{ij^\prime}}_{\vbfield{j}}
      \big(z,k_\perp^\mu,\bar p^\mu,\mathrm{L},\mathrm{L}\big)
  \bigg]^{\mu \nu}
  &\xrightarrow[m_j \to 0]{\gbfield{j} \to \higgsfield{j}}
  \bigg[
    \mathcal{P}
      {}^{\vbfield{ij^{\vphantom{\prime}}}}_{\vbfield{i}}
      {}^{\vbfield{ij^\prime}}_{\higgsfield{j}}
      \big(z,k_\perp^\mu,\bar p^\mu,\mathrm{L}\big)
  \bigg]^{\mu \nu}_{m_j=0}
  ,
  \nonumber\\
  \bigg[
    \mathcal{P}
      {}^{\vbfield{ij^{\vphantom{\prime}}}}_{\vbfield{i}}
      {}^{\vbfield{ij^\prime}}_{\vbfield{j}}
      \big(z,k_\perp^\mu,\bar p^\mu,\mathrm{L},\mathrm{L}\big)
  \bigg]^{\mu \nu}
  &\xrightarrow[m_{i/j} \to 0]{\gbfield{i/j} \to \higgsfield{i/j}}
  \bigg[
    \mathcal{P}
      {}^{\vbfield{ij^{\vphantom{\prime}}}}_{\higgsfield{i}}
      {}^{\vbfield{ij^\prime}}_{\higgsfield{j}}
      \big(z,k_\perp^\mu,\bar p^\mu\big)
  \bigg]^{\mu \nu}_{m_i=m_j=0}
  ,
\end{align}
where the splitting functions on the r.h.s.\ can be found in 
Sections~\ref{sect:VHV} and \ref{sect:VHH} below.

\paragraph{Unpolarized case.}
Combining the above results, we obtain the unpolarized~$V^\ast \to VV$
splitting function.
Denoting the summation over all polarization configurations
with~$\polvb_i=\polvb_j=\mathrm{U}$, we have
\begin{alignat}{3}
  &
  \lefteqn{
    \bigg[
      \mathcal{P}
        {}^{\vbfield{ij^{\vphantom{\prime}}}}_{\vbfield{i}}
        {}^{\vbfield{ij^\prime}}_{\vbfield{j}}
        \big(z,k_\perp^\mu,\bar p^\mu,\mathrm{U},\mathrm{U}\big)
    \bigg]^{\mu \nu}
  }
  \nonumber\\
  &=
  I_{\vbfield{i} \vbantifield{j}}^{\vbfield{ij}} 
  I_{\vbfield{i} \vbantifield{j}}^{\vbfield{ij^\prime} \ast}
  \bigg\{
    &&
    \lefteqn{
      \frac{1}{2}
      \big(\rho (m_j,m_i,m_{ij}^{\prime }) \rho (m_i,m_j,m_{ij})+8\big)
      \big(2 z-\rho (m_{ij}^{\prime },m_i,m_j)\big)
      \bar{p}^{\nu } k_{\perp }^{\mu }
    }
    \nonumber\\
    &&&
    \lefteqn{
      {}+
      \frac{1}{2}
      \big(\rho (m_j,m_i,m_{ij}) \rho (m_i,m_j,m_{ij}^{\prime })+8\big)
      \big(2 z-\rho (m_{ij},m_i,m_j)\big)
      \bar{p}^{\mu } k_{\perp }^{\nu }
    }
    \nonumber\\
    &&&
    {}+
    \bigg[
      &&
      -k_\perp^2
      \left(
        \frac{
            \rho (m_{ij},m_j,m_i) \rho (m_{ij}^{\prime },m_j,m_i)
          }{
            z^2 m_j^2
          }
        +
        \frac{
            \rho (m_{ij},m_i,m_j) \rho (m_{ij}^{\prime },m_i,m_j)
          }{
            (z-1)^2 m_i^2
          }
      \right)
    \nonumber\\
        &&&&& {} +
        \frac{
          \chi (z,m_i,m_j,m_{ij}) \chi (z,m_i,m_j,m_{ij}^{\prime })
        }{
          4 (z-1)^2 z^2 m_i^2 m_{ij}^2 m_j^2 m_{ij}^{\prime\,2 }
        }
        \nonumber\\
        &&&&& {} +
        2 \big(2 z-\rho (m_{ij},m_i,m_j)\big) \big(2 z-\rho (m_{ij}^{\prime },m_i,m_j)\big)
    \bigg]
    \bar{p}^{\mu } \bar{p}^{\nu }
    \nonumber\\
    &&&
    \lefteqn{
      {}+
      \big(\rho (m_j,m_i,m_{ij}) \rho (m_i,m_j,m_{ij}^{\prime })+8\big)
      k_{\perp }^{\mu } k_{\perp }^{\nu }
    }
    \nonumber\\
    &&&
    {}+
    \bigg[
      &&
      {} -
      \bigg(\frac{2}{z}-\rho (m_i,m_{ij},m_j)\bigg)
      \bigg(\frac{2}{z}-\rho (m_i,m_{ij}^{\prime },m_j)\bigg)
      m_i^2
      \nonumber\\
      %
      &&&&&
      {} -
      \bigg(\frac{2}{1-z}-\rho (m_j,m_{ij},m_i)\bigg)
      \bigg(\frac{2}{1-z}-\rho (m_j,m_{ij}^{\prime },m_i)\bigg)
      m_j^2
      \nonumber\\
      &&&&&
      {} +
      4
      \bigg(\frac{1}{z^2}+\frac{1}{(z-1)^2}\bigg)
      k_\perp^2
    \bigg]
    g^{\mu \nu }
    \bigg\}
    .
  \label{eq:VVV-UU-split-func}
\end{alignat}

%
\subsubsection[%
  \mathinhead{H^\ast \to VV}{H* to V V}
  splitting functions%
  ]{%
  \mathinheadbold{H^\ast \to VV}{H* to V V}
  splitting functions}
\label{sect:HVV}
For the computation of $H^\ast \to V V$ splitting functions 
we require the diagrams
\drawmomtrue%
\drawvertdottrue%
\begin{alignat}{2}%
  \ftntmath{%
    \MContrExplFields{\higgsfield{ij}}{\vbfield{i}}{\vbfield{j}}^{\polvb_i \polvb_j}%
  }%
  &\ftntmath{{}={}}%
  \FSDiagGeo%
  \ftntmath{%
    \splittingdiagIndivFS{T^{(n)}_{\higgsfield{ij} X}}%
      {scalar}{photon}{photon}%
      {\higgsfield{ij}}{\vbfield{i}, \polvb_i}{\vbfield{j}, \polvb_j}%
    ,%
  }%
  \resetdiagargs%
  &\hspace{2mm}
  \ftntmath{%
    \MContrExplFields{\higgsfield{ij}}{\gbfield{i}}{\vbfield{j}}^{\mathrm{G} \polvb_j}%
  }%
  &\ftntmath{{}={}}%
  \FSDiagGeo%
  \ftntmath{%
    \splittingdiagIndivFS{T^{(n)}_{\higgsfield{ij} X}}%
      {scalar}{scalar}{photon}%
      {\higgsfield{ij}}{\gbfield{i}}{\vbfield{j}, \polvb_j}%
    ,%
  }%
  \resetdiagargs%
  \nonumber\\%
  \ftntmath{%
    \MContrExplFields{\higgsfield{ij}}{\vbfield{i}}{\gbfield{j}}^{\polvb_i \mathrm{G}}%
  }%
  &\ftntmath{{}={}}%
  \FSDiagGeo%
  \ftntmath{%
    \splittingdiagIndivFS{T^{(n)}_{\higgsfield{ij} X}}%
      {scalar}{photon}{scalar}%
      {\higgsfield{ij}}{\vbfield{i}, \polvb_i}{\gbfield{j}}%
    ,%
  }%
  \resetdiagargs%
  &%
  \ftntmath{%
    \MContrExplFields{\higgsfield{ij}}{\gbfield{i}}{\gbfield{j}}^{\mathrm{G}\mathrm{G}}%
  }%
  &\ftntmath{{}={}}%
  \FSDiagGeo%
  \ftntmath{%
    \splittingdiagIndivFS{T^{(n)}_{\higgsfield{ij} X}}%
      {scalar}{scalar}{scalar}%
      {\higgsfield{ij}}{\gbfield{i}}{\gbfield{j}}%
    .%
  }%
  \resetdiagargs%
  \label{eq:HVV-diags-FS}
\end{alignat}%

which translate, using the coupling relations~\eqref{eq:coupling-relat-GGH-VGH}
and~\eqref{eq:coupling-relat-SVV-VSG}, into
\begin{align}
  %
  %
  \MContrExplFields{\higgsfield{ij}}{\vbfield{i}}{\vbfield{j}}^{\polvb_i \polvb_j}
  ={}&
  - 2 \imagi e
  I_{\higgsantifield{ij} \gbantifield{j}}^{\vbantifield{i}}
  m_j \eta _{\vbfield{j}}^\ast
  \frac{\imagi}{p_{ij}^2-m_{ij}^2}
  T_{\higgsfield{ij} X}^{(n)}\left(p_{ij}\right) 
  \varepsilon_{\polvb_i}^{\alpha \ast}(p_i)
  \varepsilon_{\polvb_j \alpha}^{\ast}(p_j)
  ,
  \nonumber\\
  %
  %
  \MContrExplFields{\higgsfield{ij}}{\gbfield{i}}{\vbfield{j}}^{\mathrm{G} \polvb_j}
  ={}&
  - \imagi e
  I_{\higgsantifield{ij} \gbantifield{j}}^{\vbantifield{i}}
  \frac{m_j \eta _{\vbfield{j}}^\ast}{m_i \eta _{\vbfield{i}}^\ast}
  \frac{\imagi}{p_{ij}^2-m_{ij}^2}
  T_{\higgsfield{ij} X}^{(n)}\left(p_{ij}\right) 
  \big(2 p_i+p_j\big)_{\alpha}
  \varepsilon_{\polvb_j}^{\alpha \ast}(p_j) 
  ,
  \nonumber\\
  %
  %
  \MContrExplFields{\higgsfield{ij}}{\vbfield{i}}{\gbfield{j}}^{\polvb_i \mathrm{G}}
  ={}&
  - \imagi e
  I_{\higgsantifield{ij} \gbantifield{j}}^{\vbantifield{i}}
  \frac{\imagi}{p_{ij}^2-m_{ij}^2}
  T_{\higgsfield{ij} X}^{(n)}\left(p_{ij}\right) 
  \big(p_i+2 p_j\big)_{\alpha}
  \varepsilon_{\polvb_i}^{\alpha \ast}(p_i) 
  ,
  \nonumber\\
  %
  %
  \MContrExplFields{\higgsfield{ij}}{\gbfield{i}}{\gbfield{j}}^{\mathrm{G} \mathrm{G}}
  ={}&
  - \imagi e
  I_{\higgsantifield{ij} \gbantifield{j}}^{\vbantifield{i}}
  \frac{m_{ij}^2}{m_i} \eta _{\vbfield{i}}
  \frac{\imagi}{p_{ij}^2-m_{ij}^2}
  T_{\higgsfield{ij} X}^{(n)}\left(p_{ij}\right) 
  .
  \label{eq:HVV-EnhAmp}
\end{align}
None of these functions involves
gauge parameters~$\xi_{\vbfield{}}$.

\paragraph{Polarization \boldmath{$\polvb_i=\mathrm{T}$}, \boldmath{$\polvb_j=\mathrm{T}$}.}
The TT splitting function
\begin{align}
  \mathcal{P}
    {}^{\higgsfield{ij^{\vphantom{\prime}}}}_{\vbfield{i}}
    {}^{\higgsfield{ij^\prime}}_{\vbfield{j}}
    \big(z,k_\perp^\mu,\bar p^\mu,\mathrm{T},\mathrm{T}\big)
  =
  8 m_j^2
  I_{\higgsantifield{ij} \gbantifield{j}}^{\vbantifield{i}}
  I_{\higgsantifield{ij^\prime} \gbantifield{j}}^{\vbantifield{i} \ast}
  =
  8 m_i^2
  I_{\higgsantifield{ij} \gbantifield{i}}^{\vbantifield{j}}
  I_{\higgsantifield{ij^\prime} \gbantifield{i}}^{\vbantifield{j} \ast}
  \label{eq:HVV-TT-split-func}
\end{align}
is symmetric under the substitution of $i\leftrightarrow j$, see \eqref{eq:i-j-symmru}.
This is due to the coupling relation~\eqref{eq:coupling-relat-VSG-VSG}, which was used in the
second equality.
In the \gls{ewsm}, we have~$\higgsfield{ij} = \higgsfield{ij^\prime} = H$
and~$m_{ij} = m_{ij}^\prime$, as well as~$m_i=m_j$.
Inserting~\eqref{eq:VGH-cpl-vb-mass}, we obtain
\begin{align}
  \mathcal{P}
    {}^{\higgsfield{}}_{\vbfield{i}}
    {}^{\higgsfield{}}_{\vbfield{j}}
    \big(z,k_\perp^\mu,\bar p^\mu,\mathrm{T},\mathrm{T}\big)
  \bigg|_\mathrm{EWSM}
  =
  \frac{8 m_i^4}{(e v)^2}
  \delta_{\vbfield{i} \vbantifield{j}}
  .
  \label{eq:HVV-TT-split-func-EWSM}
\end{align}

\paragraph{Polarization \boldmath{$\polvb_i=\mathrm{L}$}, \boldmath{$\polvb_j=\mathrm{T}$}.}
The LT splitting function
\begin{align}
  \mathcal{P}
    {}^{\higgsfield{ij^{\vphantom{\prime}}}}_{\vbfield{i}}
    {}^{\higgsfield{ij^\prime}}_{\vbfield{j}}
    \big(z,k_\perp^\mu,\bar p^\mu,\mathrm{L},\mathrm{T}\big)
  &=
  -\frac{4 k_\perp^2 m_j^2}{m_i^2 (z-1)^2}
  I_{\higgsantifield{ij} \gbantifield{j}}^{\vbantifield{i}}
  I_{\higgsantifield{ij^\prime} \gbantifield{j}}^{\vbantifield{i} \ast}
  =
  -\frac{4 k_\perp^2}{(z-1)^2}
  I_{\higgsantifield{ij} \gbantifield{i}}^{\vbantifield{j}}
  I_{\higgsantifield{ij^\prime} \gbantifield{i}}^{\vbantifield{j} \ast}
  .
  \label{eq:HVV-LT-split-func}
\end{align}
is well behaved in the limit of a massless vector
boson~$\vbfield{i}$, and its relation to the $H^\ast \to H V$
splitting function~\eqref{eq:HHV-T-split-func} is obvious.
The \gls{ewsm} result reads
\begin{align}
  \mathcal{P}
    {}^{\higgsfield{}}_{\vbfield{i}}
    {}^{\higgsfield{}}_{\vbfield{j}}
    \big(z,k_\perp^\mu,\bar p^\mu,\mathrm{L},\mathrm{T}\big)
  \bigg|_\mathrm{EWSM}
  &=
  -\frac{4 k_\perp^2 m_i^2}{(e v)^2 (z-1)^2}
  \delta_{\vbfield{i} \vbantifield{j}}
  .
  \label{eq:HVV-LT-split-func-EWSM}
\end{align}

\paragraph{Polarization \boldmath{$\polvb_i=\mathrm{T}$}, \boldmath{$\polvb_j=\mathrm{L}$}.}
As for the symmetry of the TT splitting function, the relation between the TL result
\begin{align}
  \mathcal{P}
    {}^{\higgsfield{ij^{\vphantom{\prime}}}}_{\vbfield{i}}
    {}^{\higgsfield{ij^\prime}}_{\vbfield{j}}
    \big(z,k_\perp^\mu,\bar p^\mu,\mathrm{T},\mathrm{L}\big)
  &=
  -\frac{4 k_\perp^2 }{z^2}
  I_{\higgsantifield{ij} \gbantifield{j}}^{\vbantifield{i}}
  I_{\higgsantifield{ij^\prime} \gbantifield{j}}^{\vbantifield{i} \ast}
  =
  -\frac{4 k_\perp^2 m_i^2}{m_j^2 z^2}
  I_{\higgsantifield{ij} \gbantifield{i}}^{\vbantifield{j}}
  I_{\higgsantifield{ij^\prime} \gbantifield{i}}^{\vbantifield{j} \ast}
  .
  \label{eq:HVV-TL-split-func}
\end{align}
and the LT counterpart~\eqref{eq:HVV-LT-split-func}
via application of~\eqref{eq:i-j-symmru-splfunc},
which expresses the substitution $i\leftrightarrow j$,
requires the coupling relation~\eqref{eq:coupling-relat-VSG-VSG}.
The \gls{ewsm} result reads
\begin{align}
  \mathcal{P}
    {}^{\higgsfield{}}_{\vbfield{i}}
    {}^{\higgsfield{}}_{\vbfield{j}}
    \big(z,k_\perp^\mu,\bar p^\mu,\mathrm{T},\mathrm{L}\big)
  \bigg|_\mathrm{EWSM}
  &=
  -\frac{4 k_\perp^2 m_i^2}{(e v)^2 z^2}
  \delta_{\vbfield{i} \vbantifield{j}}
  .
  \label{eq:HVV-TL-split-func-EWSM}
\end{align}

\paragraph{Polarization \boldmath{$\polvb_i=\mathrm{L}$}, \boldmath{$\polvb_j=\mathrm{L}$}.}
In the LL result
\begin{align}
  \mathcal{P}
      {}^{\higgsfield{ij^{\vphantom{\prime}}}}_{\vbfield{i}}
      {}^{\higgsfield{ij^\prime}}_{\vbfield{j}}
    \big(z,k_\perp^\mu,\bar p^\mu,\mathrm{L},\mathrm{L}\big)
  ={}&
  m_j^2
  I_{\higgsantifield{ij} \gbantifield{j}}^{\vbantifield{i}}
  I_{\higgsantifield{ij^\prime} \gbantifield{j}}^{\vbantifield{i} \ast}
    \bigg(
      \frac{m_i \bar{\rho}(m_i,m_j,m_{ij})}{m_j}
      -
      \frac{2 m_i}{z m_j}
      +
      \frac{2 m_j}{(z-1) m_i}
    \bigg)
    \nonumber\\
    & {} \cdot
    \bigg(
      \frac{m_i \bar{\rho}(m_i,m_j,m_{ij}^\prime)}{m_j}
      -
      \frac{2 m_i}{z m_j}
      +
      \frac{2 m_j}{(z-1) m_i}
    \bigg)
    ,
  \label{eq:HVV-LL-split-func}
\end{align}
we introduce the auxiliary function
\begin{align}
  \bar{\rho}(m_1,m_2,m_3)
  \equiv
  \frac{m_1^2 + m_2^2 + m_3^2}{m_1^2}
  .
  \label{eq:def-rhobar-abb}
\end{align}
Again, the symmetry of the result under
substitution~\eqref{eq:i-j-symmru} follows
from the coupling relation~\eqref{eq:coupling-relat-VSG-VSG}.
The latter also implies that~\eqref{eq:HVV-LL-split-func} is well
behaved in the individual limits~$m_i \to 0$ and~$m_j \to 0$, 
which follows from arguments similar to the ones given 
for the factors~\eqref{eq:rho-ml-limit}.
Replacing~$\gbfield{i} \to \higgsfield{i}$ and~$\gbfield{j} \to \higgsfield{j}$, we get 
\begin{align}
  \mathcal{P}
    {}^{\higgsfield{ij^{\vphantom{\prime}}}}_{\vbfield{i}}
    {}^{\higgsfield{ij^\prime}}_{\vbfield{j}}
    \big(z,k_\perp^\mu,\bar p^\mu,\mathrm{L},\mathrm{L}\big)
  &\xrightarrow[m_i \to 0]{\gbfield{i} \to \higgsfield{i}}
  \mathcal{P}
    {}^{\higgsfield{ij^{\vphantom{\prime}}}}_{\higgsfield{i}}
    {}^{\higgsfield{ij^\prime}}_{\vbfield{j}}
    \big(z,k_\perp^\mu,\bar p^\mu,\mathrm{L}\big)
  \bigg|_{m_i=0}
  ,
  \nonumber\\
  \mathcal{P}
    {}^{\higgsfield{ij^{\vphantom{\prime}}}}_{\vbfield{i}}
    {}^{\higgsfield{ij^\prime}}_{\vbfield{j}}
    \big(z,k_\perp^\mu,\bar p^\mu,\mathrm{L},\mathrm{L}\big)
  &\xrightarrow[m_j \to 0]{\gbfield{j} \to \higgsfield{j}}
  \mathcal{P}
    {}^{\higgsfield{ij^{\vphantom{\prime}}}}_{\vbfield{i}}
    {}^{\higgsfield{ij^\prime}}_{\higgsfield{j}}
    \big(z,k_\perp^\mu,\bar p^\mu,\mathrm{L}\big)
  \bigg|_{m_j=0}
  .
\end{align}
In the simultaneous limit~$m_i,m_j \to 0$,~\eqref{eq:HVV-LL-split-func}
remains finite due to coupling 
relation~\eqref{eq:coupling-relat-GGH-VGH} and reproduces the
splitting function for~$H^\ast \to H H$,
\begin{align}
  \mathcal{P}
    {}^{\higgsfield{ij^{\vphantom{\prime}}}}_{\vbfield{i}}
    {}^{\higgsfield{ij^\prime}}_{\vbfield{j}}
    \big(z,k_\perp^\mu,\bar p^\mu,\mathrm{L},\mathrm{L}\big)
   &\xrightarrow[m_{i/j} \to 0]{\gbfield{i/j} \to \higgsfield{i/j}}
   \mathcal{P}
     {}^{\higgsfield{ij^{\vphantom{\prime}}}}_{\higgsfield{i}}
     {}^{\higgsfield{ij^\prime}}_{\higgsfield{j}}
     \big(z,k_\perp^\mu,\bar p^\mu\big)
  \bigg|_{m_i=m_j=0}
  .
\end{align}
The \gls{ewsm} result reads
\begin{align}
  \mathcal{P}
    {}^{\higgsfield{}}_{\vbfield{i}}
    {}^{\higgsfield{}}_{\vbfield{j}}
    \big(z,k_\perp^\mu,\bar p^\mu,\mathrm{L},\mathrm{L}\big)
  \bigg|_\mathrm{EWSM}
  &=
  \frac{
    \delta_{\vbfield{i} \vbantifield{j}}
  }{(ev)^2}
  \bigg(
    2 \bigg(1+\frac{1}{(z-1)z}\bigg) m_i^2
    +
    m_{ij}^2
  \bigg)^2
  .
  \label{eq:HVV-LL-split-func-EWSM}
\end{align}

\paragraph{Unpolarized case.}
The above results combine into the unpolarized splitting function
\begin{align}
  \lefteqn{\mathcal{P}
      {}^{\higgsfield{ij^{\vphantom{\prime}}}}_{\vbfield{i}}
      {}^{\higgsfield{ij^\prime}}_{\vbfield{j}}
    \big(z,k_\perp^\mu,\bar p^\mu,\mathrm{U},\mathrm{U}\big) }
  \phantom{{}={}}
  &
  \nonumber\\*
  ={} &
  m_j^2
  I_{\higgsantifield{ij} \gbantifield{j}}^{\vbantifield{i}}
  I_{\higgsantifield{ij^\prime} \gbantifield{j}}^{\vbantifield{i} \ast}
      \bigg\{
      \frac{4 (m_i^2-k_\perp^2)}{z^2 m_j^2}
      +
      \frac{4 (m_j^2-k_\perp^2)}{(z-1)^2 m_i^2}
      +
      \bar{\rho}(m_j,m_i,m_{ij}) \bar{\rho}(m_i,m_j,m_{ij}^{\prime })
      +
      8
      \nonumber\\
      &{}-
      \frac{
        2 m_i^2 \big(
          \rho (m_i,m_{ij}^{\prime },m_j)
          +
          \rho (m_i,m_{ij},m_j)
        \big)
      }{z m_j^2}
      +
      \frac{
        2 m_j^2 \big(
          \rho (m_j,m_{ij}^{\prime },m_i)
          +
          \rho (m_j,m_{ij},m_i)
        \big)
      }{(z-1) m_i^2}
    \bigg\}
    ,
    \label{eq:HVV-UU-split-func}
\end{align}
which in the \gls{ewsm} reads
\begin{align}
  &\mathcal{P}
      {}^{\higgsfield{}}_{\vbfield{i}}
      {}^{\higgsfield{}}_{\vbfield{j}}
    \big(z,k_\perp^\mu,\bar p^\mu,\mathrm{U},\mathrm{U}\big)
  \bigg|_\mathrm{EWSM}
  \nonumber\\*
  &=
  \frac{
    \delta_{\vbfield{i} \vbantifield{j}}
  }{(ev)^2}
  \bigg\{
    4 m_i^2 (m_i^2 - k_\perp^2) \left(\frac{1}{z^2}+\frac{1}{(z-1)^2}\right)
    +
    \frac{4 m_i^2 m_{ij}^2}{(z-1) z}
    +
    12 m_i^4 + 4 m_i^2 m_{ij}^2 + m_{ij}^4
  \bigg\}
  .
  \label{eq:HVV-UU-split-func-EWSM}
\end{align}

%
\subsubsection[%
  \mathinhead{V^\ast/H^\ast \to VV}{V*/H* to V V}
  splitting functions%
  ]{%
  \mathinheadbold{V^\ast/H^\ast \to VV}{V*/H* to V V}
  splitting functions}
\label{sect:VHVV}
The functions for $V^\ast/H^\ast \to V V$ splittings resulting from
interferences
are obtained using
the ingredients from the previous sections.
Regarding the massless limits of the following result, compare to
Sections~\ref{sect:VVV} and~\ref{sect:HVV}.

\paragraph{Polarization \boldmath{$\polvb_i=\mathrm{T}$}, \boldmath{$\polvb_j=\mathrm{T}$}.}
The symmetry of the TT splitting function
\begin{align}
  &\bigg[
    \mathcal{P}
      {}^{\vbfield{ij^{\vphantom{\prime}}}}_{\vbfield{i}}
      {}^{\higgsfield{ij^\prime}}_{\vbfield{j}}
      \big(z,k_\perp^\mu,\bar p^\mu,\mathrm{T},\mathrm{T}\big)
  \bigg]^\mu
  \nonumber\\*
  &=
  - 4 m_j \eta_{\vbfield{j}}
  I_{\vbfield{i} \vbantifield{j}}^{\vbfield{ij}} 
  I_{\higgsantifield{ij^\prime} \gbantifield{j}}^{\vbantifield{i} \ast}
  \bigg\{
    \bigg(
      2+
      \frac{      
        1
      }{(z-1) z}
    \bigg)
    k_{\perp }^{\mu }
    -
    \big(
      \rho(m_{ij},m_i,m_j)-2 z
    \big)
    \bar{p}^{\mu } 
  \bigg\}
  \label{eq:VHVV-TT-split-func}
\end{align}
under~\eqref{eq:i-j-symmru} is somewhat hidden and 
can be made explicit as described in
Section~\ref{sect:HVV}.
In the \gls{ewsm}, where~$m_i=m_j$ and $m_{ij}=M_{\mathrm{H}}$, we obtain
\begin{align}
  \bigg[
    \mathcal{P}
      {}^{\vbfield{ij}}_{\vbfield{i}}
      {}^{\higgsfield{\vphantom{ij}}}_{\vbfield{j}}
      \big(z,k_\perp^\mu,\bar p^\mu,\mathrm{T},\mathrm{T}\big)
    \bigg]^\mu_\mathrm{EWSM}
  =
  4 m_i^2
  \frac{
    I_{\vbfield{i} \vbantifield{j}}^{\vbfield{ij}}
  }{ev}
  \delta_{\vbfield{i} \vbantifield{j}}
  \bigg\{
    \bigg(
      2+
      \frac{      
        1
      }{(z-1) z}
    \bigg)
    k_{\perp }^{\mu }
    -
    (1-2 z) \bar{p}^{\mu } 
  \bigg\}
  .
  \label{eq:VHVV-TT-split-func-EWSM}
\end{align}

\paragraph{Polarization \boldmath{$\polvb_i=\mathrm{L}$}, \boldmath{$\polvb_j=\mathrm{T}$}.}
The LT result reads
\begin{align}
  &\bigg[
    \mathcal{P}
      {}^{\vbfield{ij^{\vphantom{\prime}}}}_{\vbfield{i}}
      {}^{\higgsfield{ij^\prime}}_{\vbfield{j}}
      \big(z,k_\perp^\mu,\bar p^\mu,\mathrm{L},\mathrm{T}\big)
  \bigg]^\mu
  \nonumber\\*
  &=
  2 m_j \eta_{\vbfield{j}}
  I_{\vbfield{i} \vbantifield{j}}^{\vbfield{ij}} 
  I_{\higgsantifield{ij^\prime} \gbantifield{j}}^{\vbantifield{i} \ast}
  \bigg\{
    \bigg(
      \frac{\rho(m_i,m_j,m_{ij})}{z-1}
      -
      \frac{2}{z}
    \bigg)
    k_{\perp }^{\mu }
    -
    \bigg(
      \frac{k_\perp^2  \rho(m_{ij},m_i,m_j)}{(z-1)^2 m_i^2}
    \bigg)
    \bar{p}^{\mu }
  \bigg\}
  \label{eq:VHVV-LT-split-func}
\end{align}
and takes the following form in the \gls{ewsm},
\begin{align}
  \bigg[
    \mathcal{P}
      {}^{\vbfield{ij}}_{\vbfield{i}}
      {}^{\higgsfield{\phantom{ij}}}_{\vbfield{j}}
      \big(z,k_\perp^\mu,\bar p^\mu,\mathrm{L},\mathrm{T}\big)
    \bigg]^\mu_\mathrm{EWSM}
  &=
  -
  \frac{2}{z-1}
  \frac{
    I_{\vbfield{i} \vbantifield{j}}^{\vbfield{ij}}
  }{ev}
  \delta_{\vbfield{i} \vbantifield{j}}
  \bigg\{
    \bigg( \frac{2 m_i^2}{z}-m_{ij}^2 \bigg) k_{\perp }^{\mu }
    -
    \frac{k_\perp^2}{z-1} \bar{p}^{\mu }
  \bigg\}
  .
  \label{eq:VHVV-LT-split-func-EWSM}
\end{align}

\paragraph{Polarization \boldmath{$\polvb_i=\mathrm{T}$}, \boldmath{$\polvb_j=\mathrm{L}$}.}
The connection between the TL result
\begin{align}
  &\bigg[
    \mathcal{P}
      {}^{\vbfield{ij^{\vphantom{\prime}}}}_{\vbfield{i}}
      {}^{\higgsfield{ij^\prime}}_{\vbfield{j}}
      \big(z,k_\perp^\mu,\bar p^\mu,\mathrm{T},\mathrm{L}\big)
  \bigg]^\mu
  \nonumber\\*
  &=
  - 2 m_j \eta_{\vbfield{j}}
  I_{\vbfield{i} \vbantifield{j}}^{\vbfield{ij}} 
  I_{\higgsantifield{ij^\prime} \gbantifield{j}}^{\vbantifield{i} \ast}
  \bigg\{
    \bigg(
      \frac{\rho(m_j,m_i,m_{ij})}{z}
      -
      \frac{2}{z-1}
    \bigg)
    k_{\perp }^{\mu }
    -
    \bigg(
      \frac{k_\perp^2  \rho(m_{ij},m_j,m_i)}{z^2 m_j^2}
    \bigg)
    \bar{p}^{\mu }
  \bigg\}
  .
  \label{eq:VHVV-TL-split-func}
\end{align}
and the LT result~\eqref{eq:VHVV-LT-split-func}
via~\eqref{eq:i-j-symmru-splfunc} is elucidated as described in
Section~\ref{sect:HVV}.
The \gls{ewsm} result reads
\begin{align}
  \bigg[
    \mathcal{P}
      {}^{\vbfield{ij}}_{\vbfield{i}}
      {}^{\higgsfield{\vphantom{ij}}}_{\vbfield{j}}
      \big(z,k_\perp^\mu,\bar p^\mu,\mathrm{T},\mathrm{L}\big)
    \bigg]^\mu_\mathrm{EWSM}
  &=
  -
  \frac{2}{z}
  \frac{
    I_{\vbfield{i} \vbantifield{j}}^{\vbfield{ij}}
  }{ev}
  \delta_{\vbfield{i} \vbantifield{j}}
  \bigg\{
    \bigg( \frac{2 m_i^2}{z-1} + m_{ij}^2 \bigg) k_{\perp }^{\mu }
    +
    \frac{k_\perp^2}{z} \bar{p}^{\mu }
  \bigg\}
  .
  \label{eq:VHVV-TL-split-func-EWSM}
\end{align}

\paragraph{Polarization \boldmath{$\polvb_i=\mathrm{L}$}, \boldmath{$\polvb_j=\mathrm{L}$}.}
We get the LL splitting function
\begin{align}
  \bigg[
    \mathcal{P}
      {}^{\vbfield{ij^{\vphantom{\prime}}}}_{\vbfield{i}}
      {}^{\higgsfield{ij^\prime}}_{\vbfield{j}}
      \big(z,k_\perp^\mu,\bar p^\mu,\mathrm{L},\mathrm{L}\big)
  \bigg]^\mu
  {}={}&
  - m_j \eta_{\vbfield{j}}
  I_{\vbfield{i} \vbantifield{j}}^{\vbfield{ij}} 
  I_{\higgsantifield{ij^\prime} \gbantifield{j}}^{\vbantifield{i} \ast}
    \bigg(
      \frac{m_i \bar{\rho}(m_i,m_j,m_{ij}^\prime)}{m_j}
      -
      \frac{2 m_i}{z m_j}
      +
      \frac{2 m_j}{(z-1) m_i}
    \bigg)
    \nonumber\\*
    & {} \cdot
    \bigg(
      \frac{
        m_i
        \rho (m_i,m_j,m_{ij})
      }{
        m_j
      }
      k_{\perp }^{\mu }
      +
      \frac{
        \chi (z,m_i,m_j,m_{ij})
      }{
        2 (z-1) z m_i m_j m_{ij}^2
      }
      \bar{p}^{\mu }
    \bigg)
    ,
    \label{eq:VHVV-LL-split-func}
\end{align}
with the \gls{ewsm} result
\begin{align}
  \bigg[
    \mathcal{P}
      {}^{\vbfield{ij}}_{\vbfield{i}}
      {}^{\higgsfield{\vphantom{ij}}}_{\vbfield{j}}
      \big(z,k_\perp^\mu,\bar p^\mu,\mathrm{L},\mathrm{L}\big)
    \bigg]^\mu_\mathrm{EWSM}
  ={}&
  \frac{
    I_{\vbfield{i} \vbantifield{j}}^{\vbfield{ij}}
  }{e v}
  \delta_{\vbfield{i} \vbantifield{j}}
  \bigg(
    2 \bigg(1+\frac{1}{(z-1)z}\bigg) m_i^2 + m_{ij}^{\prime \, 2}
  \bigg)
  \nonumber\\
  &{}
  \cdot
  \bigg\{
    \rho(m_i,m_i,m_{ij})
    k_{\perp }^{\mu } 
    +
    \frac{\chi(z,m_i,m_i,m_{ij})}{2 (z-1) z m_i^2 m_{ij}^2}
    \bar{p}^{\mu }
  \bigg\}
  .
  \label{eq:VHVV-LL-split-func-EWSM}
\end{align}

\paragraph{Unpolarized case.}
The combination
of~\eqref{eq:VHVV-LT-split-func}--\eqref{eq:VHVV-LL-split-func} leads to
the unpolarized splitting function
\begin{align}
  \lefteqn{
    \bigg[
    \mathcal{P}
        {}^{\vbfield{ij^{\vphantom{\prime}}}}_{\vbfield{i}}
        {}^{\higgsfield{ij^\prime}}_{\vbfield{j}}
        \big(z,k_\perp^\mu,\bar p^\mu,\mathrm{U},\mathrm{U}\big)
    \bigg]^\mu
  }
  \phantom{{}={}}
  \nonumber\\*
  ={}&
  m_j \eta_{\vbfield{j}}
  I_{\vbfield{i} \vbantifield{j}}^{\vbfield{ij}} 
  I_{\higgsantifield{ij^\prime} \gbantifield{j}}^{\vbantifield{i} \ast}
  \bigg\{
    \bigg[
      \frac{
        \chi (z,m_i,m_j,m_{ij})
      }{2 (1-z) z m_i m_{ij}^2 m_j}
      \bigg(
        \frac{m_i \bar{\rho}(m_i,m_j,m_{ij}^{\prime })}{m_j}
        -
        \frac{2 m_i}{z m_j}
        +
        \frac{2 m_j}{(z-1) m_i}
      \bigg)
      \nonumber\\
     & \quad {}
      +
      \frac{2 k_\perp^2 \rho (m_{ij},m_j,m_i)}{z^2 m_j^2}
      -
      \frac{2 k_\perp^2 \rho (m_{ij},m_i,m_j)}{(z-1)^2 m_i^2}
      + 4 \big(\rho (m_{ij},m_i,m_j)-2 z\big)
    \bigg]
    \bar{p}^{\mu }
    \nonumber\\
    & \quad {}
    -
    \big(
      \rho (m_i,m_j,m_{ij}) \bar{\rho}(m_j,m_i,m_{ij}^{\prime })+8
    \big)
    k_{\perp }^{\mu }
  \bigg\}
  ,
  \label{eq:VHVV-UU-split-func}
\end{align}
with \gls{ewsm} form
\begin{alignat}{2}
  &
  \lefteqn{
    \bigg[
      \mathcal{P}
        {}^{\vbfield{ij}}_{\vbfield{i}}
        {}^{\higgsfield{\vphantom{ij}}}_{\vbfield{j}}
        \big(z,k_\perp^\mu,\bar p^\mu,\mathrm{U},\mathrm{U}\big)
      \bigg]^\mu_\mathrm{EWSM}
  }
  \nonumber\\*
  &=
  \frac{
    I_{\vbfield{i} \vbantifield{j}}^{\vbfield{ij}}
  }{e v}
  \delta_{\vbfield{i} \vbantifield{j}}
  \bigg\{
    &&
    \bigg(
      2 \frac{m_i^2 - k_\perp^2}{(z-1)^2 z^2}
      +
      \frac{m_{ij}^2 + m_{ij}^{\prime\, 2}}{(z-1)z}
    \bigg)
    (1-2z)
    \bar{p}^{\mu }
    \nonumber\\
    &&&
    +
    m_i^2 
    \big(
      \rho (m_i,m_i,m_{ij}) \bar{\rho}(m_i,m_i,m_{ij}^{\prime })+8
    \big)
    \bigg(
      k_{\perp }^{\mu }
      -
      \frac{(1-2z)}{2}
      \bar{p}^{\mu }
    \bigg)
  \bigg\}
  .
  \label{eq:VHVV-UU-split-func-EWSM}
\end{alignat}

%
\subsubsection[%
  \mathinhead{H^\ast/V^\ast \to VV}{H*/V* to V V}
  splitting functions%
  ]{%
  \mathinheadbold{H^\ast/V^\ast \to VV}{H*/V* to V V}
  splitting functions}
\label{sect:HVVV}
The splitting functions for $H^\ast/V^\ast \to V V$ are related to those
for $V^\ast/H^\ast \to V V$ via application
of~\eqref{eq:ij-ijpr-symmru-splfunc},
where~$\genfield{ij} \leftrightarrow \genfield{ij^\prime}$
corresponds to 
$\vbfield{ij}\leftrightarrow \higgsfield{ij^\prime}$.
We get
\begin{align}
  \bigg[
    \mathcal{P}
      {}^{\higgsfield{ij^{\vphantom{\prime}}}}_{\vbfield{i}}
      {}^{\vbfield{ij^\prime}}_{\vbfield{j}}
      \big(z,k_\perp^\mu,\bar p^\mu,\polvb_i,\polvb_j\big)
  \bigg]^{\mu}
  =
  \bigg[
    \mathcal{P}
      {}^{\vbfield{ij^\prime}}_{\vbfield{i}}
      {}^{\higgsfield{ij^{\vphantom{\prime}}}}_{\vbfield{j}}
      \big(z,k_\perp^\mu,\bar p^\mu,\polvb_i,\polvb_j\big)
  \bigg]^{\mu \ast}
  _{m_{ij} \leftrightarrow m_{ij}^\prime}
  .
\end{align}

%
\subsubsection[%
  \mathinhead{V^\ast \to HV}{V* to H V}
  splitting functions%
  ]{%
  \mathinheadbold{V^\ast \to HV}{V* to H V}
  splitting functions}
\label{sect:VHV}
For $V^\ast \to H V$ splittings, the issue regarding longitudinal
polarization of the \gls{fs} vector boson is treated as discussed above.
In analogy to~\eqref{eq:LT-goldstone-strat}, we write
\begin{align}
  \mathcal{M}^{(n+1)\, \mathrm{L}}_{\higgsfield{i} \vbfield{j} X}
  =
  \left[
    T^{(n+1)}_{\higgsfield{i} \vbfield{j} X}
  \right]_{\mu_j}
  \varepsilon_{\mathrm{L}}^{\mu_j \ast}(p_j)
  =
  \eta_{\vbfield{j}}
  \mathcal{M}^{(n+1)\, \mathrm{G}}_{\higgsfield{i} \gbfield{j} X}
  +
  \mathcal{M}^{(n+1)\, \mathrm{R}}_{\higgsfield{i} \vbfield{j} X}
  \label{eq:VHV-L-goldstone-strat}
\end{align}
and select the diagrams of type~\diagtype{1} separately for the two
terms on the r.h.s., amounting to
\drawmomtrue%
\drawvertdottrue%
\begin{alignat}{2}%
  \ftntmath{%
    \MContrExplFields{\vbfield{ij}}{\higgsfield{i}}{\vbfield{j}}^{\polvb_j}%
  }%
  &\ftntmath{{}={}}%
  \FSDiagGeo%
  \ftntmath{%
    \splittingdiagIndivFS{T^{(n)}_{\vbfield{ij} X}}%
      {photon}{scalar}{photon}%
      {\vbfield{ij}}{\higgsfield{i}}{\vbfield{j}, \polvb_j}%
    ,%
  }%
  \resetdiagargs%
  &\hspace{3mm}
  \ftntmath{%
    \MContrExplFields{\gbfield{ij}}{\higgsfield{i}}{\vbfield{j}}^{\polvb_j}%
  }%
  &\ftntmath{{}={}}%
  \FSDiagGeo%
  \ftntmath{%
    \splittingdiagIndivFS{T^{(n)}_{\gbfield{ij} X}}%
      {scalar}{scalar}{photon}%
      {\gbfield{ij}}{\higgsfield{i}}{\vbfield{j}, \polvb_j}%
    ,%
  }%
  \nonumber\\%
  \ftntmath{%
    \MContrExplFields{\vbfield{ij}}{\higgsfield{i}}{\gbfield{j}}^{\mathrm{G}}%
  }%
  &\ftntmath{{}={}}%
  \FSDiagGeo%
  \ftntmath{%
    \splittingdiagIndivFS{T^{(n)}_{\vbfield{ij} X}}%
      {photon}{scalar}{scalar}%
      {\vbfield{ij}}{\higgsfield{i}}{\gbfield{j}}%
    ,%
  }%
  \resetdiagargs%
  &%
  \ftntmath{%
    \MContrExplFields{\gbfield{ij}}{\higgsfield{i}}{\gbfield{j}}^{\mathrm{G}}%
  }%
  &\ftntmath{{}={}}%
  \FSDiagGeo%
  \ftntmath{%
    \splittingdiagIndivFS{T^{(n)}_{\gbfield{ij} X}}%
      {scalar}{scalar}{scalar}%
      {\gbfield{ij}}{\higgsfield{i}}{\gbfield{j}}%
    .%
  }%
  \resetdiagargs%
  \label{eq:VHV-diags-FS}
\end{alignat}%
After insertion of the coupling
relations~\eqref{eq:coupling-relat-SVV-VSG}
and~\eqref{eq:coupling-relat-GGH-VGH}, we obtain
\begin{align}
  %
  %
  \MContrExplFields{\vbfield{ij}}{\higgsfield{i}}{\vbfield{j}}^{\polvb_j}
  ={}&
  - 2 \imagi e
  I_{\higgsfield{i} \gbantifield{j}}^{\vbfield{ij}}
  m_j \eta_{\vbfield{j}}^\ast
  T^{(n)\, \sigma}_{\vbfield{ij} X}\left(p_{ij}\right)
  G^{\vbantifield{ij} \vbfield{ij}}_{\sigma \gamma}(p_{ij},-p_{ij})
  \varepsilon_{\polvb_j}^{\gamma  \ast}(p_j)
  ,
  \nonumber\\
  %
  %
  \MContrExplFields{\gbfield{ij}}{\higgsfield{i}}{\vbfield{j}}^{\polvb_j}
  ={}&
  - \imagi e
  I_{\higgsfield{i} \gbantifield{j}}^{\vbfield{ij}}
  \frac{m_j \eta_{\vbfield{j}}^\ast}{m_{ij}^2}
  G^{\gbantifield{ij} \gbfield{ij}} (p_{ij},-p_{ij})
  T_{\vbfield{ij} X}^{(n)\, \alpha }\left(p_{ij}\right)
  p_{ij,\alpha }
  \varepsilon_{\polvb_j}^{\gamma  \ast}(p_j)
  \left(2p_i+p_j\right)_{\gamma }
  ,
  \nonumber\\
  %
  %
  \MContrExplFields{\vbfield{ij}}{\higgsfield{i}}{\gbfield{j}}^{\mathrm{G}}
  ={}&
  - \imagi e
  I_{\higgsfield{i} \gbantifield{j}}^{\vbfield{ij}}
  T^{(n)\, \sigma}_{\vbfield{ij} X}\left(p_{ij}\right)
  G^{\vbantifield{ij} \vbfield{ij}}_{\sigma \alpha}(p_{ij},-p_{ij})
  \left(p_j-p_i\right)_{\alpha }
  ,
  \nonumber\\
  %
  %
  \MContrExplFields{\gbfield{ij}}{\higgsfield{i}}{\gbfield{j}}^{\mathrm{G}}
  ={}&
  + \imagi e
  I_{\higgsfield{i} \gbantifield{j}}^{\vbfield{ij}}
  \frac{m_i^2}{m_{ij}^2}
  G^{\gbantifield{ij} \gbfield{ij}} (p_{ij},-p_{ij})
  T_{\vbfield{ij} X}^{(n)\, \alpha }\left(p_{ij}\right)
  p_{ij,\alpha }
  \label{eq:VHV-EnhAmp}
\end{align}
and arrive at the splitting functions quoted below.
%
%
The results for $V^\ast \to V H$ splittings can be computed similarly,
or alternatively obtained via~\eqref{eq:i-j-symmru-splfunc}, i.e.\
\begin{align}
  \bigg[
    \mathcal{P}
      {}^{\vbfield{ij^{\vphantom{\prime}}}}_{\vbfield{i}}
      {}^{\vbfield{ij^\prime}}_{\higgsfield{j}}
      \big(z,k_\perp^\mu,\bar p^\mu,\polvb\big)
  \bigg]^{\mu \nu}
  =
  \bigg[
    \mathcal{P}
      {}^{\vbfield{ij^{\vphantom{\prime}}}}_{\higgsfield{j}}
      {}^{\vbfield{ij^\prime}}_{\vbfield{i}}
      \big(1-z,-k_\perp^\mu,\bar p^\mu,\polvb\big)
  \bigg]^{\mu \nu}
  _{m_i \leftrightarrow m_j}
  .
\end{align}

\paragraph{Polarization \boldmath{$\polvb_j=\mathrm{T}$}.}
The transverse-polarization result
\begin{align}
  &
  \bigg[
    \mathcal{P}
      {}^{\vbfield{ij^{\vphantom{\prime}}}}_{\higgsfield{i}}
      {}^{\vbfield{ij^\prime}}_{\vbfield{j}}
      \big(z,k_\perp^\mu,\bar p^\mu,\mathrm{T}\big)
  \bigg]^{\mu \nu}
  \nonumber\\*
  &=
  - 4 m_j^2
  I_{\higgsfield{i} \gbantifield{j}}^{\vbfield{ij}}
  I_{\higgsfield{i} \gbantifield{j}}^{\vbfield{ij^\prime} \ast}
  \bigg\{
    g^{\mu \nu }
    +
    \frac{
      \bar{p}^{\nu } k_{\perp }^{\mu }
    }{(z-1) m_{ij}^{\prime\, 2}}
    +
    \frac{
      \bar{p}^{\mu } k_{\perp }^{\nu }
    }{(z-1) m_{ij}^2}
    +
    \frac{k_\perp^2}{(z-1)^2 m_{ij}^2 m_{ij}^{\prime\, 2}}
    \bar{p}^{\mu } \bar{p}^{\nu }
  \bigg\}
  \label{eq:VHV-T-split-func}
\end{align}
has well-defined limits~$m_{ij} \to 0$ and~$m_{ij} \to 0$ due
to \gls{wi}~\eqref{eq:VVV-ward-id-ul-me} and coupling
relation~\eqref{eq:coupling-relat-VSG-VSG}.
In the \gls{ewsm}, we have $m_{ij}=m_{ij}^\prime=m_j$ and get the result
\begin{align}
  &
  \bigg[
    \mathcal{P}
      {}^{\vbfield{ij^{\vphantom{\prime}}}}_{H}
      {}^{\vbfield{ij^\prime}}_{\vbfield{j}}
      \big(z,k_\perp^\mu,\bar p^\mu,\mathrm{T}\big)
    \bigg]^{\mu \nu}_\mathrm{EWSM}
  \nonumber\\*
  &=
  -
  \frac{
    4
  }{(e v)^2}
  \delta_{\vbfield{ij} \vbfield{j}}
  \delta_{\vbfield{ij^\prime} \vbfield{j}}
  \bigg\{
    m_j^4
    g^{\mu \nu }
    +
    \frac{m_j^2  }{z-1}
    \big(
      \bar{p}^{\mu } k_{\perp }^{\nu } 
      +
      \bar{p}^{\nu } k_{\perp }^{\mu } 
    \big)
    +
    \frac{k_\perp^2}{(z-1)^2}
    \bar{p}^{\mu } \bar{p}^{\nu } 
  \bigg\}
  .
  \label{eq:VHV-T-split-func-EWSM}
\end{align}

\paragraph{Polarization \boldmath{$\polvb_j=\mathrm{L}$}.}
The result for longitudinal polarization reads
\begin{alignat}{2}
  &
  \lefteqn{
    \bigg[
      \mathcal{P}
        {}^{\vbfield{ij^{\vphantom{\prime}}}}_{\higgsfield{i}}
        {}^{\vbfield{ij^\prime}}_{\vbfield{j}}
        \big(z,k_\perp^\mu,\bar p^\mu,\mathrm{L}\big)
    \bigg]^{\mu \nu}
  }
  \nonumber\\*
  &={}
  &
  I_{\higgsfield{i} \gbantifield{j}}^{\vbfield{ij}}
  I_{\higgsfield{i} \gbantifield{j}}^{\vbfield{ij^\prime} \ast}
    &
    \bigg\{
      \bigg(
        \rho (m_{ij},m_j,m_i)
        +
        \frac{2 m_j^2}{(z-1) m_{ij}^2}
        +
        2 (z-1)
      \bigg)
      \bar{p}^{\mu }
      +
      2 k_{\perp }^{\mu } 
    \bigg\}
    \nonumber\\
    &&
    \cdot
    &
    \bigg\{
      \bigg(
        \rho (m_{ij}^\prime,m_j,m_i)
        +
        \frac{2 m_j^2}{(z-1) m_{ij}^{\prime\, 2}}
        +
        2 (z-1)
      \bigg)
      \bar{p}^{\nu }
      +
      2 k_{\perp }^{\nu }
    \bigg\}
    .
  \label{eq:VHV-L-split-func}
\end{alignat}
The limits~$m_{ij} \to 0$ and~$m_{ij} \to 0$ behave as described as
for~$\polvb_j=\mathrm{T}$.
In the limit~$m_j \to 0$, this result reproduces the~$V^\ast \to H H$
splitting function when taking~$\gbfield{j} \to \higgsfield{j}$,
\begin{align}
  \bigg[
    \mathcal{P}
      {}^{\vbfield{ij^{\vphantom{\prime}}}}_{\higgsfield{i}}
      {}^{\vbfield{ij^\prime}}_{\vbfield{j}}
      \big(z,k_\perp^\mu,\bar p^\mu,\mathrm{L}\big)
  \bigg]^{\mu \nu}
  \xrightarrow[m_j \to 0]{\gbfield{j} \to \higgsfield{j}}
  \bigg[
    \mathcal{P}
      {}^{\vbfield{ij^{\vphantom{\prime}}}}_{\higgsfield{i}}
      {}^{\vbfield{ij^\prime}}_{\higgsfield{j}}
      \big(z,k_\perp^\mu,\bar p^\mu\big)
    \bigg]^{\mu \nu}_{m_j = 0}
    .
\end{align}
The \gls{ewsm} result reads
\begin{alignat}{2}
  \bigg[
    \mathcal{P}
      {}^{\vbfield{ij^{\vphantom{\prime}}}}_{H}
      {}^{\vbfield{ij^\prime}}_{\vbfield{j}}
      \big(z,k_\perp^\mu,\bar p^\mu,\mathrm{L}\big)
    \bigg]^{\mu \nu}_\mathrm{EWSM}
  ={}&
  \frac{
    1
  }{(e v)^2}
  \delta_{\vbfield{ij} \vbfield{j}}
  \delta_{\vbfield{ij^\prime} \vbfield{j}}
  \bigg\{
    2
    m_j
    k_{\perp }^{\mu }
    +
    \bigg(
      2
      \bigg(z+\frac{1}{z-1}\bigg)
      m_j^2
      -
      m_i^2
    \bigg)
    \frac{\bar{p}^{\mu }}{m_j}
  \bigg\}
  \nonumber\\
  &
  \cdot
  \bigg\{
    2
    m_j
    k_{\perp }^{\mu }
    +
    \bigg(
      2
      \bigg(z+\frac{1}{z-1}\bigg)
      m_j^2
      -
      m_i^2
    \bigg)
    \frac{\bar{p}^{\mu }}{m_j}
  \bigg\}
  .
  \label{eq:vhv-L-split-func-EWSM}
\end{alignat}

%
\subsubsection[%
  \mathinhead{H^\ast \to HV}{H* to H V}
  splitting functions%
  ]{%
  \mathinheadbold{H^\ast \to HV}{H* to H V}
  splitting functions}
\label{sect:HHV}
To compute the splitting functions~$H^\ast \to H V$, we need the diagrams
\drawmomtrue%
\drawvertdottrue%
\begin{alignat}{2}%
  \ftntmath{%
    \MContrExplFields{\higgsfield{ij}}{\higgsfield{i}}{\vbfield{j}}^{\polvb_j}%
  }%
  &\ftntmath{{}={}}%
  \FSDiagGeo%
  \ftntmath{%
    \splittingdiagIndivFS{T^{(n)}_{\higgsfield{ij} X}}%
      {scalar}{scalar}{photon}%
      {\higgsfield{ij}}{\higgsfield{i}}{\vbfield{j}, \polvb_j}%
    ,%
  }%
  \resetdiagargs%
  &\hspace{3mm}
  \ftntmath{%
    \MContrExplFields{\higgsfield{ij}}{\higgsfield{i}}{\gbfield{j}}^{\mathrm{G}}%
  }%
  &\ftntmath{{}={}}%
  \FSDiagGeo%
  \ftntmath{%
    \splittingdiagIndivFS{T^{(n)}_{\higgsfield{ij} X}}%
      {scalar}{scalar}{scalar}%
      {\higgsfield{ij}}{\higgsfield{i}}{\gbfield{j}}%
    ,%
  }%
  \resetdiagargs%
  \label{eq:HHV-diags-FS}
\end{alignat}%
translating into the expressions
\begin{align}
  %
  %
  \MContrExplFields{\higgsfield{ij}}{\higgsfield{i}}{\vbfield{j}}^{\polvb_j}
  ={}&
  - \imagi e
  I_{\higgsantifield{ij} \higgsantifield{i}}^{\vbantifield{j}}
  \frac{\imagi}{p_{ij}^2-m_{ij}^2}
  T_{\higgsfield{ij} X}^{(n)}\left(p_{ij}\right) 
  \varepsilon_{\polvb_j}^{\gamma  \ast}(p_j)
  \left(2p_i+p_j\right)_{\gamma }
  ,
  \nonumber\\
  %
  %
  \MContrExplFields{\higgsfield{ij}}{\higgsfield{i}}{\gbfield{j}}^{\mathrm{G}}
  ={}&
  - \imagi e
  I_{\higgsantifield{ij} \higgsantifield{i}}^{\vbantifield{j}}
  \frac{m_{ij}^2 - m_i^2}{m_j \eta_{\vbfield{j}}^\ast}
  \frac{\imagi}{p_{ij}^2-m_{ij}^2}
  T_{\higgsfield{ij} X}^{(n)}\left(p_{ij}\right) 
  ,
  \label{eq:HHV-EnhAmp}
\end{align}
where we used coupling relation~\eqref{eq:coupling-relat-VHH-GHH}.
In the \gls{ewsm}, the couplings between two Higgs bosons and one vector boson vanish.
Therefore, the following splitting functions are absent in the \gls{ewsm}.

\paragraph{Polarization \boldmath{$\polvb_j=\mathrm{T}$}.}
For a transversely polarized vector boson, we find
\begin{align}
  &
  \mathcal{P}
    {}^{\higgsfield{ij^{\vphantom{\prime}}}}_{\higgsfield{i}}
    {}^{\higgsfield{ij^\prime}}_{\vbfield{j}}
    \big(z,k_\perp^\mu,\bar p^\mu,\mathrm{T}\big)
  =
  - \frac{4 k_\perp^2}{(z-1)^2}
  I_{\higgsantifield{ij} \higgsantifield{i}}^{\vbantifield{j}}
  I_{\higgsantifield{ij^\prime} \higgsantifield{i}}^{\vbantifield{j} \ast}
  .
  \label{eq:HHV-T-split-func}
\end{align}

\paragraph{Polarization \boldmath{$\polvb_j=\mathrm{L}$}.}
In the longitudinal case, we get
\begin{align}
  \mathcal{P}
    {}^{\higgsfield{ij^{\vphantom{\prime}}}}_{\higgsfield{i}}
    {}^{\higgsfield{ij^\prime}}_{\vbfield{j}}
    \big(z,k_\perp^\mu,\bar p^\mu,\mathrm{L}\big)
  =
  m_j^2
  I_{\higgsantifield{ij} \higgsantifield{i}}^{\vbantifield{j}}
  I_{\higgsantifield{ij^\prime} \higgsantifield{i}}^{\vbantifield{j} \ast}
  \bigg\{
    \rho(m_j,m_{ij},m_i) + \frac{2}{z-1}
  \bigg\}
  \bigg\{
    \rho(m_j,m_{ij}^\prime,m_i) + \frac{2}{z-1}
  \bigg\}
  .
  \label{eq:HHV-L-split-func}
\end{align}
Taking~$m_j \to 0$ and~$\gbfield{j} \to \higgsfield{j}$, we obtain
\begin{align}
  \mathcal{P}
    {}^{\higgsfield{ij^{\vphantom{\prime}}}}_{\higgsfield{i}}
    {}^{\higgsfield{ij^\prime}}_{\vbfield{j}}
    \big(z,k_\perp^\mu,\bar p^\mu,\mathrm{L}\big)
  \xrightarrow[m_j \to 0]{\gbfield{j} \to \higgsfield{j}}
  \mathcal{P}
    {}^{\higgsfield{ij^{\vphantom{\prime}}}}_{\higgsfield{i}}
    {}^{\higgsfield{ij^\prime}}_{\higgsfield{j}}
    \big(z,k_\perp^\mu,\bar p^\mu\big)
  .
\end{align}

%
\subsubsection[%
  \mathinhead{V^\ast/H^\ast \to HV}{V*/H* to H V}
  splitting functions%
  ]{%
  \mathinheadbold{V^\ast/H^\ast \to HV}{V*/H* to H V}
  splitting functions}
\label{sect:VHHV}
The functions for the interference of the splittings
$V^\ast/H^\ast \to H V$ are computed with
the ingredients from Sections~\ref{sect:VHV} and~\ref{sect:HHV}.
They vanish in the \gls{ewsm}, and
the various massless limits behave as discussed above.
%
%
The case $V^\ast/H^\ast \to V H$ is obtained via application
of the $i\leftrightarrow j$ symmetry relation~\eqref{eq:i-j-symmru-splfunc}
\begin{align}
  \bigg[
    \mathcal{P}
      {}^{\vbfield{ij^{\vphantom{\prime}}}}_{\vbfield{i}}
      {}^{\higgsfield{ij^\prime}}_{\higgsfield{j}}
      \big(z,k_\perp^\mu,\bar p^\mu,\mathrm{T}\big)
  \bigg]^{\mu}
  =
  \bigg[
    \mathcal{P}
      {}^{\vbfield{ij^{\vphantom{\prime}}}}_{\higgsfield{j}}
      {}^{\higgsfield{ij^\prime}}_{\vbfield{i}}
      \big(1-z,-k_\perp^\mu,\bar p^\mu,\mathrm{T}\big)
  \bigg]^{\mu}
  _{m_i \leftrightarrow m_j, \eta_i \leftrightarrow \eta_j}
  .
\end{align}
Note that we do not only swap masses, but also phases~$\eta$.

\paragraph{Polarization \boldmath{$\polvb_j=\mathrm{T}$}.}
\begin{align}
  &
  \bigg[
    \mathcal{P}
      {}^{\vbfield{ij^{\vphantom{\prime}}}}_{\higgsfield{i}}
      {}^{\higgsfield{ij^\prime}}_{\vbfield{j}}
      \big(z,k_\perp^\mu,\bar p^\mu,\mathrm{T}\big)
  \bigg]^{\mu}
  =
  - 4
  m_j \eta_{\vbfield{j}}^\ast
  I_{\higgsfield{i} \gbantifield{j}}^{\vbfield{ij}}
  I_{\higgsantifield{ij^\prime} \higgsantifield{i}}^{\vbantifield{j} \ast}
  \bigg\{
    \frac{k_\perp^2}{(z-1)^2 m_{ij}^2}
    \bar{p}^{\mu }
    +
    \frac{1}{z-1}
    k_{\perp }^{\mu }
  \bigg\}
  .
  \label{eq:VHHV-T-split-func}
\end{align}

\paragraph{Polarization \boldmath{$\polvb_j=\mathrm{L}$}.}
\begin{align}
    \bigg[
      \mathcal{P}
        {}^{\vbfield{ij^{\vphantom{\prime}}}}_{\higgsfield{i}}
        {}^{\higgsfield{ij^\prime}}_{\vbfield{j}}
        \big(z,k_\perp^\mu,\bar p^\mu,\mathrm{L}\big)
    \bigg]^{\mu}
  ={}&
  m_j \eta_{\vbfield{j}}^\ast
  I_{\higgsfield{i} \gbantifield{j}}^{\vbfield{ij}}
  I_{\higgsantifield{ij^\prime} \higgsantifield{i}}^{\vbantifield{j} \ast}
    \,\bigg\{
      \rho(m_j,m_{ij}^\prime,m_i) + \frac{2}{z-1}
    \bigg\}
    \nonumber\\
    & {} \cdot
    \bigg\{
      \bigg(
        \rho (m_{ij},m_j,m_i)
        +
        \frac{2 m_j^2}{(z-1) m_{ij}^2}
        +
        2 (z-1)
      \bigg)
      \bar{p}^{\mu }
      +
      2 k_{\perp }^{\mu } 
    \bigg\}
    .
    \label{eq:VHHV-L-split-func}
\end{align}

%
\subsubsection[%
  \mathinhead{V^\ast \to HH}{V* to H H}
  splitting function%
  ]{%
  \mathinheadbold{V^\ast \to HH}{V* to H H}
  splitting function}
\label{sect:VHH}
The diagrams
\drawmomtrue%
\drawvertdottrue%
\begin{alignat}{2}%
  \ftntmath{%
    \MContrExplFields{\vbfield{ij}}{\higgsfield{i}}{\higgsfield{j}}%
  }%
  &\ftntmath{{}={}}%
  \FSDiagGeo%
  \ftntmath{%
    \splittingdiagIndivFS{T^{(n)}_{\vbfield{ij} X}}%
      {photon}{scalar}{scalar}%
      {\vbfield{ij}}{\higgsfield{i}}{\higgsfield{j}}%
    ,%
  }%
  \resetdiagargs%
  &\quad
  \ftntmath{%
    \MContrExplFields{\gbfield{ij}}{\higgsfield{i}}{\higgsfield{j}}%
  }%
  &\ftntmath{{}={}}%
  \FSDiagGeo%
  \ftntmath{%
    \splittingdiagIndivFS{T^{(n)}_{\gbfield{ij} X}}%
      {scalar}{scalar}{scalar}%
      {\gbfield{ij}}{\higgsfield{i}}{\higgsfield{j}}%
  }%
  \resetdiagargs%
  \label{eq:VHH-diags-FS}
\end{alignat}%
translate, after inserting the coupling
relation~\eqref{eq:coupling-relat-VHH-GHH}, into the expressions
\begin{align}
  %
  %
  \MContrExplFields{\vbfield{ij}}{\higgsfield{i}}{\higgsfield{j}}
  ={}&
  -\imagi e
  I_{\higgsfield{i} \higgsantifield{j}}^{\vbfield{ij}}
  T^{(n)\, \sigma}_{\vbfield{ij} X}\left(p_{ij}\right)
  G^{\vbantifield{ij} \vbfield{ij}}_{\sigma \alpha}(p_{ij},-p_{ij})
  \left(p_j-p_i\right)^{\alpha }
  ,
  \nonumber\\
  %
  %
  \MContrExplFields{\gbfield{ij}}{\higgsfield{i}}{\higgsfield{j}}
  ={}&
  \imagi e
  I_{\higgsfield{i} \higgsantifield{j}}^{\vbfield{ij}}
  \frac{m_i^2-m_j^2}{m_{ij}^2}
  T^{(n)\, \alpha}_{\vbfield{ij} X}\left(p_{ij}\right)
  G^{\gbantifield{ij} \gbfield{ij}}(p_{ij},-p_{ij})
  p_{ij,\alpha }
  .
  \label{eq:VHH-EnhAmp}
\end{align}
We obtain the $V \to H H$ splitting function
\begin{align}
  &
  \bigg[
    \mathcal{P}
      {}^{\vbfield{ij^{\vphantom{\prime}}}}_{\higgsfield{i}}
      {}^{\vbfield{ij^\prime}}_{\higgsfield{j}}
      \big(z,k_\perp^\mu,\bar p^\mu\big)
  \bigg]^{\mu \nu}
  \nonumber\\*
  &=
  I_{\higgsfield{i} \higgsantifield{j}}^{\vbfield{ij}}
  I_{\higgsfield{i} \higgsantifield{j}}^{\vbfield{ij^\prime} \ast}
  \big\{
    \big(2 z-\rho (m_{ij},m_i,m_j)\big)
    \bar{p}^{\mu }
    +
    2 k_{\perp }^{\mu }
  \big\}
  \big\{
    \big(2 z-\rho (m_{ij}^{\prime },m_i,m_j)\big)
    \bar{p}^{\nu }
    +
    2 k_{\perp }^{\nu }
  \big\}
  .
  \label{eq:VHH-split-func}
\end{align}
The limits~$m_{ij} \to 0$ and~$m_{ij}^\prime \to 0$ are well-defined due
to \gls{wi}~\eqref{eq:VVV-ward-id-ul-me}
and coupling relation~\eqref{eq:coupling-relat-VHH-GHH}.
Again, this splitting function does not exist in the \gls{ewsm}.

%
\subsubsection[%
  \mathinhead{H^\ast \to HH}{H* to H H}
  splitting function%
  ]{%
  \mathinheadbold{H^\ast \to HH}{H* to H H}
  splitting function}
\label{sect:HHH}
The splitting function $H^\ast \to H H$ is trivially obtained as
\begin{align}
  \mathcal{P}
    {}^{\higgsfield{ij^{\vphantom{\prime}}}}_{\higgsfield{i}}
    {}^{\higgsfield{ij^\prime}}_{\higgsfield{j}}
    \big(z,k_\perp^\mu,\bar p^\mu\big)
  =  
  C_{\higgsfield{ij} \higgsfield{i} \higgsfield{j}}
  C^\ast_{\higgsfield{ij^\prime} \higgsfield{i} \higgsfield{j}}
  .
  \label{eq:HHH-split-func}
\end{align}
Note that dimensional analysis of the Lagrangian yields for the
pure-scalar coupling
\begin{align}
  C_{H_{ij} H_{i} H_{j}}=\orderof{m}
  .
\end{align}
This implies that the above expression is of~$\orderof{m^2}$, just like
all other splitting functions.
Inserting the \gls{ewsm} triple-Higgs coupling~\eqref{eq:app-HHH-cpl}, we get
\begin{align}
  \mathcal{P}
    {}^{\higgsfield{}}_{\higgsfield{}}
    {}^{\higgsfield{}}_{\higgsfield{}}
    \big(z,k_\perp^\mu,\bar p^\mu\big)
  \bigg|_\mathrm{EWSM}
  =  
  \frac{9 m_i^4}{(e v)^2}
  .
  \label{eq:HHH-split-func-EWSM}
\end{align}

%
\subsubsection[%
  \mathinhead{V^\ast/H^\ast \to HH}{V*/H* to H H}
  splitting function%
  ]{%
  \mathinheadbold{V^\ast/H^\ast \to HH}{V*/H* to H H}
  splitting function}
\label{sect:VHHH}
Combining the results of Sections~\ref{sect:VHH} and~\ref{sect:HHH}, we get
\begin{align}
  &
  \bigg[
    \mathcal{P}
      {}^{\vbfield{ij^{\vphantom{\prime}}}}_{\higgsfield{i}}
      {}^{\higgsfield{ij^\prime}}_{\higgsfield{j}}
      \big(z,k_\perp^\mu,\bar p^\mu\big)
  \bigg]^{\mu}
  =
  -
  I_{\higgsfield{i} \higgsantifield{j}}^{\vbfield{ij}}
  C^\ast_{\higgsfield{ij^\prime} \higgsfield{i} \higgsfield{j}}
  \big\{
    \big(2 z-\rho (m_{ij},m_i,m_j)\big)
    \bar{p}^{\mu }
    +
    2 k_{\perp }^{\mu }
  \big\}
  .
  \label{eq:VHHH-split-func}
\end{align}
This splitting function again vanishes in the~\gls{ewsm}.

\subsection{Fermionic FS splitting functions}
\label{sect:expl-splfuncs-FS-fermionic}
Next, we switch to the derivation of \gls{fs} splitting functions that involve \emph{fermions}.
On top of the symmetries from Section~\ref{sect:splfunc-symm}, these
results are subject to relations under charge conjugation,
which is discussed in some detail below.
Again, splitting functions that are not considered explicitly may be
obtained via symmetry relations and charge conjugation.

%
\subsubsection[%
  \mathinhead{V^\ast \to f \bar f}{V* to f fbar}
  splitting functions%
  ]{%
  \mathinheadbold{V^\ast \to f \bar f}{V* to f fbar}
  splitting functions}
\label{sect:VFFbar}
The relevant diagrams for the computation of the $V^\ast \to f \bar f$
splitting functions read
\drawmomtrue%
\drawvertdottrue%
\begin{alignat}{2}%
  \ftntmath{%
    \MContrExplFields{\vbfield{ij}}{\fermfield{i}}{\antifermfield{j}}^{\polferm_i \polferm_j}%
  }%
  &\ftntmath{{}={}}%
  \FSDiagGeo%
  \ftntmath{%
    \splittingdiagIndivFS{T^{(n)}_{\vbfield{ij} X}}%
      {photon}{fermion}{antifermion}%
      {\vbfield{ij}}{\fermfield{i}, \polferm_i}{\antifermfield{j}, \polferm_j}%
    ,%
  }%
  \resetdiagargs%
  &\quad
  \ftntmath{%
    \MContrExplFields{\gbfield{ij}}{\fermfield{i}}{\antifermfield{j}}^{\polferm_i \polferm_j}%
  }%
  &\ftntmath{{}={}}%
  \FSDiagGeo%
  \ftntmath{%
    \splittingdiagIndivFS{T^{(n)}_{\gbfield{ij} X}}%
      {scalar}{fermion}{antifermion}%
      {\gbfield{ij}}{\fermfield{i}, \polferm_i}{\antifermfield{j}, \polferm_j}%
    ,%
  }%
  \resetdiagargs%
  \label{eq:VFFbar-diags-FS}
\end{alignat}%
and correspond, upon insertion of~\eqref{eq:coupling-relat-Vff-Gff}, to
the expressions
\begin{align}
  %
  %
  \MContrExplFields{\vbfield{ij}}{\fermfield{i}}{\antifermfield{j}}^{\polferm_i \polferm_j}
  ={}&
  - \imagi e
  T^{(n)\, \alpha }_{\vbfield{ij} X}(p_{ij})
  G^{\vbantifield{ij} \vbfield{ij}}_{\alpha \beta}(p_{ij},-p_{ij})
  \bar{u}_{\polferm_i}(p_i)
  \Big(
    \gamma ^{\beta }
    \omega_-
    I_{\fermfield{i}^\mathrm{L} \fermfield{j}^\mathrm{L}}^{\vbfield{ij}}
    +
    \gamma ^{\beta }
    \omega_+ 
    I_{\fermfield{i}^\mathrm{R} \fermfield{j}^\mathrm{R}}^{\vbfield{ij}}
  \Big)
  v_{\polferm_j}(p_j)
  ,
  \nonumber\\
  %
  %
  \MContrExplFields{\gbfield{ij}}{\fermfield{i}}{\antifermfield{j}}^{\polferm_i \polferm_j}
  ={}&
  - \imagi e
  T^{(n)\, \alpha}_{\vbfield{ij} X}(p_{ij})
  p_{ij,\alpha}
  G^{\gbantifield{ij} \gbfield{ij}}(p_{ij},-p_{ij})
\, \frac{1}{m_{ij}^2}
  \nonumber\\
  &
  \cdot
  \bar{u}_{\polferm_i}(p_i)
  \left\{
    \omega_+
    \left(
      m_i I_{\fermfield{i}^\mathrm{R} \fermfield{j}^\mathrm{R}}^{\vbfield{ij}}
      -
      m_j I_{\fermfield{i}^\mathrm{L} \fermfield{j}^\mathrm{L}}^{\vbfield{ij}}
    \right)
    +
    \omega_-
    \left(
      m_i I_{\fermfield{i}^\mathrm{L} \fermfield{j}^\mathrm{L}}^{\vbfield{ij}}
      -
      m_j I_{\fermfield{i}^\mathrm{R} \fermfield{j}^\mathrm{R}}^{\vbfield{ij}}
    \right)
  \right\}
  v_{\polferm_j}(p_j)
  .
  \label{eq:VFFbar-EnhAmp}
\end{align}
The projectors~$\omega_\pm$ onto right- and left-handed chirality,
respectively, are defined in~\eqref{eq:chir-proj-op}.
Combining the above equations leads us to
\begin{alignat}{2}  
\lefteqn{  \MContrExplFields
    {\vbfield{ij},\mathrm{tHF}}
    {\fermfield{i}}
    {\antifermfield{j}}^{\polferm_i \polferm_j}
  +
  \MContrExplFields
    {\gbfield{ij},\mathrm{tHF}}
    {\fermfield{i}}
    {\antifermfield{j}}^{\polferm_i \polferm_j} }
\phantom{{}={}}
  \nonumber\\*
  ={}&
  \frac{1}{2}
  \frac{e T^{(n)\, \mu}_{\vbfield{ij} X}(p_{ij})}{ p_{ij}^2 - m_{ij}^2 }
  \bigg\{
  - \frac{m_i+m_j}{m_{ij}^2}
  p_{ij,\mu} 
  \bar{u}_{\polferm_i}(p_i) \gamma^5 v_{\polferm_j}(p_j)
  \Big(
    I_{\fermfield{i}^\mathrm{L} \fermfield{j}^\mathrm{L}}^{\vbfield{ij}}
    -
    I_{\fermfield{i}^\mathrm{R} \fermfield{j}^\mathrm{R}}^{\vbfield{ij}}
  \Big)
  \nonumber\\
  & \quad {} + \frac{m_i-m_j}{m_{ij}^2}
  p_{ij, \mu} 
  \bar{u}_{\polferm_i}(p_i) v_{\polferm_j}(p_j)
  \Big(
    I_{\fermfield{i}^\mathrm{L} \fermfield{j}^\mathrm{L}}^{\vbfield{ij}}
    +
    I_{\fermfield{i}^\mathrm{R} \fermfield{j}^\mathrm{R}}^{\vbfield{ij}}
  \Big)
  \nonumber\\
  & \quad {} - \bar{u}_{\polferm_i}(p_i)
  \gamma_\mu v_{\polferm_j}(p_j)
  \Big(
    I_{\fermfield{i}^\mathrm{L} \fermfield{j}^\mathrm{L}}^{\vbfield{ij}}
    +
    I_{\fermfield{i}^\mathrm{R} \fermfield{j}^\mathrm{R}}^{\vbfield{ij}}
  \Big)
+ \bar{u}_{\polferm_i}(p_i)
  \gamma_\mu \gamma^5 v_{\polferm_j}(p_j)
  \Big(
    I_{\fermfield{i}^\mathrm{L} \fermfield{j}^\mathrm{L}}^{\vbfield{ij}}
    -
    I_{\fermfield{i}^\mathrm{R} \fermfield{j}^\mathrm{R}}^{\vbfield{ij}}
  \Big)
  \bigg\}
  .
  \label{eq:VFF-indiv-amp-fullL}
\end{alignat}
After squaring this expression, we plug in the spin-projected
completeness relations~\eqref{eq:spinproj-complrelat-ferm}, keeping the
polarization-defining gauge vectors $\restpol^\mu_i$ and $\restpol^\mu_j$ separate.
We again stress that we are not restricted to helicity in some
specific frame, as one might intrinsically assume when
employing explicit parametrizations of spinors.\footnotemark\
\footnotetext{%
  Recall that helicity is not a Lorentz-invariant property.
}%
Specifically, given some reference frame, helicity is recovered with
$\restpol^\mu$ that fulfills $\boldsymbol{\restpol}\parallel\boldsymbol{p}$.
Despite the similarity of the vector~$s^\mu$
in~\eqref{eq:spin-projection-op} with the longitudinal polarization
vector~\eqref{eq:long-pol-vec}, its mass-singular factor does not
require special care.
Indeed, it is already fully cancelled
in~\eqref{eq:spinproj-complrelat-ferm} due to~$\slashed p^2=p^2=m_f^2$.


We are left with evaluating Dirac traces.
Apart from traces with even and odd numbers of slashed four-vectors,
which can be straightforwardly evaluated as chains of scalar products,
this also involves traces with~$\gamma^5$.
As we work in~$D=4$ dimensions, there are no problems regarding the
definition of~$\gamma^5$.
However, the evaluation of the latter type of traces requires the
contraction of four-vectors with the antisymmetric
tensor~$\epsilon^{\mu\nu\sigma\rho}$.
For example, with the four-momenta~$p_1^\mu,p_2^\mu,p_3^\mu$, and
$p_4^\mu$, we get
\begin{align}
  \mathrm{Tr}
  \{\slashed p_1 \slashed p_2 \slashed p_3 \slashed p_4 \gamma^5\}
  =
  - 4 \imagi \epsilon_{\mu\nu\sigma\rho} p_1^\mu p_2^\nu p_3^\sigma p_4^\rho
  .
  \label{eq:example-gam5-epsilon-trace}
\end{align}
To facilitate the evaluation of the mass order of these expressions, we
decompose such contractions into individual scalar products, as
described in the following.
A similar procedure has been employed in~\citep{dittmaier2008polarized}
for the evaluation of \gls{is}~$f \to f \gamma^\ast$ splitting
functions.
We consistently use the convention~$\epsilon^{0123}=1$.
First, we write
\begin{align}
  \epsilon_{\mu\nu\sigma\rho}
  p_1^\mu p_2^\nu p_3^\sigma p_4^\rho
  =
  \epsilon^{\mu^\prime\nu^\prime\sigma^\prime\rho^\prime}
  g_{\mu^\prime \mu}
  g_{\nu^\prime \nu}
  g_{\sigma^\prime \sigma}
  g_{\rho^\prime \rho}
  p_1^\mu p_2^\nu p_3^\sigma p_4^\rho
  \label{eq:eps-tens-replace}
\end{align}
and express the metric tensor as an outer product of four-vectors
\begin{align}
  g^{\mu^\prime \mu}
  =
  -
  \frac{\tilde m_{ij}^2 n^{\mu^\prime} n^{\mu}}{(\bar p n)^2}
  +
  \frac{\bar p^{\mu^\prime} n^{\mu} + n^{\mu^\prime} \bar p^{\mu}}{\bar p n}
  -
  s_\varepsilon
  \big(
    \bar \varepsilon_+^{\mu^\prime}
    \bar \varepsilon_-^\mu
    +
    \bar \varepsilon_-^{\mu^\prime}
    \bar \varepsilon_+^\mu
  \big)
  ,
  \label{eq:gmunu-decomp}
\end{align}
where $\bar p^2=\tilde m_{ij}^2$ (see comment after (\ref{eq:notation-stripped-ubornme-antifermion})).
Here, we have taken~$\bar p^\mu$ and~$n^\mu$
from~\eqref{eq:sudakov-param-fs} and defined 
polarization vectors
$\bar\varepsilon_\polvb^\mu = \varepsilon_\polvb^\mu(\bar p)$
with $\polvb=\pm$
for the momentum~$\bar p^\mu$,
fulfilling both
\begin{align}
  \bar p \bar \varepsilon_\polvb = 0
  ,\qquad
  \bar \varepsilon_{-\polvb}
  =
  s_{\varepsilon}^\ast
  \bar \varepsilon_\polvb^\ast
  ,\qquad
  \bar \varepsilon_{\polvb} \bar \varepsilon_{-\polvb^\prime}
  =
  s_{\varepsilon}^\ast \bar \varepsilon_{\polvb} \bar \varepsilon_{\polvb^\prime}^\ast
  =
  - s_{\varepsilon}^\ast \delta_{\polvb \polvb^\prime}
  ,
  \label{eq:eps-bar-conditions}
\end{align}
and the gauge condition
\begin{align}
  n \bar \varepsilon_\polvb = 0
  .
  \label{eq:eps-bar-gauge-condition}
\end{align}
We support arbitrary phases~$s_\varepsilon$ to account for different
conventions used in the literature.\footnotemark\
\footnotetext{%
  For example,~\cite{Cuomo:2019siu} works with the
  convention~$s_\varepsilon=-1$, while~\cite{dittmaier2008polarized}
  implements~$s_\varepsilon=+1$.
}%
The polarization vectors~$\bar \varepsilon_\polvb^\mu$ 
define transverse \emph{helicity states} in
frames where $\bar{\boldsymbol{p}} \parallel \boldsymbol n$.
In any frame,~$k_\perp^\mu$ can be expressed as linear combination
of $\bar \varepsilon_\pm^\mu$,
\begin{align}
  k_\perp^\mu
  =
  -
  s_\varepsilon
  (k_\perp \bar \varepsilon_+)
  \bar \varepsilon_-^\mu
  -
  s_\varepsilon
  (k_\perp \bar \varepsilon_-)
  \bar \varepsilon_+^\mu
  ,
  \qquad
  k_\perp^2
  =
  -
  2
  s_\varepsilon
  (k_\perp \bar \varepsilon_+)
  (k_\perp \bar \varepsilon_-)
  .
  \label{eq:kt-from-eps-via-sp}
\end{align}
Inserting the decomposition~\eqref{eq:gmunu-decomp}
into~\eqref{eq:eps-tens-replace}, only one non-zero contraction
of~$\varepsilon^{\mu\nu\rho\sigma}$ with four-vectors remains, which can
be expressed explicitly as
\begin{align}
  \epsilon_{\mu\nu\rho\sigma}
  \bar p^\mu n^\nu
  \bar \varepsilon_-^\rho
  \bar \varepsilon_+^\sigma
  =
  \imagi s_\varepsilon^\ast \bar p n
  .
  \label{eq:remaining-eps-tens-contraction}
\end{align}
This Lorentz-invariant expression can be derived from
\begin{align}
\epsilon^{\mu\nu\rho\sigma} \bar p_\mu  n_\nu   
= \imagi s_\varepsilon (\bar pn) 
\left( \bar\varepsilon_+^\rho \bar\varepsilon_-^\sigma
      -\bar\varepsilon_-^\rho \bar\varepsilon_+^\sigma \right),
\label{eq:eps_barp_n}
\end{align}
which follows from an analogous manipulation as described in Footnote~6
of \cite{dittmaier2008polarized}
(although with massive momentum $\bar{p}^\mu$), 
or more directly
in a frame where
\begin{align}
  n^\mu
  =
  \left(1,0,0,-1\right)
  ,\quad
  \bar p^\mu
  =
  \left(\sqrt{m_{ij}^2 + |\bar{\boldsymbol{p}}|^2},0,0,|\bar{\boldsymbol{p}}|\right)
  ,\quad
  \bar{\varepsilon}_\pm^\mu
  =
  s_\pm
  \left(0,\frac{1}{\sqrt{2}},\frac{\pm \imagi}{\sqrt{2}},0\right)
  ,
  \label{eq:frame-n-parr-pb}
\end{align}
with $s_+ s_- = s_\varepsilon^\ast$.
Henceforth, the tensor~$\epsilon^{\mu\nu\sigma\rho}$ is completely
eliminated from the r.h.s.\ of~\eqref{eq:eps-tens-replace}.


Finally, we insert the Sudakov
parametrization~\eqref{eq:sudakov-param-fs} and neglect terms of higher
orders in the quasi-collinear rescaling parameter~$\lambda$.
Once more, we only assume~\eqref{eq:oc-summary} in this procedure and
impose no further restrictions on the gauge vectors $\restpol^\mu$.
We arrive at the splitting functions presented below.
As anticipated, these results are independent of the gauge vectors and
thereby of the precise definition of polarization.


It turns out that all types of \gls{fs} splitting functions involving
two (anti-)fermions in the corresponding splitting
process from~\eqref{eq:spl-proc} are built from similar analytic structures.
Therefore, it is convenient to introduce appropriate abbreviations.
The third 
bosonic field that participates in the splitting process besides the
two (anti-)fermions is denoted as~$\varrho$.
Employing chirality labels $\chir_1, \chir_2 \in
\{\mathrm{L},\mathrm{R}\}$ and the notation~\eqref{eq:generic-couplings-expl-def}
for generic couplings~$\mathcal{C}$, we define
\begin{align}
  a^{\varrho \chir_1 \chir_2}_{\mathcal{X} \mathcal{I} \bar{\mathcal{I}}}
  (\pm 1)
  &=
  \zeta_{\mathcal{I}} m_{\mathcal{I}}
  \mathcal{C}^{\chir_1}_{
    \nonfermtildefield{\mathcal{X}}
    \antifermfield{\mathcal{I}}
    \fermfield{\bar{\mathcal{I}}}
    }
  \pm
  \zeta_{\bar{\mathcal{I}}} m_{\bar{\mathcal{I}}}
  \mathcal{C}^{\chir_2}_{
    \nonfermtildefield{\mathcal{X}}
    \antifermfield{\mathcal{I}}
    \fermfield{\bar{\mathcal{I}}}
    }
  ,
  \nonumber
  \\
  b^{\varrho \chir_1 \chir_2}_{\mathcal{X} \mathcal{I} \bar{\mathcal{I}}}
  (\pm 1)
  &=
  m_{\mathcal{I}}
  \mathcal{C}^{\chir_1}_{
    \nonfermtildefield{\mathcal{X}}
    \antifermfield{\mathcal{I}}
    \fermfield{\bar{\mathcal{I}}}
    }
  \pm
  m_{\bar{\mathcal{I}}}
  \mathcal{C}^{\chir_2}_{
    \nonfermtildefield{\mathcal{X}}
    \antifermfield{\mathcal{I}}
    \fermfield{\bar{\mathcal{I}}}
    }
  ,
  \nonumber
  \\
  c^{\varrho \chir_1 \chir_2}_{\mathcal{X} \mathcal{I} \bar{\mathcal{I}}}
  &=
  -
  \big(
    2 \zeta_{\mathcal{X}} m_{\mathcal{X}}^2
    + \zeta_{\mathcal{I}} m_{\mathcal{I}}^2
    + \zeta_{\bar{\mathcal{I}}} m_{\bar{\mathcal{I}}}^2
  \big)
  \mathcal{C}^{\chir_1}_{
    \nonfermtildefield{\mathcal{X}}
    \antifermfield{\mathcal{I}}
    \fermfield{\bar{\mathcal{I}}}
    }
  +
  m_{\mathcal{I}} m_{\bar{\mathcal{I}}}
  \big(
    \zeta_{\mathcal{I}}
    +
    \zeta_{\bar{\mathcal{I}}}
  \big)
  \mathcal{C}^{\chir_2}_{
    \nonfermtildefield{\mathcal{X}}
    \antifermfield{\mathcal{I}}
    \fermfield{\bar{\mathcal{I}}}
    }
  ,
  \nonumber
  \\
  d^{\varrho}_{\mathcal{X} \mathcal{I} \bar{\mathcal{I}}}
  (\pm 1)
  &=
  \mathcal{C}^{\mathrm{L}}_{
    \nonfermtildefield{\mathcal{X}}
    \antifermfield{\mathcal{I}}
    \fermfield{\bar{\mathcal{I}}}
    }
  \mathcal{C}^{\mathrm{L} \ast}_{
    \nonfermtildefield{\mathcal{X}^\prime}
    \antifermfield{\mathcal{I}^\prime}
    \fermfield{\bar{\mathcal{I}}^\prime}
  }
  \pm
  \mathcal{C}^{\mathrm{R}}_{
    \nonfermtildefield{\mathcal{X}}
    \antifermfield{\mathcal{I}}
    \fermfield{\bar{\mathcal{I}}}
    }
  \mathcal{C}^{\mathrm{R} \ast}_{
    \nonfermtildefield{\mathcal{X}^\prime}
    \antifermfield{\mathcal{I}^\prime}
    \fermfield{\bar{\mathcal{I}}^\prime}
    }
  ,
  \nonumber\\
  e^{\varrho}_{\mathcal{X} \mathcal{I} \bar{\mathcal{I}}}
  (\pm 1)
  &=
  \mathcal{C}^{\mathrm{L}}_{
    \nonfermtildefield{\mathcal{X}}
    \antifermfield{\mathcal{I}}
    \fermfield{\bar{\mathcal{I}}}
    }
  \mathcal{C}^{\mathrm{R} \ast}_{
    \nonfermtildefield{\mathcal{X}^\prime}
    \antifermfield{\mathcal{I}^\prime}
    \fermfield{\bar{\mathcal{I}}^\prime}
    }
  \pm
  \mathcal{C}^{\mathrm{R}}_{
    \nonfermtildefield{\mathcal{X}}
    \antifermfield{\mathcal{I}}
    \fermfield{\bar{\mathcal{I}}}
    }
  \mathcal{C}^{\mathrm{L} \ast}_{
    \nonfermtildefield{\mathcal{X}^\prime}
    \antifermfield{\mathcal{I}^\prime}
    \fermfield{\bar{\mathcal{I}}^\prime}
    }
  .
  \label{eq:gen-ferm-abbs}
\end{align}
These generic abbreviations are assigned to specific fermionic splitting
processes by choosing explicit indices~$\mathcal{I}, \bar{\mathcal{I}}
\in \{i,j,ij^{(\prime)}\}$ of the corresponding {\it outgoing} fermion and
antifermion, respectively, as well as the index $\mathcal{X} \in
\{i,j,ij^{(\prime)}\}$ of~$\varrho$;
recall that the field labels on the symbols
$\mathcal{C}^{\dots}_{\dots}$ are considered {\it incoming}, whereas the
considered splitting process may involve {\it incoming} and {\it
outgoing} particles.
In
$V^*_{ij}\to f_i \bar{f}_j$
splittings,
for example, the (anti-)fermions are both {\it outgoing} and correspond
to $\bar{\mathcal{I}}=j$ and $\mathcal{I}=i$, while the vector boson is
{\it incoming}, $\mathcal{X}=ij$.
Below, we explicitly discuss the abbreviations that arise for this situation.
In the case
$f^*_{ij}\to V_i f_j$, on the other hand,
the fermion with the index $\bar{\mathcal{I}} = ij$ is {\it incoming} and corresponds to an
{\it outgoing} antifermion.
Similarly,
we write
\begin{align}
  \nonfermtildefield{ij^{(\prime)}} = \nonfermfield{ij^{(\prime)}}
  ,\qquad
  \nonfermtildefield{i} = \nonfermantifield{i}
  ,\qquad
  \nonfermtildefield{j} = \nonfermantifield{j}
  ,
\end{align}
because $\nonfermfield{ij^{(\prime)}}$ is incoming w.r.t.\ the
considered splitting
process, while $\nonfermfield{i}$ and $\nonfermfield{j}$ are outgoing.
For the symbols~$d^\varrho$ and~$e^\varrho$, we specify~$i^\prime=i$,~$j^\prime=j$.
The object~$\zeta$ is defined via
\begin{align}
  \zeta_i = 1-z
  ,\qquad
  \zeta_j = z
  ,\qquad
  \zeta_{ij} = \zeta_{ij^\prime} = - (1-z) z
  .
  \label{eq:gen-mom-sharing-symbol}
\end{align}
For each possible assignment of $\mathcal{I}$, $\bar{\mathcal{I}}$,
and~$\mathcal{X}$ to the indices $i,j$, and $ij^{(\prime)}$, we have
\begin{align}
  \zeta_{\mathcal{X}}
  =
  -
  \frac{
    \zeta_{\mathcal{I}}
    \zeta_{\bar{\mathcal{I}}}
  }{
    \zeta_{\mathcal{I}}
    +
    \zeta_{\bar{\mathcal{I}}}
  }
  \label{eq:generic-relat-zeta}
  .
\end{align}
Similar generic abbreviations are defined for the \gls{is} case
in~\eqref{eq:gen-ferm-abbs-IS} below.


As described above, each specific splitting process corresponds to a
choice of~$\varrho$ and the indices~$\mathcal{I}$, $\bar{\mathcal{I}}$,
$\mathcal{X}$.
This turns the generic
abbreviations~\eqref{eq:gen-ferm-abbs} into explicit expressions.
For \gls{fs} $V^* \to f \bar f$ splittings, which are the subject of this
section, we get
\begin{align}
  a^{V \chir_1 \chir_2}_{ij, i, j}
  (\pm 1)
  &=
  (1-z) m_i
  I_{\fermfield{i}^{\chir_1} \fermfield{j}^{\chir_1}}^{\vbfield{ij}}
  \pm
  z m_j
  I_{\fermfield{i}^{\chir_2} \fermfield{j}^{\chir_2}}^{\vbfield{ij}}
  ,
  \nonumber
  \\
  b^{V \chir_1 \chir_2}_{ij, i, j}
  (\pm 1)
  &=
  m_i
  I_{\fermfield{i}^{\chir_1} \fermfield{j}^{\chir_1}}^{\vbfield{ij}}
  \pm
  m_j
  I_{\fermfield{i}^{\chir_2} \fermfield{j}^{\chir_2}}^{\vbfield{ij}}
  ,
  \nonumber
  \\
  c^{V \chir_1 \chir_2}_{ij, i, j}
  &=
  \Big(
    2 (1-z) z m_{ij}^2
    - (1-z) m_i^2
    - z m_j^2
  \Big)
  I_{\fermfield{i}^{\chir_1} \fermfield{j}^{\chir_1}}^{\vbfield{ij}}
  +
  m_i m_j
  I_{\fermfield{i}^{\chir_2} \fermfield{j}^{\chir_2}}^{\vbfield{ij}}
  ,
  \nonumber\\
  d^{V}_{ij, i, j}
  (\pm 1)
  &=
  I_{\fermfield{i}^\mathrm{L} \fermfield{j}^\mathrm{L}}^{\vbfield{ij}}
  I_{\fermfield{i}^\mathrm{L} \fermfield{j}^\mathrm{L}}^{\vbfield{ij^\prime} \ast}
  \pm
  I_{\fermfield{i}^\mathrm{R} \fermfield{j}^\mathrm{R}}^{\vbfield{ij}}
  I_{\fermfield{i}^\mathrm{R} \fermfield{j}^\mathrm{R}}^{\vbfield{ij^\prime} \ast}
  ,
  \nonumber\\
  e^V_{ij, i, j}
  (\pm 1)
  &=
  I_{\fermfield{i}^\mathrm{L} \fermfield{j}^\mathrm{L}}^{\vbfield{ij}}
  I_{\fermfield{i}^\mathrm{R} \fermfield{j}^\mathrm{R}}^{\vbfield{ij^\prime} \ast}
  \pm
  I_{\fermfield{i}^\mathrm{R} \fermfield{j}^\mathrm{R}}^{\vbfield{ij}}
  I_{\fermfield{i}^\mathrm{L} \fermfield{j}^\mathrm{L}}^{\vbfield{ij^\prime} \ast}
  .
  \label{eq:VFFbar-abbrev}
\end{align}
In order to reduce notational complexity
with these kind of symbols,
we often drop their indices
and use the shorthand notations~$a_{\chir_1 \chir_2}^{V}$,
$b_{\chir_1 \chir_2}^{V}$, $c_{\chir_1 \chir_2}^{V}$, $d^{V}$,
$e^{V}$, respectively.
The proper index assignments are always understood
w.r.t.\ the splitting process that is discussed in the given context.
Similarly, replacing~$ij$
by~$ij^\prime$ in the first three symbols
from~\eqref{eq:VFFbar-abbrev} defines~$a_{\chir_1 \chir_2}^{V
\prime}$, $b_{\chir_1 \chir_2}^{V \prime}$, and $c_{\chir_1
\chir_2}^{V \prime}$.


Wherever two splitting functions are related by symmetries, the
corresponding mappings also relate the respective sets of abbreviations, 
as discussed in detail below.
Writing our results in terms of the expressions~\eqref{eq:gen-ferm-abbs}
therefore highlights their symmetry properties.
This also applies to the crossing relations between \gls{is} and
\gls{fs} splitting functions from
Section~\ref{sect:splfunc-symm-crossing}, with the definition of the
generic \gls{is} abbreviations from~\eqref{eq:gen-ferm-abbs-IS}.


We conclude with a comment on the various massless limits of the
abbreviations.
First, we focus on~$m_{\mathcal{X}} \to 0$.
To this end, we rewrite
\begin{align}
  c^{\varrho \chir (-\chir)}_{\mathcal{X} \mathcal{I} \bar{\mathcal{I}}}
  =
  -
  \zeta_{\mathcal{I}} m_{\mathcal{I}}
  b^{\varrho \chir (-\chir)}_{\mathcal{X} \mathcal{I} \bar{\mathcal{I}}}
  (- 1)
  +
  \zeta_{\bar{\mathcal{I}}} m_{\bar{\mathcal{I}}}
  b^{\varrho (-\chir) \chir}_{\mathcal{X} \mathcal{I} \bar{\mathcal{I}}}
  (- 1)
  -
  2 \zeta_{\mathcal{X}} m_{\mathcal{X}}^2 
  \mathcal{C}^{\chir}_{
    \nonfermtildefield{\mathcal{X}}
    \antifermfield{\mathcal{I}}
    \fermfield{\bar{\mathcal{I}}}
    }
  .
  \label{eq:c-abb-via-b-abb-massless}
\end{align}
Furthermore, the coupling relation~\eqref{eq:coupling-relat-Vff-Gff} implies
\begin{align}
  b^{V \chir (-\chir)}_{\mathcal{X} \mathcal{I} \bar{\mathcal{I}}}
  (-1)
  =
  \eta_{\vbtildefield{\mathcal{X}}}
  m_{\mathcal{X}}
  \mathcal{C}^{\chir}_{
    \gbtildefield{\mathcal{X}}
    \antifermfield{\mathcal{I}}
    \fermfield{\bar{\mathcal{I}}}
  }
  .
  \label{eq:bV-abb-via-scalar-cpl}
\end{align}
Similar to the discussion for the terms~\eqref{eq:rho-ml-limit}, this means
that~$b^{V \chir (-\chir)}_{\mathcal{X} \mathcal{I} \bar{\mathcal{I}}}/m_{\mathcal{X}}$
and~%
$c^{V \chir (-\chir)}_{\mathcal{X} \mathcal{I} \bar{\mathcal{I}}}/m_{\mathcal{X}}$
are well behaved for~$m_{\mathcal{X}} \to 0$.
In fact, we find
\begin{align} 
  \frac{1}{m_{\mathcal{X}}}
  c^{V \chir (-\chir)}_{\mathcal{X} \mathcal{I} \bar{\mathcal{I}}}
\;
  \xrightarrow[m_{\mathcal{X}} \to 0]{\gbfield{\mathcal{X}} \to \higgsfield{\mathcal{X}}}
\;
  -
  \eta_{\vbtildefield{\mathcal{X}}}
  a^{H \chir (-\chir)}_{\mathcal{X} \mathcal{I} \bar{\mathcal{I}}}
  (- 1)
  ,
  \label{eq:cV-abb-via-aH-abb-massless}
\end{align}
upon relabeling the fields~$\gbfield{\mathcal{X}} \to \higgsfield{\mathcal{X}}$.
For massless fermions, $m_{\mathcal{I}},m_{\mathcal{\bar{I}}} \to 0$,
all symbols from \eqref{eq:gen-ferm-abbs} are obviously well behaved,
and we get
\begin{align} 
  a^{\varrho \chir_1 \chir_2}_{\mathcal{X} \mathcal{I} \bar{\mathcal{I}}}
  (\pm 1)
  ,
  b^{\varrho \chir_1 \chir_2}_{\mathcal{X} \mathcal{I} \bar{\mathcal{I}}}
  (\pm 1)
\;  \xrightarrow[]{m_{\mathcal{I}},m_{\mathcal{\bar{I}}} \to 0} \;
  0
  ,\qquad
  c^{\varrho \chir_1 \chir_2}_{\mathcal{X} \mathcal{I} \bar{\mathcal{I}}}
\; \xrightarrow[]{m_{\mathcal{I}},m_{\mathcal{\bar{I}}} \to 0} \;
  -
  2 \zeta_{\mathcal{X}} m_{\mathcal{X}}^2
  \mathcal{C}^{\chir_1}_{
    \nonfermtildefield{\mathcal{X}}
    \antifermfield{\mathcal{I}}
    \fermfield{\bar{\mathcal{I}}}
    }
  .
  \label{eq:gen-ferm-abbs-mlferm}
\end{align}

\paragraph{Polarization \boldmath{$\polferm_i=+$}, \boldmath{$\polferm_j=+$}.}
We obtain
\begin{align}
\lefteqn{  \bigg[
    \mathcal{P}
      {}^{\vbfield{ij^{\vphantom{\prime}}}}_{\fermfield{i}}
      {}^{\vbfield{ij^\prime}}_{\antifermfield{j}}
      \big(z,k_\perp^\mu,\bar p^\mu,+,+\big)
    \bigg]^{\mu \nu} }
  \phantom{{}={}}
  \nonumber\\*
  ={}&
  \frac{1}{(z-1)z}
      \bigg\{
      a_{\mathrm{L}\mathrm{R}}^{V}(+1) a_{\mathrm{L}\mathrm{R}}^{V \prime \ast}(+1)
      \Big[
        g^{\mu \nu }
        +
        s_\varepsilon
        \bar{\varepsilon }_-^{\nu }
        \bar{\varepsilon }_+^{\mu }
        -
        s_\varepsilon
        \bar{\varepsilon }_-^{\mu }
        \bar{\varepsilon }_+^{\nu }
      \Big]
      +
      k_\perp^2
      \frac{
        b_{\mathrm{L}\mathrm{R}}^{V}(-1) b_{\mathrm{L}\mathrm{R}}^{V \prime \ast}(-1)
      }{m_{ij}^2 m_{ij}^{\prime \, 2}}
      \bar{p}^{\mu } \bar{p}^{\nu }
      \nonumber\\
      & \quad {}+
      2
      s_\varepsilon
      (k_{\perp }  \bar{\varepsilon }_-)
      \frac{
      a_{\mathrm{L}\mathrm{R}}^{V \prime \ast}(+1) b_{\mathrm{L}\mathrm{R}}^{V}(-1)
      }{m_{ij}^2}
      \bar{p}^{\mu }
      \bar{\varepsilon}_+^{\nu }      
      +
      2
      s_\varepsilon
      (k_{\perp }  \bar{\varepsilon }_+)
      \frac{
        a_{\mathrm{L}\mathrm{R}}^{V}(+1) b_{\mathrm{L}\mathrm{R}}^{V \prime \ast}(-1)
      }{m_{ij}^{\prime \, 2}}
      \bar{p}^{\nu }
      \bar{\varepsilon }_-^{\mu }      
    \bigg\}
  ,
  \label{eq:VFF-PP-split-func}
\end{align}
This result is well behaved in the limit of massless vector bosons since
$b_{\mathrm{L}\mathrm{R}}^{V}(-1)/m_{ij}$ is well behaved, and because
of \gls{wi}~\eqref{eq:VVV-ward-id-ul-me}.
It is easily seen via~\eqref{eq:gen-ferm-abbs-mlferm} that the
splitting function~\eqref{eq:VFF-PP-split-func} vanishes in the limit of
massless fermions.
This aligns with the expectation from the chirality structure of
the~$V\bar ff$ vertex~\eqref{eq:app-Vff-vertex}.

\paragraph{Polarization \boldmath{$\polferm_i=+$}, \boldmath{$\polferm_j=-$}.}
We find
\begin{alignat}{2}
  &
  \lefteqn{
    \bigg[
      \mathcal{P}
        {}^{\vbfield{ij^{\vphantom{\prime}}}}_{\fermfield{i}}
        {}^{\vbfield{ij^\prime}}_{\antifermfield{j}}
        \big(z,k_\perp^\mu,\bar p^\mu,+,-\big)
    \bigg]^{\mu \nu}
  }
  \nonumber\\*
  &=
  \frac{1}{(z-1)z}
  \bigg\{
    &&{}
    2
    I_{\fermfield{i}^{\mathrm{R}} \fermfield{j}^{\mathrm{R}}}^{\vbfield{ij^\prime} \ast}
    \frac{
      c_{\mathrm{R}\mathrm{L}}^{V}
    }{m_{ij}^2}
    \bar{p}^{\mu }
    \Big(
      s_\varepsilon
      (k_{\perp }  \bar{\varepsilon }_+)
      \bar{\varepsilon }_-^{\nu }
      +
      z
      k_{\perp }^{\nu }
    \Big)
    +
    2
    I_{\fermfield{i}^{\mathrm{R}} \fermfield{j}^{\mathrm{R}}}^{\vbfield{ij}}
    \frac{
      c_{\mathrm{R}\mathrm{L}}^{V \prime \ast}
    }{m_{ij}^{\prime \, 2}}
    \bar{p}^{\nu }
    \Big(
      s_\varepsilon
      (k_{\perp }  \bar{\varepsilon }_-)
      \bar{\varepsilon }_+^{\mu }
      +
      z
      k_{\perp }^{\mu }
    \Big)
    \nonumber\\
    &&&{}+
    I_{\fermfield{i}^{\mathrm{R}} \fermfield{j}^{\mathrm{R}}}^{\vbfield{ij}}
    I_{\fermfield{i}^{\mathrm{R}} \fermfield{j}^{\mathrm{R}}}^{\vbfield{ij^\prime} \ast}
    \Big(
      -
      k_\perp^2
      g^{\mu \nu }
      -
      4
      (z-1) z
      k_{\perp }^{\mu } k_{\perp }^{\nu }
      +
      s_\varepsilon
      (2 z-1)
      k_\perp^2
      \big(
        \bar{\varepsilon }_-^{\mu } \bar{\varepsilon }_+^{\nu }
        -
        \bar{\varepsilon }_-^{\nu } \bar{\varepsilon }_+^{\mu }
      \big)
    \Big)
    \nonumber\\
    &&&{}-
    \frac{
      c_{\mathrm{R}\mathrm{L}}^{V}
      c_{\mathrm{R}\mathrm{L}}^{V \prime \ast}
    }{m_{ij}^2 m_{ij}^{\prime \, 2}}
    \bar{p}^{\mu } \bar{p}^{\nu }
  \bigg\}
  ,
  \label{eq:VFF-PM-split-func}
\end{alignat}
which remains finite in the limits~$m_{ij} \to 0$
and~$m_{ij}^\prime \to 0$ due to~\eqref{eq:VVV-ward-id-ul-me}
and~\eqref{eq:cV-abb-via-aH-abb-massless}.
For massless fermions,~\eqref{eq:gen-ferm-abbs-mlferm} implies
that all left-handed couplings $I_{\fermfield{i}^{\mathrm{L}}
\fermfield{j}^{\mathrm{L}}}^{\vbfield{ij}}$ drop out
of~\eqref{eq:VFF-PM-split-func}.
This is, again, expected form the structure of the $V\bar ff$
vertex~\eqref{eq:app-Vff-vertex}.


\paragraph{Flipping the fermion polarizations.}
Flipping the (anti-)fermion polarizations in the $V^*\to f \bar f$ splitting
functions amounts to flipping the signs of the $\gamma^5$ matrices inside
all spin projection matrices~\eqref{eq:spin-projection-op}.
This, in turn, is equivalent to a global replacement 
$\gamma^5 \to -\gamma^5$, followed by
$I_{\fermfield{i}^{\mathrm{L}} \fermfield{j}^{\mathrm{L}}}^{\vbfield{ij}} \leftrightarrow
 I_{\fermfield{i}^{\mathrm{R}} \fermfield{j}^{\mathrm{R}}}^{\vbfield{ij}}$
to reverse this 
chirality flip in the couplings of the splitting vertices.
Because of~\eqref{eq:example-gam5-epsilon-trace},
$\gamma^5 \to -\gamma^5$ 
goes along with a sign flip of the $\epsilon$-tensor and, thus, with
exchanging $\bar \varepsilon_-^\mu$ and $\bar
\varepsilon_+^\mu$,
according to~\eqref{eq:remaining-eps-tens-contraction} and \eqref{eq:eps_barp_n}.
With this at hand, we obtain
\begin{align}
  \bigg[
    \mathcal{P}
      {}^{\vbfield{ij^{\vphantom{\prime}}}}_{\fermfield{i}}
      {}^{\vbfield{ij^\prime}}_{\antifermfield{j}}
      \big(z,k_\perp^\mu,\bar p^\mu,-,\mp\big)
  \bigg]^{\mu \nu}
  &=
  \bigg[
    \mathcal{P}
      {}^{\vbfield{ij^{\vphantom{\prime}}}}_{\fermfield{i}}
      {}^{\vbfield{ij^\prime}}_{\antifermfield{j}}
      \big(z,k_\perp^\mu,\bar p^\mu,+,\pm\big)
  \bigg]^{\mu \nu}
  _{\mathrm{L} \leftrightarrow \mathrm{R}, \bar \varepsilon_+ \leftrightarrow \bar \varepsilon_-}
  .
  \label{eq:VFF-MM-MP-split-func}
\end{align}

\paragraph{Unpolarized case.}
Summing the above results, we get
\begin{alignat}{2}
  &
  \lefteqn{
    \bigg[
      \mathcal{P}
        {}^{\vbfield{ij^{\vphantom{\prime}}}}_{\fermfield{i}}
        {}^{\vbfield{ij^\prime}}_{\antifermfield{j}}
        \big(z,k_\perp^\mu,\bar p^\mu,\mathrm{U},\mathrm{U}\big)
    \bigg]^{\mu \nu}
  }
  \nonumber\\*
  &{}={}
  &&
  \frac{
    k_\perp^2
    \Big(
      (m_i^2 + m_j^2)
      d^V(+1)
      -
      2 m_i m_j
      e^V(+1)
    \Big)
    -
    c_{\mathrm{L}\mathrm{R}}^{V} c_{\mathrm{L}\mathrm{R}}^{V \prime  \ast}
    -
    c_{\mathrm{R}\mathrm{L}}^{V} c_{\mathrm{R}\mathrm{L}}^{V \prime  \ast}
  }{
    m_{ij}^2 m_{ij}^{\prime\, 2}
    (z-1) z
  }
  \bar{p}^{\mu } \bar{p}^{\nu }
  \nonumber\\
  &&&
  +
  \bigg(
    d^V(+1)
    \frac{
      - k_\perp^2
      + m_i^2 (z-1)^2
      + m_j^2 z^2
    }{(z-1) z}
    -
    2 m_i m_j e^V(+1)
  \bigg)
  g^{\mu \nu }
  -
  4
  d^V(+1)
  k_{\perp }^{\mu } k_{\perp }^{\nu }
  \nonumber\\
  &&&
  +
  2
  \bigg(
    d^V(+1)
    \frac{
      m_i^2 + m_{ij}^2 (1-2z) - m_j^2
    }{m_{ij}^2} 
    +
    d^V(-1)
  \bigg)
  \bar{p}^{\mu } k_{\perp }^{\nu }
  \nonumber\\
  &&&
  +
  2
  \bigg(
    d^V(+1)
    \frac{
      m_i^2 + m_{ij}^{\prime\, 2} (1-2z) - m_j^2
    }{m_{ij}^{\prime\, 2}} 
    +
    d^V(-1)
  \bigg)
  \bar{p}^{\nu } k_{\perp }^{\mu }
  \nonumber\\
  &&&
  +
  d^V(-1)
  \frac{
    (2z-1)
    k_\perp^2
    +
    m_i^2 (z-1)^2
    -
    m_j^2 z^2
  }{(z-1) z}
  s_{\varepsilon }
  \Big(
    \bar{\varepsilon }_-^{\nu } \bar{\varepsilon }_+^{\mu }
    -
    \bar{\varepsilon }_-^{\mu } \bar{\varepsilon }_+^{\nu }
  \Big)
  \nonumber\\
  &&&
  +
  4
  s_{\varepsilon }
  d^V(-1)
  \Big(
    (k_{\perp }  \bar{\varepsilon }_-)
    \bar{p}^{\nu } \bar{\varepsilon }_+^{\mu }
    +
    (k_{\perp }  \bar{\varepsilon }_+)
    \bar{p}^{\mu } \bar{\varepsilon }_-^{\nu }
  \Big)
  .
  \label{eq:VFF-UU-split-func-simp}
\end{alignat}

%
\subsubsection[%
  \mathinhead{V^\ast \to \bar f f}{V* to f barf}
  splitting functions%
  ]{%
  \mathinheadbold{V^\ast \to \bar f f}{V* to f barf}
  splitting functions}
\label{sect:VFbarF}
The splitting functions for $V^\ast \to \bar f f$ may be obtained
via~\eqref{eq:i-j-symmru-splfunc} from those for $V^\ast \to f
\bar f$,
\begin{align}
  \bigg[
    \mathcal{P}
      {}^{\vbfield{ij^{\vphantom{\prime}}}}_{\antifermfield{i}}
      {}^{\vbfield{ij^\prime}}_{\fermfield{j}}
      \big(z,k_\perp^\mu,\bar p^\mu,\polferm_i,\polferm_j\big)
  \bigg]^{\mu \nu}
  =
  \bigg[
    \mathcal{P}
      {}^{\vbfield{ij^{\vphantom{\prime}}}}_{\fermfield{j}}
      {}^{\vbfield{ij^\prime}}_{\antifermfield{i}}
      \big(1-z,-k_\perp^\mu,\bar p^\mu,\polferm_j,\polferm_i\big)
  \bigg]^{\mu \nu}
  _{m_i \leftrightarrow m_j}
  .
  \label{eq:VFF-i-j-symm}
\end{align}
More precisely, the couplings on the r.h.s.\ are related to the ones of
$V^\ast \to f\bar f$ of Section~\ref{sect:VFFbar} by
\begin{align}
  I_{\fermfield{i}^{\chir} \fermfield{j}^{\chir}}^{\vbfield{ij}^{(\prime)}}
  \to
  I_{\fermfield{j}^{\chir} \fermfield{i}^{\chir}}^{\vbfield{ij}^{(\prime)}}
  .
  \label{eq:VFF-swap-cpl-wo-chir}
\end{align}
Of course, this is reflected by the fact that the diagrams for~$V^\ast
\to \bar f f$ splittings,
\drawmomtrue%
\drawvertdottrue%
\begin{alignat}{2}%
  \ftntmath{%
    \MContrExplFields{\vbfield{ij}}{\antifermfield{i}}{\fermfield{j}}^{\polferm_i \polferm_j}%
  }%
  &\ftntmath{{}={}}%
  \FSDiagGeo%
  \ftntmath{%
    \splittingdiagIndivFS{T^{(n)}_{\vbfield{ij} X}}%
      {photon}{antifermion}{fermion}%
      {\vbfield{ij}}{\antifermfield{i}, \polferm_i}{\fermfield{j}, \polferm_j}%
    ,%
  }%
  \resetdiagargs%
  &\quad
  \ftntmath{%
    \MContrExplFields{\gbfield{ij}}{\antifermfield{i}}{\fermfield{j}}^{\polferm_i \polferm_j}%
  }%
  &\ftntmath{{}={}}%
  \FSDiagGeo%
  \ftntmath{%
    \splittingdiagIndivFS{T^{(n)}_{\gbfield{ij} X}}%
      {scalar}{antifermion}{fermion}%
      {\gbfield{ij}}{\antifermfield{i}, \polferm_i}{\fermfield{j}, \polferm_j}%
    ,%
  }%
  \resetdiagargs%
  \label{eq:VFbarF-diags-FS}
\end{alignat}%
and their corresponding analytical expressions,
\begin{align}
  %
  %
  \MContrExplFields{\vbfield{ij}}{\antifermfield{i}}{\fermfield{j}}^{\polferm_i \polferm_j}
  ={}&
  \imagi e
  T^{(n)\, \alpha }_{\vbfield{ij} X}(p_{ij})
  G^{\vbantifield{ij} \vbfield{ij}}_{\alpha \beta}(p_{ij},-p_{ij})
  \bar{u}_{\polferm_j}(p_j)
  \Big(
    \gamma ^{\beta }
    \omega_-
    I_{\fermfield{j}^\mathrm{L} \fermfield{i}^\mathrm{L}}^{\vbfield{ij}}
    +
    \gamma ^{\beta }
    \omega_+ 
    I_{\fermfield{j}^\mathrm{R} \fermfield{i}^\mathrm{R}}^{\vbfield{ij}}
  \Big)
  v_{\polferm_i}(p_i)
  ,
  \nonumber\\
  %
  %
  \MContrExplFields{\gbfield{ij}}{\antifermfield{i}}{\fermfield{j}}^{\polferm_i \polferm_j}
  ={}&
  \imagi e
  T^{(n)\, \alpha}_{\vbfield{ij} X}(p_{ij})
  p_{ij,\alpha}
  G^{\gbantifield{ij} \gbfield{ij}}(p_{ij},-p_{ij})
  \frac{1}{m_{ij}^2}
  \nonumber\\
  &
  \cdot
  \bar{u}_{\polferm_j}(p_j)
  \left\{
    \omega_+
    \left(
      m_j I_{\fermfield{j}^\mathrm{R} \fermfield{i}^\mathrm{R}}^{\vbfield{ij}}
      -
      m_i I_{\fermfield{j}^\mathrm{L} \fermfield{i}^\mathrm{L}}^{\vbfield{ij}}
    \right)
    +
    \omega_-
    \left(
      m_j I_{\fermfield{j}^\mathrm{L} \fermfield{i}^\mathrm{L}}^{\vbfield{ij}}
      -
      m_i I_{\fermfield{j}^\mathrm{R} \fermfield{i}^\mathrm{R}}^{\vbfield{ij}}
    \right)
  \right\}
  v_{\polferm_i}(p_i)
  ,
  \label{eq:VFbarF-EnhAmp}
\end{align}
are obtained from the $V^\ast \to f \bar f$ counterparts by swapping~$i
\leftrightarrow j$ and multiplying with a conventional
sign~$-1$.\footnotemark\
\footnotetext{%
  The relative sign factor between \eqref{eq:VFbarF-EnhAmp}
  and~\eqref{eq:VFFbar-EnhAmp} accounts for the interchange of 
  fermionic fields that is involved in the symmetry relation between the two
  splitting processes.
}%
This relation can, alternatively, be expressed in terms of charge
conjugation.
The charge conjugation matrix~$C$ is defined via its action on Dirac matrices,
\begin{align}
  C \gamma^\mu C^{-1}
  =
  - (\gamma^{\mu})^\mathrm{T}
  ,\qquad
  C \gamma^5 C^{-1}
  =
+ (\gamma^{5})^\mathrm{T}
  .
  \label{eq:ch-conj-def}
\end{align}
In the Dirac and the chiral representations of the $\gamma^\mu$ matrices, we get
\begin{align}
  C=\imagi \gamma^2 \gamma^0
  ,\qquad
  C^{-1}=C^\dagger=C^\mathrm{T}=-C
  ,
\end{align}
relating spinor fields~$\psi$, $\bar{\psi}$
to their
respective charge-conjugated counterparts~$\psi^\mathrm{c}$,
$\bar{\psi}^\mathrm{c}$ via
\begin{align}
  \psi^\mathrm{c} = C \bar \psi^\mathrm{T}
  ,\qquad
  \bar \psi^\mathrm{c} = \psi^\mathrm{T} C
  ,
  \label{eq:ch-conj-gen-spinors}
\end{align}
where trivial charge parity phases are assumed.
In particular, we can choose the phases of the Dirac spinors $u,v$ so that
\begin{align}
  v_\polferm(p) = C \bar u_\polferm^\mathrm{T}(p)
  ,\quad
  u_\polferm(p) = C \bar v_\polferm^\mathrm{T}(p)
  ,\quad
  \bar v_\polferm(p) = u_\polferm^\mathrm{T}(p) C
  ,\quad
  \bar u_\polferm(p) = v_\polferm^\mathrm{T}(p) C
  .
  \label{eq:ch-conj-expl-spinors}
\end{align}
By transposition, insertions of~$1 = C^\dagger C$, and identification
of~\eqref{eq:ch-conj-expl-spinors}, it can be seen
that~\eqref{eq:VFbarF-EnhAmp} is obtained from~\eqref{eq:VFFbar-EnhAmp}
by multiplication with the conventional sign~$-1$ and swapping the
couplings
\begin{align}
  I_{\fermfield{i}^{\mathrm{L}} \fermfield{j}^{\mathrm{L}}}^{\vbfield{ij}^{(\prime)}}
  \to
  I_{\fermfield{j}^{\mathrm{R}} \fermfield{i}^{\mathrm{R}}}^{\vbfield{ij}^{(\prime)}}
  ,
  \qquad
  I_{\fermfield{i}^{\mathrm{R}} \fermfield{j}^{\mathrm{R}}}^{\vbfield{ij}^{(\prime)}}
  \to
  I_{\fermfield{j}^{\mathrm{L}} \fermfield{i}^{\mathrm{L}}}^{\vbfield{ij}^{(\prime)}}
  .
  \label{eq:VFF-swap-cpl-w-chir}
\end{align}
In contrast to~\eqref{eq:VFF-swap-cpl-wo-chir}, this flips
the assignment of chiralities to the couplings
(without changing $m_i$, $m_j$, $z$).
Altogether,
\begin{align}
  \bigg[
    \mathcal{P}
      {}^{\vbfield{ij^{\vphantom{\prime}}}}_{\antifermfield{i}}
      {}^{\vbfield{ij^\prime}}_{\fermfield{j}}
      \big(z,k_\perp^\mu,\bar p^\mu,\polferm_i,\polferm_j\big)
  \bigg]^{\mu \nu}
  &=
  \bigg[
    \mathcal{P}
      {}^{\vbfield{ij^{\vphantom{\prime}}}}_{\fermfield{i}}
      {}^{\vbfield{ij^\prime}}_{\antifermfield{j}}
      \big(z,k_\perp^\mu,\bar p^\mu,\polferm_i,\polferm_j\big)
  \bigg]^{\mu \nu}
  _{\eqref{eq:VFF-swap-cpl-w-chir}}
  ,
  \label{eq:VFFbar-VFbarF-symm}
\end{align}
where the polarizations~$\polferm_i$ and~$\polferm_j$ remain fixed.
Note that this derivation does not assume CP invariance of the model,
i.e.\ this relation remains valid in the presence of CKM-like mixing.


For $V^\ast \to \bar f f$ splittings, we obtain the following
abbreviations from~\eqref{eq:gen-ferm-abbs},
\begin{align}
  a^{V \chir_1 \chir_2}_{ij, j, i}
  (\pm 1)
  &=
  z m_j
  I_{\fermfield{j}^{\chir_1} \fermfield{i}^{\chir_1}}^{\vbfield{ij}}
  \pm
  (1-z) m_i
  I_{\fermfield{j}^{\chir_2} \fermfield{i}^{\chir_2}}^{\vbfield{ij}}
  ,
  \nonumber
  \\
  b^{V \chir_1 \chir_2}_{ij, j, i}
  (\pm 1)
  &=
  m_j
  I_{\fermfield{j}^{\chir_1} \fermfield{i}^{\chir_1}}^{\vbfield{ij}}
  \pm
  m_i
  I_{\fermfield{j}^{\chir_2} \fermfield{i}^{\chir_2}}^{\vbfield{ij}}
  ,
  \nonumber
  \\
  c^{V \chir_1 \chir_2}_{ij, j, i}
  &=
  \big[
    2 (1-z) z m_{ij}^2
    - (1-z) m_i^2
    - z m_j^2)
  \big]
  I_{\fermfield{j}^{\chir_1} \fermfield{i}^{\chir_1}}^{\vbfield{ij}}
  +
  m_i m_j
  I_{\fermfield{j}^{\chir_2} \fermfield{i}^{\chir_2}}^{\vbfield{ij}}
  ,
  \nonumber\\
  d^{V}_{ij, j, i}
  (\pm 1)
  &=
  I_{\fermfield{j}^\mathrm{L} \fermfield{i}^\mathrm{L}}^{\vbfield{ij}}
  I_{\fermfield{j}^\mathrm{L} \fermfield{i}^\mathrm{L}}^{\vbfield{ij^\prime} \ast}
  \pm
  I_{\fermfield{j}^\mathrm{R} \fermfield{i}^\mathrm{R}}^{\vbfield{ij}}
  I_{\fermfield{j}^\mathrm{R} \fermfield{i}^\mathrm{R}}^{\vbfield{ij^\prime} \ast}
  ,
  \nonumber\\
  e^V_{ij, j, i}
  (\pm 1)
  &=
  I_{\fermfield{j}^\mathrm{L} \fermfield{i}^\mathrm{L}}^{\vbfield{ij}}
  I_{\fermfield{j}^\mathrm{R} \fermfield{i}^\mathrm{R}}^{\vbfield{ij^\prime} \ast}
  \pm
  I_{\fermfield{j}^\mathrm{R} \fermfield{i}^\mathrm{R}}^{\vbfield{ij}}
  I_{\fermfield{j}^\mathrm{L} \fermfield{i}^\mathrm{L}}^{\vbfield{ij^\prime} \ast}
  .
  \label{eq:VFbarF-abbrev}
\end{align}
As anticipated in Section~\ref{sect:VFFbar}, the combination of 
the substitutions 
$m_i \leftrightarrow m_j$, $z \to 1-z$, and~\eqref{eq:VFF-swap-cpl-wo-chir}
from the $i\leftrightarrow j$ symmetry rule~\eqref{eq:VFF-i-j-symm} maps
the abbreviations~\eqref{eq:VFFbar-abbrev} for $V^\ast \to f \bar f$
splittings into these expressions.
Thus, when written in terms of the respective abbreviations, the
$i\leftrightarrow j$ symmetry relation between $V^\ast \to \bar f f$ and
$V^\ast \to f \bar f$ splitting functions becomes particularly
transparent.\footnotemark\
\footnotetext{%
  \label{ftnt:charge-conj-abbreviation-vff}
  The rule~\eqref{eq:VFF-swap-cpl-w-chir} from charge conjugation 
  (without changing $m_i$, $m_j$, and $z$)
  also
  induces mappings between the two sets of abbreviations,
  \begin{align}
    a^{V \mathrm{L} \mathrm{R}}_{ij, i, j}(\pm 1)
    &\to
    \pm
    a^{V \mathrm{L} \mathrm{R}}_{ij, j, i}(\pm 1)
    ,
    &
    a^{V \mathrm{R} \mathrm{L}}_{ij, i, j}(\pm 1)
    &\to
    \pm
    a^{V \mathrm{R} \mathrm{L}}_{ij, j, i}(\pm 1)
    ,
    \nonumber\\
    b^{V \mathrm{L} \mathrm{R}}_{ij, i, j}(\pm 1)
    &\to
    \pm
    b^{V \mathrm{L} \mathrm{R}}_{ij, j, i}(\pm 1)
    ,
    &
    b^{V \mathrm{R} \mathrm{L}}_{ij, i, j}(\pm 1)
    &\to
    \pm
    b^{V \mathrm{R} \mathrm{L}}_{ij, j, i}(\pm 1)
    ,
    \nonumber\\
    c^{V \mathrm{L} \mathrm{R}}_{ij, i, j}
    &\to
    c^{V \mathrm{R} \mathrm{L}}_{ij, j, i}
    ,
    &
    c^{V \mathrm{R} \mathrm{L}}_{ij, i, j}
    &\to
    c^{V \mathrm{L} \mathrm{R}}_{ij, j, i}
    ,
    \nonumber\\
    d^{V}_{ij, i, j}(\pm 1)
    &\to
    \pm
    d^{V}_{ij, j, i}(\pm 1)
    ,
    &
    e^{V}_{ij, i, j}(\pm 1)
    &\to
    \pm
    e^{V}_{ij, j, i}(\pm 1)
    .
    \label{eq:VFFbar-to-VFbarF-abbrev}
  \end{align}
  Although slightly more complicated, these relations highlight the
  symmetry between the associated splitting functions as well.
}%


As an explicit example, we consider the special case of $V^\ast \to \bar
f f$ splittings with massless vector bosons $\vbfield{ij}$,
$\vbfield{ij^\prime}$.
The corresponding $V^\ast \to f \bar f$ splitting
functions are obtained from the results of Section~\ref{sect:VFFbar} by
taking the limit
\begin{align}
  m_{ij}, m_{ij}^{\prime} \to 0
  ,\qquad
  \frac{\bar p^\mu}{m_{ij}} \to 0
  ,\qquad
  \frac{\bar p^\nu}{m_{ij}^\prime} \to 0
  ,
  \label{eq:VFF-mV0-limit}
\end{align}
leading to 
\begin{align}
  %
  %
  \bigg[
    {\mathcal{P}}
        {}^{\vbfield{ij^{\vphantom{\prime}}}}_{\fermfield{i}}
        {}^{\vbfield{ij^\prime}}_{\antifermfield{j}}
      \big(z,k_\perp^\mu,\bar p^\mu,\pm,\mp\big)
  \bigg]^{\mu \nu}_{m_{ij}^{(\prime)}=0}
  &=
  I_{\fermfield{i}^\mathrm{R/L} \fermfield{j}^\mathrm{R/L}}^{\vbfield{ij}}
  \Big(I_{\fermfield{i}^\mathrm{R/L} \fermfield{j}^\mathrm{R/L}}^{\vbfield{ij^\prime}}\Big)^\ast
  \begin{aligned}[t]
    \bigg\{
      &\frac{-k_\perp^2}{(z-1)z}
      g^{\mu \nu}
      -
      4
      k_\perp^\mu
      k_\perp^\nu
      \\
      &{}
      \pm
      s_\varepsilon
      \frac{2z-1}{(z-1)z}
      k_\perp^2
      \big(
        \bar{\varepsilon }_-^{\mu }
        \bar{\varepsilon }_+^{\nu }
        -
        \bar{\varepsilon }_+^{\mu }
        \bar{\varepsilon }_-^{\nu }
      \big)
    \bigg\}
    ,
  \end{aligned}
  \nonumber\\
  %
  %
  \bigg[
    {\mathcal{P}}
        {}^{\vbfield{ij^{\vphantom{\prime}}}}_{\fermfield{i}}
        {}^{\vbfield{ij^\prime}}_{\antifermfield{j}}
      \big(z,k_\perp^\mu,\bar p^\mu,\pm,\pm\big)
  \bigg]^{\mu \nu}_{m_{ij}^{(\prime)}=0}
  &=
  \frac{
    a^{V \, \mathrm{LR/RL}}_{ij, i, j}(+1)
    a^{V \, \mathrm{LR/RL}}_{ij^\prime, i, j}(+1)
  }{(z-1)z}
  \bigg\{
    g^{\mu \nu}
    \mp
    s_\varepsilon
    \big(
      \bar{\varepsilon }_-^{\mu }
      \bar{\varepsilon }_+^{\nu }
      -
      \bar{\varepsilon }_+^{\mu }
      \bar{\varepsilon }_-^{\nu }
    \big)
  \bigg\}
  .
  \label{eq:FS-VFFbar-splfunc-specialcase}
\end{align}
First, applying~\eqref{eq:VFF-i-j-symm} and~\eqref{eq:VFF-swap-cpl-wo-chir} via
\begin{align}
  z \to 1-z
  ,\quad
  k_\perp^\mu \to - k_\perp^\mu
  ,\quad
  \kappa_i \leftrightarrow \kappa_j
  ,\quad
  m_i \leftrightarrow m_j
  ,\quad
  I_{\fermfield{i}^\sigma \fermfield{j}^\sigma}^{\vbfield{ij\smash{{}^{(\prime)}}}}
  \to
  I_{\fermfield{j}^\sigma \fermfield{i}^\sigma}^{\vbfield{ij\smash{{}^{(\prime)}}}}
  \label{eq:VFFbar-VFbarF-crossing-cpl}
\end{align}
yields the $V^\ast \to \bar f f$ splitting functions
\begin{align}
  %
  %
  \bigg[
    {\mathcal{P}}
        {}^{\vbfield{ij^{\vphantom{\prime}}}}_{\antifermfield{i}}
        {}^{\vbfield{ij^\prime}}_{\fermfield{j}}
      \big(z,k_\perp^\mu,\bar p^\mu,\mp,\pm\big)
  \bigg]^{\mu \nu}_{m_{ij}^{(\prime)}=0}
  &=
  I_{\fermfield{j}^\mathrm{R/L} \fermfield{i}^\mathrm{R/L}}^{\vbfield{ij}}
  \Big(I_{\fermfield{j}^\mathrm{R/L} \fermfield{i}^\mathrm{R/L}}^{\vbfield{ij^\prime}}\Big)^\ast
  \begin{aligned}[t]
    \bigg\{
      &
      \frac{-k_\perp^2}{(z-1)z}
      g^{\mu \nu}
      -
      4
      k_\perp^\mu
      k_\perp^\nu
      \\
      &{}
      \mp
      s_\varepsilon
      \frac{2z-1}{(z-1)z}
      k_\perp^2
      \big(
        \bar{\varepsilon }_-^{\mu }
        \bar{\varepsilon }_+^{\nu }
        -
        \bar{\varepsilon }_+^{\mu }
        \bar{\varepsilon }_-^{\nu }
      \big)
    \bigg\}
    ,
  \end{aligned}
  \nonumber\\
  %
  %
  \bigg[
    {\mathcal{P}}
        {}^{\vbfield{ij^{\vphantom{\prime}}}}_{\antifermfield{i}}
        {}^{\vbfield{ij^\prime}}_{\fermfield{j}}
      \big(z,k_\perp^\mu,\bar p^\mu,\pm,\pm\big)
  \bigg]^{\mu \nu}_{m_{ij}^{(\prime)}=0}
  &=
  \frac{
    a^{V \, \mathrm{LR/RL}}_{ij, j, i}(+1)
    a^{V \, \mathrm{LR/RL}}_{ij^\prime, j, i}(+1)
  }{(z-1)z}
  \bigg\{
    g^{\mu \nu}
    \mp
    s_\varepsilon
    \big(
      \bar{\varepsilon }_-^{\mu }
      \bar{\varepsilon }_+^{\nu }
      -
      \bar{\varepsilon }_+^{\mu }
      \bar{\varepsilon }_-^{\nu }
    \big)
  \bigg\}
  .
  \label{eq:FS-VFbarF-splfunc-specialcase}
\end{align}
Second, these results may also be obtained from~\eqref{eq:FS-VFFbar-splfunc-specialcase}
by simply applying~\eqref{eq:VFF-swap-cpl-w-chir}.

%
\subsubsection[%
  \mathinhead{H^\ast \to f \bar f}{H* to f fbar}
  splitting functions%
  ]{%
  \mathinheadbold{H^\ast \to f \bar f}{H* to f fbar}
  splitting functions}
\label{sect:HFFbar}
The only relevant diagram for the $H^\ast \to f \bar f$ splitting
functions,
\drawmomtrue%
\drawvertdottrue%
\begin{alignat}{2}%
  \ftntmath{%
    \MContrExplFields{\higgsfield{ij}}{\fermfield{i}}{\antifermfield{j}}^{\polferm_i \polferm_j}%
  }%
  &\ftntmath{{}={}}%
  \FSDiagGeo%
  \ftntmath{%
    \splittingdiagIndivFS{T^{(n)}_{\higgsfield{ij} X}}%
      {scalar}{fermion}{antifermion}%
      {\higgsfield{ij}}{\fermfield{i}, \polferm_i}{\antifermfield{j}, \polferm_j}%
    ,%
  }%
  \resetdiagargs%
  \label{eq:HFFbar-diags-FS}
\end{alignat}%
translates into
\begin{align}
  %
  %
  \MContrExplFields{\higgsfield{ij}}{\fermfield{i}}{\antifermfield{j}}^{\polferm_i \polferm_j}
  =
  \frac{
    e T^{(n)}_{\higgsfield{ij} X}(p_{ij})
  }{
    p_{ij}^2-m_{ij}^2
  }
  \bar{u}_{\polferm_i}(p_i)
  \left(
    \omega_+
    C^{\mathrm{R}}_{\higgsfield{ij} \antifermfield{i} \fermfield{j}}
    +
    \omega_-
    C^{\mathrm{L}}_{\higgsfield{ij} \antifermfield{i} \fermfield{j}}
  \right)
  v_{\polferm_j}(p_j)
  .
  \label{eq:HFFbar-EnhAmp}
\end{align}
In the derivation of the splitting functions given below, we use the abbreviations
\begin{align}
  a^{H \chir_1 \chir_2}_{ij, i, j}
  (\pm 1)
  &=
  (1-z) m_i
  C^{\chir_1}_{\higgsfield{ij} \antifermfield{i} \fermfield{j}}
  \pm
  z m_j
  C^{\chir_2}_{\higgsfield{ij} \antifermfield{i} \fermfield{j}}
  ,
  \nonumber
  \\
  b^{H \chir_1 \chir_2}_{ij, i, j}
  (\pm 1)
  &=
  m_i
  C^{\chir_1}_{\higgsfield{ij} \antifermfield{i} \fermfield{j}}
  \pm
  m_j
  C^{\chir_2}_{\higgsfield{ij} \antifermfield{i} \fermfield{j}}
  ,
  \nonumber\\
  d^{H}_{ij, i, j}
  (\pm 1)
  &=
  C^{\mathrm{L}}_{\higgsfield{ij} \antifermfield{i} \fermfield{j}}
  C^{\mathrm{L} \ast}_{\higgsfield{ij^\prime} \antifermfield{i} \fermfield{j}}
  \pm
  C^{\mathrm{R}}_{\higgsfield{ij} \antifermfield{i} \fermfield{j}}
  C^{\mathrm{R} \ast}_{\higgsfield{ij^\prime} \antifermfield{i} \fermfield{j}}
  ,
  \nonumber\\
  e^{H}_{ij, i, j}
  (\pm 1)
  &=
  C^{\mathrm{L}}_{\higgsfield{ij} \antifermfield{i} \fermfield{j}}
  C^{\mathrm{R} \ast}_{\higgsfield{ij^\prime} \antifermfield{i} \fermfield{j}}
  \pm
  C^{\mathrm{R}}_{\higgsfield{ij} \antifermfield{i} \fermfield{j}}
  C^{\mathrm{L} \ast}_{\higgsfield{ij^\prime} \antifermfield{i} \fermfield{j}}
  ,
  \label{eq:HFFbar-abbrev}
\end{align}
as obtained from~\eqref{eq:gen-ferm-abbs}.
The symbols~$c^H$ are not relevant and hence omitted.

\paragraph{Polarization \boldmath{$\polferm_i=+$}, \boldmath{$\polferm_j=+$}.}
The result
\begin{align}
  \mathcal{P}
    {}^{\higgsfield{ij^{\vphantom{\prime}}}}_{\fermfield{i}}
    {}^{\higgsfield{ij^\prime}}_{\antifermfield{j}}
    \big(z,k_\perp^\mu,\bar p^\mu,+,+\big)
  &=
  \frac{k_\perp^2}{(z-1) z}
  C^{\mathrm{L}}_{\higgsfield{ij} \antifermfield{i} \fermfield{j}}
  C^{\mathrm{L} \ast}_{\higgsfield{ij^\prime} \antifermfield{i} \fermfield{j}}
  \label{eq:HFF-PP-split-func}
\end{align}
only involves left-handed couplings.
In the massless limit, this reproduces the expectation from the
chirality structure of the 
$H\bar ff$ vertex~\eqref{eq:app-Sff-vertex}.
In the \gls{ewsm},~\eqref{eq:HFF-PP-split-func} becomes
\begin{align}
  \mathcal{P}
    {}^{\higgsfield{}}_{\fermfield{i}}
    {}^{\higgsfield{}}_{\antifermfield{j}}
    \big(z,k_\perp^\mu,\bar p^\mu,+,+\big)
  \bigg|_{\mathrm{EWSM}}
  =
  \frac{m_i^2}{(e v)^2}
  \frac{k_\perp^2}{(z-1) z}
  \delta_{\fermfield{i} \fermfield{j}}
  .
  \label{eq:HFF-PP-split-func-EWSM}
\end{align}

\paragraph{Polarization \boldmath{$\polferm_i=+$}, \boldmath{$\polferm_j=-$}.}
Consistently with the structure of the~$H\bar ff$
vertex~\eqref{eq:app-Sff-vertex}, the properties
\eqref{eq:gen-ferm-abbs-mlferm} imply that the splitting
function
\begin{align}
  \mathcal{P}
    {}^{\higgsfield{ij^{\vphantom{\prime}}}}_{\fermfield{i}}
    {}^{\higgsfield{ij^\prime}}_{\antifermfield{j}}
    \big(z,k_\perp^\mu,\bar p^\mu,+,-\big)
  &=
  -
  \frac{
    a_{\mathrm{R}\mathrm{L}}^{H}(-1)
    a_{\mathrm{R}\mathrm{L}}^{H \prime \ast}(-1)
  }{
    (z-1) z
  }
  \label{eq:HFF-PM-split-func}
\end{align}
vanishes for massless fermions.
In the \gls{ewsm}, we obtain
\begin{align}
  \mathcal{P}
    {}^{\higgsfield{}}_{\fermfield{i}}
    {}^{\higgsfield{}}_{\antifermfield{j}}
    \big(z,k_\perp^\mu,\bar p^\mu,+,-\big)
  \bigg|_{\mathrm{EWSM}}
  =
  -
  \frac{m_i^4}{(e v)^2}
  \frac{(1-2 z)^2}{(z-1) z}
  \delta_{\fermfield{i} \fermfield{j}}
  .
  \label{eq:HFF-PM-split-func-EWSM}
\end{align}

\paragraph{Flipping the fermion polarizations.}
This proceeds as described in~\eqref{eq:VFF-MM-MP-split-func}, with $\mathrm{L}
\leftrightarrow \mathrm{R}$ meaning
\begin{align}
  C^{\mathrm{L}}_{\higgsfield{ij} \antifermfield{i} \fermfield{j}}
  \leftrightarrow
  C^{\mathrm{R}}_{\higgsfield{ij} \antifermfield{i} \fermfield{j}}
  .
  \label{eq:SFF-swap-cpl-w-chir} 
\end{align}

\paragraph{Unpolarized case.}
The above expressions, together with their polarization-flipped
equivalents, combine into
\begin{align}
  \mathcal{P}
    {}^{\higgsfield{ij^{\vphantom{\prime}}}}_{\fermfield{i}}
    {}^{\higgsfield{ij^\prime}}_{\antifermfield{j}}
    \big(z,k_\perp^\mu,\bar p^\mu,\mathrm{U},\mathrm{U}\big)
  &=
  \frac{
    k_\perp^2 d^H(+1)
    -
    a_{\mathrm{L}\mathrm{R}}^{H}(-1)
    a_{\mathrm{L}\mathrm{R}}^{H \prime \ast}(-1)
    -
    a_{\mathrm{R}\mathrm{L}}^{H}(-1)
    a_{\mathrm{R}\mathrm{L}}^{H \prime \ast}(-1)
  }{(z-1) z}
  .
  \label{eq:HFF-UU-split-func}
\end{align}

%
\subsubsection[%
  \mathinhead{H^\ast \to \bar f f}{H* to f fbar}
  splitting functions%
  ]{%
  \mathinheadbold{H^\ast \to \bar f f}{H* to f fbar}
  splitting functions}
\label{sect:HFbarF}
Again, the splitting functions for $H^\ast \to \bar f f$ and $H^\ast \to
f \bar f$ are related via~\eqref{eq:i-j-symmru-splfunc},
\begin{align}
  \mathcal{P}
    {}^{\higgsfield{ij^{\vphantom{\prime}}}}_{\antifermfield{i}}
    {}^{\higgsfield{ij^\prime}}_{\fermfield{j}}
    \big(z,k_\perp^\mu,\bar p^\mu,\polferm_i,\polferm_j\big)
  &=
  \mathcal{P}
    {}^{\higgsfield{ij^{\vphantom{\prime}}}}_{\fermfield{j}}
    {}^{\higgsfield{ij^\prime}}_{\antifermfield{i}}
    \big(1-z,-k_\perp^\mu,\bar p^\mu,\polferm_j,\polferm_i\big)
  \Big|
  _{m_i \leftrightarrow m_j}
  \label{eq:HFFbar-split-func-i-j-symmru}
  ,
\end{align}
where the ones on the r.h.s.\ are obtained from the results of
Section~\ref{sect:HFFbar} through
\begin{align}
  C^{\chir}_{\higgsfield{ij}^{(\prime)} \antifermfield{i} \fermfield{j}}
  \to
  C^{\chir}_{\higgsfield{ij}^{(\prime)} \antifermfield{j} \fermfield{i}}
  .
  \label{eq:HFF-swap-cpl-wo-chir}
\end{align}
Inspecting the r.h.s.\ of~\eqref{eq:HFFbar-split-func-i-j-symmru}, we observe
that~$m_i \leftrightarrow m_j$ is equivalent to~$z\to1-z$,
$k_\perp \to - k_\perp$, and $\polferm_i \leftrightarrow \polferm_j$.
Thus, the splitting functions for $H^\ast \to \bar f f$ are recovered
from those for~$H^\ast \to f \bar f$ by
applying~\eqref{eq:HFF-swap-cpl-wo-chir} alone.


This is confirmed when studying the relation of
\drawmomtrue%
\drawvertdottrue%
\begin{alignat}{2}%
  \ftntmath{%
    \MContrExplFields{\higgsfield{ij}}{\antifermfield{i}}{\fermfield{j}}^{\polferm_i \polferm_j}%
  }%
  &\ftntmath{{}={}}%
  \FSDiagGeo%
  \ftntmath{%
    \splittingdiagIndivFS{T^{(n)}_{\higgsfield{ij} X}}%
      {scalar}{antifermion}{fermion}%
      {\higgsfield{ij}}{\antifermfield{i}, \polferm_i}{\fermfield{j}, \polferm_j}%
  }%
  \resetdiagargs%
  \label{eq:HFbarF-diags-FS}
\end{alignat}%

and
\begin{align}
  %
  %
  \MContrExplFields{\higgsfield{ij}}{\antifermfield{i}}{\fermfield{j}}^{\polferm_i \polferm_j}
  =
  \frac{
    - e T^{(n)}_{\higgsfield{ij} X}(p_{ij})
  }{
    p_{ij}^2-m_{ij}^2
  }
  \bar{u}_{\polferm_j}(p_j)
  \left(
    \omega_+
    C^{\mathrm{R}}_{\higgsfield{ij} \antifermfield{j} \fermfield{i}}
    +
    \omega_-
    C^{\mathrm{L}}_{\higgsfield{ij} \antifermfield{j} \fermfield{i}}
  \right)
  v_{\polferm_i}(p_i)
  ,
  \label{eq:HFbarF-EnhAmp}
\end{align}
to~\eqref{eq:HFFbar-diags-FS} and~\eqref{eq:HFFbar-EnhAmp} via charge
conjugation, along the lines of Section~\ref{sect:VFFbar}.
Specifically, due to the absence of a~$\gamma^\mu$-matrix in the
scalar--fermion coupling, it is found that~\eqref{eq:HFFbar-EnhAmp}
differs from~\eqref{eq:HFbarF-EnhAmp} only
by~\eqref{eq:HFF-swap-cpl-wo-chir}, in agreement with the results given above.
In contrast to~\eqref{eq:VFF-swap-cpl-w-chir} for the~$V^\ast \to \bar f
f$ case, the chirality labels in the couplings are \emph{not} swapped
in~\eqref{eq:HFFbar-EnhAmp}, and charge conjugation compensates the
conventional sign between the diagrams~\eqref{eq:HFFbar-diags-FS}
and~\eqref{eq:HFbarF-diags-FS}.


For~$H^\ast \to \bar f f$ splittings, the
abbreviations~\eqref{eq:gen-ferm-abbs} read
\begin{align}
  a^{H \chir_1 \chir_2}_{ij, j, i}
  (\pm 1)
  &=
  z m_j
  C^{\chir_1}_{\higgsfield{ij} \antifermfield{j} \fermfield{i}}
  \pm
  (1-z) m_i
  C^{\chir_2}_{\higgsfield{ij} \antifermfield{j} \fermfield{i}}
  ,
  \nonumber
  \\
  b^{H \chir_1 \chir_2}_{ij, j, i}
  (\pm 1)
  &=
  m_j
  C^{\chir_1}_{\higgsfield{ij} \antifermfield{j} \fermfield{i}}
  \pm
  m_i
  C^{\chir_2}_{\higgsfield{ij} \antifermfield{j} \fermfield{i}}
  ,
  \nonumber\\
  d^{H}_{ij, j, i}
  (\pm 1)
  &=
  C^{\mathrm{L}}_{\higgsfield{ij} \antifermfield{j} \fermfield{i}}
  C^{\mathrm{L} \ast}_{\higgsfield{ij^\prime} \antifermfield{j} \fermfield{i}}
  \pm
  C^{\mathrm{R}}_{\higgsfield{ij} \antifermfield{j} \fermfield{i}}
  C^{\mathrm{R} \ast}_{\higgsfield{ij^\prime} \antifermfield{j} \fermfield{i}}
  ,
  \nonumber\\
  e^{H}_{ij, j, i}
  (\pm 1)
  &=
  C^{\mathrm{L}}_{\higgsfield{ij} \antifermfield{j} \fermfield{i}}
  C^{\mathrm{R} \ast}_{\higgsfield{ij^\prime} \antifermfield{j} \fermfield{i}}
  \pm
  C^{\mathrm{R}}_{\higgsfield{ij} \antifermfield{j} \fermfield{i}}
  C^{\mathrm{L} \ast}_{\higgsfield{ij^\prime} \antifermfield{j} \fermfield{i}}
  .
  \label{eq:HFbarF-abbrev}
\end{align}
Similar to the situation for $V^\ast \to \bar f f$ splittings, the
combination of $m_i \leftrightarrow m_j$ and $z \to 1-z$
with~\eqref{eq:HFF-swap-cpl-wo-chir} maps the
abbreviations~\eqref{eq:HFFbar-abbrev} for $H^\ast \to f \bar f$ splittings
into these expressions.
Again, this emphasizes the $i \leftrightarrow j$ symmetry between the
associated splitting functions.\footnotemark\
\footnotetext{%
  As in the $V\to \bar f f$ case, the symmetry relation between $H^\ast
  \to \bar f f$ and $H^\ast \to f \bar f$ splitting functions that is
  suggested by charge conjugation, i.e.\ the application
  of~\eqref{eq:HFF-swap-cpl-wo-chir} \emph{alone}, yields abbreviation
  mappings
  \begin{align*}
    a^{H \chir_1 \chir_2}_{ij, i, j}(\pm 1)
    &\to
    \pm
    a^{H \chir_2 \chir_1}_{ij, j, i}(\pm 1)
    ,
    &
    b^{H \chir_1 \chir_2}_{ij, i, j}(\pm 1)
    &\to
    \pm
    b^{H \chir_2 \chir_1}_{ij, j, i}(\pm 1)
    ,
    \nonumber\\
    d^{H}_{ij, i, j}(\pm 1)
    &\to
    d^{H}_{ij, j, i}(\pm 1)
    ,
    &
    e^{H}_{ij, i, j}(\pm 1)
    &\to
    e^{H}_{ij, j, i}(\pm 1)
    ,
  \end{align*}
  which are slightly more complicated.
  Nevertheless, these substitutions highlight the symmetries of the
  splitting functions, too.
  Compare also to Footnote~\ref{ftnt:charge-conj-abbreviation-vff}.
}%

%
\subsubsection[%
  \mathinhead{V^\ast/H^\ast \to f \bar f}{V*/H* to f fbar}
  splitting functions%
  ]{%
  \mathinheadbold{V^\ast/H^\ast \to f \bar f}{V*/H* to f fbar}
  splitting functions}
\label{sect:VHFFbar}
Utilizing the ingredients from above, we obtain splitting functions
for the interference~$V^\ast/H^\ast \to f \bar{f}$.
Because they are proportional to the product of ``spin-preserving''
$V\bar ff$ and ``spin-flipping'' $H\bar ff$ couplings, they vanish for massless
fermions.
Results for flipped polarizations and the case $V^\ast/H^\ast \to \bar f
f$ may be obtained as before.
Specifically, charge conjugation corresponds
to~\eqref{eq:VFF-swap-cpl-w-chir}, \eqref{eq:HFF-swap-cpl-wo-chir}, and
multiplication by~$-1$.
When recovering $H^\ast/V^\ast \to f \bar f$ splitting functions
via~\eqref{eq:ij-ijpr-symmru-splfunc}, complex conjugation entails
$\bar{\varepsilon}^\mu_\pm \to \bar{\varepsilon}^{\ast \mu}_\pm =
s_\varepsilon \bar{\varepsilon}^\mu_\mp$.

\paragraph{Polarization \boldmath{$\polferm_i=+$}, \boldmath{$\polferm_j=+$}.}
We get
\begin{align}
  &
  \bigg[
    \mathcal{P}
      {}^{\vbfield{ij^{\vphantom{\prime}}}}_{\fermfield{i}}
      {}^{\higgsfield{ij^\prime}}_{\antifermfield{j}}
      \big(z,k_\perp^\mu,\bar p^\mu,+,+\big)
  \bigg]^{\mu}
  =
  \frac{
    C^{\mathrm{L} \ast}_{\higgsfield{ij^\prime} \antifermfield{i} \fermfield{j}}
  }{(z-1)z}
  \bigg\{
    \frac{
      k_\perp^2
      b_{\mathrm{L}\mathrm{R}}^{V}(-1)
    }{
      m_{ij}^2
    }
    \bar{p}^{\mu }
    +
    2
    s_{\varepsilon }
    a_{\mathrm{L}\mathrm{R}}^{V}(+1)
    (k_{\perp }  \bar{\varepsilon }_+)
    \bar{\varepsilon }_-^{\mu }
  \bigg\}
  ,
  \label{eq:VHFF-PP-split-func}
\end{align}
with the \gls{ewsm} special case
\begin{align}
\bigg[
    \mathcal{P}
      {}^{\vbfield{ij}}_{\fermfield{i}}
      {}^{\higgsfield{\vphantom{ij}}}_{\antifermfield{j}}
      \big(z,k_\perp^\mu,\bar p^\mu,+,+\big)
  \bigg]^{\mu}_{\mathrm{EWSM}} 
  ={}&
  \frac{1}{e v}
  \frac{m_i^2}{(z-1)z}
      \bigg\{
      2
      s_\varepsilon
      \Big(
        (z-1)
        I_{\fermfield{i}^\mathrm{L} \fermfield{j}^\mathrm{L}}^{\vbfield{ij}}
        -
        z
        I_{\fermfield{i}^\mathrm{R} \fermfield{j}^\mathrm{R}}^{\vbfield{ij}}
      \Big)
      (k_{\perp }  \bar{\varepsilon }_+)
      \bar{\varepsilon }_-^{\mu }
      \nonumber\\
      & \quad {} 
      - \frac{
        k_\perp^2
      }{m_{ij}^2}
      \Big(
        I_{\fermfield{i}^\mathrm{L} \fermfield{j}^\mathrm{L}}^{\vbfield{ij}}
        -
        I_{\fermfield{i}^\mathrm{R} \fermfield{j}^\mathrm{R}}^{\vbfield{ij}}
      \Big)
      \bar{p}^{\mu }
    \bigg\}
  \delta_{\fermfield{i} \fermfield{j}}
  .
  \label{eq:VHFF-PP-split-func-EWSM}
\end{align}

\paragraph{Polarization \boldmath{$\polferm_i=+$}, \boldmath{$\polferm_j=-$}.}
For
\begin{align}
  &\bigg[
    \mathcal{P}
      {}^{\vbfield{ij^{\vphantom{\prime}}}}_{\fermfield{i}}
      {}^{\higgsfield{ij^\prime}}_{\antifermfield{j}}
      \big(z,k_\perp^\mu,\bar p^\mu,+,-\big)
  \bigg]^{\mu}
  =
  2
  s_{\varepsilon }
  I_{\fermfield{i}^{\mathrm{R}} \fermfield{j}^{\mathrm{R}}}^{\vbfield{ij}}
    a_{\mathrm{R}\mathrm{L}}^{H \prime \ast}(-1)
  \bigg(
    \frac{
      (k_{\perp }  \bar{\varepsilon }_-)
    }{z}
    \bar{\varepsilon }_+^{\mu }
    +
    \frac{
    (k_{\perp } \bar{\varepsilon }_+)
    }{z-1}
    \bar{\varepsilon }_-^{\mu }
  \bigg)
  +
  \frac{
    a_{\mathrm{R}\mathrm{L}}^{H \prime \ast}(-1)
    c_{\mathrm{R}\mathrm{L}}^{V}
  }{(z-1) z m_{ij}^2}
  \bar{p}^{\mu }
  ,
  \label{eq:VHFF-PM-split-func}
\end{align}
the \gls{ewsm} limit becomes
\begin{alignat}{2}
  &
  \lefteqn{
    \bigg[
      \mathcal{P}
        {}^{\vbfield{ij}}_{\fermfield{i}}
        {}^{\higgsfield{\vphantom{ij}}}_{\antifermfield{j}}
        \big(z,k_\perp^\mu,\bar p^\mu,+,-\big)
    \bigg]^{\mu}_{\mathrm{EWSM}}
  }
  \nonumber\\*
  &=
  \frac{(2 z-1) m_i^2}{e v}
  \bigg\{
    &&
    2 s_\varepsilon
    I_{\fermfield{i}^\mathrm{R} \fermfield{j}^\mathrm{R}}^{\vbfield{ij}}
    \biggl(
    \frac{(k_{\perp }\bar{\varepsilon }_-)}{z}
    \bar{\varepsilon }_+^{\mu }
    +
    \frac{(k_{\perp }\bar{\varepsilon }_+)}{z-1}
    \bar{\varepsilon }_-^{\mu }
    \biggr)
    \nonumber\\
    &&&{}-
    \frac{1}{(z-1) z m_{ij}^2}
    \Big(
      \big(m_i^2+2 (z-1) z m_{ij}^2\big)
      I_{\fermfield{i}^\mathrm{R} \fermfield{j}^\mathrm{R}}^{\vbfield{ij}}
      -
      m_i^2
      I_{\fermfield{i}^\mathrm{L} \fermfield{j}^\mathrm{L}}^{\vbfield{ij}}
    \Big)
    \bar{p}^{\mu }
  \bigg\}
  \delta_{\fermfield{i} \fermfield{j}}
  .
  \label{eq:VHFF-PM-split-func-EWSM}
\end{alignat}

\paragraph{Unpolarized case.}
The combination of all polarization configurations yields
\begin{align}
  &\bigg[
    \mathcal{P}
      {}^{\vbfield{ij^{\vphantom{\prime}}}}_{\fermfield{i}}
      {}^{\higgsfield{ij^\prime}}_{\antifermfield{j}}
      \big(z,k_\perp^\mu,\bar p^\mu,\mathrm{U},\mathrm{U}\big)
  \bigg]^{\mu}
  \nonumber\\*
  &=
  \frac{
    a_{\mathrm{L}\mathrm{R}}^{H \prime \ast}(-1) c_{\mathrm{L}\mathrm{R}}^{V}
    +
    a_{\mathrm{R}\mathrm{L}}^{H \prime \ast}(-1) c_{\mathrm{R}\mathrm{L}}^{V}
    +
    k_\perp^2
    \Big(
      I_{\fermfield{i}^{\mathrm{L}} \fermfield{j}^{\mathrm{L}}}^{\vbfield{ij}}
      b_{\mathrm{L}\mathrm{R}}^{H \prime \ast}(-1)
      +
      I_{\fermfield{i}^{\mathrm{R}} \fermfield{j}^{\mathrm{R}}}^{\vbfield{ij}}
      b_{\mathrm{R}\mathrm{L}}^{H \prime \ast}(-1)
    \Big)
  }{(z-1) z m_{ij}^2}
  \bar{p}^{\mu }
  \nonumber\\
  &\phantom{={}}
  +
  2
  \bigg(
    I_{\fermfield{i}^{\mathrm{L}} \fermfield{j}^{\mathrm{L}}}^{\vbfield{ij}}
    b_{\mathrm{L}\mathrm{R}}^{H \prime \ast}(+1)
    +
    I_{\fermfield{i}^{\mathrm{R}} \fermfield{j}^{\mathrm{R}}}^{\vbfield{ij}}
    b_{\mathrm{R}\mathrm{L}}^{H \prime \ast}(+1)
  \bigg)
  k_{\perp }^{\mu }
  .
  \label{eq:VHFF-UU-split-func}
\end{align}

%
\subsubsection[%
  \mathinhead{\bar f^\ast \to V \bar f}{fbar* to V fbar}
  splitting functions%
  ]{%
  \mathinheadbold{\bar f^\ast \to V \bar f}{fbar* to V fbar}
  splitting functions}
\label{sect:FbarVFbar}
The computation of~$\bar f^\ast \to V \bar f$ splitting functions
involves the diagrams
\drawmomtrue%
\drawvertdottrue%
\begin{alignat}{2}%
  \ftntmath{%
    \MContrExplFields{\antifermfield{ij}}{\vbfield{i}}{\antifermfield{j}}^{\polvb_i \polferm_j}%
  }%
  &\ftntmath{{}={}}%
  \FSDiagGeo%
  \ftntmath{%
    \splittingdiagIndivFS{\bar T^{(n)}_{\antifermfield{ij} X}}%
      {antifermion}{photon}{antifermion}%
      {\antifermfield{ij}}{\vbfield{i}, \polvb_i}{\antifermfield{j}, \polferm_j}%
    ,%
  }%
  \resetdiagargs%
   &\quad
  \ftntmath{%
    \MContrExplFields{\antifermfield{ij}}{\gbfield{i}}{\antifermfield{j}}^{\mathrm{G} \polferm_j}%
  }%
  &\ftntmath{{}={}}%
  \FSDiagGeo%
  \ftntmath{%
    \splittingdiagIndivFS{\bar T^{(n)}_{\antifermfield{ij} X}}%
      {antifermion}{scalar}{antifermion}%
      {\antifermfield{ij}}{\gbfield{i}}{\antifermfield{j}, \polferm_j}%
    ,%
  }%
  \resetdiagargs%
  \label{eq:FbarVFbar-diags-FS}
\end{alignat}%
corresponding to the analytical expressions
\begin{align}
  %
  %
  \MContrExplFields{\antifermfield{ij}}{\vbfield{i}}{\antifermfield{j}}^{\polvb_i \polferm_j}%
  &=
  - \imagi e
  \bar T^{(n)}_{\antifermfield{ij} X} (p_{ij})
  G^{\fermfield{ij} \antifermfield{ij}}(p_{ij},-p_{ij})
  \slashed \varepsilon_{\polvb_i}^{\ast} (p_{i})
  \left(
    \omega_-
    I_{\fermfield{ij}^\mathrm{L} \fermfield{j}^\mathrm{L}}^{\vbantifield{i}}
    +
    \omega_+
    I_{\fermfield{ij}^\mathrm{R} \fermfield{j}^\mathrm{R}}^{\vbantifield{i}}
  \right)
  v_{\polferm_j}(p_j)
  ,
  \nonumber\\
  %
  %
  \MContrExplFields{\antifermfield{ij}}{\gbfield{i}}{\antifermfield{j}}^{\mathrm{G} \polferm_j}%
  &=
  \imagi e
  \eta _{\vbfield{i}}
  \bar T^{(n)}_{\antifermfield{ij} X} (p_{ij})
  G^{\fermfield{ij} \antifermfield{ij}}(p_{ij},-p_{ij})
  \frac{1}{m_i}
  \nonumber\\
  &\phantom{{}={}}
  \cdot
  \left\{
    \omega_+
    \left(
      m_{ij}
      I_{\fermfield{ij}^\mathrm{R} \fermfield{j}^\mathrm{R}}^{\vbantifield{i}}
      -
      m_j
      I_{\fermfield{ij}^\mathrm{L} \fermfield{j}^\mathrm{L}}^{\vbantifield{i}}
    \right)
    +
    \omega_-
    \left(
      m_{ij}
      I_{\fermfield{ij}^\mathrm{L} \fermfield{j}^\mathrm{L}}^{\vbantifield{i}}
      -
      m_j
      I_{\fermfield{ij}^\mathrm{R} \fermfield{j}^\mathrm{R}}^{\vbantifield{i}}
    \right)
  \right\}
  v_{\polferm_j}(p_j)
  ,
  \label{eq:FbarVFbar-EnhAmp}
\end{align}
where \eqref{eq:coupling-relat-Vff-Gff} was used for the ${\bar G_i \bar f_{ij} f_j}$ coupling.
Combining the techniques introduced in the previous sections,
we obtain the results quoted below.
In particular, we employ an identity similar
to the decomposition~\eqref{eq:LT-goldstone-strat} for the treatment of mass-singular
parts in the splitting functions involving longitudinally polarized vector bosons.
On top of this, we insert the spin-projected completeness
relation~\eqref{eq:spinproj-complrelat-ferm}, the Sudakov
parametrization~\eqref{eq:sudakov-param-fs}, reduce the occurring Dirac
chains into a minimal set, and evaluate the quasi-collinear limit
using the power-counting~\eqref{eq:oc-summary}.
The adaption of
our alternative strategy for the treatment of mass-singular factors,
developed in
Appendix~\ref{sect:VVV-alt-approach} for $V^\ast \to V V$ splittings,
to the $\bar f^\ast \to V \bar f$ case is
discussed in Appendix~\ref{sect:FbarVFbar-alt-approach}.
We arrive at results featuring the following Dirac structures,
sandwiched between stripped underlying amplitudes
$\bar T^{(n)}_{\bar f X}$ and
$T^{(n)}_{\bar f X}$,
\begin{align}
  \slashed{\bar{p}}
  ,\qquad
  \slashed{\bar{p}}
  \gamma^5
  ,\qquad
  \slashed{k}_{\perp }
  \slashed{\bar{p}}
  ,\qquad
  \slashed{k}_{\perp }
  \slashed{\bar{p}}
  \gamma^5
  .
  \label{eq:ferm-splittings-dirac-structures}
\end{align}
As we detail in Section~\ref{sect:id-hard-me},
these contractions 
may be expressed in terms of matrix elements
of definite polarization,
\begin{align}
  \mathcal{M}^{(n)\, \pm}_{\bar f X}
  &=
  \bar T^{(n)}_{\bar f X}
  v_\pm
  ,
  \qquad
  \mathcal{M}^{(n)\, \pm\, \ast}_{\bar f X}
  =
  \bar v_\pm
  T^{(n)}_{\bar f X}
  .
  \label{eq:pol-ul-me-antiferm}
\end{align}
Recall that $\Sigma_\pm v_\pm=v_\pm$ correspond to antifermions with spin
$\pm1/2$ in the
reference direction defined by a vector $r^\mu$, as discussed
below~\eqref{eq:spin-projection-op}.
In particular, for massless
antifermions $\antifermfield{ij}$, $\antifermfield{ij'}$
with on-shell momentum $p^\mu$,
we can identify matrix elements of definite helicity via
\begin{align}
  \bar T^{(n)}_{\antifermfield{ij} X}(p)
  \,
  \slashed p
  \frac{1 \pm \gamma^5}{2}
  \,
  T^{(n)}_{\antifermfield{ij^\prime} X}(p)
  =
  \bar T^{(n)}_{\antifermfield{ij} X}(p)
  \,
  \frac{1 \mp \gamma^5}{2}
  \slashed p
  \,
  T^{(n)}_{\antifermfield{ij^\prime} X}(p)
  =
  \mathcal{M}^{(n)\, \pm}_{\antifermfield{ij} X}
  \mathcal{M}^{(n)\, \pm\, \ast}_{\antifermfield{ij^\prime} X}
  .
  \label{eq:id-sq-ulme-massless-antiferm}
\end{align}
Terms involving~$\slashed{k}_{\perp }$ typically receive an additional
suppression in the
quasi-collinear limit, as detailed in Section~\ref{sect:ktslash-terms}.
According to~\eqref{eq:ij-ijpr-symmru-splfunc}, our results are
symmetric under
\begin{align}
  m_{ij} \leftrightarrow m_{ij}^\prime
  ,\quad
  \fermfield{ij} \leftrightarrow \fermfield{ij^\prime}
  ,
  \label{eq:FVF-swap-fermions-ij-ijpr}
\end{align}
and complex conjugation.
Combining complex conjugation with the transposition of Dirac chains
and subsequent reordering of $\gamma_0$ from
$T^\dagger\dots\gamma_0 T$ to $T^\dagger\gamma_0\dots T$
implies
\begin{align}
  \slashed{\bar p}
  \to
  \slashed{\bar p}
  ,\qquad
  \slashed{\bar p} \gamma^5
  \to
  \slashed{\bar p} \gamma^5
  ,\qquad
  \slashed{k}_\perp \slashed{\bar p}
  \to
  - \slashed{k}_\perp \slashed{\bar p}
  ,\qquad
  \slashed{k}_\perp \slashed{\bar p} \gamma^5
  \to
  \slashed{k}_\perp \slashed{\bar p} \gamma^5
  \label{eq:cc-of-dirac-chains}
\end{align}
for the Dirac structures from~\eqref{eq:ferm-splittings-dirac-structures}.
In the \gls{ewsm} with diagonal \gls{ckm} matrix, we have $m_{ij}=m_{ij}^\prime$,
and our splitting functions simplify accordingly.


Finally, for~$\bar f^\ast \to V \bar f$ splittings, the abbreviations 
introduced in~\eqref{eq:gen-ferm-abbs} read
\begin{align}
  a^{V \chir_1 \chir_2}_{i, ij, j}
  (\pm 1)
  &=
  - (1-z) z m_{ij}
  I_{\fermfield{ij}^{\chir_1} \fermfield{j}^{\chir_1}}^{\vbantifield{i}}
  \pm
  z m_j
  I_{\fermfield{ij}^{\chir_2} \fermfield{j}^{\chir_2}}^{\vbantifield{i}}
  ,
  \nonumber
  \\
  b^{V \chir_1 \chir_2}_{i, ij, j}
  (\pm 1)
  &=
  m_{ij}
  I_{\fermfield{ij}^{\chir_1} \fermfield{j}^{\chir_1}}^{\vbantifield{i}}
  \pm
  m_j
  I_{\fermfield{ij}^{\chir_2} \fermfield{j}^{\chir_2}}^{\vbantifield{i}}
  ,
  \nonumber
  \\
  c^{V \chir_1 \chir_2}_{i, ij, j}
  &=
  \big(
    (1-z) z m_{ij}^2
    - 2 (1-z) m_i^2
    - z m_j^2)
  \big)
  I_{\fermfield{ij}^{\chir_1} \fermfield{j}^{\chir_1}}^{\vbantifield{i}}
  +
  m_{ij} m_j z^2
  I_{\fermfield{ij}^{\chir_2} \fermfield{j}^{\chir_2}}^{\vbantifield{i}}
  ,
  \nonumber
  \\
  d^{V}_{i, ij, j}
  (\pm 1)
  &=
  I_{\fermfield{ij}^{\mathrm{L}} \fermfield{j}^{\mathrm{L}}}^{\vbantifield{i}}
  I_{\fermfield{ij^\prime}^{\mathrm{L}} \fermfield{j}^{\mathrm{L}}}^{\vbantifield{i} \ast}
  \pm
  I_{\fermfield{ij}^{\mathrm{R}} \fermfield{j}^{\mathrm{R}}}^{\vbantifield{i}}
  I_{\fermfield{ij^\prime}^{\mathrm{R}} \fermfield{j}^{\mathrm{R}}}^{\vbantifield{i} \ast}
  \nonumber
  \\
  e^{V}_{i, ij, j}
  (\pm 1)
  &=
  I_{\fermfield{ij}^{\mathrm{L}} \fermfield{j}^{\mathrm{L}}}^{\vbantifield{i}}
  I_{\fermfield{ij^\prime}^{\mathrm{R}} \fermfield{j}^{\mathrm{R}}}^{\vbantifield{i} \ast}
  \pm
  I_{\fermfield{ij}^{\mathrm{R}} \fermfield{j}^{\mathrm{R}}}^{\vbantifield{i}}
  I_{\fermfield{ij^\prime}^{\mathrm{L}} \fermfield{j}^{\mathrm{L}}}^{\vbantifield{i} \ast}
  .
  \label{eq:FbarVFbar-abbrev}
\end{align}

\paragraph{Polarization \boldmath{$\polvb_i=\mathrm{T}$}, \boldmath{$\polferm_j=+$}.}
We find
\begin{align}
  &
  \lefteqn{
    \mathcal{P}
      {}^{\antifermfield{ij^{\vphantom{\prime}}}}_{\vbfield{i}}
      {}^{\antifermfield{ij^\prime}}_{\antifermfield{j}}
      \big(z,k_\perp^\mu,\bar p^\mu,\mathrm{T},+\big)
  }
  \nonumber\\*
  &=
  \frac{1}{z^2}
  \bigg(
    I_{\fermfield{ij}^\mathrm{L} \fermfield{j}^\mathrm{L}}^{\vbantifield{i}}
    a_{\mathrm{L}\mathrm{R}}^{V \prime \ast}(+1)
    +
    I_{\fermfield{ij^\prime}^\mathrm{L} \fermfield{j}^\mathrm{L}}^{\vbantifield{i} \ast}
    a_{\mathrm{L}\mathrm{R}}^{V}(+1)
  \bigg)
  \slashed{k}_{\perp }
  \slashed{\bar{p}}
  \gamma^5
  \nonumber\\
  &\phantom{={}}
  -
  \frac{1}{z^2}
  \bigg(
    I_{\fermfield{ij}^\mathrm{L} \fermfield{j}^\mathrm{L}}^{\vbantifield{i}}
    a_{\mathrm{L}\mathrm{R}}^{V \prime \ast}(+1)
    -
    I_{\fermfield{ij^\prime}^\mathrm{L} \fermfield{j}^\mathrm{L}}^{\vbantifield{i} \ast}
    a_{\mathrm{L}\mathrm{R}}^{V}(+1)
  \bigg)
  \slashed{k}_{\perp }
  \slashed{\bar{p}}
  \nonumber\\
  &\phantom{={}}
  +
  \frac{1}{(z-1) z^2}
  \bigg(
    I_{\fermfield{ij}^\mathrm{L} \fermfield{j}^\mathrm{L}}^{\vbantifield{i}}
    I_{\fermfield{ij^\prime}^\mathrm{L} \fermfield{j}^\mathrm{L}}^{\vbantifield{i} \ast}
    k_\perp^2 (z^2-2 z+2)
    -
    a_{\mathrm{L}\mathrm{R}}^{V}(+1)
    a_{\mathrm{L}\mathrm{R}}^{V \prime \ast}(+1)
  \bigg)
  \slashed{\bar{p}}
  \nonumber\\
  &\phantom{={}}
  +
  \frac{1}{(z-1) z^2}
  \bigg(
    I_{\fermfield{ij}^\mathrm{L} \fermfield{j}^\mathrm{L}}^{\vbantifield{i}}
    I_{\fermfield{ij^\prime}^\mathrm{L} \fermfield{j}^\mathrm{L}}^{\vbantifield{i} \ast}
    k_\perp^2 (z^2-2 z+2)
    +
    a_{\mathrm{L}\mathrm{R}}^{V}(+1)
    a_{\mathrm{L}\mathrm{R}}^{V \prime \ast}(+1)
  \bigg)
  \slashed{\bar{p}}
  \gamma^5
  .
  \label{eq:FbarVFbar-TP-split-func}
\end{align}
The limit of vanishing fermion masses can be taken
via~\eqref{eq:gen-ferm-abbs-mlferm} and yields
\begin{align}
  &
  \mathcal{P}
    {}^{\antifermfield{ij^{\vphantom{\prime}}}}_{\vbfield{i}}
    {}^{\antifermfield{ij^\prime}}_{\antifermfield{j}}
    \big(z,k_\perp^\mu,\bar p^\mu,\mathrm{T},+\big)
  \bigg|_{m_{ij} = m_{ij}^\prime = m_j = 0}
  =
  2 k_\perp^2
  \frac{z^2-2 z+2}{(z-1) z^2}
  \,
  I_{\fermfield{ij}^\mathrm{L} \fermfield{j}^\mathrm{L}}^{\vbantifield{i}}
  I_{\fermfield{ij^\prime}^\mathrm{L} \fermfield{j}^\mathrm{L}}^{\vbantifield{i} \ast}
  \,
  \slashed{\bar{p}}
  \frac{1+\gamma^5}{2}
  .
  \label{eq:FbarVFbar-TP-split-func-mlferm}
\end{align}
Using~\eqref{eq:id-sq-ulme-massless-antiferm}, we can associate this
result to the hard matrix element for a massless antifermion of
helicity~$+\frac{1}{2}$.
Once more, the spin structure matches the anticipation from the
``spin-preserving'' nature of the~$V\bar ff$
vertex~\eqref{eq:app-Vff-vertex}.
Similarly, only left-handed couplings remain.

\paragraph{Polarization \boldmath{$\polvb_i=\mathrm{L}$}, \boldmath{$\polferm_j=+$}.}
We get
\begin{align}
  \mathcal{P}
      {}^{\antifermfield{ij^{\vphantom{\prime}}}}_{\vbfield{i}}
      {}^{\antifermfield{ij^\prime}}_{\antifermfield{j}}
      \big(z,k_\perp^\mu,\bar p^\mu,\mathrm{L},+\big)
  ={} &
  \frac{1}{2 m_i^2 (z-1) z}
      \bigg\{
      \Big(
        b_{\mathrm{L}\mathrm{R}}^{V}(-1)
        c_{\mathrm{L}\mathrm{R}}^{V \prime \ast}
        -
        b_{\mathrm{L}\mathrm{R}}^{V \prime \ast}(-1)
        c_{\mathrm{L}\mathrm{R}}^{V})
      \Big)
      \slashed{k}_{\perp }
      \slashed{\bar{p}}
      \nonumber\\
      & \quad {}
      +
      \Big(
        b_{\mathrm{L}\mathrm{R}}^{V}(-1)
        c_{\mathrm{L}\mathrm{R}}^{V \prime \ast}
        +
        b_{\mathrm{L}\mathrm{R}}^{V \prime \ast}(-1)
        c_{\mathrm{L}\mathrm{R}}^{V})
      \Big)
      \slashed{k}_{\perp }
      \slashed{\bar{p}}
      \gamma^5
      \nonumber\\
      & \quad {}
      -
      \bigg(
        z
        k_\perp^2
        b_{\mathrm{L}\mathrm{R}}^{V}(-1)
        b_{\mathrm{L}\mathrm{R}}^{V \prime \ast}(-1)
        +
        \frac{
           c_{\mathrm{L}\mathrm{R}}^{V} c_{\mathrm{L}\mathrm{R}}^{V \prime \ast}
        }{z}
      \bigg)
      \slashed{\bar{p}}
      \gamma^5
      \nonumber\\
      & \quad {}
      +
      \bigg(
        z
        k_\perp^2
        b_{\mathrm{L}\mathrm{R}}^{V}(-1)
        b_{\mathrm{L}\mathrm{R}}^{V \prime \ast}(-1)
        -
        \frac{
           c_{\mathrm{L}\mathrm{R}}^{V} c_{\mathrm{L}\mathrm{R}}^{V \prime \ast}
        }{z}
      \bigg)
      \slashed{\bar{p}}
    \bigg\}
    .
  \label{eq:FbarVFbar-LP-split-func}
\end{align}
This result is well behaved for~$m_i \to 0$ due to the fact that
$c_{\mathrm{L}\mathrm{R}}^{V (\prime)}/m_{i}$,
$b_{\mathrm{L}\mathrm{R}}^{V}(-1)/m_{i}$, and
$b_{\mathrm{R}\mathrm{L}}^{V}(-1)/m_{i}$ 
all have proper small-mass limits.
More specifically,~\eqref{eq:bV-abb-via-scalar-cpl}
and~\eqref{eq:cV-abb-via-aH-abb-massless} yield
\begin{align}
  \frac{b_{\chir (-\chir)}^{V}(-1)}{m_{i}}
  \,\xrightarrow[m_i \to 0]{\gbfield{i} \to \higgsfield{i}}\,
  -
  \eta_{\vbfield{i}}^\ast
  C^\chir_{\higgsantifield{i} \antifermfield{ij} \fermfield{j}}
  ,\qquad
  \frac{c_{\chir (-\chir)}^{V}}{m_i}
  \,\xrightarrow[m_i \to 0]{\gbfield{i} \to \higgsfield{i}}\,
  \eta_{\vbfield{i}}^\ast
  a_{\chir (-\chir)}^{H}(-1)
  ,
  \label{eq:FbarVFbar-FbarHFbar-abbrev-mlVH}
\end{align}
which implies that we recover the corresponding splitting
function~\eqref{eq:FbarHFbar-P-split-func} for~$\bar{f}^\ast \to H
\bar{f}$ given below,
\begin{align}
  &
  \mathcal{P}
      {}^{\antifermfield{ij^{\vphantom{\prime}}}}_{\vbfield{i}}
      {}^{\antifermfield{ij^\prime}}_{\antifermfield{j}}
      \big(z,k_\perp^\mu,\bar p^\mu,\mathrm{L},\polferm\big)
  \,\xrightarrow[m_i \to 0]{\gbfield{i} \to \higgsfield{i}}\,
  \mathcal{P}
    {}^{\antifermfield{ij^{\vphantom{\prime}}}}_{\higgsfield{i}}
    {}^{\antifermfield{ij^\prime}}_{\antifermfield{j}}
    \big(z,k_\perp^\mu,\bar p^\mu,\polferm\big)
  \bigg|_{m_i=0}
  .
\end{align}
For vanishing fermion masses, \eqref{eq:gen-ferm-abbs-mlferm} implies
\begin{align}
  &
  \mathcal{P}
    {}^{\antifermfield{ij^{\vphantom{\prime}}}}_{\vbfield{i}}
    {}^{\antifermfield{ij^\prime}}_{\antifermfield{j}}
    \big(z,k_\perp^\mu,\bar p^\mu,\mathrm{L},+\big)
  \bigg|_{m_{ij} = m_{ij}^\prime = m_j = 0}
  =
  -
  \frac{
    2 m_i^2 (z-1)
  }{z^2}
  \,
  I_{\fermfield{ij}^\mathrm{L} \fermfield{j}^\mathrm{L}}^{\vbantifield{i}}
  I_{\fermfield{ij^\prime}^\mathrm{L} \fermfield{j}^\mathrm{L}}^{\vbantifield{i} \ast}
  \,
  \slashed{\bar{p}}
  \frac{1+\gamma^5}{2}
  ,
  \label{eq:FbarVFbar-LP-split-func-mlferm}
\end{align}
again consistent with the ``spin-preserving'' $V\bar ff$
vertex~\eqref{eq:app-Vff-vertex}.

\paragraph{Flipping the antifermion polarization.}
This proceeds as described in Section~\ref{sect:VFFbar}, with the
exception that the~$\gamma^5$ matrices appear explicitly in our results.
We get
\begin{align}
  \mathcal{P}
    {}^{\antifermfield{ij^{\vphantom{\prime}}}}_{\vbfield{i}}
    {}^{\antifermfield{ij^\prime}}_{\antifermfield{j}}
    \big(z,k_\perp^\mu,\bar p^\mu,\mathrm{T}/\mathrm{L},\mp\big)
  &=
  \mathcal{P}
    {}^{\antifermfield{ij^{\vphantom{\prime}}}}_{\vbfield{i}}
    {}^{\antifermfield{ij^\prime}}_{\antifermfield{j}}
    \big(z,k_\perp^\mu,\bar p^\mu,\mathrm{T}/\mathrm{L},\pm\big)
  \bigg|_{\mathrm{L} \leftrightarrow \mathrm{R}, \gamma^5 \to -\gamma^5}
  .
  \label{eq:FbarVFbar-TM-LM-split-func}
\end{align}

\paragraph{Unpolarized cases.}
Summing over the polarizations of the external
antifermion while keeping the vector boson transverse,
we get
\begin{align}
  &
  \lefteqn{
    \mathcal{P}
      {}^{\antifermfield{ij^{\vphantom{\prime}}}}_{\vbfield{i}}
      {}^{\antifermfield{ij^\prime}}_{\antifermfield{j}}
      \big(z,k_\perp^\mu,\bar p^\mu,\mathrm{T},\mathrm{U}\big)
  }
  \nonumber\\*
  &=
  \frac{1}{(z-1) z^2}
  \begin{aligned}[t]
    \Big\{
      &\Big(
        k_\perp^2
        (z^2-2 z+2)
        d^V(-1)
        +
        a_{\mathrm{L}\mathrm{R}}^{V}(+1)
        a_{\mathrm{L}\mathrm{R}}^{V \prime \ast}(+1)
        -
        a_{\mathrm{R}\mathrm{L}}^{V}(+1)
        a_{\mathrm{R}\mathrm{L}}^{V \prime \ast}(+1)
      \Big)
      \slashed{\bar{p}}
      \gamma^5
      \nonumber\\
      &{}
      +
      \Big(
        k_\perp^2
        (z^2-2 z+2)
        d^V(+1)
        -
        a_{\mathrm{L}\mathrm{R}}^{V}(+1)
        a_{\mathrm{L}\mathrm{R}}^{V \prime \ast}(+1)
        -
        a_{\mathrm{R}\mathrm{L}}^{V}(+1)
        a_{\mathrm{R}\mathrm{L}}^{V \prime \ast}(+1)
      \Big)
      \slashed{\bar{p}}
    \Big\}
  \end{aligned}
  \nonumber\\
  &\phantom{={}}
  +
  \frac{
    z-1
  }{z}
  \Big\{
    (
      m_{ij}
      +
      m_{ij}^{\prime}
    )
    d^V(-1)
    \slashed{k}_{\perp }
    \slashed{\bar{p}}
    \gamma^5
    +
    (
      m_{ij}
      -
      m_{ij}^{\prime}
    )
    d^V(+1)
  \slashed{k}_{\perp }
  \slashed{\bar{p}}
  \Big\}
  .
  \label{eq:FbarVFbar-TU-split-func-EWSM}
\end{align}
For a longitudinal vector boson, this procedure yields
\begin{align}
  &
  \lefteqn{
    \mathcal{P}
      {}^{\antifermfield{ij^{\vphantom{\prime}}}}_{\vbfield{i}}
      {}^{\antifermfield{ij^\prime}}_{\antifermfield{j}}
      \big(z,k_\perp^\mu,\bar p^\mu,\mathrm{L},\mathrm{U}\big)
  }
  \nonumber\\*
  &=  
  \frac{1}{2 (z-1) m_i^2}
  \Big\{
    \frac{1}{z}
    \Big(
      b_{\mathrm{L}\mathrm{R}}^{V}(-1) c_{\mathrm{L}\mathrm{R}}^{V \prime \ast}
      -
      b_{\mathrm{L}\mathrm{R}}^{V \prime \ast}(-1) c_{\mathrm{L}\mathrm{R}}^{V}
      +
      b_{\mathrm{R}\mathrm{L}}^{V}(-1) c_{\mathrm{R}\mathrm{L}}^{V \prime \ast}
      -
      b_{\mathrm{R}\mathrm{L}}^{V \prime \ast}(-1) c_{\mathrm{R}\mathrm{L}}^{V}
    \Big)
    \slashed{k}_{\perp }
    \slashed{\bar{p}}
    \nonumber\\
    &\phantom{={}}
    +
    \frac{1}{z}
    \Big(
      b_{\mathrm{L}\mathrm{R}}^{V}(-1) c_{\mathrm{L}\mathrm{R}}^{V \prime \ast}
      +
      b_{\mathrm{L}\mathrm{R}}^{V \prime \ast}(-1) c_{\mathrm{L}\mathrm{R}}^{V}
      -
      b_{\mathrm{R}\mathrm{L}}^{V}(-1) c_{\mathrm{R}\mathrm{L}}^{V \prime \ast}
      -
      b_{\mathrm{R}\mathrm{L}}^{V \prime \ast}(-1) c_{\mathrm{R}\mathrm{L}}^{V}
    \Big)
    \slashed{k}_{\perp }
    \slashed{\bar{p}}
    \gamma^5
    \nonumber\\
    &\phantom{={}}
    +
    \bigg(
      -
      \frac{
         c_{\mathrm{R}\mathrm{L}}^{V} c_{\mathrm{R}\mathrm{L}}^{V \prime \ast}
      }{z^2}
      -
      \frac{
         c_{\mathrm{L}\mathrm{R}}^{V} c_{\mathrm{L}\mathrm{R}}^{V \prime \ast}
      }{z^2}
      +
      k_\perp^2
      \Big(
        b_ {\mathrm{L} \mathrm{R}}^{V}(-1) b_ {\mathrm{L} \mathrm{R}}^{V \prime \ast}(-1)
        +
        b_ {\mathrm{R} \mathrm{L}}^{V}(-1) b_ {\mathrm{R} \mathrm{L}}^{V \prime \ast}(-1)
      \Big)
    \bigg)
    \slashed{\bar{p}}
    \nonumber\\
    &\phantom{={}}
    +
    \bigg(
      \frac{
         c_{\mathrm{R}\mathrm{L}}^{V} c_{\mathrm{R}\mathrm{L}}^{V \prime \ast}
      }{z^2}
      -
      \frac{
         c_{\mathrm{L}\mathrm{R}}^{V} c_{\mathrm{L}\mathrm{R}}^{V \prime \ast}
      }{z^2}
      -
      k_\perp^2
      \Big(
        b_ {\mathrm{L} \mathrm{R}}^{V}(-1) b_ {\mathrm{L} \mathrm{R}}^{V \prime \ast}(-1)
        -
        b_ {\mathrm{R} \mathrm{L}}^{V}(-1) b_ {\mathrm{R} \mathrm{L}}^{V \prime \ast}(-1)
      \Big)
    \bigg)
    \slashed{\bar{p}}
    \gamma^5
  \Big\}
  .
  \label{eq:FbarVFbar-LU-split-func-EWSM}
\end{align}

%
\subsubsection[%
  \mathinhead{f^\ast \to V f}{f* to V f}
  splitting functions%
  ]{%
  \mathinheadbold{f^\ast \to V f}{f* to V f}
  splitting functions}
\label{sect:FVF}
The splitting functions~$f^\ast \to V f$ can be computed similarly, from the diagrams
\drawmomtrue%
\drawvertdottrue%
\begin{alignat}{2}%
  \ftntmath{%
    \MContrExplFields{\fermfield{ij}}{\vbfield{i}}{\fermfield{j}}^{\polvb_i \polferm_j}%
  }%
  &\ftntmath{{}={}}%
  \FSDiagGeo%
  \ftntmath{%
    \splittingdiagIndivFS{T^{(n)}_{\fermfield{ij} X}}%
      {fermion}{photon}{fermion}%
      {\fermfield{ij}}{\vbfield{i}, \polvb_i}{\fermfield{j}, \polferm_j}%
    ,%
  }%
  \resetdiagargs%
  &\quad
  \ftntmath{%
    \MContrExplFields{\fermfield{ij}}{\gbfield{i}}{\fermfield{j}}^{\mathrm{G} \polferm_j}%
  }%
  &\ftntmath{{}={}}%
  \FSDiagGeo%
  \ftntmath{%
    \splittingdiagIndivFS{T^{(n)}_{\fermfield{ij} X}}%
      {fermion}{scalar}{fermion}%
      {\fermfield{ij}}{\gbfield{i}}{\fermfield{j}, \polferm_j}%
    ,%
  }%
  \resetdiagargs%
  \label{eq:FVF-diags-FS}
\end{alignat}%
with the analytical expressions
\begin{align}
  %
  %
  \MContrExplFields{ \fermfield{ij}}{\vbfield{i}}{ \fermfield{j}}^{\polvb_i \polferm_j}%
  &=
  \imagi e
  \bar u_{\polferm_j}(p_j) 
  \slashed \varepsilon_{\polvb_i}^{\ast}(p_j)
  \left(
    \omega_-
    I_{\fermfield{j}^\mathrm{L} \fermfield{ij}^\mathrm{L}}^{\vbantifield{i}}
    +
    \omega_+
    I_{\fermfield{j}^\mathrm{R} \fermfield{ij}^\mathrm{R}}^{\vbantifield{i}}
  \right)
  G^{\fermfield{ij} \antifermfield{ij}}(-p_{ij},p_{ij})
  T^{(n)}_{ \fermfield{ij} X} (p_{ij})
  ,
  \nonumber\\
  %
  %
  \MContrExplFields{ \fermfield{ij}}{\gbfield{i}}{ \fermfield{j}}^{\mathrm{G} \polferm_j}%
  &=
  - \imagi e
  \eta _{\vbfield{i}}
  \bar u_{\polferm_j}(p_j)
  \frac{1}{m_i}
  \left\{
    \omega_+
    \left(
      m_{j}
      I_{\fermfield{j}^\mathrm{R} \fermfield{ij}^\mathrm{R}}^{\vbantifield{i}}
      -
      m_{ij}
      I_{\fermfield{j}^\mathrm{L} \fermfield{ij}^\mathrm{L}}^{\vbantifield{i}}
    \right)
    +
    \omega_-
    \left(
      m_{j}
      I_{\fermfield{j}^\mathrm{L} \fermfield{ij}^\mathrm{L}}^{\vbantifield{i}}
      -
      m_{ij}
      I_{\fermfield{j}^\mathrm{R} \fermfield{ij}^\mathrm{R}}^{\vbantifield{i}}
    \right)
  \right\}
  \nonumber\\*
  &\phantom{{}={}}
  \cdot
  G^{\fermfield{ij} \antifermfield{ij}}(-p_{ij},p_{ij})
  T^{(n)}_{\fermfield{ij} X} (p_{ij})
  .
  \label{eq:FVF-EnhAmp}
\end{align}
Alternatively, we may recover these results from the $\bar f^\ast \to V \bar f$
case by studying the effect of charge conjugation.
Specifically, we find that~\eqref{eq:FVF-EnhAmp} is obtained
from~\eqref{eq:FbarVFbar-EnhAmp} by transposition, multiplication
with~$-1$, replacing
\begin{align}
  I_{\fermfield{ij^{(\prime)}}^\mathrm{L} \fermfield{j}^\mathrm{L}}^{\vbantifield{i}}
  \to
  I_{\fermfield{j}^\mathrm{R} \fermfield{ij^{(\prime)}}^\mathrm{R}}^{\vbantifield{i}}
  , \qquad
  I_{\fermfield{ij^{(\prime)}}^\mathrm{R} \fermfield{j}^\mathrm{R}}^{\vbantifield{i}}
  \to
  I_{\fermfield{j}^\mathrm{L} \fermfield{ij^{(\prime)}}^\mathrm{L}}^{\vbantifield{i}}
  ,
  \label{eq:FVF-swap-cpl-w-chir}
\end{align}
and inserting the spinor relations~\eqref{eq:ch-conj-expl-spinors} as well as
\begin{align}
  \bar T^{(n)}_{\antifermfield{ij} X} (p_{ij})
  \to
  T^{(n)\, T}_{\fermfield{ij} X} (p_{ij}) C
  ,
  \label{eq:ch-conj-transpos}
\end{align}
with $C$ denoting the charge-conjugation matrix defined in~\eqref{eq:ch-conj-def}.
The object $T^{(n)}_{\antifermfield{ij} X}$
occurs in the $\bar f^\ast \to V \bar f$ splitting functions
in Dirac chains of the form 
\begin{align}
  \bar T^{(n)}_{\antifermfield{ij} X} (p_{ij}) \Gamma T^{(n)}_{\antifermfield{ij} X} (p_{ij})
  \quad
  \mathrm{and}
  \quad
  \bar T^{(n)}_{\antifermfield{ij} X} (p_{ij}) \Gamma \gamma^5 T^{(n)}_{\antifermfield{ij} X} (p_{ij})
  ,
  \label{eq:dirac-sandwich-in-FVF}
\end{align}
where~$\Gamma=\slashed{v}_1 \dots \slashed{v}_m$ for some
real four-vectors~$v_i^\mu$.
To be precise, we encounter
\begin{align}
  \Gamma = \slashed{\bar p}
  \quad
  \mathrm{and}
  \quad
  \Gamma = \slashed{k}_\perp \slashed{\bar p}
  .
  \label{eq:dirac-chains-in-FFV}
\end{align}
The substitution~\eqref{eq:ch-conj-transpos} transforms
the Dirac chains from~\eqref{eq:dirac-sandwich-in-FVF}
in the following way:
\begin{align}
  \bar T^{(n)}_{\antifermfield{ij} X} (p_{ij}) \Gamma T^{(n)}_{\antifermfield{ij} X} (p_{ij})
  &\to
  (-1)^{m+1}
  \bar T^{(n)}_{\fermfield{ij} X} (p_{ij}) \tilde \Gamma T^{(n)}_{\fermfield{ij} X} (p_{ij})
  ,
  \nonumber\\
  \bar T^{(n)}_{\antifermfield{ij} X} (p_{ij}) \Gamma \gamma^5 T^{(n)}_{\antifermfield{ij} X} (p_{ij})
  &\to
  -\bar T^{(n)}_{\fermfield{ij} X} (p_{ij}) \tilde \Gamma \gamma^5 T^{(n)}_{\fermfield{ij} X} (p_{ij})
  ,
  \label{eq:ch-conj-transpos-effect-on-dirac-chains}
\end{align}
where~$\tilde \Gamma=\slashed{v}_m \dots \slashed{v}_1$.
Utilization of commutation relations and~$\bar p k_\perp =0$ shows that
the application of~\eqref{eq:ch-conj-transpos-effect-on-dirac-chains}
and therefore of~\eqref{eq:ch-conj-transpos} to the
Dirac chains involving~\eqref{eq:dirac-chains-in-FFV} is equivalent to
\begin{align}
  \slashed{\bar p}
  \to
  \slashed{\bar p}
  ,\qquad
  \slashed{\bar p} \gamma^5
  \to
  - \slashed{\bar p} \gamma^5
  ,\qquad
  \slashed{k}_\perp \slashed{\bar p}
  \to
  \slashed{k}_\perp \slashed{\bar p}
  ,\qquad
  \slashed{k}_\perp \slashed{\bar p} \gamma^5
  \to
  \slashed{k}_\perp \slashed{\bar p} \gamma^5
  .
  \label{eq:ch-conj-transpos-effect-on-dirac-chains-equiv}
\end{align}
Combining the above, we have
\begin{align}
  \mathcal{P}
    {}^{\fermfield{ij^{\vphantom{\prime}}}}_{\vbfield{i}}
    {}^{\fermfield{ij^\prime}}_{\fermfield{j}}
    \big(z,k_\perp^\mu,\bar p^\mu,\polvb_i,\polferm_j\big)
  =
  \mathcal{P}
    {}^{\antifermfield{ij^{\vphantom{\prime}}}}_{\vbfield{i}}
    {}^{\antifermfield{ij^\prime}}_{\antifermfield{j}}
    \big(z,k_\perp^\mu,\bar p^\mu,\polvb_i,\polferm_j\big)
  \bigg|_{
    \eqref{eq:FVF-swap-cpl-w-chir}, \,
    \eqref{eq:ch-conj-transpos-effect-on-dirac-chains-equiv}
  }
  .
  \label{eq:FbarVFbar-to-FVF-FS-1}
\end{align}
Finally, similar to the $\bar f^\ast \to V \bar f$ case, the Dirac
structures~\eqref{eq:ferm-splittings-dirac-structures} are related to
matrix elements of definite polarization,
\begin{align}
  \mathcal{M}^{(n)\, \pm}_{f X}
  &=
  \bar u_\pm 
  T^{(n)}_{f X}
  ,
  \qquad
  \mathcal{M}^{(n)\, \pm\, \ast}_{f X}
  =
  \bar T^{(n)}_{f X}
  u_\pm
  ,
  \label{eq:pol-ul-me-ferm}
\end{align}
as we discuss in Appendix~\ref{sect:id-hard-me}.
Again, the spinors $\Sigma_\pm u_\pm=u_\pm$ correspond to fermions with
spin $\pm1/2$ in the reference direction defined by some vector $r^\mu$,
as detailed below~\eqref{eq:spin-projection-op}.
For massless
fermions $\fermfield{ij}$, $\fermfield{ij'}$
with on-shell momentum $p^\mu$,
we get
\begin{align}
  \bar T^{(n)}_{\fermfield{ij} X}(p)
  \,
  \slashed p
  \frac{1 \mp \gamma^5}{2}
  \,
  T^{(n)}_{\fermfield{ij^\prime} X}(p)
  =
  \bar T^{(n)}_{\fermfield{ij} X}(p)
  \,
  \frac{1 \pm \gamma^5}{2}
  \slashed p
  \,
  T^{(n)}_{\fermfield{ij^\prime} X}(p)
  =
  \mathcal{M}^{(n)\, \pm}_{\fermfield{ij} X}
  \mathcal{M}^{(n)\, \pm\, \ast}_{\fermfield{ij^\prime} X}
  ,
  \label{eq:id-sq-ulme-massless-ferm}
\end{align}
with matrix elements of definite helicity.

%
\subsubsection[%
  \mathinhead{\bar f^\ast \to H \bar f}{fbar* to H fbar}
  splitting functions%
  ]{%
  \mathinheadbold{\bar f^\ast \to H \bar f}{fbar* to H fbar}
  splitting functions}
\label{sect:FbarHFbar}
The diagram required for the computation of the $\bar f^\ast \to H \bar
f$ splitting functions,
\drawmomtrue%
\drawvertdottrue%
\begin{alignat}{2}%
  \ftntmath{%
    \MContrExplFields{\antifermfield{ij}}{\higgsfield{i}}{\antifermfield{j}}^{\polferm_j}%
  }%
  &\ftntmath{{}={}}%
  \FSDiagGeo%
  \ftntmath{%
    \splittingdiagIndivFS{\bar T^{(n)}_{\antifermfield{ij} X}}%
      {antifermion}{scalar}{antifermion}%
      {\antifermfield{ij}}{\higgsfield{i}}{\antifermfield{j}, \polferm_j}%
    ,
  }%
  \resetdiagargs%
  \label{eq:FbarHFbar-diags-FS}
\end{alignat}%
is evaluated to
\begin{align}
  %
  %
  \MContrExplFields{\antifermfield{ij}}{\higgsfield{i}}{\antifermfield{j}}^{\polferm_j}%
  &=
  - \imagi e
  \bar T^{(n)}_{\antifermfield{ij} X} (p_{ij})
  G^{\fermfield{ij} \antifermfield{ij}}(p_{ij},-p_{ij})
  \Big(
    \omega_+
    C^{\mathrm{R}}_{\higgsantifield{i} \antifermfield{ij} \fermfield{j}} 
    +
    \omega_-
    C^{\mathrm{L}}_{\higgsantifield{i} \antifermfield{ij} \fermfield{j}} 
  \Big)
  v_{\polferm_j}(p_j)
  .
  \label{eq:FbarHFbar-EnhAmp}
\end{align}
Our results on the splitting functions, 
quoted below, are symmetric
under~\eqref{eq:FVF-swap-fermions-ij-ijpr}, combined
with~\eqref{eq:cc-of-dirac-chains}.
They are expressed in terms of the abbreviations
\begin{align}
  a^{H \chir_1 \chir_2}_{i, ij, j}
  (\pm 1)
  &=
  - (1-z) z m_{ij}
  C^{\chir_1}_{\higgsantifield{i} \antifermfield{ij} \fermfield{j}}
  \pm
  z m_j
  C^{\chir_2}_{\higgsantifield{i} \antifermfield{ij} \fermfield{j}}
  ,
  \nonumber
  \\
  b^{H \chir_1 \chir_2}_{i, ij, j}
  (\pm 1)
  &=
  m_{ij}
  C^{\chir_1}_{\higgsantifield{i} \antifermfield{ij} \fermfield{j}}
  \pm
  m_j
  C^{\chir_2}_{\higgsantifield{i} \antifermfield{ij} \fermfield{j}}
  \nonumber
  \\
  d^{H}_{i, ij, j}
  (\pm 1)
  &=
  C^{\mathrm{L}}_{\higgsantifield{i} \antifermfield{ij} \fermfield{j}}
  C^{\mathrm{L} \ast}_{\higgsantifield{i} \antifermfield{ij^\prime} \fermfield{j}}
  \pm
  C^{\mathrm{R}}_{\higgsantifield{i} \antifermfield{ij} \fermfield{j}}
  C^{\mathrm{R} \ast}_{\higgsantifield{i} \antifermfield{ij^\prime} \fermfield{j}}
  \nonumber
  \\
  e^{H}_{i, ij, j}
  (\pm 1)
  &=
  C^{\mathrm{L}}_{\higgsantifield{i} \antifermfield{ij} \fermfield{j}}
  C^{\mathrm{R} \ast}_{\higgsantifield{i} \antifermfield{ij^\prime} \fermfield{j}}
  \pm
  C^{\mathrm{R}}_{\higgsantifield{i} \antifermfield{ij} \fermfield{j}}
  C^{\mathrm{L} \ast}_{\higgsantifield{i} \antifermfield{ij^\prime} \fermfield{j}}
  ,
  \label{eq:FbarHFbar-abbrev}
\end{align}
which are implied by~\eqref{eq:gen-ferm-abbs}.

\paragraph{Polarization \boldmath{$\polferm_j=+$}.}
We get
\begin{align}
    \mathcal{P}
      {}^{\antifermfield{ij^{\vphantom{\prime}}}}_{\higgsfield{i}}
      {}^{\antifermfield{ij^\prime}}_{\antifermfield{j}}
      \big(z,k_\perp^\mu,\bar p^\mu,+\big)
  =
  \frac{1}{2 (z-1) z}
      \bigg\{
      &
      \bigg(
        z
        C^\mathrm{L}_{\higgsantifield{i} \antifermfield{ij} \fermfield{j}}
        C^{\mathrm{L} \ast}_{\higgsantifield{i} \antifermfield{ij^\prime} \fermfield{j}}
        k_\perp^2
        -
        \frac{
          a_{\mathrm{L}\mathrm{R}}^{H}(-1)
          a_{\mathrm{L}\mathrm{R}}^{H \prime \ast}(-1)
        }{z}
      \bigg)
      \slashed{\bar{p}}
      \nonumber\\
      &
      -
      \bigg( 
        z
        C^\mathrm{L}_{\higgsantifield{i} \antifermfield{ij} \fermfield{j}}
        C^{\mathrm{L} \ast}_{\higgsantifield{i} \antifermfield{ij^\prime} \fermfield{j}}
        k_\perp^2
        +
        \frac{
          a_{\mathrm{L}\mathrm{R}}^{H}(-1)
          a_{\mathrm{L}\mathrm{R}}^{H \prime \ast}(-1)
        }{z}
      \bigg)
      \slashed{\bar{p}}
      \gamma^5
      \nonumber\\
      &
      -
      \bigg(
        C^\mathrm{L}_{\higgsantifield{i} \antifermfield{ij} \fermfield{j}}
        a_{\mathrm{L}\mathrm{R}}^{H \prime \ast}(-1)
        -
        C^{\mathrm{L} \ast}_{\higgsantifield{i} \antifermfield{ij^\prime} \fermfield{j}}
        a_{\mathrm{L}\mathrm{R}}^{H}(-1)
      \bigg)
      \slashed{k}_{\perp }
      \slashed{\bar{p}}
      \nonumber\\
      &
      -
      \bigg(
        C^\mathrm{L}_{\higgsantifield{i} \antifermfield{ij} \fermfield{j}}
        a_{\mathrm{L}\mathrm{R}}^{H \prime \ast}(-1)
        +
        C^{\mathrm{L} \ast}_{\higgsantifield{i} \antifermfield{ij^\prime} \fermfield{j}}
        a_{\mathrm{L}\mathrm{R}}^{H}(-1)
      \bigg)
      \slashed{k}_{\perp }
      \slashed{\bar{p}}
      \gamma^5
    \bigg\}
    ,
    \label{eq:FbarHFbar-P-split-func}
\end{align}
with the \gls{ewsm} form
\begin{align}
  &
  \mathcal{P}
    {}^{\antifermfield{ij^{\vphantom{\prime}}}}_{H}
    {}^{\antifermfield{ij^\prime}}_{\antifermfield{j}}
    \big(z,k_\perp^\mu,\bar p^\mu,+\big)
  \bigg|_{\mathrm{EWSM}}
  \nonumber\\*
  &=
  \frac{1}{(e v)^2}\frac{m_j^2}{2 (z-1)}
  \bigg\{  
    \big(
      k_\perp^2
      -
      (z-2)^2 m_j^2
    \big)
    \slashed{\bar{p}}
    -
    \big(
      k_\perp^2
      +
      (z-2)^2 m_j^2
    \big)
    \slashed{\bar{p}}
    \gamma^5
    -
    2
    (z-2)
    m_j
    \slashed{k}_{\perp }
    \slashed{\bar{p}}
    \gamma^5
  \bigg\}
  \delta_{\fermfield{ij} \fermfield{j}}
  \delta_{\fermfield{ij}^\prime \fermfield{j}}
  .
  \label{eq:FbarHFbar-P-split-func-EWSM}
\end{align}
For massless fermions, \eqref{eq:gen-ferm-abbs-mlferm} implies
\begin{align}
  &
  \mathcal{P}
    {}^{\antifermfield{ij^{\vphantom{\prime}}}}_{\higgsfield{i}}
    {}^{\antifermfield{ij^\prime}}_{\antifermfield{j}}
    \big(z,k_\perp^\mu,\bar p^\mu,+\big)
  \bigg|_{m_{ij} = m_{ij}^\prime = m_j = 0}
  =
  \frac{k_\perp^2}{z-1}
  \,
  C^\mathrm{L}_{\higgsantifield{i} \antifermfield{ij} \fermfield{j}}
  C^{\mathrm{L} \ast}_{\higgsantifield{i} \antifermfield{ij^\prime} \fermfield{j}}
  \,
  \slashed{\bar{p}}
  \frac{1-\gamma^5}{2}
  ,
  \label{eq:FbarHFbar-P-split-func-mlferm}
\end{align}
with a Dirac structure that is associated to a hard matrix element with
a massless antifermion of 
helicity~$-\frac{1}{2}$, reflecting the
``spin-flipping'' nature of the 
$H\bar ff$ vertex~\eqref{eq:app-Sff-vertex}.
Consistently, only left-handed couplings remain
in~\eqref{eq:FbarHFbar-P-split-func-mlferm}.

\paragraph{Unpolarized case.}
\begin{alignat}{2}
  \mathcal{P}
    {}^{\antifermfield{ij^{\vphantom{\prime}}}}_{\higgsfield{i}}
    {}^{\antifermfield{ij^\prime}}_{\antifermfield{j}}
    \big(z,k_\perp^\mu,\bar p^\mu,+\big)
  ={}&
  \frac{1}{2}
      \bigg\{
      &&
      \frac{1}{z-1}
      \bigg(
        k_\perp^2 d^H(+1)
        -
        \frac{
           a_{\mathrm{L}\mathrm{R}}^{H}(-1)
           a_{\mathrm{L}\mathrm{R}}^{H \prime \ast}(-1)
           +
           a_{\mathrm{R}\mathrm{L}}^{H}(-1)
           a_{\mathrm{R}\mathrm{L}}^{H \prime \ast}(-1)
         }{z^2}
      \bigg)
      \slashed{\bar{p}}
      \nonumber\\
      &&&
      -
      \frac{1}{z-1}
      \bigg(
        k_\perp^2 d^H(-1)
        +
        \frac{
           a_{\mathrm{L}\mathrm{R}}^{H}(-1)
           a_{\mathrm{L}\mathrm{R}}^{H \prime \ast}(-1)
           -
           a_{\mathrm{R}\mathrm{L}}^{H}(-1)
           a_{\mathrm{R}\mathrm{L}}^{H \prime \ast}(-1)
         }{z^2}
      \bigg)
      \slashed{\bar{p}}
      \gamma^5 
      \nonumber\\
      &&&
      +
      \big(
        m_{ij}
        -
        m_{ij}^{\prime }
      \big)
      d^H(+1) 
      \slashed{k}_{\perp }
      \slashed{\bar{p}}
      -
      \big(
        m_{ij}
        +
        m_{ij}^{\prime }
      \big)
      d^H(-1) 
      \slashed{k}_{\perp }
      \slashed{\bar{p}}
      \gamma^5
    \bigg\}
    .
    \label{eq:FbarHFbar-U-split-func}
\end{alignat}

%
\subsubsection[%
  \mathinhead{f^\ast \to Hf}{f* to H f}
  splitting functions%
  ]{%
  \mathinheadbold{f^\ast \to Hf}{f* to H f}
  splitting functions}
\label{sect:FHF}
The diagram corresponding to the $f^\ast \to H f$ splitting functions reads
\drawmomtrue%
\drawvertdottrue%
\begin{alignat}{2}%
  \ftntmath{%
    \MContrExplFields{\fermfield{ij}}{\higgsfield{i}}{\fermfield{j}}^{\polferm_j}%
  }%
  &\ftntmath{{}={}}%
  \FSDiagGeo%
  \ftntmath{%
    \splittingdiagIndivFS{T^{(n)}_{\fermfield{ij} X}}%
      {fermion}{scalar}{fermion}%
      {\fermfield{ij}}{\higgsfield{i}}{\fermfield{j}, \polferm_j}%
  }%
  \resetdiagargs%
  \label{eq:FHF-diags-FS}
\end{alignat}%
and translates into
\begin{align}
  %
  %
  \MContrExplFields{ \fermfield{ij}}{\higgsfield{i}}{ \fermfield{j}}^{\mathrm{G} \polferm_j}%
  &=
  \imagi e
  \bar u_{\polferm _j}(p_j)
  \Big(
    \omega_+
    C^{\mathrm{R}}_{\higgsantifield{i} \antifermfield{j} \fermfield{ij}} 
    +                                    
    \omega_-                             
    C^{\mathrm{L}}_{\higgsantifield{i} \antifermfield{j} \fermfield{ij}} 
  \Big)
  G^{\fermfield{ij} \antifermfield{ij}}(-p_{ij},p_{ij})
  T^{(n)}_{\fermfield{ij} X} (p_{ij})
  ,
  \label{eq:FHF-EnhAmp}
\end{align}
which is recovered from~\eqref{eq:FbarHFbar-EnhAmp} by transposition
and
insertion of~\eqref{eq:ch-conj-expl-spinors}, \eqref{eq:ch-conj-transpos},
and
\begin{align}
  C^{\chir}_{\higgsantifield{i} \antifermfield{ij^{(\prime)}} \fermfield{j}}
  \to
  C^{\chir}_{\higgsantifield{i} \antifermfield{j} \fermfield{ij^{(\prime)}}}
  .
  \label{eq:FHF-swap-cpl-wo-chir}
\end{align}
In contrast to~\eqref{eq:FVF-swap-cpl-w-chir}, the chiralities of the couplings
are not swapped in~\eqref{eq:FHF-swap-cpl-wo-chir}.
Altogether, the $f^\ast \to H f$ splitting functions are obtained from the
$\bar f^\ast \to H \bar f$ results via
\begin{align}
  \mathcal{P}
    {}^{\fermfield{ij^{\vphantom{\prime}}}}_{\higgsfield{i}}
    {}^{\fermfield{ij^\prime}}_{\fermfield{j}}
    \big(z,k_\perp^\mu,\bar p^\mu,\polferm\big)
  =
  \mathcal{P}
    {}^{\antifermfield{ij^{\vphantom{\prime}}}}_{\higgsfield{i}}
    {}^{\antifermfield{ij^\prime}}_{\antifermfield{j}}
    \big(z,k_\perp^\mu,\bar p^\mu,\polferm\big)
  \bigg|_{
    \eqref{eq:FHF-swap-cpl-wo-chir}, \,
    \eqref{eq:ch-conj-transpos-effect-on-dirac-chains-equiv}
  }
  .
  \label{eq:FbarHFbar-to-FHF-FS-1}
\end{align}

\subsection{Computing the azimuthal average of FS splitting functions}
\label{sect:FS-az-avg}
\subsubsection{General procedure}
The tensor-valued structure of the splitting functions from
Sections~\ref{sect:expl-splfuncs-FS-bosonic} and~\ref{sect:expl-splfuncs-FS-fermionic}
encodes spin correlations between
the matrix elements for the 
$(n+1)$- and $n$-particle process
in~\eqref{eq:realme-sq-fact-nonsudakov}.
Via $k_\perp^\mu$, these correlations depend on the plane of branching,
and the azimuthal average around the collinear axis 
corresponds to taking the spin average 
of the branching particle $\varphi_{ij}$
over the polarizations transverse 
to the collinear axis.
This average is of interest 
for experiments that are not sensitive to the
orientation of the splitting plane and significantly simplifies the
structure of the splitting functions.
Moreover, when
splitting functions are
used for the construction of subtraction algorithms,
the azimuthal average constitutes an essential step in the computation
of integrated counterterms.


Examples
for the computation of the azimuthal average
are provided
in~\cite{catani1996general,catani2002massivedipole} for the unpolarized
branching processes~$\mathrm{g}^\ast \to \mathrm{g} \mathrm{g}$,
$\mathrm{g}^\ast \to q \bar q$, and in~\cite{dittmaier2008polarized} for
unpolarized~$\gamma^\ast \to f \bar f$.
In these cases, the azimuthal average turns the respective Lorentz structures into
the transverse polarization sums
of the gluon or photon.           


The precise definition of the azimuthal angle requires a choice of
frame, which we consider to be distinguished by~$\bar{\boldsymbol{p}}
\parallel \boldsymbol{n}$, with
$\bar{p}^\mu$ and $n^\mu$ from the
Sudakov parametrization~\eqref{eq:sudakov-param-fs}.
Aligning~$\bar{\boldsymbol{p}}$ with the 3-axis, we get the explicit
component-wise expressions for~$\bar p^\mu$, $n^\mu$,
and~$\bar{\varepsilon}^\mu_\pm$ from~\eqref{eq:frame-n-parr-pb}.
Additionally, the conditions~$\bar p k_\perp = n k_\perp =0$ imply
\begin{align}
  k_\perp^\mu
  &=
  (0,|\boldsymbol k_\perp| \cos \phi,|\boldsymbol k_\perp| \sin \phi,0)
  \label{eq:azavg-sudkt-form}
  ,
\end{align}
where~$k_\perp^2=-|\boldsymbol k_\perp|^2$ and~$\phi$ defines the
azimuthal angle of the splitting plane.
Taking the average with respect to~$\phi$, denoted by
\begin{align}
  \langle \dots \rangle_\phi
  =
  \int_0^{2 \pi} \frac{\D \phi}{2 \pi} \dots
  ,
  \label{eq:azavg-def}
\end{align}
yields
\begin{align}
  \langle k_{\perp}^\mu \rangle_\phi
  =0
  .
  \label{eq:azavg-kt}
\end{align}
All remaining four-vectors~$\bar p^\mu$, $n^\mu$,
and~$\bar{\varepsilon}_\pm^\mu$ are independent of~$\phi$ and do not
change under~$\langle \dots \rangle_\phi$.
The scalar product of~$k_\perp^\mu$ with any of these
four-vectors vanishes
upon integrating over~$\phi$.
In particular,
\begin{align}
  \langle k_{\perp} \bar{\varepsilon}_\pm \rangle_\phi
  =0.
  \label{eq:azavg-kteps}
\end{align}
Explicitly, we find, in the chosen frame of reference,
\begin{align}
  k_\perp \bar{\varepsilon}_\pm
  =
  - s_\pm
  \frac{
    |\boldsymbol k_\perp|
  }{
    \sqrt{2}
  }
  \expe^{\pm \imagi \phi}
  ,
  \label{eq:azavg-kteps-sp}
\end{align}
so that~\eqref{eq:kt-from-eps-via-sp} implies
\begin{align}
  k_\perp^\mu k_\perp^\nu
  =
  -\frac{k_\perp^2}{2}
  \left(
    \epsepssum
    +
    s_+
    s_-^\ast
    \expe^{+ 2 \imagi \phi}
    \,
    \bar{\varepsilon}_-^{\mu}
    \bar{\varepsilon}_+^{\nu \ast}
    +
    s_-
    s_+^\ast
    \expe^{- 2 \imagi \phi}
    \,
    \bar{\varepsilon}_+^{\mu}
    \bar{\varepsilon}_-^{\nu \ast}
  \right)
  \label{eq:azavg-ktmuktnu-nonav}
\end{align}
and
\begin{align}
  \langle
    k_\perp^\mu k_\perp^\nu
  \rangle_\phi
  =
  -\frac{k_\perp^2}{2}
  \left(
    \epsepssum
  \right)
  \label{eq:azavg-ktmuktnu-av}
  ,
\end{align}
which agrees with $k_\perp^2 g^{\mu\nu}/2$ in leading mass order
according to \eqref{eq:gmunu-decomp}.
All other outer products~$k_\perp^\mu v^\nu$ between $k_\perp^\mu$ and
four-vectors~$v^\nu = \bar p^\nu, n^\nu, \bar{\varepsilon}_\pm^\nu$
vanish under the azimuthal integration, while the outer
products~$v_1^\mu v_2^\nu$ that do not involve~$k_\perp^\mu$
and~$k_\perp^\nu$ are unaffected.
This, of course, also applies to~$g^{\mu \nu}$.


Combining these results, the computation of the azimuthal average of our
\gls{fs} splitting functions becomes straightforward.
For the individual splitting types, this is discussed in more detail
below.
Throughout, we perform our discussion in view of the construction of
subtraction functions for subtraction algorithms.
Specifically, this means that we are interested in the \emph{exact}
azimuthal average of our
splitting functions, including potential
mass-suppressed terms.

\subsubsection{Splittings of vector bosons}
\label{sect:V-az-avg}
We start with the splittings $\vbfield{ij}^\ast \to \genfield{i}
\genfield{j}$, discussing separately the case of massless and massive
vector bosons.
In the basic formulas, we keep the general case of fields
$\vbfield{ij}$, $\vbfield{ij^\prime}$. Note, however, that 
in the non-diagonal case $\vbfield{ij} \ne \vbfield{ij^\prime}$
the evaluation of products
$\mathcal{M}^{(n)\, \polvb}_{\vbfield{ij} X}
\big(\mathcal{M}^{(n)\, \polvb}_{\vbfield{ij^\prime} X}\big)^*$
of matrix elements,
where $\lambda$ is the polarization of $\vbfield{ij}$ and $\vbfield{ij^\prime}$,
deserve particular care if $m_{ij} \ne m_{ij}^\prime$,
i.e.\ in one of the two factors some distortion of momenta 
is required to obtain the relevant on-shell kinematics
(which is admitted in leading mass order).
In the diagonal case $\vbfield{ij} = \vbfield{ij^\prime}$, some of the
emerging tensor structures can be identified
with spin-summed squared matrix elements of the underlying hard $n$-particle process.

\paragraph{%
  Examples: Unpolarized
  \mathinheadbold{\mathrm{g}^\ast \to \mathrm{g} \mathrm{g}}{g* to gg},
  \mathinheadbold{\mathrm{g}^\ast \to q \bar q}{g* to qqbar},
  \mathinheadbold{\gamma^\ast \to f \bar f}{P* to ffbar}
  .
  }%
As described in Appendices~\ref{sect:VVV-litcheck}
and~\ref{sect:VFFbar-litcheck}, the splitting functions
from~\cite{catani1996general,catani2002massivedipole,dittmaier2008polarized}
with \emph{massless} vector bosons exhibit structures 
proportional to $k_\perp^\mu k_\perp^\nu$ or $g^{\mu \nu}$.
Because of~\eqref{eq:azavg-ktmuktnu-av} and~\eqref{eq:gmunu-decomp},
combined with \eqref{eq:VVV-ggg-limit} and~\eqref{eq:VFF-gqq-limit},
both structures become proportional to the transverse polarization sum
(up to irrelevant terms $\propto\bar p^\mu$ or $\propto\bar p^\nu$),
as claimed.

\paragraph{Massless intermediate vector bosons.}
The same procedure applies to all other splittings of a massless vector
boson, resulting in
\begin{align}
  \bigg\langle
  \bigg[
    \mathcal{P}
      {}^{\vbfield{ij^{\vphantom{\prime}}}}_{\genfield{i}}
      {}^{\vbfield{ij^\prime}}_{\genfield{j}}
      \big(z,k_\perp^\mu,\bar p^\mu,\pol_i,\pol_j\big)
  \bigg]^{\mu \nu}
  \bigg\rangle_\phi
  &=
  C
  \big(\epsepssum\big)
  ,
  \nonumber\\
  \bigg\langle
  T^{(n)}_{\vbfield{ij} X, \mu}
  \bigg[
    \mathcal{P}
      {}^{\vbfield{ij^{\vphantom{\prime}}}}_{\genfield{i}}
      {}^{\vbfield{ij^\prime}}_{\genfield{j}}
      \big(z,k_\perp^\mu,\bar p^\mu,\pol_i,\pol_j\big)
  \bigg]^{\mu \nu}
  T^{(n) \ast}_{\vbfield{ij^\prime} X, \nu}
  \bigg\rangle_\phi
  &=
  C
  \sum_{\polvb = \pm}
  \mathcal{M}^{(n)\, \polvb}_{\vbfield{ij} X}
  \left(\mathcal{M}^{(n)\, \polvb}_{\vbfield{ij^\prime} X}\right)^*
  .
  \label{eq:azavg-splfunc-VXY-massless-coeffs}
\end{align}
The coefficient $C$ depends on the specific splitting and
polarizations.
We have
\begin{align}
  C_{V^\ast \to V V}
  &=
  - 4 k_\perp^2
  \bigg(
    \frac{1}{z^2}
    +
    \frac{1}{(z-1)^2}
    +
    1
  \bigg)
  I_{\vbfield{i} \vbantifield{j}}^{\vbfield{ij}} 
  I_{\vbfield{i} \vbantifield{j}}^{\vbfield{ij^\prime} \ast}
  ,
  \nonumber\\
  C_{V^\ast \to f \bar f}
  &=
  \bigg(
    -
    \frac{2 \big(m_i^2 - k_\perp^2\big)}{z (z-1)}
    +
    4
    k_\perp^2
  \bigg)
  I_{\fermfield{i} \fermfield{j}}^{\vbfield{ij}}
  I_{\fermfield{i} \fermfield{j}}^{\vbfield{ij^\prime} \ast}
  \label{eq:azavg-Ccoeff-examples-massless}
\end{align}
for entirely massless, unpolarized $V^\ast \to V V$ splittings and
unpolarized $V^\ast \to f \bar f$ splittings with 
non-chiral fermions of mass
$m_i$, respectively, corresponding to the above examples.


An exception arises for~$V^\ast \to f \bar f$ with
\emph{polarized fermions} or \emph{chiral couplings}.
The corresponding splitting functions contain additional terms
$\propto (\epsepsdiff)$, which remain after azimuthal averaging.
We obtain
\begin{align}
  \bigg\langle
  T^{(n)}_{\vbfield{ij} X, \mu}
  \bigg[
    \mathcal{P}
      {}^{\vbfield{ij^{\vphantom{\prime}}}}_{\fermfield{i}}
      {}^{\vbfield{ij^{\prime}}}_{\antifermfield{j}}
      \big(z,k_\perp^\mu,\bar p^\mu,\polferm_i,\polferm_j\big)
  \bigg]^{\mu \nu}
  T^{(n) \ast}_{\vbfield{ij^\prime} X, \nu}
  \bigg\rangle_\phi
  &=
  \sum_{\polvb = \pm}
  C_\polvb
    \mathcal{M}^{(n)\, \polvb}_{\vbfield{ij} X}
    \left(\mathcal{M}^{(n)\, \polvb}_{\vbfield{ij^\prime} X}\right)^*
  ,
  \label{eq:azavg-splfunc-VFF-massless-contracted-coeffs}
\end{align}
with two coefficients $C_\polvb$, encoding correlations with hard matrix
elements of different polarization.
A similar situation is detailed for the azimuthal average of polarized
IS $f \to f \gamma^\ast$ splittings in \cite{dittmaier2008polarized}.


\paragraph{%
  Massive intermediate vector bosons.
}%

The case of \emph{massive} vector bosons is more involved, as terms
$\propto \bar p^\mu$ remain and $-g^{\mu \nu}$ cannot
simply be replaced by~\epsepssum\
in the azimuth-averaged splitting functions.
Their tensor structure
remains intricate, and the identification of spin-summed products of matrix elements
for the hard process similar to~\eqref{eq:azavg-splfunc-VXY-massless-coeffs} is not directly possible.
Keeping mass-suppressed terms emerging from the averaging process,
much simpler results, however, are obtained for splitting functions emerging from
the following
\emph{modification} of the $\vbfield{ij}^\ast \to \genfield{i}
\genfield{j}$ results from
Sections~\ref{sect:expl-splfuncs-FS-bosonic}
and \ref{sect:expl-splfuncs-FS-fermionic},
\begin{align}
  k_\perp^\mu k_\perp^\nu
  \to
  k_\perp^\mu k_\perp^\nu
  +
  \frac{k_\perp^2}{2}
  \left(
    \frac{
      \bar p^\mu
      n^\nu
      +
      n^\mu
      \bar p^\nu
    }{\bar p n}
    -
    \frac{
      \tilde{m}_{ij}^2
      n^\mu
      n^\nu
    }{(\bar p n)^2}
  \right)
  ,
  \qquad
  \bar{p}^2 = \tilde{m}_{ij}^2
  ,
  \label{eq:azavg-ktkt-redef}
\end{align}
which only affects terms beyond the leading mass order.
Because the corresponding adaption of the azimuth-averaged
results amounts to
\begin{align}
  \langle
    k_\perp^\mu k_\perp^\nu
  \rangle_\phi
  \to
  \frac{k_\perp^2}{2}
  g^{\mu\nu}
  ,
  \label{eq:azavg-ktkt-redef-avg}
\end{align}
only $\bar p^\mu \bar p^\nu$ and $g^{\mu \nu}$
are left after taking $\langle \dots \rangle_\phi$.
This allows us to write
\begin{align}
  \bigg\langle
  \bigg[
    \mathcal{P}
      {}^{\vbfield{ij}}_{\genfield{i}}
      {}^{\vbfield{ij^\prime}}_{\genfield{j}}
      \big(z,k_\perp^\mu,\bar p^\mu,\pol_i,\pol_j\big)
  \bigg]^{\mu \nu}
  \bigg\rangle_\phi
  &=
  C_V
  \left(
    -g^{\mu \nu}
    +\frac{\bar p^\mu \bar p^\nu}{m_{ij} m_{ij}^\prime}
  \right)
  +
  C_G\,
  \frac{\bar p^\mu \bar p^\nu}{m_{ij} m_{ij}^\prime}
  ,
\end{align}
with suitable coefficients~$C_V$ and~$C_G$.
For $\vbfield{ij}=\vbfield{ij^\prime}$
this facilitates the identification of spin-summed squared matrix elements.
Specifically, employing~\eqref{eq:VVV-ward-id-ul-me} we find
\begin{align}
  \bigg\langle
  T^{(n)}_{\vbfield{ij} X, \mu}
  \bigg[
    \mathcal{P}
      {}^{\vbfield{ij}}_{\genfield{i}}
      {}^{\vbfield{ij}}_{\genfield{j}}
      \big(z,k_\perp^\mu,\bar p^\mu,\pol_i,\pol_j\big)
  \bigg]^{\mu \nu}
  T^{(n) \ast}_{\vbfield{ij} X, \nu}
  \bigg\rangle_\phi
  &=
  C_V
  \smashoperator[l]{\sum_{\polvb = \pm,\mathrm{L}}}
  \Big|
    \mathcal{M}^{(n)\, \polvb}_{\vbfield{ij} X}
  \Big|^2
  +
  C_G\,
  \Big|
    \mathcal{M}^{(n)}_{\gbfield{ij} X}
  \Big|^2
  .
  \label{eq:azavg-splfunc-VVV-coeffs}
\end{align}
In contrast to the massless case, we are left
with a sum that does not only involve
a squared,
unpolarized matrix element 
for the vector boson,
but also a squared would-be Goldstone-boson matrix
element.
We note that
the specific frame in which the transverse and longitudinal
polarization vectors
are defined does not matter for the evaluation of the r.h.s.\
of~\eqref{eq:azavg-splfunc-VVV-coeffs}.
This is a key property, for example, when applying splitting
functions in subtraction algorithms.
Because the latter require the exact correspondence between splitting functions
and their azimuth-averaged counterparts,
we stress that this conclusion is only reached \emph{after} performing
the redefinition~\eqref{eq:azavg-ktkt-redef}, which introduces terms beyond the leading mass order.


Once more, this does not apply to $V^\ast \to f \bar f$ splittings with
polarized fermions or chiral
couplings, where additional structures
$\propto \epsepsdiff$ remain after the integration over $\phi$.
Even with the modification~\eqref{eq:azavg-ktkt-redef}, the integrated
splitting functions cannot be expressed in the
form~\eqref{eq:azavg-splfunc-VVV-coeffs}.
Rather, separate coefficients~$C_{V,\polvb}$ for the longitudinal and
the two transverse polarizations are found.
A simpler result is obtained by performing, instead
of~\eqref{eq:azavg-ktkt-redef}, the substitution
\begin{align}
  g^{\mu\nu}
  \to
  -
  \big(\epsepssum\big)
  ,
  \label{eq:azavg-gmunu-redef}
\end{align}
which is valid in leading mass order, according to \eqref{eq:gmunu-decomp}.
After averaging over~$\phi$,
we find
\begin{align}
  \bigg\langle
  T^{(n)}_{\vbfield{ij} X, \mu}
  \bigg[
    \mathcal{P}
      {}^{\vbfield{ij^{\vphantom{\prime}}}}_{\fermfield{i}}
      {}^{\vbfield{ij^\prime}}_{\antifermfield{j}}
      \big(z,k_\perp^\mu,\bar p^\mu,\polferm_i,\polferm_j\big)
  \bigg]^{\mu \nu}
  T^{(n) \ast}_{\vbfield{ij^\prime} X, \nu}
  \bigg\rangle_\phi
  =
  {}&
  \sum_{\polvb = \pm}
  C_{V,\polvb} \,
  \mathcal{M}^{(n)\, \polvb}_{\vbfield{ij} X}
  \left(\mathcal{M}^{(n)\, \polvb}_{\vbfield{ij^\prime} X}\right)^*
  \nonumber\\
  &{}
  +
  C_G \,
  \mathcal{M}^{(n)}_{\gbfield{ij} X}
  \left(\mathcal{M}^{(n)}_{\gbfield{ij^\prime} X}\right)^*
  .
  \label{eq:azavg-splfunc-VFF-contracted-coeffs}
\end{align}
Unlike~\eqref{eq:azavg-splfunc-VVV-coeffs},
the result from the evaluation of this
expression \emph{does}
retain dependence on the frame used for the definition of the
polarization vectors of the \emph{massive} vector
bosons $\vbfield{ij}$ and~$\vbfield{ij^\prime}$.

\subsubsection{Splittings of vector-boson--scalar interferences}
\label{sect:VH-az-avg}
All terms of the splitting functions
for~$\vbfield{ij}^\ast/\higgsfield{ij^\prime}^\ast \to \genfield{i}
\genfield{j}$ and~$\higgsfield{ij}^\ast/\vbfield{ij^\prime}^\ast \to
\genfield{i} \genfield{j}$ are proportional to the four-vectors~$\bar
p^\mu$, $k_\perp^\mu$, 
or $(k_\perp \bar{\varepsilon}_\mp)
\bar{\varepsilon}_\pm^\mu$.
Only the terms~$\propto \bar p^\mu$
survive upon integrating over $\phi$.
With suitable coefficients~$C$, we arrive at
\begin{align}
  \bigg\langle
  T^{(n)}_{\vbfield{ij} X, \mu}
  \bigg[
    \mathcal{P}
      {}^{\vbfield{ij^{\vphantom{\prime}}}}_{\genfield{i}}
      {}^{\higgsfield{ij^{\prime}}}_{\genfield{j}}
      \big(z,k_\perp^\mu,\bar p^\mu,\pol_i,\pol_j\big)
  \bigg]^{\mu}
  T^{(n) \ast}_{\higgsfield{ij^\prime} X}
  \bigg\rangle_\phi
  &=
  C\,
  \mathcal{M}^{(n)}_{\gbfield{ij} X}
  \mathcal{M}^{(n)\, \ast}_{\higgsfield{ij^\prime} X}
  .
  \label{azavg-splfunc-VHFF-contracted-coeffs}
\end{align}
This result is free from gauge-boson matrix elements and therefore not
subject to issues regarding the frame-dependent definition of
polarization vectors.

\subsubsection{Splittings of fermions and antifermions}
\label{sect:F-az-avg}
All coefficients of the
structures~\eqref{eq:ferm-splittings-dirac-structures} in the
$\fermfield{ij}^\ast \to \genfield{i} \genfield{j}$ and
$\antifermfield{ij}^\ast \to \genfield{i} \genfield{j}$ splitting functions are
azimuthal-angle independent.
Thus, the terms~$\propto \slashed{k}_\perp$ vanish after azimuthal
integration.
The remaining contributions can be collected into
terms~$\propto \Sigma_\kappa (\slashed{\bar{p}} \pm \tilde m_{ij})$
after performing the redefinitions
\begin{align}
  \slashed{\bar{p}}
  \to
  \slashed{\bar{p}} \pm \tilde m_{ij}
  ,\qquad
  \slashed{\bar{p}}
  \gamma^5
  \to
  \mp\gamma^5
  \slashed{\bar{s}}
  (
  \slashed{\bar{p}} \pm \tilde m_{ij}
  )
  ,
  \label{eq:ferm-splittings-dirac-structures-redefinitions}
\end{align}
where the signs apply to fermion and antifermion splittings,
respectively.
We write $\bar s^\mu = \bar p^\mu/\tilde m_{ij} + \restpol^\mu$
with some unspecified
gauge vector~$\restpol^\mu$, according to~\eqref{eq:spin-projection-op}.
Again, these substitutions are valid at leading mass order.
Assuming $\fermfield{ij} = \fermfield{ij^\prime}$ and
$m_{ij} = m_{ij}^\prime = \tilde m_{ij}$,
we get,
via~\eqref{eq:spinproj-complrelat-ferm},
\begin{align}
  \bigg\langle
  \bar T^{(n)}_{\fermfield{ij} X}
  \bigg[
    \mathcal{P}
      {}^{\fermfield{ij}}_{\genfield{i}}
      {}^{\fermfield{ij}}_{\genfield{j}}
      \big(z,k_\perp^\mu,\bar p^\mu,\pol_i,\pol_j)
      \bigg]
  T^{(n)}_{\fermfield{ij} X}  
  \bigg\rangle_\phi
  &=
  C_+\,
  \Big|\mathcal{M}^{(n)\, +}_{\fermfield{ij} X}\Big|^2
  +
  C_-\,
  \Big|\mathcal{M}^{(n)\, -}_{\fermfield{ij} X}\Big|^2
  ,
  \label{azavg-splfunc-FXY-contracted-coeffs}
\end{align}
with coefficients~$C_\pm$, where $\pm$ indicates the two 
possible polarizations of $\fermfield{ij}$.
A similar relation is found for $\antifermfield{ij}^\ast \to \genfield{i}
\genfield{j}$.

\subsection{IS splitting functions}
\label{sect:expl-splfuncs-IS}
The computation of \gls{is} splitting functions directly carries over
from the \gls{fs} case, including the treatment of mass-singular terms
from longitudinal polarization states for vector bosons.
However, care w.r.t.\ signs and phases 
has to be taken when
employing~\glspl{wi} and coupling relations.
For example, the \gls{wi} for the underlying amplitude, with incoming
momentum~$p_{ai}^\mu$ and incoming~$\vbfield{ai}$, reads
\begin{align}
  T^{(n)}_{\gbfield{ij} X}(p_{ai})
  =
  \eta_{\vbfield{ai}}^\ast
  \frac{p_{ai}^\mu}{m_{ai}}
  T^{(n)}_{\vbfield{ai} X, \mu}(p_{ai})
  .
  \label{eq:VVV-ward-id-ul-me-IS}
\end{align}
Similar to~\eqref{eq:oc-summary}, the quasi-collinear limit is evaluated
entirely via
\begin{align}
  k_\perp^\mu
    = \orderof{\lambda},
  \quad
  p_a p_i
    = \orderof{\lambda^2},
  \quad
  \bar p^\mu \, T^{(n)}_{\vbfield{ai} X, \mu}(p_{ai})
    = \orderof{\lambda},
  \quad
  \restpol^\mu_{a/i}
  = \orderof{\lambda}
  .
  \label{eq:oc-summary-IS}
\end{align}
Potentially remaining scalar products~$\bar p \restpol_a$ and $\bar p \restpol_i$ are
eliminated via
\begin{align}
  0
  &=
  p_a \varepsilon_{\mathrm{L}}(p_a)
  =
  m_a
  +
  \bigg(\frac{1}{x} \bar p - B_a n \bigg) \restpol_a
  =
  m_a + \frac{1}{x} \, \big(\bar p \restpol_a\big) + \orderof{\lambda^2}
  ,
  \nonumber\\
  0
  &=
  p_i \varepsilon_{\mathrm{L}}(p_i)
  =
  m_i
  +
  \bigg( \frac{1-x}{x} \bar p - B_i n + k_\perp \bigg) \restpol_i
  =
  m_i + \frac{1-x}{x} \, \big(\bar p \restpol_j\big) + \orderof{\lambda^2}
  ,
\end{align}
with $B_a$ and $B_i$ defined in \eqref{eq:sudakov-param-is}.
As in the \gls{fs} case, we observe that our results obey the symmetries from
Sections~\ref{sect:splfunc-symm-i-j-and-a-i},
\ref{sect:splfunc-symm-ij-ijpr-and-ai-aipr}, and charge conjugation.
Moreover, as discussed in Section~\ref{sect:splfunc-symm-crossing}, all
\gls{is} splitting functions can be recovered from \gls{fs} counterparts
via crossing symmetry.
For this reason, we do not explicitly report our results for the
\gls{is} case and refer to \eqref{eq:fs-to-is-splfunc-mapping} instead.
For example, instead of deriving the~$\bar f \to \bar f V^\ast$
splitting functions from the diagrams
\newcommand{\ISDiagGeo}{%
  \renewcommand{\sclfac}{1}%
  \renewcommand{\intsclfac}{1}%
  \renewcommand{\extsclfac}{1}%
  \renewcommand{\len}{1.1cm}%
  \renewcommand{\rad}{.6cm}%
  \renewcommand{\dotrad}{1mm}%
  \renewcommand{\diagang}{45}%
}%
\drawmomtrue%
\drawvertdottrue%
\begin{alignat}{2}%
  \ftntmath{%
    \MContrExplFields{\vbfield{ai}}{\antifermfield{a}}{\antifermfield{i}}^{\polferm_a \polferm_i}%
  }%
  &\ftntmath{{}={}}%
  \ISDiagGeo%
  \ftntmath{%
    \splittingdiagIndivIS{T^{(n)}_{\vbfield{ai} X}}%
      {antifermion}{antifermion}{photon}%
      {\antifermfield{a}, \polferm_a}{\antifermfield{i}, \polferm_i}{\vbfield{ai}}%
    ,%
  }%
  \resetdiagargs%
  &\quad
  \ftntmath{%
    \MContrExplFields{\gbfield{ai}}{\antifermfield{a}}{\antifermfield{i}}^{\polferm_a \polferm_i}%
  }%
  &\ftntmath{{}={}}%
  \ISDiagGeo%
  \ftntmath{%
    \splittingdiagIndivIS{T^{(n)}_{\gbfield{ai} X}}%
      {antifermion}{antifermion}{scalar}%
      {\antifermfield{a}, \polferm_a}{\antifermfield{i}, \polferm_i}{\gbfield{ai}}%
    ,%
  }%
  \resetdiagargs%
  \label{eq:FbarFbarV-diags-IS}
\end{alignat}%
the crossing formula \eqref{eq:fs-to-is-splfunc-mapping} implies the relation
\begin{align}
  \bigg[
    \bar{\mathcal{P}}
        {}^{\vbfield{ai^{\vphantom{\prime}}}}_{\antifermfield{a}}
        {}^{\vbfield{ai^\prime}}_{\antifermfield{i}}
      \bigg(x,k_\perp^\mu,\bar p^\mu,\polferm_a,\polferm_i\bigg)
  \bigg]^{\mu \nu}
  =
  -
  \bigg[
    \mathcal{P}
      {}^{\vbantifield{ai^{\vphantom{\prime}}}}_{\fermfield{a}}
      {}^{\vbantifield{ai^\prime}}_{\antifermfield{i}}
      \bigg(\frac{1}{x},-\frac{k_\perp^\mu}{x},-\hat p^\mu,-\polferm_a,\polferm_i\bigg)
  \bigg]^{\mu \nu}
  _{m_\mathrm{FS} \to m_\mathrm{IS}}
  \label{eq:VFF-fs-to-FFV-is-splfunc-mapping}
\end{align}
to the \gls{fs} case~$V^\ast \to f \bar f$ from
Section~\ref{sect:VFFbar},
with the identification
\begin{align}
  I_{\fermfield{i}^{\chir} \fermfield{j}^{\chir}}^{\vbfield{ij\smash{{}^{(\prime)}}}}
  \to
  I_{\fermfield{a}^{\chir} \fermfield{i}^{\chir}}^{\vbantifield{ai\smash{{}^{(\prime)}}}}
  .
  \label{eq:VFF-FVF-crossing-cpl}
\end{align}
Here, we introduced 
\begin{align}
  \hat{p}^\mu
  \equiv
  \bar{p}^\mu
  -
  k_\perp^\mu
  ,
  \label{eq:def-ptilde}
\end{align}
which considerably simplifies 
the results and makes both the
symmetry rule~\eqref{eq:a-i-symmru} and the crossing
relation~\eqref{eq:fs-to-is-splfunc-mapping} more obvious.\footnotemark%
\footnotetext{%
  This is also beneficial in view of the leading-mass-order expression
  of the \gls{wi}~\eqref{eq:VVV-ward-id-ul-me-IS},
  \begin{align}
    p_{ai}^\mu \,
    T^{(n)}_{\vbfield{ai} X,\mu}(p_{ai})
    =
    \left(
      \bar p
      -
      k_\perp
    \right)^\mu
    T^{(n)}_{\vbfield{ai} X,\mu}(p_{ai})
    +
    \orderof{m^2}
    .
  \end{align}
}


The crossing symmetry~\eqref{eq:fs-to-is-splfunc-mapping} suggests the
following definition of generic abbreviations for \gls{is} splittings
with two (anti-)fermions and a third field~$\varrho$,
\begin{align}
  \bar{a}^{\varrho \chir_1 \chir_2}_{\mathcal{X} \mathcal{I} \bar{\mathcal{I}}}
  (\pm 1)
  &=
  \bar{\zeta}_{\mathcal{I}} m_{\mathcal{I}}
  \mathcal{C}^{\chir_1}_{
    \nonfermtildefield{\mathcal{X}}
    \antifermfield{\mathcal{I}}
    \fermfield{\bar{\mathcal{I}}}
    }
  \pm
  \bar{\zeta}_{\bar{\mathcal{I}}} m_{\bar{\mathcal{I}}}
  \mathcal{C}^{\chir_2}_{
    \nonfermtildefield{\mathcal{X}}
    \antifermfield{\mathcal{I}}
    \fermfield{\bar{\mathcal{I}}}
    }
  ,
  \nonumber
  \\
  \bar{b}^{\varrho \chir_1 \chir_2}_{\mathcal{X} \mathcal{I} \bar{\mathcal{I}}}
  (\pm 1)
  &=
  m_{\mathcal{I}}
  \mathcal{C}^{\chir_1}_{
    \nonfermtildefield{\mathcal{X}}
    \antifermfield{\mathcal{I}}
    \fermfield{\bar{\mathcal{I}}}
    }
  \pm
  m_{\bar{\mathcal{I}}}
  \mathcal{C}^{\chir_2}_{
    \nonfermtildefield{\mathcal{X}}
    \antifermfield{\mathcal{I}}
    \fermfield{\bar{\mathcal{I}}}
    }
  ,
  \nonumber
  \\
  \bar{c}^{\varrho \chir_1 \chir_2}_{\mathcal{X} \mathcal{I} \bar{\mathcal{I}}}
  &=
  -
  \big(
    2 \bar{\zeta}_{\mathcal{X}} m_{\mathcal{X}}^2
    + \bar{\zeta}_{\mathcal{I}} m_{\mathcal{I}}^2
    + \bar{\zeta}_{\bar{\mathcal{I}}} m_{\bar{\mathcal{I}}}^2
  \big)
  \mathcal{C}^{\chir_1}_{
    \nonfermtildefield{\mathcal{X}}
    \antifermfield{\mathcal{I}}
    \fermfield{\bar{\mathcal{I}}}
    }
  +
  m_{\mathcal{I}} m_{\bar{\zeta}_{\mathcal{I}}}
  \big(
    \bar{\zeta}_{\mathcal{I}}
    +
    \bar{\zeta}_{\bar{\mathcal{I}}}
  \big)
  \mathcal{C}^{\chir_2}_{
    \nonfermtildefield{\mathcal{X}}
    \antifermfield{\mathcal{I}}
    \fermfield{\bar{\mathcal{I}}}
    }
  ,
  \label{eq:gen-ferm-abbs-IS}
\end{align}
where $\mathcal{I}, \bar{\mathcal{I}}, \mathcal{X} \in
\{a,i,ai^{(\prime)}\}$ denote the indices of the outgoing fermion,
outgoing antifermion, and~$\varrho$, respectively, in the
processes~$\genfield{a} \to \genfield{i}
[\genfield{ai}^\ast/\genfield{ai^\prime}^\ast]$.
We write~$\nonfermtildefield{a} = \nonfermfield{a}$,
$\nonfermtildefield{ai^{(\prime)}} = \nonfermantifield{ai^{(\prime)}}$,
and~$\nonfermtildefield{i} = \nonfermantifield{i}$.
These abbreviations are obtained from the \gls{fs}
counterparts~\eqref{eq:gen-ferm-abbs} simply by replacing the
symbols~$\zeta$ with~$\bar{\zeta}$, defined as
\begin{align}
  \bar{\zeta}_a = \frac{x-1}{x}
  ,\qquad
  \bar{\zeta}_i = \frac{1}{x}
  ,\qquad
  \bar{\zeta}_{ai} = \bar{\zeta}_{ai^\prime} = - \frac{x-1}{x^2}
  .
  \label{eq:gen-mom-sharing-symbol-IS}
\end{align}
Similar to the identity~\eqref{eq:generic-relat-zeta} for $\zeta$, this variable obeys
\begin{align}
  \bar{\zeta}_{\mathcal{X}}
  =
  -
  \frac{
    \bar{\zeta}_{\mathcal{I}}
    \bar{\zeta}_{\bar{\mathcal{I}}}
  }{
    \bar{\zeta}_{\mathcal{I}}
    +
    \bar{\zeta}_{\bar{\mathcal{I}}}
  }
  \label{eq:generic-relat-zeta-IS}
  .
\end{align}
Specifically, the \gls{is} abbreviations are constructed such that they
are related to the \gls{fs} counterparts by crossing.
For example, the \gls{is} abbreviations for $\bar f \to \bar f V^\ast$, i.e.\
\begin{align}
  \bar{a}^{V \chir_1 \chir_2}_{ai, a, i}
  (\pm 1)
  &=
  \frac{x-1}{x}
  m_a
  I_{\fermfield{a}^{\chir_1} \fermfield{i}^{\chir_1}}^{\vbantifield{ai}}
  \pm
  \frac{1}{x}
  m_i
  I_{\fermfield{a}^{\chir_2} \fermfield{i}^{\chir_2}}^{\vbantifield{ai}}
  ,
  \nonumber
  \\
  \bar{b}^{V \chir_1 \chir_2}_{ai, a, i}
  (\pm 1)
  &=
  m_a
  I_{\fermfield{a}^{\chir_1} \fermfield{i}^{\chir_1}}^{\vbantifield{ai}}
  \pm
  m_i
  I_{\fermfield{a}^{\chir_2} \fermfield{i}^{\chir_2}}^{\vbantifield{ai}}
  ,
  \nonumber
  \\
  \bar{c}^{V \chir_1 \chir_2}_{ai, a, i}
  &=
  \bigg(
    2 \frac{x-1}{x^2} m_{ai}^2
    - \frac{x-1}{x} m_a^2
    - \frac{1}{x} m_i^2
  \bigg)
  I_{\fermfield{a}^{\chir_1} \fermfield{i}^{\chir_1}}^{\vbantifield{ai}}
  +
  m_a m_i
  I_{\fermfield{a}^{\chir_2} \fermfield{i}^{\chir_2}}^{\vbantifield{ai}}
  ,
  \label{eq:FbarFbarV-abbrev-IS}
\end{align}
are obtained from the corresponding \gls{fs} $V^\ast \to f \bar f$
symbols \eqref{eq:VFFbar-abbrev} via the combination of the mappings
$m_a \leftrightarrow m_i$, $x \to 1/x$,
and~\eqref{eq:VFF-FVF-crossing-cpl} from the crossing
rule~\eqref{eq:VFF-fs-to-FFV-is-splfunc-mapping}.
Thus, when recovering \gls{is} $\bar f \to \bar f V^\ast$ splitting
functions from \gls{fs} $V^\ast \to f \bar f$ results, the associated
abbreviation symbols~\eqref{eq:VFFbar-abbrev} simply need to be
exchanged with those from~\eqref{eq:FbarFbarV-abbrev-IS}.


To make the above example even more explicit, we consider the special case of
\gls{is} $\bar f \to \bar f V^\ast$ splittings with
massless vector bosons $\vbfield{ai}$, $\vbfield{ai^\prime}$.
The corresponding \gls{fs} $V^\ast \to f \bar f$ splitting functions can
be found in~\eqref{eq:FS-VFFbar-splfunc-specialcase}.
Multiplication of these results with $-1$ and application of
\begin{align}
  & z \to \frac{1}{x}
  ,\quad
  k_\perp^\mu \to \frac{- k_\perp^\mu}{x}
  ,\quad
  \bar{p}^\mu \to - \bar{p}^\mu + k_\perp^\mu
  ,\quad
  \kappa_i \to -\kappa_a
  ,\quad
  \kappa_j \to \kappa_i
  ,
\nonumber\\
  & m_i \to m_a
  ,\quad
  m_j \to m_i
  ,\quad
  I_{\fermfield{i}^{\sigma} \fermfield{j}^{\sigma}}^{\vbfield{ij\smash{{}^{(\prime)}}}}
  \to
  I_{\fermfield{a}^{\sigma} \fermfield{i}^{\sigma}}^{\vbantifield{ai\smash{{}^{(\prime)}}}}
  \label{eq:IS-VFF-FS-FVF-crossing-cpl}
\end{align}
leads us to the \gls{is} $\bar f \to \bar f V^\ast$ splitting functions
\begin{align}
  %
  %
  \bigg[
    \bar{\mathcal{P}}
        {}^{\vbfield{ai^{\vphantom{\prime}}}}_{\antifermfield{a}}
        {}^{\vbfield{ai^\prime}}_{\antifermfield{i}}
      \big(x,k_\perp^\mu,\bar p^\mu,\mp,\mp\big)
  \bigg]^{\mu \nu}_{m_{ai}^{(\prime)}=0}
  &=
  I_{\fermfield{a}^\mathrm{R/L} \fermfield{i}^\mathrm{R/L}}^{\vbantifield{ai}}
  \Big(I_{\fermfield{a}^\mathrm{R/L} \fermfield{i}^\mathrm{R/L}}^{\vbantifield{ai^\prime}}\Big)^\ast
  \begin{aligned}[t]
    \bigg\{
      &
      \frac{-k_\perp^2}{x-1}
      g^{\mu \nu}
      +
      \frac{4}{x^2}
      k_\perp^\mu
      k_\perp^\nu
      \\
      &{}
      \pm
      s_\varepsilon
      \frac{2-x}{(x-1)x}
      k_\perp^2
      \big(
        \bar{\varepsilon }_-^{\mu }
        \bar{\varepsilon }_+^{\nu }
        -
        \bar{\varepsilon }_+^{\mu }
        \bar{\varepsilon }_-^{\nu }
      \big)
    \bigg\}
    ,
  \end{aligned}
  \nonumber\\
  %
  %
  \bigg[
    \bar{\mathcal{P}}
        {}^{\vbfield{ai^{\vphantom{\prime}}}}_{\antifermfield{a}}
        {}^{\vbfield{ai^\prime}}_{\antifermfield{i}}
      \big(x,k_\perp^\mu,\bar p^\mu,\mp,\pm\big)
  \bigg]^{\mu \nu}_{m_{ai}^{(\prime)}=0}
  &=
  \frac{
    x^2
    \bar{a}^{V \, \mathrm{LR/RL}}_{ai, a, i}(+1)
    \bar{a}^{V \, \mathrm{LR/RL}}_{ai^\prime, a, i}(+1)
  }{x-1}
  \bigg\{
    g^{\mu \nu}
    \mp
    s_\varepsilon
    \big(
      \bar{\varepsilon }_-^{\mu }
      \bar{\varepsilon }_+^{\nu }
      -
      \bar{\varepsilon }_+^{\mu }
      \bar{\varepsilon }_-^{\nu }
    \big)
  \bigg\}
  .
\end{align}

\subsection{Computing the azimuthal average of IS splitting functions}
\label{sect:IS-az-avg}
The azimuthal average of \gls{is} splitting functions
(w.r.t.\ the kinematics defined by the \gls{is} version~\eqref{eq:sudakov-param-fs} of the Sudakov parametrization)
may be evaluated
similarly to the \gls{fs} case discussed in
Section~\ref{sect:FS-az-avg}.
Once more defining the azimuthal angle $\phi$ of the transverse momentum
$k_\perp^\mu$ in the frame distinguished by~$\bar{\boldsymbol{p}}
\parallel \boldsymbol{n}$ (where $k_\perp^\mu$, $\bar{p}^\mu$, and
$n^\mu$ are now taken from~\eqref{eq:sudakov-param-fs}), we get
the form~\eqref{eq:azavg-sudkt-form} of $k_\perp^\mu$.
Again, this allows for the straightforward computation of
the average $\langle \dots \rangle_\phi$
of the \gls{is} splitting functions by application of the
rules~\eqref{eq:azavg-kt} and \eqref{eq:azavg-ktmuktnu-av}.
With this at hand, the functions for the various splittings of vector
bosons, vector-boson--scalar interferences, and (anti-)fermions may be
treated similarly to the discussion from
Sections~\ref{sect:V-az-avg}--\ref{sect:F-az-avg}.

\subsection{\revision{Outlook towards an electroweak parton shower 
and subtraction formalisms}}
\label{sect:outlook}
\revision{The derivation of all EW splitting functions represents
a major step towards the construction of an EW parton shower or 
a subtraction formalism for soft/collinear singularities from EW
particle emissions at high energies. Nevertheless, further steps
are needed or desirable to tackle these tasks:
\begin{itemize}
\item
The most important step concerns the inclusion of the limit
of {\it soft particle emission} where the energy $p_s$ of any emitted 
EW particle is of the order of the generic mass scale $m$, i.e.\
$p_s^0\sim m$. In the FS splitting functions, this limit corresponds to
$z\to0$ or $z\to1$ for $s=i$ or $s=j$, respectively;
for IS splittings, it corresponds to $x\to1$ and $s=i$.
Soft singularities result from propagator enhancement in diagrams with
the soft particle $\genfield{s}$ directly coupled to 
any external field
$\genfield{r}$ acting as radiator. However, unlike collinear singularities,
soft singularities in the squared matrix element
$|{\cal M}^{(n+1)}_{\genfield{s} X}|^2$
arise from all possible interferences of such diagrams,
not only from ``squared contributions'' featuring the same emitter.
This leads to angular and charge correlations for soft emission that do
not exist for collinear singularities.
For instance, soft singularities in
$|{\cal M}^{(n+1)}_{\genfield{s} X}|^2$
from soft
gauge-boson emission factorize from the
underlying hard processes in the form 
$|{\cal M}^{(n)}_{\hat X}\otimes J^\mu_{\mathrm{eik,}\genfield{s}}|^2$,
where
${\cal M}^{(n)}_{\hat X}$
is the hard matrix element for the process
without $\genfield{s}$ in the final state and $J^\mu_{\mathrm{eik,}\genfield{s}}$
is the {\it eikonal current} for soft $\genfield{s}$ emission
(see, e.g., Refs.~\cite{Peskin:1995ev,catani1996general,dittmaier1999photonrad,%
bohm2001gauge,catani2002massivedipole,Dissertori:2009,Schwartz:2014sze} for QCD/QED).
Note that soft EW emission in general changes the nature of the radiating particle.
Taking the emission of a possibly massive gauge boson off a gauge-boson line
as an example, the soft limit
can be easily derived from the sum of the first two diagrams of 
\eqref{eq:VVV-diags-FS}. External longitudinal gauge bosons do not pose any
problems, so that external Goldstone fields need not considered. 
A full treatment of the emission of a soft gauge boson $V_s$ would require to include all
possibilities $V\to V'V_s$, $V\to SV_s$, $S\to VV_s$, $S\to S'V_s$, 
$f\to f'V_s$, and $\bar f\to \bar f'V_s$, with generic vectors $V^{(\prime)}$,
scalars $S^{(\prime)}$, and (anti-)fermions $f^{(\prime)}$, $\bar f^{(\prime)}$.
A comprehensive discussion of the soft limit, however, goes beyond the scope 
of this publication; further details can, e.g., be found in Ref.~\cite{Nardi:2024tce}.
\item
In the dipole subtraction formalism for soft and/or collinear singularities, such as in
QCD~\cite{catani1996general,Phaf:2001gc,catani2002massivedipole}
or QED~\cite{dittmaier1999photonrad,dittmaier2008polarized},
the angular and charge correlations of the soft singularities are taken care of by the
introduction of emitter--spectator pairs, where the emitter takes the role of the 
radiating particle and the spectator is there to accommodate the correlations.
Each splitting function for a collinear enhancement, thus, is replaced by a set of 
dipole splitting functions with a fixed emitter and a set of spectators, where each of the 
dipole splitting functions now depends on two kinematical variables instead of only one.
The generalization from QCD/QED to EW emissions seems straightforward, but some
care will be required w.r.t.\ the fact that the 
underlying hard matrix elements 
$T^{(n)\, \sigma}_{\vbfield{ij} X}\left(p_{ij}\right)$
for a massive gauge boson is not transversal, i.e.\
$T^{(n)\, \sigma}_{\vbfield{ij} X}\left(p_{ij}\right)p_{ij,\sigma} \ne 0$,
so that underlying hard matrix elements involving Goldstone fields 
appear in the dipole subtraction functions
(or likewise with longitudinal gauge bosons if the \gls{gbet} is used).

\item
\revisiontwo{We recall that the construction of a parton shower in
leading logarithmic order and of a subtraction formalism both require only
the hard matrix elements in leading quasi-collinear/soft approximation,
so that the hard functions 
$T^{(n)}_{\genfield{ij} X}\left(p_{ij}\right)$ and
$T^{(n)}_{\genfield{ai} X}\left(p_{ai}\right)$
can be approximated by
$T^{(n)}_{\genfield{ij} X}\left(\bar p\right)$ and
$T^{(n)}_{\genfield{ai} X}\left(\bar p\right)$, respectively.
This approximation is sufficient for the iterative factorization
in a parton shower, and changing for instance the recoil scheme 
in the factorization formula
leads to effects beyond leading power.
}

\item
As for parton-shower implementations, particular care is required if different
particles $\genfield{ij}$ and $\genfield{ij}^\prime$ 
\revisiontwo{(or $\genfield{ai}$ and $\genfield{ai}^\prime$ for initial-state splittings)}
can interfere in the \revisiontwo{infrared (collinear/soft)} limit,
because in this case there are different interfering hard matrix elements, so that
the on-shell condition $p_{ij}^2=m_{ij}^2$ 
\revisiontwo{(or $p_{ai}^2=m_{ai}^2$) cannot be fulfilled in the two interfering
hard matrix elements at the same time for $m_{ij}\ne m'_{ij}$ (or $m_{ai}\ne m'_{ai}$).
As explained in the previous item, a leading-power approximation
in the quasi-collinear/soft limit is sufficient in the hard matrix elements.
Thus, the hard momentum $\bar p^\mu$ from the interfering hard matrix elements
could be modified by subleading terms such that the respective mass-shell
conditions in (sub)matrix elements are fulfilled.
This allows for the consistent (gauge-invariant) evaluation of all matrix
elements.
In view of interferences between different particles types in the
branchings in the infrared (collinear/soft) limits, branching algorithms based
on matrix elements seem the ideal framework for a parton shower,
but simpler approximative approaches based on density matrices or
reweighting algorithms seem to be a promising intermediate solution,
as e.g.\ described in Refs.~\cite{Chen:2016wkt,Kleiss:2020rcg,Brooks:2021kji}.}

Another aspect relevant for EW parton showers concerns the eventual truncation of
some shower branches by particle decays, which e.g.\ set in if the particle virtuality
$p_{ij}^2-m_{ij}^2$ becomes of the order of $m_{ij}\Gamma_{ij}$ if the decay of
$\genfield{ij}$ (with total) decay width $\Gamma_{ij}$
is kinematically possible.

\item
Finally, we left considerations about the subleading behaviour in the quasi-collinear limits,
which are for instance relevant for analyses at subleading power, for future work. 
As detailed in Section~\ref{sect:quasi-coll-fact},
the leading quasi-collinear enhancement is of ${\cal O}(m^{-2})$ in $|{\cal M}^{(n+1)}|^2$ and results
from the square of all diagrams of type~\diagtype{1}.
Subleading terms are, thus, at most of ${\cal O}(m^{-1})$.
They also receive contributions from the interference
of diagrams of types~\diagtype{1} and~\diagtype{2}.
Obviously a thorough analysis of these interferences
are beyond the scope of this paper.
Subleading terms may also be introduced when the external momenta
$p_{ij}$ and $p_{ai}$ in the hard matrix elements
$T^{(n)}_{\genfield{ij} X}\left(p_{ij}\right)$
and
$T^{(n)}_{\genfield{ai} X}\left(p_{ai}\right)$
are replaced by on-shell counterparts via some explicit recoil prescription
\revisiontwo{as described above}
(shifting other external momenta in the matrix elements, as well).
In the soft limit this issues touches upon a formulation of the 
Low--Burnett--Kroll theorem~\cite{Low:1958sn,Burnett:1967km}
(for recent progress see, e.g., Ref.~\cite{Engel:2023ifn} and references therein) 
for EW particles.
\end{itemize}
}

\section{Comparison with existing results}
\label{sect:lit-review-splittings}
In this section, we briefly review the existing 
literature~\cite{Chen:2016wkt,Cuomo:2019siu,Kleiss:2020rcg,Brooks:2021kji,Masouminia:2021kne}
on electroweak quasi-collinear factorization
and comment on the compatibility with our results.
As already mentioned, these references derive factorization formulas at
the level of individual amplitudes, in contrast to our approach.
This considerably complicates the comparison with our findings.
Specifically, the partial matrix 
elements~\eqref{eq:def-partialM} are written as
products of matrix elements for the underlying hard process and
\emph{splitting amplitudes} for $1 \to 2$ splitting processes.
Both
are evaluated with on-shell polarization vectors and spinors,
while the incoming momentum of the splitting amplitudes is off-shell.
This procedure relies on the non-trivial, well-defined decomposition of
\emph{off-shell} propagator numerators into
\emph{on-shell} polarization sums of outer products of wave functions.
These steps yield remainder terms that must be identified as
subleading in the quasi-collinear limit.
The splitting amplitudes are evaluated either in explicit,
component-wise, and frame-dependent parametrizations of momenta,
spinors, and polarization
vectors~\cite{Chen:2016wkt,Cuomo:2019siu,Masouminia:2021kne}, or using
spinor--helicity methods~\cite{Kleiss:2020rcg,Brooks:2021kji}.
The issues regarding mass-singular factors in longitudinal polarization
vectors are dealt with in different ways, as detailed in the following.
Throughout this section, we focus on \gls{fs} splittings.
The~\gls{is} case can be treated analogously.

\subsection{Comparison to Cuomo, Vecchi, and Wulzer~\texorpdfstring{\cite{Cuomo:2019siu}}{}}
\label{sect:wulzer-vecchi-cuomo}
\paragraph{Details of the approach.}
The approach of~\cite{Cuomo:2019siu} as well as its further development
described in \cite{Nardi:2024tce} are
based on the redefinition of
longitudinal polarization vectors with the help of \glspl{wi}.
In the following comparison to our findings
we concentrate on~\cite{Cuomo:2019siu} only; the results of 
\cite{Nardi:2024tce} 
were found to be in agreement with the ones of~\cite{Cuomo:2019siu}
there.
We note that~\cite{Nardi:2024tce} goes beyond \cite{Cuomo:2019siu} by
employing a formalism that is not tied to a specific Lorentz frame;
moreover, the soft emission limits are considered as well, in contrast to 
\cite{Cuomo:2019siu} and our article.

While the procedure of~\cite{Cuomo:2019siu}
is defined for a general spontaneously broken gauge
theory with arbitrarily many gauge bosons and scalars,
we review it in the simpler setting with only one vector-boson
field~$V^\mu$ and its associated would-be Goldstone boson field~$G$.
As starting point, their fields are collected into a common
5-dimensional vector~$\Theta^M=(V^\mu,G)$, where the index runs over the
four Lorentz components~$M=\mu$ and a scalar component~$M=\mathrm{G}$.
In this notation, the \gls{wi}~\eqref{eq:multi-ward-id} can be written
as\footnotemark
\footnotetext{%
  %
  Here and in the following, we adapt the formulas
  from~\cite{Cuomo:2019siu} to our conventions and notation.
  This concerns, for example, the phases of would-be Goldstone bosons.
}%
\begin{align}
  \Big \langle T
    \prod_{i=0}^{m}
    \Big[
      \mathcal{K}_{M_i}(p_i) 
      \underline{\Theta^{M_i}}(p_i)
    \Big]
    \prod_{j=1}^{n} \genfield{j}^{\mathrm{phys}}
  \Big \rangle_{\mathrm{c}}
  =
  0
  .
  \label{eq:wulzer-ward-gengaugetheo}
\end{align}
At tree level, the vector and scalar components of $\mathcal{K}$ read 
\begin{align}
  \mathcal{K}^0_\mu(p)
  =
  \imagi \frac{p_{\mu}}{\sqrt{p^2}}
  ,
  \qquad
  \mathcal{K}^0_{\mathrm{G}}(p)
  =
  -
  \imagi
  \eta_{V}
  \frac{M_{V}}{\sqrt{p^2}}
  .
  \label{eq:kappa-factor-def}
\end{align}
The standard polarization vectors are extended into 5-dimensional
notation via
\begin{align}
  \varepsilon^{\mathrm{st}}_{\polvb, M}(p)
  =
  \big(\varepsilon^{\mathrm{st}}_{\polvb, \mu}(p),0\big)
  .
\end{align}
Next, longitudinal polarization vectors for a general \emph{off-shell}
momentum~$p^\mu$ are redefined.\footnotemark\
\footnotetext{%
  In the approach of~\cite{Cuomo:2019siu}, off-shell longitudinal
  polarization vectors, obtained from~\eqref{eq:long-pol-vec} by
  replacing $M \to \sqrt{p^2}$, are required for the decomposition of
  propagators.
  See below.
}%
For incoming and outgoing vector bosons~$V$, respectively,
\begin{alignat}{2}
  \varepsilon_{\mathrm{L}, M}(p)
  &=
  \varepsilon^{\mathrm{st}}_{\mathrm{L},M}(p)
  +
  \imagi \mathcal{K}_M(+p)
  &&=
  \bigg(\restpol_\mu(p), + \eta_{V} \frac{M_{V}}{\sqrt{p^2}} \bigg)
  ,\nonumber\\
  \overline \varepsilon_{\mathrm{L},M}(p)
  &=
  \varepsilon^{\mathrm{st}\, \ast}_{\mathrm{L},M}(p)
  -
  \imagi \bar{\mathcal{K}}_M(-p)
  &&=
  \bigg(\restpol_\mu(p), - \eta_{\bar V} \frac{M_{V}}{\sqrt{p^2}} \bigg)
  ,
  \label{eq:redef-long-polvec}
\end{alignat}
where $\bar{\mathcal{K}}_M(-p)$ refers to an incoming field $\bar V$ (=~outgoing~$V$).
Here, the problematic terms~$p^\mu/\sqrt{p^2}$ are canceled
between~$\varepsilon^{\mathrm{st}}_{\mathrm{L}}$ and $\mathcal{K}$,
such that~$\varepsilon_{\mathrm{L}}$ and~$\overline
\varepsilon_{\mathrm{L}}$ are well behaved.
The computation of amplitudes with redefined longitudinal polarization
vectors~$\varepsilon_{\mathrm{L}}$ now involves would-be Goldstone
amplitudes, too.
Due to the \gls{wi}~\eqref{eq:wulzer-ward-gengaugetheo},
the result agrees with the amplitude
for~$\varepsilon^{\mathrm{st}}_{\mathrm{L}}$.
For incoming vector bosons, this reads
\begin{align}
  &\prod_{i=0}^{m}
  \varepsilon^{\mathrm{st}}_{\mathrm{L},\mu_i}(p_i)
  \Big \langle T
    \prod_{i=0}^{m}
    \underline{V^{\mu_i}}(p_i)
    \prod_{j=1}^{n} \genfield{j}^{\mathrm{phys}}
  \Big \rangle
  =
  \prod_{i=0}^{m}
  \varepsilon_{\mathrm{L}, M_i}(p_i)
  \Big \langle T
    \prod_{i=0}^{m}
    \underline{\Theta^{M_i}_{\vphantom{a_i}}}(p_i)
    \prod_{j=1}^{n} \genfield{j}^{\mathrm{phys}}
  \Big \rangle
  .
  \label{eq:pol-vec-equivalence-wulzer}
\end{align}
The r.h.s.\ is now manifestly free of mass-singular factors.
Because this statement is equivalent to
the exact version~\eqref{eq:multi-gbet-exact} of the \gls{gbet}, the
redefinition \eqref{eq:redef-long-polvec} is analogous to our procedure
from~\eqref{eq:VVV-LT-partial-amp-redef}
and~\eqref{eq:VVV-LL-partial-amp-redef}.


The new
propagators for gauge bosons require special care.
First, they are written as $5\times5$ matrices $G_{M N}$.
Then, for the off-shell particle that undergoes the splitting
an \emph{equivalent} propagator~$G^{\mathrm{eq}}_{M N}$ is
defined, which is manifestly free of mass-singular factors.
At tree level,
%
\begin{align}
  G^{\mathrm{eq},0}_{M N}
  (p,-p)
  =
  \frac{1}{p^2-M_V^2}
  \sum_{\polvb = \pm,\mathrm{L}}
  \varepsilon^\polvb_{M}(p)
  \overline \varepsilon^\polvb_{N}(p)
  +
  \frac{1}{p^2}
  \mathcal{P}_{\mathrm{S}, M}
  \mathcal{P}_{\mathrm{S}, M}
  ,
  \qquad
  \mathcal{P}
  _{\mathrm{S}, M}
  =
  (0,0,0,0,1)
  .
  \label{eq:equiv-prop}
\end{align}
Similar to the redefinition of polarization vectors, the
\gls{wi}~\eqref{eq:wulzer-ward-gengaugetheo} implies that computing the
diagrams~\eqref{eq:enhanced-diag-topologies} with equivalent and
standard propagators produces the same result.
Finally, the polarization vectors in~\eqref{eq:equiv-prop} are replaced
by counterparts that are evaluated on the on-shell-projected momentum
$p^\mu_{\mathrm{on}}$, which is obtained from $p^\mu$ via modification
of the component $p^0$.
This absorbs the second, scalar part of~\eqref{eq:equiv-prop} 
into the $\varepsilon^{\mathrm{L}}$ part of the polarization sum, up to
contributions that are irrelevant in the quasi-collinear limit.
This completes the decomposition of vector-boson propagators into outer
products of (redefined) polarization vectors.


Putting everything together, quasi-collinear factorization of
\eqref{eq:def-partialM} is expressed as
\begin{align}
  \MContrExplFields
    {\genfield{ij}}
    {\genfield{i}}
    {\genfield{j}}
    ^{\pol_i \pol_j}
  &
  \xsim{\lambda \to 0}
  \sum_{\pol_{ij}}
  \frac{1}{p_{ij}^2-m_{ij}^2}
  \,
  \mathcal{M}
    ^{(n)\, \pol_{ij}}
    _{\genfield{ij}X}
  (p_{ij}^\mathrm{on})
  \,
  \mathcal{M}
    ^{(\mathrm{split})\, \pol_{ij} \pol_{i} \pol_{j}}
    _{\genfield{ij}\genfield{i} \genfield{j}}
  (p_{ij},p_i,p_j)
  ,
  \label{eq:wulzer-fact-formula}
\end{align}
where the hard matrix elements
$\mathcal{M}^{(n)}$
are evaluated with on-shell-projected momentum and wave functions.
Here, standard polarization vectors may be used.
The splitting amplitudes $\mathcal{M}^{(\mathrm{split})}$ for the
processes $\genfield{ij}^\ast \to \genfield{i} \genfield{j}$ are
evaluated with the \emph{off-shell} momentum~$p_{ij}^\mu = p_i^\mu +
p_j^\mu$, but with on-shell wave functions for the intermediate
particles.
Here, modified longitudinal polarization
vectors~\eqref{eq:redef-long-polvec} need to be used.
Similar to our approach, contributions for intermediate would-be
Goldstone bosons are not explicitly considered, as they are
already fully taken into account in the definition of the equivalent
propagator.
However, the computation of~$\mathcal{M}^{(\mathrm{split})}$ \emph{does}
require the evaluation of would-be Goldstone-boson amplitudes, due
to~\eqref{eq:redef-long-polvec}.
To this end, the employed Feynman rules are parametrized in such a way
that the coupling relations from Section~\ref{sect:generic-gb-cpl-relat}
are manifest.


The splitting matrix elements are finally evaluated at leading order in
the quasi-collinear limit, employing explicit, component-wise
parametrizations of the splitting 
3-momenta (cf.~(5.16) of \cite{Cuomo:2019siu}),
\begin{align}
  \boldsymbol{p}_{ij} &= (0,0,|\boldsymbol{p}_{ij}|),
  \nonumber
  \\
  \boldsymbol{p}_i &= (|\boldsymbol k_\perp| \cos \phi,|\boldsymbol k_\perp| \sin \phi,(1-x^\prime)|\boldsymbol{p}_{ij}|),
  \nonumber
  \\
  \boldsymbol{p}_j &= (-|\boldsymbol k_\perp| \cos \phi,-|\boldsymbol k_\perp| \sin \phi,x^\prime|\boldsymbol{p}_{ij}|)
  ,
  \label{eq:wulzer-momentum-param-fs}
\end{align}
polarization vectors (with $s_\varepsilon = -1$), and spinors.
The results are reported in Appendix~B of~\cite{Cuomo:2019siu}.
These formulas, however, appear to contain some typographical errors.
First, the denominator in the expressions for $V_0^a \to V_{\pm 1}^b
V_{\mp 1}^c$ should be $m_a$ instead of $m_1$.
In the numerator, $m_c^3$ should be replaced by $m_c^2$.
Likewise, in the numerator for $V_0^a \to V_{\pm 1}^b V_{0}^c$, the
exponent of $m_a$ should be replaced by~$2$.

\paragraph{Compatibility of results.}
Matching the Sudakov 
decomposition~\eqref{eq:sudakov-param-fs} to the explicit
pa\-ra\-me\-tri\-za\-tion~\eqref{eq:wulzer-momentum-param-fs},
our quasi-collinear factorization formula~\eqref{eq:realme-sq-fact-nonsudakov}
agrees with the 
counterpart of factorization~\eqref{eq:wulzer-fact-formula} in~\cite{Cuomo:2019siu}
(up to a trivial factor $e^2$).
For this, we set
\begin{align}
  n^\mu &= (1,0,0,-1)
  \label{eq:wulzer-choice-n}
\end{align}
and get,
via $\bar p n = p_{ij} n$, that
\begin{align}
  z=(p_i n)/(\bar p n)  = (1-x^\prime)+\orderof{m^2}
  .
  \label{eq:z-from-wulzer-param}
\end{align}
After identifying~$k_\perp^2=-|\boldsymbol k_\perp|^2$
from~\eqref{eq:prop-denom-sudakov-massive}, the coefficients $A_i$,
$A_j$ are expressed completely in terms of the parameters
from~\eqref{eq:wulzer-momentum-param-fs}, giving us an explicit expression for
\begin{align}
  \bar p^\mu =
  p_{ij}^\mu
  + (A_i + A_j) n^\mu
  .
\end{align}
Furthermore, with~$k_\perp^\mu = p_i^\mu - z \bar p^\mu + A_i n^\mu$, we obtain
\begin{align}
  k_\perp^\mu &= (0,|\boldsymbol k_\perp| \cos \phi,|\boldsymbol k_\perp| \sin \phi,0)
  \label{eq:sudkt-from-wulzer-param}
  .
\end{align}
Fixing $s_\pm = \mp 1$ in~\eqref{eq:frame-n-parr-pb},
the specific choice of $n^\mu$ corresponds to
$\bar{\varepsilon}_+^\mu = - (0,1,\imagi,0)/\sqrt{2}$
and
$\bar{\varepsilon}_-^\mu = (0,1,-\imagi,0)/\sqrt{2}$.
Thus,
the parameters~$|\boldsymbol k _\perp|$ and phases~$\expe^{\pm \imagi
\phi}$ can be computed via the scalar products between~$k_\perp^\mu$
and~$\bar{\varepsilon}_\pm^\mu$,
\begin{align}
  k_\perp
  \bar{\varepsilon}_\pm
  =
  \pm
  \frac{
    |\boldsymbol k _\perp|
  }{\sqrt{2}}
  \expe^{\pm \imagi \phi}
  ,\qquad
  |\boldsymbol k _\perp|^2
  =
  - k_\perp^2
  =
  - 2
  \big(
    k_\perp
    \bar{\varepsilon}_+
  \big)
  \big(
    k_\perp
    \bar{\varepsilon}_-
  \big)
  .
  \label{eq:wulzer-phases-via-sp}
\end{align}


The relation between our splitting functions from
Sections~\ref{sect:expl-splfuncs-FS-bosonic} and \ref{sect:expl-splfuncs-FS-fermionic}
and the splitting matrix elements
from~\cite{Cuomo:2019siu} is now found by
plugging~\eqref{eq:wulzer-fact-formula}
into~\eqref{eq:decomp-sq-realme-by-mother} and comparing
with~\eqref{eq:realme-sq-fact-nonsudakov}.
This yields
\begin{align}
  &
  e^2
  \left[
    \mathcal{P}
        {}^{
          \vbfield{ij^{\vphantom{\prime}}}
        }_{\genfield{i}}
        {}^{
          \vbfield{ij^\prime}
        }_{\genfield{j}}
      \big(z,k_\perp^\mu,\bar p^\mu,
      \pol_i,\pol_j\big)
  \right]^{\mu \nu}
  \nonumber\\*
  &=
  \sum_{\polvb_{ij}, \polvb_{ij}^\prime \in \{\pm,\mathrm{L}\}}
  \varepsilon^{\ast \mu}_{\polvb_{ij}}
  (p_{ij}^\mathrm{on})
  \varepsilon^{\nu}_{\polvb_{ij}^\prime}
  (p_{ij}^{\mathrm{on}\prime})
  \,
  \mathcal{M}
    ^{(\mathrm{split})\, \polvb_{ij} \pol_{i} \pol_{j}}
    _{\vbfield{ij} \genfield{i} \genfield{j}}
  \Big[
  \mathcal{M}
    ^{(\mathrm{split})\, \polvb_{ij}^\prime \pol_{i} \pol_{j}}
    _{\vbfield{ij^\prime} \genfield{i} \genfield{j}}
  \Big]^\ast
  \label{eq:wulzer-contact-main-formula-vec}
\end{align}
for two intermediate vector bosons
and
\begin{align}
  &
  e^2 \,
  \mathcal{P}
      {}^{
        \fermfield{ij^{\vphantom{\prime}}}
      }_{\genfield{i}}
      {}^{
        \fermfield{ij^\prime}
      }_{\genfield{j}}
    \big(z,k_\perp^\mu,\bar p^\mu,
    \pol_i,\pol_j\big)
  \nonumber\\*
  &=
  \sum_{\polferm_{ij}, \polferm_{ij}^\prime \in \{\pm\}}
  u_{\polferm_{ij}^\prime}
  (p_{ij}^{\mathrm{on}\prime})
  \bar u_{\polferm_{ij}}
  (p_{ij}^\mathrm{on})
  \,
  \mathcal{M}
    ^{(\mathrm{split})\, \polferm_{ij} \pol_{i} \pol_{j}}
    _{\fermfield{ij} \genfield{i} \genfield{j}}
  \Big[
  \mathcal{M}
    ^{(\mathrm{split})\, \polferm_{ij}^\prime \pol_{i} \pol_{j}}
    _{\fermfield{ij^\prime} \genfield{i} \genfield{j}}
  \Big]^\ast
  \label{eq:wulzer-contact-main-formula-ferm}
\end{align}
for two intermediate fermions.
For antifermions, we need to exchange $u \to \bar v$ and $\bar u \to v$.
The situations for intermediate scalars and vector--scalar interference
contributions are obvious.
Here, $p_{ij}^{\mathrm{on} (\prime)}$ represents the on-shell projection
of $p_{ij}$ w.r.t.\ the mass $m_{ij}^{(\prime)}$.
In~\eqref{eq:wulzer-contact-main-formula-vec}, we use \emph{standard}
polarization vectors, because they belong to the hard process.


To align the parametrizations of the r.h.s.\ with the l.h.s., we
express~$x^\prime$ through~$z$, and rewrite $|\boldsymbol k_\perp|$ and
$\expe^{\pm \imagi \phi}$ in terms of the scalar products~$k_\perp
\bar{\varepsilon}_\pm$, as discussed above.
Because the mass parameters~$m_a,m_b,m_c$ in the splitting amplitudes
of~\cite{Cuomo:2019siu} are not consistently assigned to the labels of
the momenta in~\eqref{eq:wulzer-momentum-param-fs}, they must be
appropriately re-mapped to~$m_{ij}^{(\prime)},m_i,m_j$.
Moreover, the outer products of polarization vectors from the r.h.s.\
need to be expressed in the basis of the tensor structures used in
Sections~\ref{sect:expl-splfuncs-FS-bosonic} and~\ref{sect:expl-splfuncs-FS-fermionic}.
We start with the case of two intermediate vector bosons.
Vector--scalar interference contributions are treated similarly.
First, we employ $\varepsilon_\mathrm{L}^\mu \big(p_{ij}^{\mathrm{on}
(\prime)}\big) = \bar p^\mu / m_{ij}^{(\prime)} + \orderof{m}$.
Second, the choice of $n^\mu$ implies that the arguments of the
transverse polarization vectors $\varepsilon_\pm^\nu
\big(p_{ij}^{\mathrm{on} (\prime)}\big)$ can be replaced by~$\bar p$.
This allows us to use~\eqref{eq:kt-from-eps-via-sp}, with
$s_\varepsilon=-1$, to reduce
$\bar p^\mu k_\perp^\nu$,
$\bar p^\mu \bar{\varepsilon}^\nu_+$,
$\bar p^\mu \bar{\varepsilon}^\nu_-$
into at most two terms.
Third, the four structures
$\bar{\varepsilon}^\mu_\pm\bar{\varepsilon}^\nu_\pm$,
$\bar{\varepsilon}^\mu_\pm\bar{\varepsilon}^\nu_\mp$
are addressed by inserting
\begin{align}
  \bar{\varepsilon}^\mu_-
  \bar{\varepsilon}^\nu_-
  =
  &{}-
  \left(
  \frac{
      k_\perp
      \bar{\varepsilon}_-
  }{
      k_\perp
      \bar{\varepsilon}_+
  }
  \right)^2
  \bar{\varepsilon}^\mu_+
  \bar{\varepsilon}^\nu_+
  -
  \frac{
      k_\perp
      \bar{\varepsilon}_-
  }{
      k_\perp
      \bar{\varepsilon}_+
  }
  \left(
    \bar{\varepsilon}^\mu_+
    \bar{\varepsilon}^\nu_-
    +
    \bar{\varepsilon}^\mu_+
    \bar{\varepsilon}^\nu_-
  \right)
  +
  \frac{
    k_\perp^\mu
    k_\perp^\nu
  }{
    \big(
      k_\perp
      \bar{\varepsilon}_+
    \big)^2
  }
  ,
  \label{eq:wulzer-epseps-via-ktkt}
\end{align}
which simply follows from \eqref{eq:kt-from-eps-via-sp} as well.
We observe that this also eliminates
$\bar{\varepsilon}^\mu_+ \bar{\varepsilon}^\nu_+$, leaving us with
$\bar{\varepsilon}^\mu_\pm\bar{\varepsilon}^\nu_\mp$
and
$k_\perp^\mu k_\perp^\nu$.
Last, via
\begin{align}
  a\,
  \bar{\varepsilon}^\mu_+\bar{\varepsilon}^\nu_-
  +
  b\,
  \bar{\varepsilon}^\mu_-\bar{\varepsilon}^\nu_+
  =
  \frac{a+b}{2}
  \big(
    g^{\mu \nu}
    +
    \orderof{m}
  \big)
  +
  \frac{a-b}{2}
  \big(
    \bar{\varepsilon}^\mu_+\bar{\varepsilon}^\nu_-
    -
    \bar{\varepsilon}^\mu_-\bar{\varepsilon}^\nu_+
  \big)
  ,
\end{align}
we identify the metric tensor~$g^{\mu\nu}$.


For intermediate fermions or antifermions, on the other hand, we
use~\eqref{eq:helproj-spin-sum-leadingmass} to write
\begin{align}
  u_\pm(\bar p)  
  \bar u_\pm(\bar p)
  &=
  \slashed{\bar{p}}
  \omega_\mp
  + \orderof{m}
  ,&
  v_\pm(\bar p)  
  \bar v_\pm(\bar p)
  &=
  \slashed{\bar{p}}
  \omega_\pm
  + \orderof{m}
  ,
  \label{eq:id-spin-projcompl-relat}
\end{align}
and identify, similar to~\eqref{eq:ktslash-terms-amplitudes-antiferm}
and~\eqref{eq:ktslash-terms-amplitudes-ferm}, the corresponding
structures $\propto \slashed{k}_\perp$,
\begin{align}
    (k_\perp \bar{\varepsilon}_\mp) \,
  u_\pm(\bar p)  
  \bar u_\mp(\bar p)
  &=
  \mp
  \frac{1}{\sqrt{2}}
    \slashed{k}_\perp
    \slashed{\bar{p}}
\omega_\pm
  + \orderof{m}
  ,&
(k_\perp \bar{\varepsilon}_\pm) \,
  v_\pm(\bar p)  
  \bar v_\mp(\bar p)
  &=
  \pm
  \frac{1}{\sqrt{2}}
    \slashed{k}_\perp
    \slashed{\bar{p}}
\omega_\mp
  + \orderof{m}
  .
  \label{eq:id-ktslash-terms}
\end{align}

\subsection{Comparison to Chen, Han, and Tweedie~\texorpdfstring{\cite{Chen:2016wkt}}{}}
\label{sect:chen-han-tweedie}
\paragraph{Details of the approach.}
In~\cite{Chen:2016wkt}, mass-singular factors from longitudinal
polarization vectors
are circumvented by the construction of an
appropriate axial gauge, dubbed \gls{geg}.
It is defined via the momentum-space gauge-fixing Lagrangian
\begin{align}
  \mathcal{L}_\mathrm{fix}^\mathrm{GEG}
  =
  - \frac{1}{2 \xi}
  \sum_{a}
  \left[n_{\mu}(p) V^\mu_a(p) \right]
  \left[n_{\nu}(p) V^\nu_a(-p) \right]
  ,
  \label{eq:chen-gauge-fixing}
\end{align}
in the limit $\xi \to 0$.
Here, the sum is understood to run over the vector-boson fields~$V_a$ in
the real, symmetric basis.
With the choices~$n^\mu(p) =
(1,-\mathrm{sign}(p^0) \hat{\boldsymbol{p}})$, $\bar{n}^\mu(p) =
\left(1, \mathrm{sign}(p^0) \hat{\boldsymbol{p}}\right)$
and the definitions
$\varepsilon_{n}^\mu \propto n^\mu$, $\varepsilon_{\bar n}^\mu \propto \bar n^\mu$,
the vector-boson fields are parametrized as
\begin{align}
  V^\mu(p)
  =
  V_+(p) \varepsilon_{+}^\mu(p)
  +
  V_-(p) \varepsilon_{-}^\mu(p)
  +
  V_n(p) \varepsilon_{n}^\mu(p)
  +
  V_{\bar n}(p) \varepsilon_{\bar n}^\mu(p)
  \label{eq:chen-VB-param}
  .
\end{align}
The mass-singular components of the longitudinal polarizations are tied
to the~$\bar n$ polarizations.
Sending $\xi \to 0$ endows $V_{\bar n}$ with an infinite mass, so that it
ceases to propagate.
Because the would-be Goldstone-boson fields~$G$ are not involved in the
gauge-fixing Lagrangian~\eqref{eq:chen-gauge-fixing}, the propagators
retain tree-level mixing,
\begin{alignat}{2}
  G^{V_\pm V_\pm}_{0} (p,-p)
  &=
  \frac{\imagi}{p^2-M_V^2}
  ,
  &\qquad
  G^{V_n V_n}_{0} (p,-p)
  &=
  \frac{\imagi}{p^2-M_V^2} \mathrm{sign}(p^2)
  ,
  \nonumber
  \\
  G^{\mathrm{G} \mathrm{G}}_{0} (p,-p)
  &=
  \frac{\imagi}{p^2-M_V^2}
  ,
  &
  G^{V_n G}_{0} (p,-p)
  &=
  \frac{\imagi}{p^2-M_V^2} \frac{M_V}{\sqrt{|p^2|}}
  .
  \label{eq:chen-prop-mixing}
\end{alignat}
These contributions share a common pole at~$p^2=M_V^2$.
To compute a matrix element with an external longitudinal vector boson~$V_\mathrm{L}$,
the corresponding matrix elements for~$V_n$ and~$G$ have to
be combined.
With the phase choices from~\cite{Chen:2016wkt}, this reads
\begin{align}
  \mathcal{M}
    _{V_\mathrm{L} X}
  =
  \mathcal{M}
    _{V_n X}
  +
  \mathcal{M}
    _{G X}
    ,
  \label{eq:chen-me-comp-long-pol}
\end{align}
where the r.h.s.\ is manifestly free of mass-singular contributions.


In diagrams like~\eqref{eq:enhanced-diag-topologies}, all propagator
terms from~\eqref{eq:chen-prop-mixing} need to be taken into account.
Adapting the arguments from~\cite{Chen:2016wkt} to our notation, this
means
\begin{align}
  \MContrExplFields
    {\vbfield{ij}^\mathrm{L}}
    {\genfield{i}}
    {\genfield{j}}
    ^{\pol_i \pol_j}
  &
  \xsim{\lambda \to 0}
  \,
  \smashoperator[r]{
    \sum_{\genfield{},\genfield{}^\prime \in \{\vbfield{ij}^n,\gbfield{ij}\}}
  }
  \,
  -\imagi
  G^{
    \genfield{}
    \genfield{}^\prime
  }_{0} (p_{ij},-p_{ij})
  \hat{\mathcal{M}}
    ^{(n)}
    _{\genfield{} X}
  (p_{ij})
  \,
  \hat{\mathcal{M}}
    ^{(\mathrm{split})\, \pol_{i} \pol_{j}}
    _{\genfield{}^\prime \genfield{i} \genfield{j}}
  (p_{ij},p_i,p_j)
  ,
  \label{eq:chen-fact-formula-raw}
\end{align}
where the subscript of $\mathcal{M}$ indicates that we omit the
contributions from transverse~$V$.
The momentum $p_{ij}^\mu$ is off shell, and the hats denote that
$\hat{\mathcal{M}}^{(n)}$ and $\hat{\mathcal{M}}^{(\mathrm{split})}$ are
evaluated with the corresponding off-shell polarization vectors.
We also absorbed the polarization label for $ij$ into~$\genfield{ij}$.
Finally, at leading mass order, replacing the off-shell polarization
vectors with on-shell counterparts cancels the factors
$M_V/\sqrt{|p^2|}$ from the mixed propagators
in~\eqref{eq:chen-prop-mixing}, leading to
\begin{align}
  \MContrExplFields
    {\vbfield{ij}^\mathrm{L}}
    {\genfield{i}}
    {\genfield{j}}
    ^{\pol_i \pol_j}
  \xsim{\lambda \to 0}
  \frac{1}{p_{ij}^2-m_{ij}^2}
  &\left[
    \mathcal{M}
      ^{(n)}
      _{\vbfield{ij}X}
    (p_{ij}^\mathrm{on})
    +
    \mathcal{M}
      ^{(n)}
      _{\gbfield{ij}X}
    (p_{ij}^\mathrm{on})
  \right]
  \nonumber\\\cdot&
  \left[
    \mathcal{M}
      ^{(\mathrm{split})\, \pol_{i} \pol_{j}}
      _{\vbfield{ij}\genfield{i} \genfield{j}}
    (p_{ij},p_i,p_j)
    +
    \mathcal{M}
      ^{(\mathrm{split})\, \pol_{i} \pol_{j}}
      _{\gbfield{ij}\genfield{i} \genfield{j}}
    (p_{ij},p_i,p_j)
  \right]
  .
  \label{eq:chen-fact-formula}
\end{align}
Together with the transverse contributions, this provides the
factorization of~\eqref{eq:def-partialM} in terms of hard matrix elements
and splitting amplitudes, similar to~\eqref{eq:wulzer-fact-formula}.
The latter are evaluated at leading order in the quasi-collinear limit,
using explicit parametrizations of momenta and wave functions.
The Feynman rules for 3-point vertices in \gls{geg} essentially
agree with those from~$R_\xi$ gauges, with the exception
that~\eqref{eq:chen-me-comp-long-pol} is used whenever external
longitudinal polarization vectors occur.
Thereby, the practical computation of splitting matrix elements
from~\cite{Cuomo:2019siu} and~\cite{Chen:2016wkt} essentially agrees.
However, the results of~\cite{Chen:2016wkt} are reported in the form
of splitting functions, which correspond to the product of two splitting
matrix elements,
\begin{align}
  \mathcal{M}
    ^{(\mathrm{split})\, \pol_{ij} \pol_{i} \pol_{j}}
    _{\genfield{ij} \genfield{i} \genfield{j}}
  \Big[
  \mathcal{M}
    ^{(\mathrm{split})\,\pol_{ij}^\prime \pol_{i} \pol_{j}}
    _{\genfield{ij^\prime} \genfield{i} \genfield{j}}
  \Big]^\ast
  .
  \label{eq:chen-split-func}
\end{align}
Flavor and spin interference effects regarding the mother particle are
not always considered, with the exception of the pairs
$\gamma/\mathrm{Z}_\mathrm{T}$ and $\mathrm{H}/\mathrm{Z}_\mathrm{L}$.


In \cite{Chen:2016wkt}
the splitting processes are organized into two distinct classes,
labeled \emph{conventional} and \emph{ultra-collinear}.
Conventional splitting functions are non-zero in the unbroken phase of
the \gls{ewsm}, with $v=0$ and massless particles.
They are provided in Tables~1--3 of~\cite{Chen:2016wkt} and do not
receive corrections in the broken phase of the \gls{ewsm}.
Conventional splittings that involve scalars directly correspond to
counterparts with longitudinal vector bosons.
For non-conventional processes, ultra-collinear splitting
functions emerge in the broken phase of the \gls{ewsm}.
They are proportional to~$v^2$ and quoted in Tables~4--6
of~\cite{Chen:2016wkt}.\footnotemark\
\footnotetext{%
  Ultra-collinear splittings are identified systematically by
  considering all effects of~$v \neq 0$.
  First, fermion spinors of helicity~$\pm\frac{1}{2}$ receive
  left/right-chiral components proportional to the fermion mass.
  Second, the on-shell polarization vector~$\varepsilon_{n}^\mu$ is
  proportional to the vector-boson mass.
  Third, replacing a Higgs field in quartic Lagrangian terms by~$v$
  yields trilinear couplings that have no correspondence in the unbroken
  phase.
  For dimensional reasons, the factor $v^2$ is accompanied by an
  uncanceled propagator factor $1/\big(p_{ij}^2 - m_{ij}^2\big)$.
  This
  provides the ultra-collinear splitting functions with a more
  pronounced scaling in the collinear regime, which is the reason for their name.
}%
The combination of conventional and ultra-collinear results
provides the complete set of broken-phase splitting functions.

\paragraph{Compatibility of results.}
Because not all interference contributions~\eqref{eq:chen-split-func}
are quoted, a direct comparison to our results, e.g.\
via~\eqref{eq:wulzer-contact-main-formula-vec}, is not feasible.
The splitting functions from~\cite{Chen:2016wkt} can, however, be
matched to products of the splitting matrix elements from~\cite{Cuomo:2019siu}.
We performed this check explicitly for all polarization configurations
in~$V^\ast \to V V$ and~$V^\ast \to f \bar f$,
finding agreement.

\subsection{Comparison to Kleiss, Verheyen, Brooks, and Skands~\texorpdfstring{\cite{Kleiss:2020rcg,Brooks:2021kji}}{}}
\label{sect:kleiss-verheyen}
\paragraph{Details of the approach.}
The computation of splitting amplitudes is performed using the
spinor--helicity method of~\cite{Kleiss:1985yh}.
For this, massive spinors~$u_\polferm(p)$, $v_\polferm(p)$ and transverse
polarization vectors~$\varepsilon_\pm^\mu(p)$ are expressed in terms of
massless spinors~$u_\polferm(n)$
with light-like reference vectors~$n^\mu(p)=(1,-\hat{\boldsymbol{p}})$.
This expresses the splitting matrix elements in terms of ordinary scalar
products between four-vectors and structures like
\begin{align}
  S_\polferm(n_a,p_i,p_j,\dots,n_b)
  =
  \bar{u}_\polferm(n_a)
  \slashed{p}_i
  \slashed{p}_j
  \dots
  u_{\pm\polferm}(n_b),
  \label{eq:kleiss-general-spinor-product}
\end{align}
where $n_a$ and $n_b$ are light-like.
The mass-singular part $p^\mu/M_V$ of the longitudinal polarization
vector~$\varepsilon_{\mathrm{L}}^\mu(p)$ is treated similarly to our
procedure described in 
Appendix~\ref{sect:alt-approach}.
Specifically, this term generates contributions proportional to the
propagator denominator of the diagrams in~\eqref{eq:def-partialM}.
These components are manually removed from the splitting amplitudes.
In practice, this is realized via the replacements
\begin{align}
  2 p_i p_j &\to m_{ij}^2 - m_i^2 - m_j^2,
  \nonumber
  \\
  2 p_{ij} p_i &\to m_{ij}^2 + m_i^2 - m_j^2, 
  \nonumber
  \\
  2 p_{ij} p_j &\to m_{ij}^2 - m_i^2 + m_j^2.
\end{align}
For example, in \gls{fs}~$f^\ast \to f V$ splittings, this proceeds via
\begin{align}
  \frac{\slashed{p}_i\slashed{p}_j\slashed{p}_{ij}}{m_j}
  =
  \frac{1}{m_j}
  \left(
    p_{ij}^2 \slashed{p}_i -
    m_i^2 \slashed{p}_{ij}
  \right)
  \to
  \frac{1}{m_j}
  \left(
    m_{ij}^2 \slashed{p}_i -
    m_i^2 \slashed{p}_{ij}
  \right)
  .
  \label{eq:kleiss-verheyen-removal-strat}
\end{align}
Taking all reference vectors collinear, only one non-trivial spinor
product remains.
At leading order in the quasi-collinear limit, it becomes, up to a phase,
\begin{align}
  S_{-\polferm} (n,p_i,p_j,n)
  \propto
  2 (p_{ij} n)
  \sqrt{z (1-z)}
  \sqrt{\tilde Q^2}
  ,
  \qquad
  \tilde Q^2
  =
  (p_i+p_j)^2 - \frac{m_j^2}{1-z} - \frac{m_i^2}{z}
  ,
  \label{eq:kleiss-coll-limit-spinorprod}
\end{align}
where $z$ is defined as the energy fraction of $p_{ij}^\mu$ carried by
$p_i^\mu$.
The results for the splitting amplitudes are quoted in Table~1--7
of~\cite{Kleiss:2020rcg}, where the phase factors
from~\eqref{eq:kleiss-coll-limit-spinorprod} are neglected.
In contrast to~\cite{Chen:2016wkt,Cuomo:2019siu,Masouminia:2021kne}, no
component-wise parametrizations of momenta, spinors, and polarization
vectors are used.
On the other hand, no arguments for the decomposition of internal
propagators into products of on-shell wave functions are
presented.
The results from~\cite{Kleiss:2020rcg} have been restated
in~\cite{Brooks:2021kji} in squared form, using antenna
collinear momentum fractions~$x_i$ and~$x_j$.

\paragraph{Compatibility of results.}
The splitting matrix elements can directly be compared to the
corresponding expressions from~\cite{Cuomo:2019siu}.
We performed this check explicitly for all polarization configurations
in~$V^\ast \to V V$ and~$V^\ast \to f \bar f$.
Up to the neglected phase factors, the results agree, apart from an
apparent typographical mistake in the fifth row of Table~2.
Here, the sign in front of~$m_j/m_{ij}$ should be~$-$ instead of~$+$.
In~\cite{Kleiss:2020rcg}, agreement with the results
from~\cite{Chen:2016wkt} is reported.
Due to the missing phase factors, however, the results
from~\cite{Kleiss:2020rcg} cannot be used directly to reconstruct our
factorization formulas from this paper.

\subsection{Comparison to Masouminia and Richardson (Herwig 7)~\texorpdfstring{\cite{Masouminia:2021kne}}{}}
\label{sect:herwig}
\paragraph{Details of the approach.}
The splitting amplitudes from Tables~1--4 of~\cite{Masouminia:2021kne}
are evaluated in explicit parametrizations
of momenta, spinors, and polarization vectors, at leading order in the
quasi-collinear limit.
The results are also combined into splitting functions, but
spin interference contributions are neglected.
The factorization of~\eqref{eq:def-partialM} into hard amplitudes and
splitting matrix elements is not explicitly proven.


The mass-singular term~$p^\mu/m$ of the longitudinal polarization
vector~$\varepsilon_\mathrm{L}^\mu(p)$ is treated by simply subtracting
it.
This produces redefined polarization
vectors~$\varepsilon_{\mathrm{L}^\ast}^\mu(p)$.
Longitudinal splitting matrix elements are then computed
with~$\varepsilon^\mu_{\mathrm{L}^\ast}(p)$ instead
of~$\varepsilon^\mu_{\mathrm{L}}(p)$.
Since a similar technique has been used in~\cite{Dawson:1984gx} for the
derivation of the \gls{ewa}, this procedure is coined as ``Dawson's
approach''.
For the specific situation of the \gls{ewa}, i.e.\ for \gls{is} $f \to f
V^\ast$ splittings with massless fermions, the subtracted terms actually
vanish in the splitting amplitude, due to the \gls{wi} and zero 
Yukawa coupling.
For general splittings, however, this is \emph{not} the case.\footnotemark\
\footnotetext{%
  Already for \gls{is} $f_a \to f_i V^\ast_{ai}$ splittings with massive fermions
  and \gls{fs}~$f^\ast_{ij} \to f_i V_j$ splittings with massless fermions, this
  argument breaks down.
  In the first case, replacing the gauge boson with the associated
  would-be Goldstone boson yields a splitting amplitude that does not
  vanish.
  In the second case, the incoming fermion is off-shell, so that the 
  \gls{wi}
  for the gauge boson with index $j$
  cannot be applied.
}%
In contrast,~\cite{Cuomo:2019siu} replaces these contributions with
splitting amplitudes for would-be Goldstone bosons.\footnotemark\
\footnotetext{%
  The arguments leading to the procedure of~\cite{Cuomo:2019siu} involve
  \glspl{wi} for the full $(n+1)$ particle matrix element, not just the
  splitting amplitude, as suggested by ``Dawson's approach''.
}%
Thus, generally, the results from~\cite{Masouminia:2021kne} for
longitudinal vector bosons cannot be expected to consistently capture
all terms that are enhanced in the quasi-collinear limit.

\paragraph{Compatibility of results.}
As detailed above, the splitting amplitudes
from~\cite{Cuomo:2019siu} and~\cite{Masouminia:2021kne} can only be
compared with each other for splittings without longitudinal vector
bosons.
We observe full agreement for transversely polarized $V^\ast \to VV$
splittings.
However, for splittings with at least one longitudinal vector boson, the
results differ.
In particular, all splitting functions from~\cite{Masouminia:2021kne}
with two or more longitudinal vector bosons vanish, in conflict with
both our results and those from~\cite{Cuomo:2019siu}.

\section{Summary}
\label{sect:summary}
In this paper, we 
have derived \gls{lo} \gls{fs} and \gls{is} splitting
functions for generic 
spontaneously broken gauge theories with scalars, fermions,
and vector bosons.
These results constitute important ingredients for the construction of
\gls{ew} parton-shower simulations and subtraction algorithms
for soft singularities and collinear enhancements.
As such, they are of phenomenological interest 
for LHC physics in the TeV energy range and of great importance 
for physics at
future colliders with center-of-mass energies $\sqrt{s} \approx 100\,
\TeV$.
The main complication in the derivation of our results was
the presence of mass-singular terms in longitudinal polarization vectors,
severely complicating the process of quasi-collinear power counting.
To address this delicate issue, we have performed our computations with
two different strategies, employing gauge invariance and \glspl{wi} as
guiding principle.
We obtained full agreement between the results obtained with these methods.


We have discussed various special cases such as the specialization to the \gls{ewsm}
and massless limits,
including the reproduction of literature results in massless and massive
\gls{qed} and \gls{qcd}.
Moreover, we have thoroughly compared our strategy and results with existing literature
on \gls{ew} splitting functions.
While we find agreement with corresponding results from other approaches,
we also point out differences to some earlier publications.
We stress, in particular, that our derivation is unique in that it uses a
different general methodology and 
significantly reduces the number of underlying
assumptions.
In detail, we do not use any explicit, component-wise parametrization of
momenta or wave functions and also do not distinguish a particular frame
of reference.
Furthermore, we make no assumptions on the individual spin reference
axes that define the polarizations of the external particles.
Finally, our factorization formulas are expressed directly in terms of
squared matrix elements, which facilitates the application to
both parton-shower algorithms and
subtraction formalisms for soft singularities and collinear enhancements.
Our splitting functions fulfill the symmetry relations implied by those
of the corresponding real-emission matrix elements.
In particular, we confirmed that the \gls{fs} and \gls{is} splitting
functions are related by crossing relations. 


As an additional step, we discussed the procedure of taking the
azimuthal average of our splitting functions.
Where possible, we took special care that the final formulas can be
evaluated using squared matrix elements that are independent of the
considered frame of reference.
Apart from phenomenological motivations, this constitutes the first step
towards the computation of the full integral over the emission phase space.
This is relevant not only for the construction of subtraction
algorithms, but also for the comparison to and combination with the
virtual factorization formulas
from~\cite{denner2001oneloop1,denner2001oneloop2}.
In particular, this may be used for the assessment of cancellations
between the leading real and virtual corrections for specific
observables.
Both of these aspects are, however, left for future work.

  
\appendix

\section{The Electroweak Standard Model}
\label{sect:ewsm}
In this appendix we describe the embedding of the EWSM into the general
framework of our description of generic spontaneously broken gauge theories
given in Section~\ref{sect:generic-ym-theory}, which builds on the formulation
of \cite{denner2001oneloop1,denner2001oneloop2}.
In particular, we specialize all generic couplings to their specific
choices in the EWSM, based on the conventions of 
\cite{Bohm:1986rj,Denner:1991kt,bohm2001gauge,Denner:2019vbn}.

\subsection{Gauge structure}
\label{sect:ewsm-multiplets}
The gauge symmetry of the \gls{ewsm} is provided by the
group~$\mathrm{SU}(2)_\mathrm{w} \times \mathrm{U}(1)_Y$.
Left-handed neutrino fields~$\nu^\mathrm{L}_\ell$ and lepton
fields~$\ell^\mathrm{L}$ are organized into left-handed lepton
doublets~$L^\mathrm{L}=(\nu^\mathrm{L}_\ell,\ell^\mathrm{L})^\mathrm{T}$,
whereas left-handed up-type quark fields~$u^\mathrm{L}$ and down-type
quark fields~$d^\mathrm{L}$ are organized into left-handed quark
doublets~$Q^\mathrm{L}=(u^\mathrm{L},d^\mathrm{L})^\mathrm{T}$.
Generically, they are denoted as left-handed fermion
doublets~$\varphi=f^\mathrm{L}$ that transform in the fundamental
representation of~$\mathrm{SU}(2)_\mathrm{w}$.
All fermion fields occur in three generations, which are not
mixed by gauge transformations.
The scalar Higgs doublet~$\varphi=\cplxhiggsmult$ of the \gls{ewsm},
\begin{align}
  \cplxhiggsmult(x)
  =
  \left(
    \begin{matrix}
      \cplxhiggscpt^+(x)
      \\
      \cplxhiggscpt^0_{\phantom{+}}(x)
    \end{matrix}
  \right)
  ,
  \label{eq:higgs-doublet}
\end{align}
behaves in the same way under $\mathrm{SU}(2)_\mathrm{w}$ transformations.
We also write $\cplxhiggscpt^- = \cplxhiggscpt^{+\ast}$.
In the convention of
\citep{denner2001oneloop1,denner2001oneloop2,bohm2001gauge,Denner:2019vbn},
the generators of the fundamental representation
of~$\mathrm{SU}(2)_\mathrm{w}$ are defined, in the symmetric basis, as
\begin{align}
  I^{W_a}_{\varphi_k \varphi_l}
  =\frac{g_2}{e} T^a_{k l}
  ,
  \qquad
  T^a_{k l}
  = \frac{\tau^a_{kl}}{2}
  .
  \label{eq:cpl-mtr-Wfund}
\end{align}
Here,~$W_a$ denote the 
(real) gauge fields of~$\mathrm{SU}(2)_\mathrm{w}$,
$g_2$ is the coupling strength of $\mathrm{SU}(2)_\mathrm{w}$ gauge
interactions, and~$\tau^a$ represent the Pauli
matrices
\begin{align}
  \tau^1
  =
  \left(
  \begin{matrix}
    0&1\\
    1&0
  \end{matrix}
  \right)
  ,\qquad
  \tau^2
  =
  \left(
  \begin{matrix}
    0&-\imagi\\
    \imagi&0
  \end{matrix}
  \right)
  ,\qquad
  \tau^3
  =
  \left(
  \begin{matrix}
    1&0\\
    0&-1
  \end{matrix}
  \right)
  .
  \label{eq:pauli-matrices}
\end{align}
Right-handed leptons~$\ell^\mathrm{R}$, up-type quarks~$u^\mathrm{R}$,
and down-type quarks~$d^\mathrm{R}$ are summarized as right-handed
fermion singlets~$\varphi=f^\mathrm{R}$ and transform in the trivial
representation of~$\mathrm{SU}(2)_\mathrm{w}$.
The~$\mathrm{SU}(2)_\mathrm{w}$ doublets and singlet fields transform in
one-dimensional representations of~$\mathrm{U}(1)_Y$ with
\begin{align}
  I^{B}_{\varphi_k \varphi_l}
  =
  -\frac{g_1}{e} \frac{Y_{\varphi}}{2} \delta_{kl} 
  ,
  \label{eq:cpl-mtr-Bfund}
\end{align}
which defines couplings to the gauge field~$B$ with strength~$g_1$.
Here,~$Y_{\varphi}$ is the weak hypercharge of~$\varphi$.
As detailed below, it fixes the electric charges of the individual
components~$\varphi_l$.
To normalize $g_1$, we set~$Y_\cplxhiggsmult=1$, which gives the Gell-Mann--Nishijima
relation the form quoted in~\eqref{eq:gell-mann-nishijima}.


Since the adjoint representation of an abelian Lie group is trivial, the
gauge field~$B$ is invariant under $\mathrm{SU}(2)_\mathrm{w} \times
\mathrm{U}(1)_Y$ transformations.
In the symmetric field basis, the adjoint~$\mathrm{SU}(2)_\mathrm{w}$
generators are obtained from the commutation relation of fundamental
generators
\begin{align}
  \left[ I^{W_a}, I^{W_b} \right]_{\varphi_k \varphi_l}
  =
  \imagi \frac{g_2}{e} \epsilon^{abc} I^{W_c}_{\varphi_k \varphi_l}
  .
\end{align}
The fields~$W_a$ thus transform with
\begin{align}
  I^{W_a}_{W_b W_c}
  = - \imagi \frac{g_2}{e} \epsilon^{a b c}
  .
  \label{eq:cpl-mtr-VVV}
\end{align}

\subsection{Electroweak symmetry breaking}
\label{sect:ewsb}
The EWSM Lagrangian contains the Higgs potential
\begin{align}
  V(\cplxhiggsmult)
  =
  \frac{\lambda}{4} \big( \cplxhiggsmult^\dagger \cplxhiggsmult \big)^2
  -
  \mu^2 \cplxhiggsmult^\dagger \cplxhiggsmult,
  \qquad
  \mu^2,\lambda > 0,
  \label{eq:higgs-pot}
\end{align}
with a degenerate minimum for
$\cplxhiggsmult^\dagger \cplxhiggsmult = v^2/2 = 2 \mu^2 / \lambda$.
The latter is responsible for the non-zero \gls{vev} of \cplxhiggsmult\
which can, after a suitable gauge transformation, be written as
$\langle \cplxhiggsmult \rangle=(0,v/\sqrt{2})^\mathrm{T}$ with 
$v>0$.
Applying the generators~\eqref{eq:cpl-mtr-Wfund}
and~\eqref{eq:cpl-mtr-Bfund} to~$\langle \cplxhiggsmult \rangle$, we
obtain 
\begin{align}
  %
  %
  I^{W_1}_{\cplxhiggscpt_{i} \cplxhiggscpt_{j}}
  \langle \cplxhiggscpt_{j} \rangle
  &=
  \frac{g_2}{e}
  \frac{v}{2 \sqrt{2}}
  \left(
    \begin{matrix}
      1 \\ 0
    \end{matrix}
  \right)_i
  ,
  &
  I^{W_2}_{\cplxhiggscpt_{i} \cplxhiggscpt_{j}}
  \langle \cplxhiggscpt_{j} \rangle
  &=
  - \imagi
  \frac{g_2}{e}
  \frac{v}{2 \sqrt{2}}
  \left(
    \begin{matrix}
      1 \\ 0
    \end{matrix}
  \right)_i
  ,
  &
  I^{W_3}_{\cplxhiggscpt_{i} \cplxhiggscpt_{j}}
  \langle \cplxhiggscpt_{j} \rangle
  &=
  -
  \frac{g_2}{e}
  \frac{v}{2 \sqrt{2}}
  \left(
    \begin{matrix}
      0 \\ 1
    \end{matrix}
  \right)_i
  ,
  \nonumber\\
  %
  %
  I^{W_1}_{\cplxhiggscpt^\ast_{i} \cplxhiggscpt^\ast_{j}}
  \langle \cplxhiggscpt^\ast_{j} \rangle
  &=
  -
  \frac{g_2}{e}
  \frac{v}{2 \sqrt{2}}
  \left(
    \begin{matrix}
      1 \\ 0
    \end{matrix}
  \right)_i
  ,
  &
  I^{W_2}_{\cplxhiggscpt^\ast_{i} \cplxhiggscpt^\ast_{j}}
  \langle \cplxhiggscpt^\ast_{j} \rangle
  &=
  - \imagi
  \frac{g_2}{e}
  \frac{v}{2 \sqrt{2}}
  \left(
    \begin{matrix}
      1 \\ 0
    \end{matrix}
  \right)_i
  ,
  &
  I^{W_3}_{\cplxhiggscpt^\ast_{i} \cplxhiggscpt^\ast_{j}}
  \langle \cplxhiggscpt^\ast_{j} \rangle
  &=
  \frac{g_2}{e}
  \frac{v}{2 \sqrt{2}}
  \left(
    \begin{matrix}
      0 \\ 1
    \end{matrix}
  \right)_i
  ,
  \nonumber\\
  %
  %
  I^{B}_{\cplxhiggscpt_{i} \cplxhiggscpt_{j}}
  \langle \cplxhiggscpt_{j} \rangle
  &=
  -
  \frac{g_1}{e}
  \frac{v}{2 \sqrt{2}}
  \left(
    \begin{matrix}
      0 \\ 1
    \end{matrix}
  \right)_i
  ,
  &
  I^{B}_{\cplxhiggscpt^\ast_{i} \cplxhiggscpt^\ast_{j}}
  \langle \cplxhiggscpt^\ast_{j} \rangle
  &=
  \frac{g_1}{e}
  \frac{v}{2 \sqrt{2}}
  \left(
    \begin{matrix}
      0 \\ 1
    \end{matrix}
  \right)_i
  .
  \label{eq:sm-goldstone-dir-real}
\end{align}
Via a rotation of the generator basis
\begin{align}
  \left(
    \begin{matrix} I^Z \\ I^A \end{matrix}
  \right)
  =
  \left(
    \begin{matrix} c_\mathrm{w} & s_\mathrm{w} \\ -s_\mathrm{w} & c_\mathrm{w} \end{matrix}
  \right)
  \left(
    \begin{matrix} I^{W_3} \\ I^B \end{matrix}
  \right)
  ,\qquad
  \left(
    \begin{matrix} Z^\mu \\ A^\mu \end{matrix}
  \right)
  =
  \left(
    \begin{matrix} c_\mathrm{w} & s_\mathrm{w} \\ -s_\mathrm{w} & c_\mathrm{w} \end{matrix}
  \right)
  \left(
    \begin{matrix} W_3^\mu \\ B^\mu \end{matrix}
  \right),
  \label{eq:weinberg-rot}
\end{align}
by the weak mixing angle~$\theta_\mathrm{w}$, with
\begin{align}
  s_\mathrm{w}
  \equiv\sin(\theta_\mathrm{w})
  =\frac{g_1}{\sqrt{g_1^2 + g_2^2}}
  ,\qquad
  c_\mathrm{w}
  \equiv\cos(\theta_\mathrm{w})
  =\frac{g_2}{\sqrt{g_1^2 + g_2^2}}
  ,
\end{align}
we obtain the unbroken generator
\begin{align}
  I^{A}_{\cplxhiggscpt_{i} \cplxhiggscpt_{j}}
  =
  - I^{A}_{\cplxhiggscpt^\ast_{i} \cplxhiggscpt^\ast_{j}}
  =
  -\frac{1}{e}
  \frac{g_1 g_2}{\sqrt{g_1^2 + g_2^2}}
  \left(
    \begin{matrix}
      1 & 0 \\
      0 & 0 \\
    \end{matrix}
  \right)_{i j}
  ,
  \qquad
  I^{A}_{\cplxhiggscpt_{i} \cplxhiggscpt_{j}}
  \langle \cplxhiggscpt_{j} \rangle
  =
  I^{A}_{\cplxhiggscpt^\ast_{i} \cplxhiggscpt^\ast_{j}}
  \langle \cplxhiggscpt^\ast_{j} \rangle
  =
  0
  .
  \label{eq:unbroken-gen-IA}
\end{align}
The matrix~$Q \equiv - I^A$ generates the unbroken symmetry subgroup
$\mathrm{U}(1)_Q \subset \mathrm{SU}(2)_\mathrm{w} \times
\mathrm{U}(1)_Y$, which is thence identified as the group of
electromagnetic gauge transformations.
Thus,~$Q$ is interpreted as the electric charge operator.
The corresponding gauge field~$A_\mu$ is massless and identified as the
photon field.
Since $Q$ is unbroken, the \gls{vev} of the scalar doublet remains
electrically neutral.
Comparing the photon--fermion couplings to \gls{qed}, we identify 
\begin{align}
  e
  =
  \frac{
    g_1 g_2
  }{
    \sqrt{
    g_1^2
    +
    g_2^2
    }
  }
  ,
\end{align}
resulting in the Gell-Mann--Nishijima relation
\begin{align}
  Q
  =
  \frac{e}{g_2} I^{W_3} - \frac{e}{g_1} I^B
  =
  T^3 + \frac{Y}{2}
  .
  \label{eq:gell-mann-nishijima}
\end{align}
The weak hypercharges~$Y_\varphi$ of all \gls{ewsm} multiplets~$\varphi$
are chosen such that~\eqref{eq:gell-mann-nishijima} reproduces their
electric charges.
Diagonalizing the adjoint generator~$Q$ with respect to~$W_1^\mu$
and~$W_2^\mu$, we obtain
\begin{align}
  W_\pm^\mu = \frac{1}{\sqrt{2}} (W_1^\mu \mp \imagi W_2^\mu)
  ,
  \qquad
  I^{W_\pm} = \frac{1}{\sqrt{2}} (I^{W_1} \pm \imagi I^{W_2})
  ,
  \label{eq:diag-chg-gf}
\end{align}
while~$Z^\mu$ remains electrically neutral.
The generators $I^{W_\pm}$ and $I^{Z}$ are broken,
\begin{align}
  I^{W_+}_{\cplxhiggscpt_{i} \cplxhiggscpt_{j}}
  \langle \cplxhiggscpt_{j} \rangle
  &=
  -I^{W_-}_{\cplxhiggscpt^\ast_{i} \cplxhiggscpt^\ast_{j}}
  \langle \cplxhiggscpt^\ast_{j} \rangle
  =
  \frac{g_2}{e}
  \frac{v}{2}
  \left(
    \begin{matrix}
      1 \\ 0
    \end{matrix}
  \right)_i
  ,
  \qquad
  I^{W_-}_{\cplxhiggscpt_{i} \cplxhiggscpt_{j}}
  \langle \cplxhiggscpt_{j} \rangle
  =
  I^{W_+}_{\cplxhiggscpt^\ast_{i} \cplxhiggscpt^\ast_{j}}
  \langle \cplxhiggscpt^\ast_{j} \rangle
  =
  0
  ,
  \nonumber\\
  I^{Z}_{\cplxhiggscpt_{i} \cplxhiggscpt_{j}}
  \langle \cplxhiggscpt_{j} \rangle
  &=
  -I^{Z}_{\cplxhiggscpt^\ast_{i} \cplxhiggscpt^\ast_{j}}
  \langle \cplxhiggscpt^\ast_{j} \rangle
  =
  -
  \frac{\sqrt{g_1^2 + g_2^2}}{e}
  \frac{v}{2 \sqrt{2}}
  \left(
    \begin{matrix}
      0 \\ 1
    \end{matrix}
  \right)_i 
  .
\end{align}
Via~\eqref{eq:vb-mass-term}, we obtain the gauge-boson masses
\begin{align}
  M_\mathrm{Z}^2
  =
  \frac{v^2}{4} \big(g_1^2 +g_2^2\big)
  ,
  \qquad
  M_\mathrm{W}^2
  =
  \frac{v^2}{4} g_2^2
  ,
  \label{eq:mz-mw}
\end{align}
and identify
\begin{align}
  c_\mathrm{w} = M_\mathrm{W} / M_\mathrm{Z}
  .
  \label{eq:cw-from-mw-mz}
\end{align}
The would-be Goldstone-boson fields corresponding to $W_\pm$ and $Z$ read
\begin{align}
  G_\pm = \pm \eta_{W_\pm} \cplxhiggscpt^\pm
  ,\qquad
  G_Z = - \frac{\eta_{Z}}{\sqrt{2}} \left(\cplxhiggscpt^0 - \cplxhiggscpt^{0\ast}\right)
  .
\end{align}
With~$\eta_{W_\pm} = \pm 1$ and~$\eta_Z = \imagi$, we obtain the
standard conventions from
e.g.~\citep{denner2001oneloop1,denner2001oneloop2,bohm2001gauge,Denner:2019vbn},
where~$\cplxhiggscpt^\pm$ correspond to~$G_{\pm}$ and
$\chi=-\imagi/\sqrt{2} (\cplxhiggscpt^0-\cplxhiggscpt^{0\ast})$ corresponds to~$G_Z$.
The remaining scalar field can be parametrized as
\begin{align}
  H(x)
  =
  \frac{1}{\sqrt{2}}
  \big( \cplxhiggscpt^0 (x) + \cplxhiggscpt^{0\ast} (x) \big)
  ,
  \qquad
  \langle H \rangle
  =
  v
  ,
  \label{eq:ewsm-higgs-field}
\end{align}
and is identified with the Higgs field.
The associated bilinear term of the potential~\eqref{eq:higgs-pot}
yields the mass~$M_\mathrm{H} = \sqrt{2 \mu^2}$, 
while the cubic term
implies the triple-Higgs coupling~\eqref{eq:app-SSS-vertex},
\begin{align}
  C_{HHH}
  =
  -\frac{3 \lambda v}{2 e}
  =
  -\frac{3 M_\mathrm{H}^2}{e v}
  .
  \label{eq:app-HHH-cpl}
\end{align}
The relation~\eqref{eq:VGH-cpl-vb-mass} explicitly reads
\begin{align}
  M_\mathrm{W} = - e v \eta_{W_\mp} I^{W_\pm}_{G_\pm H}
  ,
  \qquad
  M_\mathrm{Z}     = - e v \eta_{Z} I^{Z}_{G_Z H}
  ,
  \label{eq:ewsm-vb-mass}
\end{align}
agreeing with~\eqref{eq:mz-mw} after plugging in the generators from
Appendix~\ref{sect:explicit-cpl-matrices}.


Finally, we discuss the fermion mass terms of the \gls{ewsm},
writing~$f^{\mathrm{L}}_{i \pm}$
for the left-handed fermion field of generation~$i$ and~$\mathrm{SU}(2)_\mathrm{w}$
doublet component~$\pm$, and~$f^{\mathrm{R}}_{i \pm}$ for the right-handed
$\mathrm{SU}(2)_\mathrm{w}$ singlet counterpart.\footnotemark\
\footnotetext{%
  This means we divide the set of fermion indices~$i$
  from~\eqref{eq:general-ferm-yukawa-cpl} into gauge group (singlet and
  doublet) and generation indices.
  This is convenient because gauge transformations do not mix
  generation indices.
  }%
Moreover, we define the charge conjugate scalar doublet
\begin{align}
  \cplxhiggsmult^\mathrm{c}_\doubcptone
  =
  \imagi \tau^2_{\doubcptone \doubcpttwo} \cplxhiggsmult_\doubcpttwo^\ast
  =
  \left( \begin{matrix} \cplxhiggscpt^{0\ast} \\ -\cplxhiggscpt^- \end{matrix} \right)_\doubcptone
  ,
  \label{eq:chg-conj-higgs-doublet}
\end{align}
which transforms like~$\cplxhiggsmult$
under~$\mathrm{SU}(2)_\mathrm{w}$.
Note that~$\cplxhiggscpt_\pm$ represent the upper and lower components
of the Higgs doublet, respectively, and \emph{differ}
from~$\cplxhiggscpt^\pm$ as defined in~\eqref{eq:higgs-doublet}.
The condition~\eqref{eq:general-ferm-yukawa-cpl-cond} is fulfilled both
for $V_a=W_1,W_2,W_3$ and for $V_a=B$ if
\begin{align}
  \Upsilon^{\cplxhiggscpt_\doubcptone}_
    {f^\mathrm{L}_{i \doubcpttwo} f^\mathrm{R}_{j \doubcptthree}}
  = 
  -
  \frac{\sqrt{2}}{v}
  M_{f^\mathrm{L}_{i -} f^\mathrm{R}_{j -}}
  \delta_{\doubcpttwo \doubcptone}
  \delta_{\doubcptthree -}
  ,
  \qquad
  \Upsilon^{\cplxhiggscpt_\doubcptone^\ast}_
    {f^\mathrm{L}_{i \doubcpttwo} f^\mathrm{R}_{j \doubcptthree}}
  =
  -
  \frac{\sqrt{2}}{v}
  M_{f^\mathrm{L}_{i +} f^\mathrm{R}_{j +}}
  \imagi \tau^2_{\doubcpttwo \doubcptone}
  \delta_{\doubcptthree +}
  .
\end{align}
In the conventions of \cite{Denner:1991kt,bohm2001gauge,Denner:2019vbn}, where the 
Yukawa coupling matrices are denoted $G^f$ ($f=\ell,d,u$), this
corresponds to
\begin{align}
\Upsilon^{\cplxhiggscpt_-}_{\ell^\mathrm{L}_{i} \ell^\mathrm{R}_{j}}
= \Upsilon^{\cplxhiggscpt_+}_{\nu^\mathrm{L}_{i} \ell^\mathrm{R}_{j}}
= -G^{\ell}_{ij},
\qquad
\Upsilon^{\cplxhiggscpt_-}_{d^\mathrm{L}_{i} d^\mathrm{R}_{j}}
= \Upsilon^{\cplxhiggscpt_+}_{u^\mathrm{L}_{i} d^\mathrm{R}_{j}}
= -G^{d}_{ij},
\qquad
\Upsilon^{\cplxhiggscpt_-^\ast}_{u^\mathrm{L}_{i} u^\mathrm{R}_{j}}
= -\Upsilon^{\cplxhiggscpt_+^\ast}_{d^\mathrm{L}_{i} u^\mathrm{R}_{j}}
= -G^{u}_{ij},
\end{align}
with $\Upsilon^{\dots}_{\dots}=0$ otherwise.
Diagonalizing the mass matrix, we arrive at the Yukawa interaction terms
\begin{align}
\mathcal{L}_{\mathrm{Yuk}} ={} &
-\sqrt{2}\frac{m_{\ell_{i}}}{v}\bar\ell^\mathrm{L}_{i}\ell^\mathrm{R}_{i} 
	\cplxhiggsmult_-
-\sqrt{2}\frac{m_{d_{i}}}{v}\bar d^\mathrm{L}_{i} d^\mathrm{R}_{i} 
	\cplxhiggsmult_-
-\sqrt{2}\frac{m_{u_{i}}}{v}\bar u^\mathrm{L}_{i} u^\mathrm{R}_{i} 
	\cplxhiggsmult^\mathrm{c}_+
\nonumber\\
& {}
-\sqrt{2}\frac{m_{\ell_{i}}}{v}\bar\nu^\mathrm{L}_{i}\ell^\mathrm{R}_{i} 
	\cplxhiggsmult_+
-\sqrt{2}\frac{m_{d_{i}}}{v}\bar u^\mathrm{L}_{i} d^\mathrm{R}_{j} 
	\cplxhiggsmult_+ V_{ij}
-\sqrt{2}\frac{m_{u_{i}}}{v}\bar d^\mathrm{L}_{i} u^\mathrm{R}_{j} 
	\cplxhiggsmult^\mathrm{c}_- V^\dagger_{ij}
  + \mathrm{h.c.},
\end{align}
where $(V_{ij})$ is the
\gls{ckm}
matrix
describing quark mixing.
Reducing the Higgs fields to its 
\glspl{vev}, $\langle \cplxhiggsmult_\doubcptone \rangle = \delta_{\doubcptone -} v/\sqrt{2}$
and~$\langle \cplxhiggsmult^\mathrm{c}_\doubcptone \rangle = \delta_{\doubcptone +} v/\sqrt{2}$,
produces the fermion mass terms
\begin{align}
\mathcal{L}_{m_f} =
  &-
  \Big(
    m_{f_{i -}} \bar f^\mathrm{L}_{i -} f^\mathrm{R}_{i -}
    +
    m_{f_{i +}} \bar f^\mathrm{L}_{i +} f^\mathrm{R}_{i +}
  \Big)
  +
  \mathrm{h.c.}
  =
  -
  \Big(
    m_{f_{i -}} \bar f_{i -} f_{i -}
    +
    m_{f_{i +}} \bar f_{i +} f_{i +}
  \Big)
  .
\end{align}
Correspondingly, the fermion--Higgs coupling~\eqref{eq:app-Sff-vertex} reads
\begin{align}
  C^\chir_{H \bar f_k f_l}
  =C_{H \bar f_k f_l}
  =-\frac{m_{f_l}}{e v} \delta_{f_k f_l}
  \label{eq:app-HFF-cpl}
  .
\end{align}

\subsection{Explicit coupling factors}
\label{sect:explicit-cpl-matrices}
Here, we provide explicit values for the coupling
factors~$I^{V_a}_{\varphi_j \varphi_k}$ for the various
fields~$\varphi_{j}$, $\varphi_{k}$, and~$V_a$ of the \gls{ewsm}.
These are largely taken from Appendix~B of~\cite{denner2001oneloop1}.
Using definition~\eqref{eq:inf-trf}, they can be easily read off from (23) of 
\cite{Denner:2019vbn} as well.
Coupling factors that are not given below vanish.
The couplings for conjugated fields are obtained
via~\eqref{eq:cpl-mtr-symmetry-1}
and \eqref{eq:cpl-mtr-symmetry-2}.

\paragraph{%
Charged-current 
couplings \mathinheadbold{I^{W_{\pm}}}{IW}.
  }%
For the $\mathrm{SU}(2)_\mathrm{w}$
doublets~$\varphi=L^\mathrm{L},Q^\mathrm{L},\cplxhiggsmult,\cplxhiggsmult^\mathrm{c}$,
the representations of~$I^{W_{\chgWindone}}$ with $\chgWindone = \pm$ have the
non-vanishing components
\begin{align}
  I^{W_\chgWindone}_{\varphi_{\doubcptone} \varphi_{-\doubcptone}}
  =
  \frac{g_2}{e}
  \frac{\delta_{\chgWindone \doubcptone}}{\sqrt{2}}
  =
  \frac{\delta_{\chgWindone \doubcptone}}{\sqrt{2} s_\mathrm{w}}
  .
  \label{eq:expl-IW}
\end{align}
Here, $\doubcptone=\pm$ indicate the upper and lower components of
the~$\mathrm{SU}(2)_\mathrm{w}$ doublet~$\varphi$,
respectively, and represent the eigenvalues of $\tau^3$.
The scalar fields~$S=H,\chi$ in the physical basis do not correspond
to eigenstates of~$\tau^3$, because the neutral Higgs
fields~$\cplxhiggscpt^0$ and~$\cplxhiggscpt^{0\ast}$ in the symmetric
basis possess distinct eigenvalues of~$\tau^3$.
Thus, the transformation of~\eqref{eq:expl-IW} into the physical basis
is non-trivial and results in
\begin{align}
  I^{W_\chgWindone}_{S \cplxhiggscpt^{-\chgWindtwo}}
  = - I^{W_\chgWindone}_{\cplxhiggscpt^{\chgWindtwo} S}
  = \delta_{\chgWindone \chgWindtwo} I^{W_\chgWindone}_{S} , 
\qquad \chgWindone, \chgWindtwo = \pm,
\end{align}
with
\begin{align}
  I^{W_\chgWindone}_{H}
  = - \frac{\chgWindone}{2 s_\mathrm{w}}
  ,
  \qquad
  I^{W_\chgWindone}_{\chi}
  = - \frac{\imagi}{2 s_\mathrm{w}}
  .
\end{align}
Again, we stress that~$\cplxhiggscpt_\pm$ differs
from~$\cplxhiggscpt^\pm$.
For the $\mathrm{SU}(2)_\mathrm{w}$ singlets
$\varphi=\ell^\mathrm{R},q^\mathrm{R},d^\mathrm{R}$, the couplings
$I^{W_\pm}$ vanish.
After transforming the $\mathrm{SU}(2)_\mathrm{w}$ structure constants
into the physical bases of the charged sector ($W_\pm$) and the neutral
sector ($A,Z$), the~$W_\pm$ couplings to gauge fields read
\begin{align}
  I^{W_\chgWindone}_{N W_{-\chgWindtwo}}
  = -I^{W_\chgWindone}_{W_{\chgWindtwo} N}
  = \delta_{\chgWindone \chgWindtwo} I^{W_\chgWindone}_{N} ,
\end{align}
with $\chgWindone,\chgWindtwo = \pm$, $N=A,Z$, and
\begin{align}
  I^{W_\chgWindone}_{A} = -\chgWindone,
  \qquad
  I^{W_\chgWindone}_{Z}
  = \chgWindone \frac{g_2}{g_1}
  = \chgWindone \frac{c_\mathrm{w}}{s_\mathrm{w}}
  .
\end{align}
\paragraph{%
    Neutral-current couplings \mathinheadbold{I^{A}}{IA},
    \mathinheadbold{I^{Z}}{IZ}.
  }%
The photon couplings $I^A=-Q$ are determined via the Gell-Mann--Nishijima
relation~\eqref{eq:gell-mann-nishijima} by the weak isospin~$T^3$ and
weak hypercharge~$Y$.
The couplings to the $Z$~field are obtained via~$T^3$ and~$Y$ according to
\begin{align}
  I^Z
  =
  \frac{1}{g_1 g_2}
  \left(g_2^2 T^3 - g_1^2 \frac{Y}{2}\right)
  =
  \frac{1}{s_\mathrm{w} c_\mathrm{w}}
  \left(c_\mathrm{w}^2 T^3 - s_\mathrm{w}^2 \frac{Y}{2}\right)
  =
  \frac{T^3 - s_\mathrm{w}^2 Q}{s_\mathrm{w} c_\mathrm{w}}
  .
\end{align}
For $\mathrm{SU}(2)_\mathrm{w}$ doublet components, we have
\begin{align}
  T^3_{\varphi_\doubcptone \varphi_{\doubcpttwo}}
  = \delta_{\doubcptone \doubcpttwo} \frac{\doubcptone}{2},
\end{align}
with~$\doubcptone,\doubcpttwo = \pm$ denoting the doublet
components as before.
For right-handed $\mathrm{SU}(2)_\mathrm{w}$ singlets,~$T^3$ vanishes.
In the basis of charge eigenstates, 
the only non-zero adjoint components of $T^3$ are
\begin{align}
  T^3_{W_\chgWindone W_{\chgWindtwo}}
  = \chgWindone \delta_{\chgWindone \chgWindtwo} ,
  \qquad
  \chgWindone,\chgWindtwo = \pm.
\end{align}
The weak hypercharges~$Y_{V_a}$ of all gauge fields
vanish,
while we have $Y_\cplxhiggsmult=-Y_{\cplxhiggsmult^\mathrm{c}}=1$ for
the Higgs doublet.
The isospin and hypercharge matrices~$I^3$ and~$Y$ are diagonal with respect to all fields of the \gls{ewsm} in the
physical basis, except for~$H$ and~$\chi$.
In the physical basis of the neutral Higgs sector, we get
\begin{align}
  T^3_{HH} = T^3_{\chi\chi} = 0, \qquad
  T^3_{H \chi} = - T^3_{\chi H} = - \frac{\imagi}{2}
\end{align}
and
\begin{align}
  T^3_{S S^\prime} = - \frac{Y_{S S^\prime}}{2}
  ,
  \qquad
  S,S^\prime \in \{ H,\chi \}
  ,
\end{align}
such that the electric charges corresponding to~$H$ and~$\chi$ indeed
vanish.
The couplings to the Z~boson are given by
\begin{align}
  I^Z_{H \chi} = - I^Z_{\chi H}
  = - \frac{\imagi}{2 s_\mathrm{w} c_\mathrm{w}} .
\end{align}
For the individual (left- and right-handed) lepton and quark fields, the
hypercharge eigenvalues are chosen such that the desired electric
charges~$Q$ are obtained.
The eigenvalues for all fields of the \gls{ewsm} are tabulated in
Table~\ref{tab:expl-IA-IZ}.
\begin{table}
  \center
\def\arraystretch{1.4}
  \begin{tabular}{|c||c|c|c|c|}
    \hline
    $\varphi$ & $T^3_\varphi$ &
    $\frac{Y_\varphi}{2}$ &
    $Q_\varphi=-I^A_{\varphi \varphi}$ &
    $I^Z_{\varphi \varphi}$
    \\
    \hline
    \hline
    $\nu^\mathrm{L}_\ell$ &
    $1/2$ & $-1/2$ & $0$ &
    $\frac{1}{2 s_\mathrm{w} c_\mathrm{w}}$
    \\
    $\ell^\mathrm{L}$ &
    $-1/2$ & $-1/2$ & $-1$ &
    $\frac{s_\mathrm{w}^2-c_\mathrm{w}^2}{2 s_\mathrm{w} c_\mathrm{w}}$
    \\
    $\ell^\mathrm{R}$ &
    $0$ & $-1$ & $-1$ &
    $\frac{s_\mathrm{w}}{c_\mathrm{w}}$
    \\
    $u^\mathrm{L}$ &
    $1/2$ & $1/6$ & $2/3$ &
    $\frac{3 c_\mathrm{w}^2-s_\mathrm{w}^2}{6 s_\mathrm{w} c_\mathrm{w}}$
    \\
    $d^\mathrm{L}$ &
    $-1/2$ & $1/6$ & $-1/3$ &
    $-\frac{3 c_\mathrm{w}^2+s_\mathrm{w}^2}{6 s_\mathrm{w} c_\mathrm{w}}$
    \\
    $u^\mathrm{R}$ &
    $0$ & $2/3$ & $2/3$ &
    $-\frac{2s_\mathrm{w}}{3c_\mathrm{w}}$
    \\
    $d^\mathrm{R}$ &
    $0$ & $-1/3$ & $-1/3$ &
    $\frac{s_\mathrm{w}}{3c_\mathrm{w}}$
    \\
    $W_+$ &
    $1$ & $0$ & $1$ &
    $\frac{c_\mathrm{w}}{s_\mathrm{w}}$
    \\
    $A$ &
    $0$ & $0$ & $0$ & $0$
    \\
    $Z$ &
    $0$ & $0$ & $0$ & $0$
    \\
    $\cplxhiggscpt^+$ &
    $1/2$ & $1/2$ & $1$ & $\frac{c_\mathrm{w}^2-s_\mathrm{w}^2}{2 s_\mathrm{w} c_\mathrm{w}}$
    \\
    $\chi$ &
    - & - & $0$ & -
    \\ 
    $H$ &
    - & - & $0$ & -
    \\
    \hline
  \end{tabular}
  \caption{
    Explicit eigenvalues of $T^3$, $Y/2$, $Q=-I^A$, and $I^Z$ for all
    fields of the \gls{ewsm}. The neutral scalar fields~$H$ and~$\chi$
    correspond to eigenvectors only of~$Q=-I^A$.
    To obtain eigenvalues for the charge conjugated fields, the entries
    are multiplied with $-1$.
  }
\label{tab:expl-IA-IZ}
\end{table}

\section{Derivation of generic would-be Goldstone couplings}
\label{sect:generic-gb-cpl-relat-derivations}
In this appendix, we discuss the derivation of the coupling relations
given in Section~\ref{sect:generic-gb-cpl-relat}.
\paragraph{\boldmath{$VGH$} couplings.}%
The vertex structure for couplings between one scalar and two vector
bosons is defined in~\eqref{eq:app-SVV-vertex}, while that for one
vector boson and two scalars is is provided
by~\eqref{eq:app-VSS-vertex}.
With this, we have
\begin{align}
  p_1^{\mu_1}
  G^{\underline{V_a V_b H_m}}_{0, \mu_1 \mu_2}
  (p_1,p_2,p_3)
  \varepsilon_b^{\mu_2}
  &=
  \imagi e
  C_{H_m V_a V_b}
  (p_1 \varepsilon_b)
  ,
  \nonumber \\
  \eta_{V_{a}} M_{V_{a}}
  G^{\underline{G_a V_b H_m}}_{0,\mu_2}
  (p_1,p_2,p_3)
  \varepsilon_b^{\mu_2}
  &=
  - 2 \imagi e
  \eta_{V_{a}} M_{V_{a}}
  I^{V_b}_{\bar G_a H_m}
  (p_1 \varepsilon_b)
  ,
\end{align}
as well as
\begin{align}
  p_2^{\mu_2}
  G^{\underline{V_a V_b H_m}}_{0, \mu_1 \mu_2}
  (p_1,p_2,p_3)
  \varepsilon_a^{\mu_1}
  &=
  \imagi e
  C_{H_m V_a V_b}
  (p_2 \varepsilon_a)
  ,
  \nonumber \\
  \eta_{V_{b}} M_{V_{b}}
  G^{\underline{V_a G_b H_m}}_{0,\mu_1}
  (p_1,p_2,p_3)
  \varepsilon_a^{\mu_1}
  &=
  - 2 \imagi e
  \eta_{V_{b}} M_{V_{b}}
  I^{V_a}_{\bar G_b H_m}
  (p_2 \varepsilon_a)
  ,
\end{align}
where~$\varepsilon_a(p_1)$ and~$\varepsilon_b(p_2)$ are polarization
vectors for the momenta~$p_1$ and~$p_2$, respectively.
Employing the \gls{wi}~\eqref{eq:single-ward-id}, we get the
identities~\eqref{eq:coupling-relat-SVV-VSG} between $HVV$ and $VGH$
couplings.
\paragraph{\boldmath{$GVV$} couplings.}%
With~\eqref{eq:app-SVV-vertex} and the pure-gauge vertex
from~\eqref{eq:app-VVV-vertex}, we have
\begin{align}
  p_1^{\mu_1}
  G^{\underline{V_a V_b V_c}}_{0, \mu_1 \mu_2 \mu_3}
  (p_1,p_2,p_3)
  \varepsilon_b^{\mu_2}
  \varepsilon_c^{\mu_3}
  &= \imagi e I^{V_a}_{\bar V_b V_c}
  [M_{V_c}^2 - M_{V_b}^2]
  (\varepsilon_b  \varepsilon_c )
  ,
  \nonumber
  \\
  \eta_{V_{a}} M_{V_{a}}
  G^{\underline{G_a V_b V_c}}_{0, \mu_2 \mu_3}
  (p_1,p_2,p_3)
  \varepsilon_b^{\mu_2}
  \varepsilon_c^{\mu_3}
  &=
  \imagi e C_{G_a V_b V_c}
  \eta_{V_{a}} M_{V_{a}}
  (\varepsilon_b \varepsilon_c)
  ,
  \label{eq:ingredients-coupling-relat-VVV-GVV}
\end{align}
where the momenta~$p_2$ and~$p_3$ are assumed on-shell.
Through the \gls{wi}~\eqref{eq:single-ward-id}, this yields the
relation~\eqref{eq:coupling-relat-VVV-GVV} between $VVV$ and $GVV$
couplings.
\paragraph{\boldmath{$VGG$} couplings.}%
We obtain~\eqref{eq:coupling-relat-VVV-VGG}, which expresses the
connection between $VGG$ and $VVV$ couplings, by combining
\begin{align}
  \eta_{V_{a}} M_{V_{a}}
  \eta_{V_{b}} M_{V_{b}}
  G^{\underline{G_a G_b V_c}}_{0, \mu_3}
  \varepsilon_c^{\mu_3}
  &=
  - 2 \imagi e
  \eta_{V_{a}} M_{V_{a}}
  \eta_{V_{b}} M_{V_{b}}
  I^{V_c}_{\bar G_a G_b}
  p_1 \varepsilon_c
  ,
  \nonumber\\
  p_1^{\mu_1}
  p_2^{\mu_2}
  G^{\underline{V_a V_b V_c}}_{0, \mu_1 \mu_2 \mu_3}
  (p_1,p_2,p_3)
  \varepsilon_c^{\mu_3}
  &=
  - \imagi e I^{V_a}_{\bar V_b V_c} M_{V_c}^2 p_1 \varepsilon_c
  ,
  \nonumber\\
  p_2^{\mu_2}
  \eta_{V_{a}} M_{V_{a}}
  G^{\underline{G_a V_b V_c}}_{0, \mu_2,\mu_3}
  (p_1,p_2,p_3)
  \varepsilon_c^{\mu_3}
  &=
  \imagi e
  \eta_{V_{a}} M_{V_{a}}
  C_{G_a V_b V_c} p_{2} \varepsilon_c
  =
  \imagi e
  I^{V_a}_{\bar V_b V_c}
  \big(M_{V_b}^2 - M_{V_c}^2\big)
  p_{1} \varepsilon_c
  ,
  \nonumber\\
  p_1^{\mu_1}
  \eta_{V_{b}} M_{V_{b}}
  G^{\underline{V_a G_b V_c}}_{0, \mu_1,\mu_3}
  (p_1,p_2,p_3)
  \varepsilon_c^{\mu_3}
  &=
  \imagi e
  \eta_{V_{b}} M_{V_{b}}
  C_{G_b V_a V_c} p_{1} \varepsilon_c
  =
  \imagi e
  I^{V_a}_{\bar V_b V_c}
  \big(M_{V_a}^2 - M_{V_c}^2\big)
  p_{1} \varepsilon_c
  ,
\end{align}
using the~\gls{wi}~\eqref{eq:double-ward-id}.
Here,~$p_3$ is assumed on-shell and we have
employed~\eqref{eq:coupling-relat-VVV-GVV} to replace $GVV$ couplings
with their $VVV$ counterparts.
\paragraph{\boldmath{$GGH$} couplings.}%
With the purely scalar vertex structure from~\eqref{eq:app-SSS-vertex},
we get
\begin{align}
  \eta_{V_{a}} M_{V_{a}}
  \eta_{V_{b}} M_{V_{b}}
  G^{\underline{G_a G_b H_m}}_{0}
  (p_1,p_2,p_3)
  &
  =
  \imagi e
  \eta_{V_{a}} M_{V_{a}}
  \eta_{V_{b}} M_{V_{b}}
  C_{G_a G_b H_m}
  .
\end{align}
Through the \gls{wi}~\eqref{eq:double-ward-id}, we combine this with
\begin{align}
  p_1^{\mu_1}
  p_2^{\mu_2}
  G^{\underline{V_a V_b H_m}}_{0, \mu_1 \mu_1}
  (p_1,p_2,p_3)
  =
  \imagi e C_{H_m V_a V_b} p_1 p_2
  =
  - 2 \imagi e \eta_{V_a} M_{V_a} I^{V_b}_{\bar G_a H_m} p_1 p_2
  ,
\end{align}
where~\eqref{eq:coupling-relat-SVV-VSG} was employed to replace the
$HVV$ coupling, and with
\begin{align}
  p_1^{\mu_1}
  \eta_{V_{b}} M_{V_{b}}
  G^{\underline{V_a G_b H_m}}_{0, \mu_1}
  (p_1,p_2,p_3)
  &
  =
  - \imagi e
  \eta_{V_{b}} M_{V_{b}}
  I^{V_a}_{\bar G_b H_m}
  (p_1 p_2 - p_1 p_3)
  ,
  \nonumber\\
  p_2^{\mu_2}
  \eta_{V_{a}} M_{V_{a}}
  G^{\underline{G_a V_b H_m}}_{0, \mu_2}
  (p_1,p_2,p_3)
  &
  =
  - \imagi e
  \eta_{V_{a}} M_{V_{a}}
  I^{V_b}_{\bar G_a H_m}
  (p_1 p_2 - p_2 p_3)
  .
\end{align}
Assuming the $H_m$ line is on-shell ($p_3^2=M_{H_m}^2$)
and using the
connection~\eqref{eq:coupling-relat-VSG-VSG} between the two different
$VGH$ couplings, we arrive at the
identity~\eqref{eq:coupling-relat-GGH-VGH} for $GGH$ and $VGH$
couplings.
\paragraph{\boldmath{$GHH$} couplings.}%
The relation~\eqref{eq:coupling-relat-VHH-GHH} between the $GHH$ and
$VHH$ couplings follows from
\begin{align}
  p_1^\mu
  G^{\underline{V_a H_m H_n}}_{0,\mu}
  (p_1,p_2,p_3)
  &=
  \imagi e I^{V_a}_{\bar H_m H_n}
  (M_{H_m}^2 - M_{H_n}^2)
  ,
  \nonumber
  \\
  \eta_{V_{a}} M_{V_{a}}
  G^{\underline{G_a H_m H_n}}_{0}
  (p_1,p_2,p_3)
  &=
  \imagi e
  \eta_{V_{a}} M_{V_{a}}
  C_{G_a H_m H_n}
  ,
\end{align}
where~$p_2$ and~$p_3$ are on-shell momenta.
\paragraph{\boldmath{$GGG$} couplings.}%
To show that all $GGG$ couplings vanish, we compute the individual
contributions to the \gls{wi}~\eqref{eq:triple-ward-id}, assuming
that the momenta of the $V_a$ and $V_b$ lines are on-shell
($p_1^2=M_{V_a}^2$, $p_2^2=M_{V_b}^2$).\footnotemark\
We obtain
\footnotetext{%
  Note, however, that the \gls{wi}~\eqref{eq:triple-ward-id} for
  three external vector bosons actually does not require any of the
  three momenta to be on-shell.
}%
\begin{align}
  %
  %
  p_1^{\mu_1} p_2^{\mu_2} p_3^{\mu_3}
  G_{0, \mu_1 \mu_2 \mu_3}^{\underline{V^{a}} \underline{V^{b}} \underline{V^{c}}}(p_1,p_2,p_3)
  &=
  0
  ,
  \nonumber\\
  %
  %
  -
  p_1^{\mu_1} p_2^{\mu_2}
  \eta_{V_{c}} M_{V_{c}}
  G_{0, \mu_1 \mu_2}^{\underline{V^{a}} \underline{V^{b}} \underline{G^{c}}}(p_1,p_2,p_3) 
  &=
  -
  \imagi e I^{V_a}_{\bar{V}_b V_c}
  \big(M_{V_b}^2-M_{V_a}^2\big)
  p_1 p_2
  ,
  \nonumber\\
  -
  p_2^{\mu_2} p_3^{\mu_3}
  \eta_{V_{a}} M_{V_{a}}
  G_{0, \mu_2 \mu_3}^{\underline{G^{a}} \underline{V^{b}} \underline{V^{c}}}(p_1,p_2,p_3) 
  &=
  -
  \imagi e I^{V_a}_{\bar{V}_b V_c}
  \big(M_{V_c}^2-M_{V_b}^2\big)
  \big(- p_1 p_2 - M_{V_b}^2\big)
  ,
  \nonumber\\
  -
  p_1^{\mu_1} p_3^{\mu_3}
  \eta_{V_{b}} M_{V_{b}}
  G_{0, \mu_1 \mu_3}^{\underline{V^{a}} \underline{G^{b}} \underline{V^{c}}}(p_1,p_2,p_3) 
  &=
  -
  \imagi e I^{V_a}_{\bar{V}_b V_c}
  \big(M_{V_a}^2-M_{V_c}^2\big)
  \big(- p_1 p_2 - M_{V_a}^2\big)
  ,
  \nonumber\\
  %
  %
  p_3^{\mu_3}
  \eta_{V_{a}} M_{V_{a}}
  \eta_{V_{b}} M_{V_{b}}
  G_{0, \mu_3}^{\underline{G^{a}} \underline{G^{b}} \underline{V^{c}}}(p_1,p_2,p_3)
  &=
  +
  \imagi e I^{V_a}_{\bar{V}_b V_c}
  \frac{M_{V_c}^2\big(M_{V_a}^2 - M_{V_b}^2\big) + M_{V_b}^4-M_{V_a}^4}{2}
  ,
  \nonumber\\
  p_1^{\mu_1}
  \eta_{V_{b}} M_{V_{b}}
  \eta_{V_{c}} M_{V_{c}}
  G_{0, \mu_1}^{\underline{V^{a}} \underline{G^{b}} \underline{G^{c}}}(p_1,p_2,p_3)
  &=
  -
  \imagi e I^{V_a}_{\bar{V}_b V_c}
  \frac{M_{V_a}^2 - M_{V_b}^2 - M_{V_c}^2}{2}
  \big(2 p_1 p_2 + M_{V_a}^2\big)
  ,
  \nonumber\\
  p_2^{\mu_2}
  \eta_{V_{a}} M_{V_{a}}
  \eta_{V_{c}} M_{V_{c}}
  G_{0, \mu_2}^{\underline{G^{a}} \underline{V^{b}} \underline{G^{c}}}(p_1,p_2,p_3)
  &=
  -
  \imagi e I^{V_a}_{\bar{V}_b V_c}
  \frac{M_{V_b}^2 - M_{V_c}^2 - M_{V_a}^2}{2}
  \big(- 2 p_1 p_2 - M_{V_b}^2\big)
  ,
  \nonumber\\
  %
  %
  -\eta_{V_{a}} M_{V_{a}}
  \eta_{V_{b}} M_{V_{b}}
  \eta_{V_{c}} M_{V_{c}}
  G_{0, \mu_3}^{\underline{G^{a}} \underline{G^{b}} \underline{G^{c}}}(p_1,p_2,p_3)
  &=
  -\imagi e C_{G_a G_b G_c} 
  \eta_{V_{a}} M_{V_{a}}
  \eta_{V_{b}} M_{V_{b}}
  \eta_{V_{c}} M_{V_{c}}
  .
\end{align}
Summing the r.h.s.\ of the first seven lines yields zero,
implying~\eqref{eq:coupling-GGG-is-zero}.
\paragraph{\boldmath{$G\bar ff$} couplings.}%
The vertex structures for couplings of scalars and vector bosons to
fermions are defined in~\eqref{eq:app-Sff-vertex}
and~\eqref{eq:app-Vff-vertex}, respectively.
Using the Dirac equations~\eqref{eq:dirac-equation} for the
spinors~$\bar v_k(p_2)$ and $u_l(p_3)$ with on-shell momenta~$p_2$
and~$p_3$ and the relation~$\gamma_\mu \omega_\chir = \omega_{-\chir}
\gamma_\mu$ for the chirality projection matrix~\eqref{eq:chir-proj-op},
we get
\begin{align}
  \bar v_k
  p_1^\mu
  G^{\underline{V_a \bar{f}_k f_l}}_{0,\mu}
  (p_1,p_2,p_3)
  u_l
  &=
  \imagi e I^{V_a}_{f^\chir_k f^\chir_l}
  \bar v_k \slashed p_1 \omega_\chir u_l
  = \imagi e I^{V_a}_{f^\chir_k f^\chir_l}
  \bar v_k
  \left(
    -\slashed p_2 \omega_\chir
    -\omega_{-\chir} \slashed p_3
  \right)
  u_l
  \nonumber
  \\
  &= \imagi e 
  \bar v_k
  \left(
    m_{f_k} I^{V_a}_{f^\chir_k f^\chir_l}
    - m_{f_l} I^{V_a}_{f^{-\chir}_k f^{-\chir}_l}
  \right)
  \omega_\chir u_l
  ,
  \\
  \eta_{V_{a}} M_{V_{a}}
  \bar v_k
  G^{\underline{G_a \bar{f}_k f_l}}_{0}
  (p_1,p_2,p_3)
  u_l
  &=
  \imagi e
  \eta_{V_{a}} M_{V_{a}}
  C^{\chir}_{G_a \bar f_k f_l}
  \bar v_k \omega_\chir u_l
  ,
\end{align}
where $\chir=\pm$ is summed over.
Plugging this into the \gls{wi}~\eqref{eq:single-ward-id} yields the
relation~\eqref{eq:coupling-relat-Vff-Gff} between
$G\bar ff$ and $V\bar ff$
couplings.

\section[%
  Alternative treatment of the mass-singular factor in
  \mathinhead{\varepsilon_\mathrm{L}}{epsilonL}
  ]{%
  Alternative treatment of the mass-singular factor in
  \mathinheadbold{\varepsilon_\mathrm{L}}{epsilonL}
}%
In this appendix,
we support our derivations from
Section~\ref{sect:splfunc-derivation} by presenting an
alternative, but closely related strategy for the treatment of the
contributions~$\propto p^\mu/M_V$ from longitudinal polarization
vectors~$\varepsilon_\mathrm{L}^\mu$ for a vector boson~$V$.
This approach follows a more ``pedestrian'' procedure.
In Section~\ref{sect:fact-gen-gaugetheo}, we detailed how the \gls{gbet}
implies the cancellation of 
mass-singular ``gauge contributions''
from diagrams
of type~\diagtype{1} that are not of collinear origin
(i.e.\ without factors of the collinearly enhanced propagator)
against counterparts from the remaining
diagrams of type~\diagtype{2}.
After performing this compensation manually, the residual contributions
from diagrams of types~\diagtype{1} and~\diagtype{2} are separately free
of such mass-singular gauge terms.
The diagrams of type~\diagtype{2} can now completely be considered
subleading.
In practice, this procedure requires the knowledge of the
compensating mass-singular gauge terms.
Conveniently, their structure was precisely and universally described
in Section~\ref{sect:fact-gen-gaugetheo}, allowing for their systematic
identification and removal from the diagrams of type~\diagtype{1}.
Below, we explicitly demonstrate this approach for the derivation of
$V^\ast \to VV$ and $\bar{f}^\ast \to V \bar{f}$ splitting functions.
An approach similar to this has been employed in~\cite{Kleiss:2020rcg}.

\label{sect:alt-approach}
\subsection[%
  \mathinhead{V^\ast \to VV}{V* to V V}
  splitting functions%
  ]{%
  \mathinheadbold{V^\ast \to VV}{V* to V V}
  splitting functions}
\label{sect:VVV-alt-approach}
For the LT configuration
of $V^\ast \to V V$ splittings,
we consider the tHF contributions that arise
from replacing~$\varepsilon^\mu(p_i)$ with~$p_i^\mu/m_i$ in the first
two lines of~\eqref{eq:VVV-EnhAmp}.
Denoting the contraction with~$p_i^\mu/m_i$ as ``auxiliary
polarization''~$\polvb_i=\mathrm{P}$, this yields
\begin{align}
  \MContrExplFields
    {\vbfield{ij},\mathrm{tHF}}
    {\vbfield{i}}
    {\vbfield{j}}^{\mathrm{P}\mathrm{T}}
  ={}&
  \frac{1}{m_i}
  \frac
    {e I_{\vbfield{i} \vbantifield{j}}^{\vbfield{ij}}}
    {p_{ij}^2-m_{ij}^2}
  \bigg\{
  -
  \big(p_{ij}^2-m_{ij}^2\big)
  \Big(\varepsilon_{\mathrm{T}}^{\ast}(p_j) T^{(n)}_{\vbfield{ij} X}(p_{ij})\Big)
  \nonumber\\
  &
  +
  \big(m_j^2- m_{ij}^2\big)
  \Big(\varepsilon_{\mathrm{T}}^{\ast}(p_j) T^{(n)}_{\vbfield{ij} X}(p_{ij})\Big)
  +
  \Big(p_{ij}  T^{(n)}_{\vbfield{ij} X}(p_{ij})\Big)
  \big(p_i  \varepsilon_{\mathrm{T}}^{\ast}(p_j)\big) 
  \bigg\}
  ,
  \nonumber\\
  \MContrExplFields
    {\gbfield{ij},\mathrm{tHF}}
    {\vbfield{i}}
    {\vbfield{j}}^{\mathrm{P}\mathrm{T}}
  ={}&
  \frac{
    e I_{\vbfield{i} \vbantifield{j}}^{\vbfield{ij}}
  }{
    p_{ij}^2-m_{ij}^2
  }
  \Bigg\{
    \frac{
      \big(m_i^2-m_j^2\big)
      \Big(p_{ij}  T^{(n)}_{\vbfield{ij} X}(p_{ij})\Big)
      \big(p_i  \varepsilon_{\mathrm{T}}^{\ast}(p_j)\big)
    }{
      m_i m_{ij}^2
    }
  \Bigg\}
  .
  \label{eq:VVV-PT-alt-approach-sepintGV}
\end{align}
In the first term of the first line, the factor~$(p_{ij}^2-m_{ij}^2)$
cancels with the propagator denominator, rendering it of the same
structure as corresponding mass-singular 
gauge terms from the unconsidered
diagrams of type~\diagtype{2}.
As argued in Section~\ref{sect:fact-gen-gaugetheo}, the \gls{gbet}
requires these two contributions, though stemming from the two
distinct types of diagrams, to cancel.
This instructs us to identify and drop the term $\propto
(p_{ij}^2-m_{ij}^2)$ from~\eqref{eq:VVV-PT-alt-approach-sepintGV}.
In other words, this manual but unambiguous procedure restores the gauge
cancellations tied to the mass-singular factor of
$\varepsilon^\mu_\mathrm{L}(p_i)$.
All other contributions from~\eqref{eq:VVV-PT-alt-approach-sepintGV}
are seemingly also singular in~$m_i$.
However, no further compensations with
diagrams of type~\diagtype{2} may occur, because 
the latter terms are
proportional to a propagator factor.
In fact, the outlined procedure leads us to
\begin{align}
  &\MContrExplFields
    {\vbfield{ij},\mathrm{tHF}}
    {\vbfield{i}}
    {\vbfield{j}}^{\mathrm{P}\mathrm{T}}
  +
  \MContrExplFields
    {\gbfield{ij},\mathrm{tHF}}
    {\vbfield{i}}
    {\vbfield{j}}^{\mathrm{P}\mathrm{T}}
  \nonumber
  \\*
  &\to
  \frac
    {e I_{\vbfield{i} \vbantifield{j}}^{\vbfield{ij}}}
    {p_{ij}^2-m_{ij}^2}
  \bigg\{
    \frac{m_j^2 - m_{ij}^2}{m_i}
    \Big(
      \varepsilon_{\mathrm{T}}^{\ast}(p_j)
      T^{(n)}_{\vbfield{ij} X}(p_{ij})
    \Big)
    +
    \frac{
      m_i^2+m_{ij}^2-m_j^2
    }{
      m_i m_{ij}
    }
    \frac{
      p_{ij}
      T^{(n)}_{\vbfield{ij} X}(p_{ij})
    }{m_{ij}}
    \big(p_i  \varepsilon_{\mathrm{T}}^{\ast}(p_j)\big) 
  \bigg\}
  ,
  \label{eq:VVV-PT-alt-approach}
\end{align}
which is well behaved in the massless limit~$m_i \to 0$, following the
discussion from Section~\ref{sect:generic-gb-cpl-relat}.
Moreover, this expression directly agrees with 
\begin{align}
  \eta_{\vbfield{i}}^\ast
  \MContrExplFields
    {\vbfield{ij},\mathrm{tHF}}
    {\gbfield{i}}
    {\vbfield{j}}^{\mathrm{G}\mathrm{T}}
  +
  \MContrExplFields
    {\gbfield{ij},\mathrm{tHF}}
    {\gbfield{i}}
    {\vbfield{j}}^{\mathrm{G}\mathrm{T}}
\end{align}
from \eqref{eq:VVV-LT-single-amp-GandR},
which means that we recover the LT splitting
function~\eqref{eq:VVV-LT-split-func} after adding the corresponding
R-polarized contributions and following the remaining procedure from
Section~\ref{sect:VVV}.
This illustrates two things:
First, selecting
would-be Goldstone-boson diagrams instead of vector-boson diagrams for
the P-polarized contributions
in Section~\ref{sect:VVV}
precisely removes the terms responsible
for incomplete gauge cancellations.
Second, the procedure from this section restores the exact
version~\eqref{eq:multi-gbet-exact} of the \gls{gbet} at the considered
mass order, which gets broken by the naive diagram selection procedure.


For the LL configuration, we have to consider the polarization
contributions PP, PR, and RP, corresponding to the different
combinations of the terms~$p^\mu/m$ and~$\restpol^\mu$
from~$\varepsilon_\mathrm{L}^\mu(p_i)$
and~$\varepsilon_\mathrm{L}^\mu(p_j)$.
For the PR and RP cases, we obtain results that are very similar
to~\eqref{eq:VVV-PT-alt-approach-sepintGV}
and treat them accordingly.
For the PP part, on the other hand, we get
\begin{align}
  \MContrExplFields
    {\vbfield{ij},\mathrm{tHF}}
    {\vbfield{i}}
    {\vbfield{j}}^{\mathrm{P}\mathrm{P}}
  ={}&
  \frac{1}{2 m_i m_j}
  \frac
    {e I_{\vbfield{i} \vbantifield{j}}^{\vbfield{ij}}}
    {p_{ij}^2-m_{ij}^2}
  \bigg\{
    \big(p_{ij}^2-m_{ij}^2\big)
    \Big( (p_i-p_j) T^{(n)}_{\vbfield{ij} X}(p_{ij}) \Big)
    \nonumber\\
    & \quad {}
    -m_i^2 \rho(m_i,m_j,m_{ij})
    \Big( (p_i-p_j)  T^{(n)}_{\vbfield{ij} X}(p_{ij}) \Big)
    \nonumber\\
    & \quad {}
    - 2 m_i^2 \Big( p_j  T^{(n)}_{\vbfield{ij} X}(p_{ij}) \Big)
    + 2 m_j^2 \Big( p_i  T^{(n)}_{\vbfield{ij} X}(p_{ij}) \Big)
  \bigg\},
  \nonumber\\
  \MContrExplFields
    {\gbfield{ij},\mathrm{tHF}}
    {\vbfield{i}}
    {\vbfield{j}}^{\mathrm{P}\mathrm{P}}
  ={}&
  \frac{1}{2 m_i m_{ij}^2 m_j}
  \frac
    {e I_{\vbfield{i} \vbantifield{j}}^{\vbfield{ij}}}
    {p_{ij}^2-m_{ij}^2}
  \bigg\{
    \big(p_{ij}^2-m_{ij}^2\big)
    \big(m_i^2-m_j^2\big)
    \Big( p_{ij}  T^{(n)}_{\vbfield{ij} X}(p_{ij}) \Big)
    \nonumber\\
    & \quad {}
    -
    m_i^2 \rho(m_i,m_j,m_{ij})
    \big(m_i^2-m_j^2\big)
    \Big( p_{ij}  T^{(n)}_{\vbfield{ij} X}(p_{ij}) \Big)
  \bigg\}
  .
  \label{eq:VVV-PP-alt-approach-sepintGV}
\end{align}
Again, the first term in each of
these equations must cancel against
counterparts from neglected diagrams of type~\diagtype{2} and therefore
be removed.
Combining the modified PP, PR, and RP contributions with the RR counterparts, we
obtain~\eqref{eq:VVV-LL-indiv-amp-fullL} and eventually reproduce the LL
splitting function~\eqref{eq:VVV-LL-split-func}.


We note that the
procedure described in this section is based on
individual amplitudes
before interfering them with complex conjugate amplitudes, 
since the multiplications
in~\eqref{eq:decomp-sq-realme-by-mother} obscure the structure
in which the problematic mass-singular gauge
terms contribute.
More specifically, consider the product of two diagrams of
type~\diagtype{1}.
%
Terms $\propto 1/M_V/(p_{ij}^2-m_{ij}^2)$ may be produced in two
distinct ways:
%
first, by multiplication of propagator-enhanced terms from one diagram
with terms~$\propto 1/M_V$ from 
another diagram, in which the
propagator denominator explicitly canceled against the numerator;
second, via products 
of two propagator-enhanced terms, where a
propagator factor is only canceled after combining 
the numerators.\footnotemark\
\footnotetext{%
  In the above example, this happens for the product of two terms
  proportional to $p_i \varepsilon_\mathrm{T}^\ast(p_j)$.
  Inserting the polarization sum~\eqref{eq:trv-poln-sum} into
  $\big(p_i \varepsilon_\mathrm{T}^\ast(p_j)\big) \big(p_i \varepsilon_\mathrm{T}(p_j)\big)$
  produces terms that are proportional to~$(p_{ij}^2-m_{ij}^2)$.
  On the level of individual diagrams, the separate
  factors~$p_i\varepsilon_\mathrm{T}^\ast(p_j)$, however, cannot fully
  cancel a propagator denominator by themselves.
  The corresponding terms, thus, may not cancel with contributions from
  neglected diagrams.
}%
While these two types of contributions are indistinguishable from each
other, only the first kind cancels against counterparts that involve
neglected diagrams of type~\diagtype{2}.

\subsection[%
  \mathinhead{\bar f^\ast \to V \bar f}{fbar* to V fbar}
  splitting functions%
  ]{%
  \mathinheadbold{\bar f^\ast \to V \bar f}{fbar* to V fbar}
  splitting functions}
\label{sect:FbarVFbar-alt-approach}
We now turn to the adaptation of this strategy to the $\bar f^\ast \to V
\bar f$ case.
In ~\cite{Kleiss:2020rcg}, this process is used for the explicit
discussion of the handling of mass-singular terms.
The corresponding manipulations are described
in~(20) and~(21) there.
However, the overall factorization method from~\cite{Kleiss:2020rcg} and thereby the precise
application of this treatment
differ significantly from our approach,
see also Section~\ref{sect:kleiss-verheyen}
and~\eqref{eq:kleiss-verheyen-removal-strat}.


Replacing~$\varepsilon^\mu(p_i)$ in the first line
of~\eqref{eq:FbarVFbar-EnhAmp} by~$p_i^\mu/m_i$ yields
\begin{align}
  \MContrExplFields{\antifermfield{ij}}{\vbfield{i}}{\antifermfield{j}}^{\mathrm{P} \polferm_j}%
  ={}&
  \frac{1}{2 m_i}
  \frac
    {e}
    {p_{ij}^2-m_{ij}^2}
  \bar T^{(n)}_{\antifermfield{ij} X} (p_{ij})
  \bigg\{
    \big(m_{ij}-m_j\big)
    \bigg(
      I_{\fermfield{ij}^\mathrm{L} \fermfield{j}^\mathrm{L}}^{\vbantifield{i}}
      +
      I_{\fermfield{ij}^\mathrm{R} \fermfield{j}^\mathrm{R}}^{\vbantifield{i}}
    \bigg)
    \slashed p_i
    \nonumber\\
    & \quad {}
    -
    \big(m_{ij}+m_j\big)
    \bigg(
      I_{\fermfield{ij}^\mathrm{L} \fermfield{j}^\mathrm{L}}^{\vbantifield{i}}
      -
      I_{\fermfield{ij}^\mathrm{R} \fermfield{j}^\mathrm{R}}^{\vbantifield{i}}
    \bigg)
    \slashed p_i
    \gamma^5
    \nonumber\\
    & \quad {}
    -
    \Big(
      m_{ij}^2-m_j^2+\big(p_{ij}^2-m_{ij}^2\big)
    \Big)
    \bigg(
      I_{\fermfield{ij}^\mathrm{L} \fermfield{j}^\mathrm{L}}^{\vbantifield{i}}
      +
      I_{\fermfield{ij}^\mathrm{R} \fermfield{j}^\mathrm{R}}^{\vbantifield{i}}
    \bigg)
    \nonumber\\
    & \quad {}
    +
    \Big(
      m_{ij}^2-m_j^2+\big(p_{ij}^2-m_{ij}^2\big)
    \Big)
    \bigg(
      I_{\fermfield{ij}^\mathrm{L} \fermfield{j}^\mathrm{L}}^{\vbantifield{i}}
      -
      I_{\fermfield{ij}^\mathrm{R} \fermfield{j}^\mathrm{R}}^{\vbantifield{i}}
    \bigg)
    \gamma^5
  \bigg\}
  v_{\polferm_j}(p_j)
  ,
  \label{eq:FbarVFbar-P-alt-approach}
\end{align}
where we simplified the Dirac algebra and employed the Dirac
equations~\eqref{eq:dirac-equation}.
Removing the terms~$\propto(p_{ij}^2-m_{ij}^2)$ and rearranging the
residual expression, we obtain
\begin{align}
  \MContrExplFields{\antifermfield{ij}}{\vbfield{i}}{\antifermfield{j}}^{\mathrm{P} \polferm_j}%
  \to{}&
  \frac{1}{m_i}
  \frac
    {e}
    {p_{ij}^2-m_{ij}^2}
  \bar T^{(n)}_{\antifermfield{ij} X} (p_{ij})
  \bigg\{
    \bigg(
      m_{ij} I_{\fermfield{ij}^\mathrm{R} \fermfield{j}^\mathrm{R}}^{\vbantifield{i}}
      -
      m_j I_{\fermfield{ij}^\mathrm{L} \fermfield{j}^\mathrm{L}}^{\vbantifield{i}}
    \bigg)
    \big(
      \slashed p_i \omega_+ - m_j \omega_- - m_{ij} \omega_+
    \big)
    \nonumber\\
       & \quad {}
    +
    \bigg(
      m_{ij}
      I_{\fermfield{ij}^\mathrm{L} \fermfield{j}^\mathrm{L}}^{\vbantifield{i}}
      -
      m_j
      I_{\fermfield{ij}^\mathrm{R} \fermfield{j}^\mathrm{R}}^{\vbantifield{i}}
    \bigg)
    \big(
      \slashed p_i \omega_- - m_j \omega_+ - m_{ij} \omega_-
    \big)
  \bigg\}
  v_{\polferm_j}(p_j)
  .
  \label{eq:FbarVFbar-P-alt-approach-modified}
\end{align}
Using the Dirac equations~\eqref{eq:dirac-equation} once more, this
expression can be identified with the partial matrix element
\begin{align}
  \eta_{\vbfield{i}}^\ast
  \MContrExplFields{\antifermfield{ij}}{\gbfield{i}}{\antifermfield{j}}^{\mathrm{G} \polferm_j}%
\end{align}
from~\eqref{eq:FbarVFbar-EnhAmp}.
Thus, supplementing~\eqref{eq:FbarVFbar-P-alt-approach-modified} with
the corresponding R-polarized contribution and completing the remaining
steps from Section~\ref{sect:FbarVFbar}, we recover the 
splitting functions for longitudinally polarized~$V$ from~\eqref{eq:FbarVFbar-LP-split-func}
and~\eqref{eq:FbarVFbar-TM-LM-split-func} for $\kappa_j=\pm$,
respectively.

\section{Reproduction of literature results for QED and QCD splitting functions}
\label{sect:qed-qcd-litres}
\subsection[%
  FS
  \mathinhead{\mathrm{g}^\ast \to \mathrm{g}\mathrm{g}}{g* to g g}
  splitting function%
  ]{%
  FS
  \mathinheadbold{\mathrm{g}^\ast \to \mathrm{g}\mathrm{g}}{g* to g g}
  splitting function}
\label{sect:VVV-litcheck}
Taking the limit
\begin{align}
  m_i, m_j, m_{ij}, m_{ij}^{\prime} \to 0
  \qquad
  \frac{\bar p^\mu}{m_{ij}} \to 0
  ,\qquad
  \frac{\bar p^\nu}{m_{ij}^\prime} \to 0
  ,
  \label{eq:VVV-ggg-limit}
\end{align}
and setting $\vbfield{ij}=\vbfield{ij^\prime}$,
the TT splitting function~\eqref{eq:VVV-TT-split-func} becomes
\begin{align}
  &
  \bigg[
    \mathcal{P}
      {}^{\vbfield{ij}}_{\vbfield{i}}
      {}^{\vbfield{ij}}_{\vbfield{j}}
      \big(z,k_\perp^\mu,\bar p^\mu,\mathrm{T},\mathrm{T}\big)
  \bigg]^{\mu \nu}
  \to
  \left| I_{\vbfield{i} \vbantifield{j}}^{\vbfield{ij}} \right|^2 
  \bigg\{
    4
    \bigg(\frac{1}{z^2}+\frac{1}{(z-1)^2}\bigg)
    k_\perp^2
    g^{\mu \nu }
    +
    8
    k_{\perp }^{\mu } k_{\perp }^{\nu }
  \bigg\}
  .
  \label{eq:VVV-ggg-match-1}
\end{align}
Accounting for the difference between the
conventions~\eqref{eq:realme-sq-fact-nonsudakov} from this work
and~(4.5) from~\cite{catani2002massivedipole} by multiplying with
\begin{align}
  G_\mathrm{CDST}
  \equiv
  \frac{e^2}{8 \pi \alpha_s}
  \frac{1}{p_{ij}^2 - m_{ij}^2}
  \to
  \frac{e^2}{8 \pi \alpha_s}
  \frac
    {z (z-1)}
    {k_\perp^2}
  \label{eq:VVV-ggg-convention-fac-csdt}
\end{align}
and setting
\begin{align}
  e^2
  \left| I_{\vbfield{i} \vbantifield{j}}^{\vbfield{ij}} \right|^2 
  \to
  4 \pi \alpha_s C_\mathrm{A}
  ,
  \label{eq:VVV-ggg-cpl}
\end{align}
we recover the triple-gluon \gls{qcd} splitting function for the
process~$\mathrm{g}^\ast \to \mathrm{g} \mathrm{g}$,
\begin{align}
  &
  G_\mathrm{CDST}
  \bigg[
    \mathcal{P}
      {}^{\vbfield{ij}}_{\vbfield{i}}
      {}^{\vbfield{ij}}_{\vbfield{j}}
      \big(z,k_\perp^\mu,\bar p^\mu,\mathrm{T},\mathrm{T}\big)
  \bigg]^{\mu \nu}
  \to
  2 C_\mathrm{A}
  \bigg\{
    \bigg(\frac{z-1}{z}+\frac{z}{z-1}\bigg)
    g^{\mu \nu }
    +
    2
    z (z-1)
    \frac{
      k_{\perp }^{\mu } k_{\perp }^{\nu }
    }{k_\perp^2}
  \bigg\}
  \label{eq:VVV-ggg-match-2}
\end{align}
in $D=4$ space--time dimensions,
given in~(4.9) of~\cite{catani2002massivedipole}
for arbitrary~$D$.

\subsection[%
  FS
  \mathinhead{\mathrm{g}^\ast \to q \bar q}{g* to q qbar}
  and
  \mathinhead{\gamma^\ast \to f \bar f}{gamma* to f fbar}
  splitting functions%
  ]{%
  FS
  \mathinheadbold{\mathrm{g}^\ast \to q \bar q}{g* to q qbar}
  and
  \mathinheadbold{\gamma^\ast \to f \bar f}{gamma* to f fbar}
  splitting functions%
  }%
\label{sect:VFFbar-litcheck}
Taking the limit
\begin{align}
  m_{ij}, m_{ij}^{\prime} \to 0
  ,\qquad
  \frac{\bar p^\mu}{m_{ij}} \to 0
  ,\qquad
  \frac{\bar p^\nu}{m_{ij}^\prime} \to 0
  \label{eq:VFF-gqq-limit}
\end{align}
and setting
\begin{align}
  m_i = m_j
  ,\qquad
  \vbfield{ij^\prime} = \vbfield{ij}
  ,\qquad
  I_{\fermfield{i}^{\mathrm{L}} \fermfield{j}^{\mathrm{L}}}^{\vbfield{ij}}
  =
  I_{\fermfield{i}^{\mathrm{R}} \fermfield{j}^{\mathrm{R}}}^{\vbfield{ij}}
  \equiv
  I_{\fermfield{i} \fermfield{j}}^{\vbfield{ij}}
  ,
  \label{eq:VFF-gqq-flavids}
\end{align}
the unpolarized splitting function~\eqref{eq:VFF-UU-split-func-simp} becomes
\begin{align}
  &
  \bigg[
  \mathcal{P}
      {}^{\vbfield{ij}}_{\fermfield{i}}
      {}^{\vbfield{ij}}_{\antifermfield{j}}
      \big(z,k_\perp^\mu,\bar p^\mu,\mathrm{U},\mathrm{U}\big)
  \bigg]^{\mu \nu}
  \to
  \left| I_{\vbfield{i} \vbantifield{j}}^{\vbfield{ij}} \right|^2 
  \bigg\{
    \frac{2 \big(m_i^2 - k_\perp^2\big)}{z (z-1)}
    g^{\mu \nu }
    -
    8
    k_{\perp }^{\mu } k_{\perp }^{\nu }
  \bigg\}
  .
  \label{eq:VFF-gqq-match-1}
\end{align}
Upon multiplication with the conventional factor
\begin{align}
  G_\mathrm{CDST}
  \to
  \frac{e^2}{8 \pi \alpha_s}
  \frac{1}{p_{ij}^2}
  =
  \frac{e^2}{8 \pi \alpha_s}
  \frac
    {z (z-1)}
    {k_\perp^2 - m_i^2}
    ,
  \label{eq:VFF-gqq-convention-fac-csdt}
\end{align}
and setting
\begin{align}
  e^2
  \left| I_{\vbfield{i} \vbantifield{j}}^{\vbfield{ij}} \right|^2 
  \to
  4 \pi \alpha_s T_\mathrm{R}
  \delta_{\fermfield{i} \fermfield{j}}
  ,
  \label{eq:VFF-gqq-cpl-csdt}
\end{align}
we recover the unpolarized~$\mathrm{g}^\ast \to q \bar q$ splitting
function from~(4.8) of~\cite{catani2002massivedipole},
\begin{align}
  &
  G_\mathrm{CDST}
  \bigg[
    \mathcal{P}
      {}^{\vbfield{ij}}_{\fermfield{i}}
      {}^{\vbfield{ij}}_{\antifermfield{j}}
      \big(z,k_\perp^\mu,\bar p^\mu,\mathrm{U},\mathrm{U}\big)
  \bigg]^{\mu \nu}
  \to
  \delta_{\fermfield{i} \fermfield{j}}
  T_\mathrm{R}
  \bigg\{
    -
    g^{\mu \nu }
    -
    \frac{
      4
      k_{\perp }^{\mu } k_{\perp }^{\nu }
    }{\big(p_i + p_j\big)^2}
  \bigg\}
  .
  \label{eq:VFF-gqq-match-2}
\end{align}
Similarly, we recover the unpolarized~$\gamma^\ast \to f \bar f$
splitting function~(4.3) from~\cite{dittmaier2008polarized} by setting
\begin{align}
  e^2
  \left| I_{\vbfield{i} \vbantifield{j}}^{\vbfield{ij}} \right|^2 
  \to
  N_{\mathrm{c},\fermfield{i}} Q_{\fermfield{i}}^2
  \delta_{\fermfield{i} \fermfield{j}}
  \label{eq:VFF-gqq-cpl-kab}
\end{align}
and multiplying with the conventional factor
\begin{align}
  G_\mathrm{DKK}
  \equiv
  \frac{1}{N_{\mathrm{c},{\fermfield{i}}} Q_{\fermfield{i}}^2}
  \frac{1}{\big(p_{ij}^2 - m_{ij}^2\big)^2}
  \to
  \frac{1}{N_{\mathrm{c},{\fermfield{i}}} Q_{\fermfield{i}}^2}
  \left(
    \frac
      {z (z-1)}
      {k_\perp^2 - m_i^2}
  \right)^2
  \label{eq:VFF-AFF-convention-fac-kab}
  ,
\end{align}
which is required to match
our~\eqref{eq:realme-sq-fact-nonsudakov} with~(4.2)
from~\cite{dittmaier2008polarized}.
Here,~$Q_f$ denotes the electric charge of fermion~$f$
and~$N_{\mathrm{c},f}$
denotes its number of colors.


Finally, we obtain the results for polarized~$\mathrm{g}^\ast \to q \bar
q$ splittings from~\cite{Czakon:2009ss}, by writing
\begin{align}
  &
  \bigg[
    \mathcal{P}
      {}^{\vbfield{ij}}_{\fermfield{i}}
      {}^{\vbfield{ij}}_{\antifermfield{j}}
      \big(z,k_\perp^\mu,\bar p^\mu,\polferm_i,\polferm_j\big)
  \bigg]^{\mu \nu}
  \nonumber\\*
  &=
  \delta_{\polferm_i +}
  \delta_{\polferm_j +}
  \bigg[
    \mathcal{P}
      {}^{\vbfield{ij}}_{\fermfield{i}}
      {}^{\vbfield{ij}}_{\antifermfield{j}}
      \big(z,k_\perp^\mu,\bar p^\mu,+,+\big)
  \bigg]^{\mu \nu}
  +
  \delta_{\polferm_i +}
  \delta_{\polferm_j -}
  \bigg[
    \mathcal{P}
      {}^{\vbfield{ij}}_{\fermfield{i}}
      {}^{\vbfield{ij}}_{\antifermfield{j}}
      \big(z,k_\perp^\mu,\bar p^\mu,+,-\big)
  \bigg]^{\mu \nu}
  \nonumber\\
  &\phantom{{}={}}
  +
  \delta_{\polferm_i -}
  \delta_{\polferm_j +}
  \bigg[
    \mathcal{P}
      {}^{\vbfield{ij}}_{\fermfield{i}}
      {}^{\vbfield{ij}}_{\antifermfield{j}}
      \big(z,k_\perp^\mu,\bar p^\mu,-,+\big)
  \bigg]^{\mu \nu}
  +
  \delta_{\polferm_i -}
  \delta_{\polferm_j -}
  \bigg[
    \mathcal{P}
      {}^{\vbfield{ij}}_{\fermfield{i}}
      {}^{\vbfield{ij}}_{\antifermfield{j}}
      \big(z,k_\perp^\mu,\bar p^\mu,-,-\big)
  \bigg]^{\mu \nu}
  .
  \label{eq:VFF-reco-kappa}
\end{align}
We insert~\eqref{eq:VFF-gqq-flavids}, contract~\eqref{eq:VFF-reco-kappa}
with~$\bar \varepsilon_{\polvb, \mu} \bar \varepsilon_{\polvb^\prime,\nu}^\ast$,
and evaluate the different scalar products of the polarization vectors via
\begin{align}
  \bar \varepsilon_{\polvb}
  \bar \varepsilon_{\pm}
  =
  - s_\varepsilon^\ast \delta_{\polvb \mp}
  ,\qquad
  \bar \varepsilon_{\polvb^\prime}^\ast
  \bar \varepsilon_{\pm}
  =
  - \delta_{\polvb^\prime \pm}
  ,\qquad
  \bar \varepsilon_{\polvb^\prime}^\ast
  \bar \varepsilon_{\polvb}
  =
  - \delta_{\polvb \polvb^\prime}
  .
\end{align}
To match the convention from~\cite{Czakon:2009ss}, we set~$s_\varepsilon \to -1$.
We reduce all occurring Kronecker deltas to a set involving
only~$\polvb^{(\prime)}$, $\polferm_i$, $\polferm_j$ and further minimize
their number by employing various algebraic relations.
Defining~$\delta^\prime_{\polferm_1 \polferm_2} = 1 - \delta_{\polferm_1
\polferm_2}$, this finally yields
\begin{alignat}{2}
\lefteqn{  \bar \varepsilon_{\polvb, \mu}
  \bar \varepsilon_{\polvb^\prime,\nu}^\ast
  \bigg[
  \mathcal{P}
      {}^{\vbfield{ij}}_{\fermfield{i}}
      {}^{\vbfield{ij}}_{\antifermfield{j}}
      \big(z,k_\perp^\mu,\bar p^\mu,\polferm_i,\polferm_j\big)
  \bigg]^{\mu \nu} }
  \nonumber\\*
  &\to
  2
  \left| I_{\vbfield{i} \vbantifield{j}}^{\vbfield{ij}} \right|^2 
  \bigg\{
    &&
    \delta_{\polvb \polvb^\prime}
    \bigg[
      m_i^2
      \bigg(
        \frac{
          \delta_{\polferm_i \polvb}
          -
          \delta^\prime_{\polferm_i \polferm_j}
        }{z}
        +
        \frac{
          \delta^\prime_{\polferm_i \polferm_j}
          -
          \delta_{\polferm_j \polvb}
        }{z-1}
      \bigg)
      +
      \frac{k_\perp^2 - m_i^2}{z (z-1)}
      \left(
        z \delta_{\polferm_i \polvb}
        -
        (z-1) \delta_{\polferm_j \polvb}
      \right)
      \delta^\prime_{\polferm_i \polferm_j}
    \bigg]
    \nonumber\\
    &&&
    -
    2
    \big(k_\perp \bar \varepsilon_\polvb\big)
    \big(k_\perp \bar \varepsilon_{\polvb^\prime}^\ast\big)
    \delta^\prime_{\polferm_i \polferm_j}
  \bigg\}.
  \label{eq:VFF-gqq-poldipol-match-2}
\end{alignat}
Multiplying with the factor~$G_\mathrm{CPW} = G_\mathrm{CDST}$, which is
implied by~(3.12) from~\cite{Czakon:2009ss}, and inserting the
couplings~\eqref{eq:VFF-gqq-cpl-csdt}, we obtain (3.14)
from~\cite{Czakon:2009ss}.

\subsection[%
  FS
  \mathinhead{q^\ast \to \mathrm{g} q}{q* to g q}
  and
  \mathinhead{f^\ast \to \gamma f}{f* to gamma f}
  splitting functions%
  ]{%
  FS
  \mathinheadbold{q^\ast \to \mathrm{g} q}{q* to g q}
  and
  \mathinheadbold{f^\ast \to \gamma f}{f* to gamma f}
  splitting functions%
  }%
\label{sect:FbarVFbar-litcheck}
In the TU-polarized $f^\ast \to V f$ splitting function, which can be
obtained from the corresponding $\bar{f}^\ast \to V \bar{f}$
counterpart~\eqref{eq:FbarVFbar-TU-split-func-EWSM}
via~\eqref{eq:FbarVFbar-to-FVF-FS-1}, we set
\begin{align}
  m_{ij} = m_{ij}^\prime = m_j
  ,\qquad
  \fermfield{ij^\prime} = \fermfield{ij}
  ,\qquad
  I_{\fermfield{j}^{\mathrm{L}} \fermfield{ij}^{\mathrm{L}}}^{\vbantifield{i}}
  =
  I_{\fermfield{j}^{\mathrm{R}} \fermfield{ij}^{\mathrm{R}}}^{\vbantifield{i}}
  \equiv
  I_{\fermfield{j} \fermfield{ij}}^{\vbantifield{i}}
  .
  \label{eq:FbarVFbar-qgq-flavids}
\end{align}
Taking the limit
$m_{i} \to 0$,
only the contributions~$\propto \slashed{\bar{p}}$ survive,
\begin{align}
  \mathcal{P}
    {}^{\fermfield{ij}}_{\vbfield{i}}
    {}^{\fermfield{ij}}_{\fermfield{j}}
    \big(z,k_\perp^\mu,\bar p^\mu,\mathrm{T},\mathrm{U}\big)
  \to
  \left| I_{\fermfield{j} \fermfield{ij}}^{\vbantifield{i}} \right|^2
  \left(
    \frac{2\big((z-1)^2 + 1\big)}{(z-1)z^2}
    \big(k_\perp^2 - m_j^2 z^2\big)
    -
    4 m_j^2
  \right)
  \slashed{\bar{p}}
  .
  \label{eq:FbarVFbar-qgq-csdt-match-1}
\end{align}
Upon multiplication with the conventional factor
\begin{align}
  G_\mathrm{CDST}
  \to
  \frac{e^2}{8 \pi \alpha_s}
  \frac{1}{2 p_i p_j}
  =
  \frac{e^2}{8 \pi \alpha_s}
  \frac
    {z (z-1)}
    {k_\perp^2 - z^2 m_j^2}
    ,
  \label{eq:FbarVFbar-qgq-convention-fac-csdt}
\end{align}
and setting
\begin{align}
  e^2
  \left| I_{\fermfield{j} \fermfield{ij}}^{\vbantifield{i}} \right|^2
  \to
  4 \pi \alpha_s C_\mathrm{F}
  \delta_{\fermfield{ij} \fermfield{j}}
  ,
  \label{eq:FbarVFbar-qgq-cpl-csdt}
\end{align}
we recover the unpolarized~$q^\ast \to \mathrm{g} q$ splitting function
from~(4.7) of~\cite{catani2002massivedipole}
for $D=4$ dimensions,
\begin{align}
  &
  G_\mathrm{CDST}
  \mathcal{P}
    {}^{\fermfield{ij}}_{\vbfield{i}}
    {}^{\fermfield{ij}}_{\fermfield{j}}
    \big(z,k_\perp^\mu,\bar p^\mu,\mathrm{T},\mathrm{U}\big)
  \to
  \delta_{\fermfield{ij} \fermfield{j}}
  C_\mathrm{F}
  \left(
    \frac{(z-1)^2 + 1}{z}
    -
    \frac{m_j^2}{p_i p_j}
  \right)
  \slashed{\bar{p}}
  .
  \label{eq:FbarVFbar-qgq-csdt-match-2}
\end{align}


To recover the polarized splitting functions for~$f^\ast \to \gamma f$
from~(2.5) of~\cite{dittmaier1999photonrad}, we consider
\begin{alignat}{2}
  &
  \lefteqn{
  \mathcal{P}
    {}^{\fermfield{ij}}_{\vbfield{i}}
    {}^{\fermfield{ij}}_{\fermfield{j}}
    \big(z,k_\perp^\mu,\bar p^\mu,\mathrm{T},\polferm_j\big)
  }
  \nonumber\\*
  &
  =
  \lefteqn{
    \delta_{\polferm_j +}
    \mathcal{P}
      {}^{\fermfield{ij}}_{\vbfield{i}}
      {}^{\fermfield{ij}}_{\fermfield{j}}
      \big(z,k_\perp^\mu,\bar p^\mu,\mathrm{T},+\big)
    +
    \delta_{\polferm_j -}
    \mathcal{P}
      {}^{\fermfield{ij}}_{\vbfield{i}}
      {}^{\fermfield{ij}}_{\fermfield{j}}
      \big(z,k_\perp^\mu,\bar p^\mu,\mathrm{T},-\big)
  }
  \nonumber\\
  &\to
  \left| I_{\fermfield{j} \fermfield{ij}}^{\vbantifield{i}} \right|^2
  \Bigg\{
    &&
    \left(
      \frac{2\big(k_\perp^2-z^2 m_j^2\big)}{z (z-1)}
      \frac{(z-1)^2 + 1}{z}
      -
      4 m_j^2
      +
      \frac{2 m_j^2 z^2}{z-1}
    \right)
    \left(
      \delta_{\polferm_j +} \slashed{\bar{p}} \omega_-
      +
      \delta_{\polferm_j -} \slashed{\bar{p}} \omega_+
    \right)
    \nonumber\\
    &&&
    -\frac{2 m_j^2 z^2}{z-1}
    \left(
      \delta_{\polferm_j +} \slashed{\bar{p}} \omega_+
      +
      \delta_{\polferm_j -} \slashed{\bar{p}} \omega_-
    \right)
  \Bigg\}
  ,
  \label{eq:FbarVFbar-qgq-stefan-match-1}
\end{alignat}
where we inserted~\eqref{eq:FbarVFbar-qgq-flavids}
into
the $f^\ast \to V f$
splitting functions for $\kappa_j=+$ and $\kappa_j=-$.
These two quantities are obtained from
the $\kappa_j=+$ result
for $\bar{f}^\ast \to V \bar{f}$
branchings,
shown in \eqref{eq:FbarVFbar-TP-split-func},
via application of 
\eqref{eq:FbarVFbar-to-FVF-FS-1}
and the combination of
\eqref{eq:FbarVFbar-TM-LM-split-func} with
\eqref{eq:FbarVFbar-to-FVF-FS-1}, respectively.
We also
took $m_i\to0$
and neglected terms~$\propto \slashed{k}_\perp
\slashed{\bar{p}}$ and~$\propto \slashed{k}_\perp \slashed{\bar{p}}
\gamma^5$.\footnotemark\
\footnotetext{%
  For more details regarding these structures, cf.\
  Appendix~\ref{sect:ktslash-terms}.
}%
While all terms $\propto m_j^2$ in the first parentheses of the third
line cancel, the following discussion becomes more transparent with this
form.
Replacing
\begin{align}
  e^2
  \left| I_{\fermfield{j} \fermfield{ij}}^{\vbantifield{i}} \right|^2
  \to
  Q_{\fermfield{j}}^2
  \delta_{\fermfield{ij} \fermfield{j}}
  \label{eq:FbarVFbar-qgq-cpl-stefan}
\end{align}
and multiplying with the conventional
factor~$G_\mathrm{D}$
that is
implied by~(2.4) from~\cite{dittmaier1999photonrad},
\begin{align}
  G_\mathrm{D} \equiv
  \frac{1}{Q_{\fermfield{j}}^2}
  \frac{1}{\big(p_{ij}^2 - m_{ij}^2\big)^2}
  \to
  \frac{1}{Q_{\fermfield{j}}^2}
  \left(
    \frac
      {1}
      {2 p_i p_j}
  \right)^2
  =
  \frac{1}{Q_{\fermfield{j}}^2}
  \left(
    \frac
      {z (z-1)}
      {k_\perp^2 - z^2 m_j^2}
  \right)^2
  \label{eq:VFF-AFF-convention-fac-stefan}
  ,
\end{align}
we get
\begin{alignat}{2}
  &
  \lefteqn{
    G_\mathrm{D} \mathcal{P}
      {}^{\fermfield{ij}}_{\vbfield{i}}
      {}^{\fermfield{ij}}_{\fermfield{j}}
      \big(z,k_\perp^\mu,\bar p^\mu,\mathrm{T},\polferm_j\big)
  }
  \nonumber\\*
  &\to
  \delta_{\fermfield{ij} \fermfield{j}}
  \bigg\{
    &&
    \bigg[
      \frac{1}{p_i p_j}
      \bigg(
        \frac{(z-1)^2+1}{z}
        -
        \frac{m_j^2}{p_i p_j}
      \bigg)
      +
      \frac{m_j^2}{2 (p_i p_j)^2}
      \frac{z^2}{z-1}
    \bigg]
    \left(
      \delta_{\polferm_j +} \slashed{\bar{p}} \omega_-
      +
      \delta_{\polferm_j -} \slashed{\bar{p}} \omega_+
    \right)
    \nonumber\\
    &&&
    -
    \frac{m_j^2}{2 (p_i p_j)^2}
    \frac{z^2}{z-1}
    \left(
      \delta_{\polferm_j +} \slashed{\bar{p}} \omega_+
      +
      \delta_{\polferm_j -} \slashed{\bar{p}} \omega_-
    \right)
  \bigg\}
  .
  \label{eq:FbarVFbar-qgq-stefan-match-2}
\end{alignat}
In the conventions of \cite{dittmaier1999photonrad} this corresponds to the
quantity $g^\mathrm{(out)}_{j,\pm}$, which turns into $g^\mathrm{(out)}_{i,\pm}$
given in (2.5) there by $z\to1-z$, $m_j\to m_i$.
According to~\eqref{eq:helproj-spin-sum-leadingmass}, the
structures~$\slashed{\bar{p}} \omega_\pm$ represent, at leading order in
the quasi-collinear limit, outer products of spinors for
fermions with helicities~$\mp\frac{1}{2}$, respectively.
These correspond to squared matrix elements for the hard process with
outgoing
fermions of helicities~$\mp \frac{1}{2}$.
The ``spin-preserving'' splitting function~$g^\mathrm{(out)}_{j,+}$
from~\cite{dittmaier1999photonrad} can therefore be read off as the
coefficient in the second line
of~\eqref{eq:FbarVFbar-qgq-stefan-match-2}, while the ``spin-flipping''
splitting function~$g^\mathrm{(out)}_{j,-}$ 
appears as prefactor in the third line.


Finally, the results for polarized~$q^\ast \to \mathrm{g} q$ splittings
from~(3.13) of~\cite{Czakon:2009ss}, summed over the two transverse
polarizations of the gluon, can be inferred from our results
as well.
For this, we identify in~\eqref{eq:FbarVFbar-qgq-stefan-match-1}
\begin{align}
  \slashed{\bar{p}} \omega_\pm
  \to
  \delta_{\polferm \polferm^\prime}
  \delta_{\polferm \mp}
  .
\end{align}
Here,~$\polferm$ and $\polferm^{\prime}$ denote helicities of the
fermions
in the associated underlying matrix elements, as discussed
above.
Employing algebraic relations between Kronecker deltas, we eventually find
\begin{alignat}{2}
  &
  \lefteqn{
    \mathcal{P}
      {}^{\fermfield{ij}}_{\vbfield{i}}
      {}^{\fermfield{ij}}_{\fermfield{j}}
      \big(z,k_\perp^\mu,\bar p^\mu,\mathrm{T},\polferm_j\big)
  }
  \nonumber\\*
  &\to
  \left| I_{\fermfield{j} \fermfield{ij}}^{\vbantifield{i}} \right|^2
  \delta_{\polferm \polferm^\prime}
  \left\{
    \left(
      2
      \frac
        {k_\perp^2 - z^2 m_j^2}
        {z (z-1)}
      \frac{(z-1)^2+1}{z}
      +
      4 m_j^2 
      \frac{(z-1)^2 + z}{z-1}
    \right)
    \delta_{\polferm \polferm_j}
    -
    \frac{
      2
      m_j^2
      z^2
    }{z-1}
  \right\}
  .
  \label{eq:FbarVFbar-qgq-poldipol-match-1}
\end{alignat}
Multiplying this with the factor~$G_\mathrm{CPW} = G_\mathrm{CDST}$, 
inserting the couplings~\eqref{eq:FbarVFbar-qgq-cpl-csdt}, 
and setting $z\to 1-z$, we obtain
agreement with (3.13) from~\cite{Czakon:2009ss}, summed over the
transverse gluon polarizations.

\subsection[%
  IS
  \mathinhead{q \to q \mathrm{g}^\ast}{q to q g*}
  and
  \mathinhead{f \to f \gamma^\ast}{f to f gamma*}
  splitting functions%
  ]{%
  IS
  \mathinheadbold{q \to q \mathrm{g}^\ast}{q to q g*}
  and
  \mathinheadbold{f \to f \gamma^\ast}{f to f gamma*}
  splitting functions%
  }%
\label{sect:FbarFbarV-litcheck-IS}
Since our procedure for the treatment of contributions involving~$\gamma^5$
follows the computation of the polarized \gls{is}~$f \to f \gamma^\ast$ splitting
functions from~\cite{dittmaier2008polarized}, we briefly discuss how those
results emerge as a special case from
our findings for general \gls{is}~$f \to f V^\ast$ splittings.
The latter are obtained
via the crossing relation \eqref{eq:fs-to-is-splfunc-mapping}
from their \gls{fs} $V^\ast\to \bar f f$ counterparts
(similar to \eqref{eq:VFF-fs-to-FFV-is-splfunc-mapping}),
which are themselves related to the corresponding \gls{fs} $V^\ast\to f \bar f$
results of Section~\ref{sect:VFFbar} as discussed in
Section~\ref{sect:VFbarF}.
Specifically, we set
\begin{align}
  m_a = m_i
  ,\qquad
  \vbfield{ai^\prime} = \vbfield{ai}
  ,\qquad
  I_{\fermfield{i}^{\mathrm{L}} \fermfield{a}^{\mathrm{L}}}^{\vbantifield{ai}}
  =
  I_{\fermfield{i}^{\mathrm{R}} \fermfield{a}^{\mathrm{R}}}^{\vbantifield{ai}}
  \equiv
  I_{\fermfield{i} \fermfield{a}}^{\vbantifield{ai}}
  ,\qquad
  s_\varepsilon=1
  \label{eq:FFV-FFA-flavids-IS}
\end{align}
in the \gls{is}~$f \to f V^\ast$ splitting functions
and investigate the limit
\begin{align}
  m_{ai}, m_{ai}^{\prime} \to 0
  ,\qquad
  \frac{\hat{p}^\mu}{m_{ai}}
  =
  \frac{\bar p^\mu - k_\perp^\mu}{m_{ai}}
  \to 0
  ,\qquad
  \frac{\hat{p}^\mu}{m_{ai}^\prime}
  =
  \frac{\bar p^\mu - k_\perp^\mu}{m_{ai}^\prime}
  \to 0
  .
  \label{eq:FFV-FFA-limit-IS}
\end{align}
The last two
replacements correspond to the \glspl{wi} for massless gauge bosons
$V_{ai}$ (such as $\gamma$ or~g).
Note that the limits~$\hat{p}^\mu /m _{ai} \to 0$ and~$\bar p^\mu /
m_{ai} \to 0$ differ at leading mass order in the quasi-collinear limit.
In this way, we obtain
\begin{alignat}{2}
  &
  \lefteqn{
    \bigg[
      \bar{\mathcal{P}}
          {}^{\vbfield{ai}}_{\fermfield{a}}
          {}^{\vbfield{ai}}_{\fermfield{i}}
        \big(x,k_\perp^\mu,\bar p^\mu,\polferm_a,\mathrm{U}\big)
    \bigg]^{\mu \nu}
  }
  \nonumber\\*
  &=
  \lefteqn{
    \delta_{\polferm_a +}
    \bigg[
      \bar{\mathcal{P}}
          {}^{\vbfield{ai}}_{\fermfield{a}}
          {}^{\vbfield{ai}}_{\fermfield{i}}
        \big(x,k_\perp^\mu,\bar p^\mu,+,\mathrm{U}\big)
    \bigg]^{\mu \nu}
    +
    \delta_{\polferm_a -}
    \bigg[
      \bar{\mathcal{P}}
          {}^{\vbfield{ai}}_{\fermfield{a}}
          {}^{\vbfield{ai}}_{\fermfield{a}}
        \big(x,k_\perp^\mu,\bar p^\mu,-,\mathrm{U}\big)
    \bigg]^{\mu \nu}
  }
  \nonumber\\
  &\to
  \left| I_{\fermfield{i} \fermfield{a}}^{\vbantifield{ai}} \right|^2
  \bigg\{
    &&
    \frac{m_a^2 x^2 - k_\perp^2}{x-1}
    g^{\mu \nu}
    +
    \frac{4}{x^2}
    k_\perp^\mu
    k_\perp^\nu
    \nonumber\\
    &&&
    {}+
    \frac{\polferm_a}{x}
    \bigg(
      (2-x)
      \frac{m_a^2 x^2 - k_\perp^2}{x-1}
      +
      2 x^2 m_a^2
    \bigg)
    \big(
      \bar{\varepsilon }_-^{\mu }
      \bar{\varepsilon }_+^{\nu }
      -
      \bar{\varepsilon }_+^{\mu }
      \bar{\varepsilon }_-^{\nu }
    \big)
  \bigg\}
  ,
  \label{eq:FFV-FFA-kab-match-IS}
\end{alignat}
where the polarizations of the \gls{fs} particle~$i$ have been summed
over.
Next, we replace
\begin{align}
  e^2
  \left| I_{\fermfield{i} \fermfield{a}}^{\vbantifield{ai}} \right|^2
  \to
  N_{\mathrm{c},{\fermfield{i}}} Q_{\fermfield{i}}^2
  \delta_{\fermfield{a} \fermfield{i}}
  \label{eq:FFV-gqq-cpl-kab-IS}
\end{align}
and multiply with the factor
\begin{align}
  \bar{G}_\mathrm{DKK}
  \equiv
  \frac{1}{N_{\mathrm{c},{\fermfield{i}}} Q_{\fermfield{i}}^2}
  \frac{1}{\big(p_{ai}^2 - m_{ai}^2\big)^2}
  \to
  \frac{1}{N_{\mathrm{c},{\fermfield{i}}} Q_{\fermfield{i}}^2}
  \frac{1}{p_{ai}^4}
  =
  \frac{1}{N_{\mathrm{c},{\fermfield{i}}} Q_{\fermfield{i}}^2}
  \left(
    \frac
      {x-1}
      {m_a^2 x^2 - k_\perp^2}
  \right)^2
  \label{eq:FFV-FFA-convention-fac-kab-IS}
  ,
\end{align}
which matches the conventions of~\eqref{eq:realme-sq-fact-nonsudakov}
and~(5.2) from~\cite{dittmaier2008polarized}.
Swapping~$\mu \leftrightarrow \nu$, we obtain the splitting functions
from (5.3) of~\cite{dittmaier2008polarized}.\footnotemark\
\footnotetext{%
  Equation (5.2) of~\cite{dittmaier2008polarized} associates the
  index~$\nu$ to~$T^{(n)}_{\vbfield{ai} X}$, not the index~$\mu$.
}%


Similarly, we recover the polarized~$q \to q \mathrm{g}^\ast$ splitting
functions from~\cite{Czakon:2009ss} in the limit defined
by~\eqref{eq:FFV-FFA-flavids-IS} and~\eqref{eq:FFV-FFA-limit-IS},
together with~$m_a, m_i \to 0$.
Specifically, we obtain
(setting $s_\varepsilon=-1$)
\begin{alignat}{2}
  &
  \bar \varepsilon^\ast_{\polvb, \mu}
  \bar \varepsilon_{\polvb^\prime, \nu} 
  \bigg[
    \bar{\mathcal{P}}
        {}^{\vbfield{ai}}_{\fermfield{a}}
        {}^{\vbfield{ai}}_{\fermfield{i}}
      \big(x,k_\perp^\mu,\bar p^\mu,\polferm_a,\polferm_i\big)
  \bigg]^{\mu \nu}
  \nonumber\\*
  &=
  \bar \varepsilon^\ast_{\polvb, \mu}
  \bar \varepsilon_{\polvb^\prime, \nu} 
  \Bigg\{
  \delta_{\polferm_a +}
  \delta_{\polferm_i +}
  \bigg[
    \bar{\mathcal{P}}
        {}^{\vbfield{ai}}_{\fermfield{a}}
        {}^{\vbfield{ai}}_{\fermfield{i}}
      \big(x,k_\perp^\mu,\bar p^\mu,+,+\big)
  \bigg]^{\mu \nu}
  +
  \delta_{\polferm_a +}
  \delta_{\polferm_i -}
  \bigg[
    \bar{\mathcal{P}}
        {}^{\vbfield{ai}}_{\fermfield{a}}
        {}^{\vbfield{ai}}_{\fermfield{i}}
      \big(x,k_\perp^\mu,\bar p^\mu,+,-\big)
  \bigg]^{\mu \nu}
  \nonumber\\
  &\phantom{{}={}}
  +
  \delta_{\polferm_a -}
  \delta_{\polferm_i +}
  \bigg[
    \bar{\mathcal{P}}
        {}^{\vbfield{ai}}_{\fermfield{a}}
        {}^{\vbfield{ai}}_{\fermfield{i}}
      \big(x,k_\perp^\mu,\bar p^\mu,-,+\big)
  \bigg]^{\mu \nu}
  +
  \delta_{\polferm_a -}
  \delta_{\polferm_i -}
  \bigg[
    \bar{\mathcal{P}}
        {}^{\vbfield{ai}}_{\fermfield{a}}
        {}^{\vbfield{ai}}_{\fermfield{i}}
      \big(x,k_\perp^\mu,\bar p^\mu,-,-\big)
  \bigg]^{\mu \nu}
  \Bigg\}
  \nonumber\\
  &\to
  \left| I_{\fermfield{i} \fermfield{a}}^{\vbantifield{ai}} \right|^2
  \bigg\{
    \frac{
      4
      (k_\perp \bar \varepsilon_{\polvb^\prime})
      (k_\perp \bar \varepsilon^\ast_{\polvb})
    }{x^2}
    +
    \frac{k_\perp^2}{2(x-1)}
    \delta_{\polvb^\prime \polvb}
    \left(
      4
      \delta^\prime_{\polferm_a \polvb}
      +
      \frac{
        4
        \left(
          2 \delta_{\polferm_a \polvb}
          -
          1
        \right)
      }{x}
    \right)
  \bigg\}
  \delta_{\polferm_a \polferm_i}
  ,
  \label{eq:FFV-FFA-poldip-match-IS}
\end{alignat}
similarly to the procedure described in Appendix~\ref{sect:VFFbar-litcheck}.
Replacing
\begin{align}
  e^2
  \left| I_{\fermfield{i} \fermfield{a}}^{\vbantifield{ai}} \right|^2
  \to
  4 \pi \alpha_s C_\mathrm{F}
  \delta_{\fermfield{a} \fermfield{i}}
  \label{eq:FFV-gqq-cpl-poldip-IS}
\end{align}
and multiplying with the conventional factor
\begin{align}
  \bar{G}_\mathrm{CPW}
  =
  -\frac{x e^2}{p_{ai}^2 - m_{ai}^2}
  \to
  \frac{x e^2}{2 p_a p_i}
  =
  e^2
  \frac
    {(x-1)x}
    {k_\perp^2}
    ,
  \label{eq:FbarFbarV-qgq-convention-fac-csdt-IS}
\end{align}
we obtain (4.32) from~\cite{Czakon:2009ss}, after the following modification of the latter,
\begin{align}
  \frac{u_i (1-u_i)}{p_i p_j}
  \left(
    \frac{p_i^\mu}{u_i}
    -
    \frac{p_j^\mu}{1-u_i}
  \right)
  \left(
    \frac{p_i^\nu}{u_i}
    -
    \frac{p_j^\nu}{1-u_i}
  \right)
  \to
  -
  2
  \frac{
    k_\perp^\mu
    k_\perp^\nu
  }{k_\perp^2}
  ,
\end{align}
where the notation from~\cite{Czakon:2009ss} is used.

\section{Further details on FS splittings of (anti-)fermions}
\label{sect:details-fermion-splittings}

\subsection{Identification of matrix elements with definite (anti-)fermion polarizations}
\label{sect:id-hard-me}
Here, we detail the relation of the
structures~\eqref{eq:ferm-splittings-dirac-structures} to underlying
matrix elements of definite polarization, i.e.\
\eqref{eq:pol-ul-me-antiferm} and~\eqref{eq:pol-ul-me-ferm}.
%
%
First, we focus on the forms $\slashed{\bar{p}}$ and $\slashed{\bar{p}} \gamma^5$.
The completeness relations~\eqref{eq:spinproj-complrelat-ferm} yield
\begin{align}
    u_{\pm}(p) \bar{u}_{\pm}(p)
    &=
    \frac{1 \pm \gamma^5}{2}
    \slashed p
    +\orderof{m_f}
    =
    \slashed p
    \frac{1 \mp \gamma^5}{2}
    +\orderof{m_f}
    ,\nonumber\\
    v_{\pm}(p) \bar{v}_{\pm}(p)
    &=
    \frac{1 \mp \gamma^5}{2}
    \slashed p
    +\orderof{m_f}
    =
    \slashed p
    \frac{1 \pm \gamma^5}{2}
    +\orderof{m_f}
    .
    \label{eq:helproj-spin-sum-leadingmass}
\end{align}
For matrix elements with an external antifermion this implies,
assuming~$m_{ij}=m_{ij}^\prime$,
\begin{align}
  \bar T^{(n)}_{\antifermfield{ij} X}
  \,
  \slashed{\bar{p}}
  \,
  T^{(n)}_{\antifermfield{ij^\prime} X}
  &=
  \mathcal{M}^{(n)\, +}_{\antifermfield{ij} X}
  \mathcal{M}^{(n)\, +\, \ast}_{\antifermfield{ij^\prime} X}
  +
  \mathcal{M}^{(n)\, -}_{\antifermfield{ij} X}
  \mathcal{M}^{(n)\, -\, \ast}_{\antifermfield{ij^\prime} X}
  +
  \orderof{m_f}
  ,
  \nonumber\\
  \bar T^{(n)}_{\antifermfield{ij} X}
  \,
  \slashed{\bar{p}}
  \gamma^5
  \,
  T^{(n)}_{\antifermfield{ij^\prime} X}
  &=
  \mathcal{M}^{(n)\, +}_{\antifermfield{ij} X}
  \mathcal{M}^{(n)\, +\, \ast}_{\antifermfield{ij^\prime} X}
  - 
  \mathcal{M}^{(n)\, -}_{\antifermfield{ij} X}
  \mathcal{M}^{(n)\, -\, \ast}_{\antifermfield{ij^\prime} X}
  +
  \orderof{m_f}
  ,
  \label{eq:non-ktslash-terms-amplitudes-antiferm}
\end{align}
while for those with an external fermion
\begin{align}
  \bar T^{(n)}_{\fermfield{ij^\prime} X}
  \,
  \slashed{\bar{p}}
  \,
  T^{(n)}_{\fermfield{ij} X}
  &=
  \mathcal{M}^{(n)\, +}_{\fermfield{ij} X}
  \mathcal{M}^{(n)\, +\, \ast}_{\fermfield{ij^\prime} X}
  +
  \mathcal{M}^{(n)\, -}_{\fermfield{ij} X}
  \mathcal{M}^{(n)\, -\, \ast}_{\fermfield{ij^\prime} X}
  +
  \orderof{m_f}
  ,
  \nonumber
  \\
  \bar T^{(n)}_{\fermfield{ij^\prime} X}
  \,
  \slashed{\bar{p}}
  \gamma^5
  \,
  T^{(n)}_{\fermfield{ij} X}
  &=
  \mathcal{M}^{(n)\, -}_{\fermfield{ij} X}
  \mathcal{M}^{(n)\, -\, \ast}_{\fermfield{ij^\prime} X}
  -
  \mathcal{M}^{(n)\, +}_{\fermfield{ij} X}
  \mathcal{M}^{(n)\, +\, \ast}_{\fermfield{ij^\prime} X}
  +
  \orderof{m_f}
  .
  \label{eq:non-ktslash-terms-amplitudes-ferm}
\end{align}
Note that the precise definition of the spin reference axes is
irrelevant for the considered approximation.


Next, we turn to the structures
$\slashed{k}_{\perp } \slashed{\bar{p}}$
and
$\slashed{k}_{\perp }\slashed{\bar{p}}\gamma^5$.
For transverse polarization vectors~$\varepsilon^\mu_\pm (p)$ and
spinors~$u_\pm(p)$, $v_\pm(p)$ that are defined as helicity eigenvectors in
some arbitrary, but common frame,
we find the relations
\begin{align}
  \slashed \varepsilon_\pm(p)
  (\slashed p - m_f)
  &=
  \pm
  \sqrt{2}
  u_\pm(p) \bar v_\pm(p)
  ,
  &
  \slashed \varepsilon_\pm (p)
  (\slashed p + m_f)
  &=
  \pm
  \sqrt{2}
  v_\mp(p) \bar u_\mp(p)
  ,
  \nonumber\\
  \slashed \varepsilon_\pm(p)
  (\slashed p - m_f)
  \gamma^5
  &=
  +
  \sqrt{2}
  u_\pm(p) \bar u_\mp(p)
  ,
  &
  \slashed \varepsilon_\pm (p)
  (\slashed p + m_f)
  \gamma^5
  &=
  +
  \sqrt{2}
  v_\mp(p) \bar v_\pm(p)
  ,
  \label{eq:ktslash-spinor-outer-product-uv}
\end{align}
where the conventions from~\cite{Dittmaier:1998nn} regarding
polarization vectors and Dirac matrices are followed (this involves
$s_\varepsilon=+1$).
These identities
can be shown
either
with explicit parametrizations or by using the
Weyl--van-der-Waerden 
spinor formalism.\footnotemark\
\footnotetext{%
  The Weyl--van-der-Waerden
  formalism expresses four-momenta, polarization vectors, and
  Dirac spinors in terms of
  Weyl--van-der-Waerden spinors
  as common fundamental object
  and thus provides a natural language
  for this kind of relations.
  In~\cite{Reyer:2024}, a proof of \eqref{eq:ktslash-spinor-outer-product-uv}
  in terms of the spinor formalism from~\cite{Dittmaier:1998nn} is presented
  (including some generalizations of the formalism to complex four-vectors).
}%
Specifying the polarization vectors in~\eqref{eq:ktslash-spinor-outer-product-uv} as $\bar{\varepsilon}^\mu_\pm$
from~\eqref{eq:eps-bar-conditions}, these
identities can be combined
into expressions for $\slashed{k}_{\perp }
\slashed{\bar{p}}$
and
$\slashed{k}_{\perp }\slashed{\bar{p}}\gamma^5$
via~\eqref{eq:kt-from-eps-via-sp}.
Using
\begin{align}
  u_{\pm}(\bar p)
  =
  v_{\mp}(\bar p)
  + \orderof{m_f}
  ,
  \label{eq:upm-vmp-relat}
\end{align}
and neglecting terms beyond leading mass order, we obtain
\begin{align}
  \bar T^{(n)}_{\antifermfield{ij} X}
  \,
  \slashed{k}_{\perp }
  \slashed{\bar{p}}
  \gamma^5
  \,
  T^{(n)}_{\antifermfield{ij^\prime} X}
  ={}
  &
  - \sqrt{2} (k_\perp \bar \varepsilon_-)
  \mathcal{M}^{(n)\, -}_{\antifermfield{ij} X}
  \mathcal{M}^{(n)\, +\, \ast}_{\antifermfield{ij^\prime} X}
  - \sqrt{2} (k_\perp \bar \varepsilon_+)
  \mathcal{M}^{(n)\, +}_{\antifermfield{ij} X}
  \mathcal{M}^{(n)\, -\, \ast}_{\antifermfield{ij^\prime} X}
  +
  \orderof{m_f}
  ,
  \nonumber
  \\
  \bar T^{(n)}_{\antifermfield{ij} X}
  \,
  \slashed{k}_{\perp }
  \slashed{\bar{p}}
  \,
  T^{(n)}_{\antifermfield{ij^\prime} X}
  ={}
  &
  - \sqrt{2} (k_\perp \bar \varepsilon_-)
  \mathcal{M}^{(n)\, -}_{\antifermfield{ij} X}
  \mathcal{M}^{(n)\, +\, \ast}_{\antifermfield{ij^\prime} X}
  + \sqrt{2} (k_\perp \bar \varepsilon_+)
  \mathcal{M}^{(n)\, +}_{\antifermfield{ij} X}
  \mathcal{M}^{(n)\, -\, \ast}_{\antifermfield{ij^\prime} X}
  +
  \orderof{m_f}
  \label{eq:ktslash-terms-amplitudes-antiferm}
\end{align}
and
\begin{align}
  \bar T^{(n)}_{\fermfield{ij^\prime} X}
  \,
  \slashed{k}_{\perp }
  \slashed{\bar{p}}
  \gamma^5
  \,
  T^{(n)}_{\fermfield{ij} X}
  ={}
  &
  - \sqrt{2} (k_\perp \bar \varepsilon_-)
  \mathcal{M}^{(n)\, -}_{\fermfield{ij} X}
  \mathcal{M}^{(n)\, +\, \ast}_{\fermfield{ij^\prime} X}
  - \sqrt{2} (k_\perp \bar \varepsilon_+)
  \mathcal{M}^{(n)\, +}_{\fermfield{ij} X}
  \mathcal{M}^{(n)\, -\, \ast}_{\fermfield{ij^\prime} X}
  +
  \orderof{m_f}
  ,
  \nonumber
  \\
  \bar T^{(n)}_{\fermfield{ij^\prime} X}
  \,
  \slashed{k}_{\perp }
  \slashed{\bar{p}}
  \,
  T^{(n)}_{\fermfield{ij} X}
  ={}
  &
  - \sqrt{2} (k_\perp \bar \varepsilon_-)
  \mathcal{M}^{(n)\, -}_{\fermfield{ij} X}
  \mathcal{M}^{(n)\, +\, \ast}_{\fermfield{ij^\prime} X}
  + \sqrt{2} (k_\perp \bar \varepsilon_+)
  \mathcal{M}^{(n)\, +}_{\fermfield{ij} X}
  \mathcal{M}^{(n)\, -\, \ast}_{\fermfield{ij^\prime} X}
  +
  \orderof{m_f}
  .
  \label{eq:ktslash-terms-amplitudes-ferm}
\end{align}
With the conventions from~\cite{Cuomo:2019siu},
there
is an extra sign in the relations
from~\eqref{eq:ktslash-spinor-outer-product-uv} that involve
$\varepsilon_+$.\footnotemark\
\footnotetext{%
  The adaption of the formalism from~\cite{Dittmaier:1998nn} to the
  conventions from~\cite{Cuomo:2019siu} is more involved than this
  simple result suggests.
  Specifically, this involves 
  $\gamma^i_{\text{\cite{Dittmaier:1998nn}}} = - \gamma^i_{\text{\cite{Cuomo:2019siu}}}$
  for $i=1,2,3,5$ and
  $\varepsilon^{\text{\cite{Dittmaier:1998nn}}}_\pm(p)
  = \pm \exp\{\mp 2 \imagi \phi\} \varepsilon^{\text{\cite{Cuomo:2019siu}}}_\pm(p)$,
  where $\phi$ denotes the azimuthal angle of the momentum $p^\mu$ w.r.t.\ the $z$-axis.
  While the explicit representation of the $\gamma$-matrices is not
  specified in~\cite{Cuomo:2019siu}, it may be reconstructed from the
  parametrization of the spinors in (A.6)--(A.8) therein.
}%
This is, however, compensated by $s_\varepsilon=-1$
in~\eqref{eq:kt-from-eps-via-sp}, such that
that~\eqref{eq:ktslash-terms-amplitudes-antiferm}
and~\eqref{eq:ktslash-terms-amplitudes-ferm} hold regardless.
We note that the precise definition of the fermion spin reference axis
is irrelevant for~\eqref{eq:ktslash-terms-amplitudes-antiferm}
and~\eqref{eq:ktslash-terms-amplitudes-ferm}, despite $u_\pm(p)$ and
$v_\pm(p)$ in~\eqref{eq:ktslash-spinor-outer-product-uv}
being defined as helicity spinors in frames where~$\varepsilon_\pm^\mu$
are transverse (which corresponds to frames where $\bar{\boldsymbol{p}} \parallel
\boldsymbol{n}$ for the choice $\varepsilon^\mu_\pm =
\bar{\varepsilon}^\mu_\pm$ made
in~\eqref{eq:ktslash-terms-amplitudes-antiferm}
and~\eqref{eq:ktslash-terms-amplitudes-ferm}).
This is because helicity and chirality eigenspinors agree at leading
mass order
(up to the sign change for antifermions).

\subsection[%
  Suppression of terms involving 
  \mathinhead{\slashed{k}_\perp}{\~kperpslash}%
  ]{%
  Suppression of terms involving
  \mathinheadbold{\slashed{k}_\perp}{\~kperpslash}%
  }
\label{sect:ktslash-terms}
In general, for $\bar f^\ast \to V \bar f$, $f^\ast \to V f$, $\bar
f^\ast \to H \bar f$, and~$f^\ast \to H f$ splittings, the structures
\begin{align}
  \slashed{k}_{\perp }
  \slashed{\bar{p}}
  ,\qquad
  \slashed{k}_{\perp }
  \slashed{\bar{p}}
  \gamma^5
  .
  \label{eq:ktslash-terms}
\end{align}
contribute to our unpolarized splitting functions if the couplings have
a non-trivial chiral structure or if the two fermion masses are not
equal, and to our polarized results even for non-chiral couplings and
equal fermion masses.
Similar terms have also been obtained and commented on
in~\cite{Kleiss:1986ct} for the computation of \gls{fs} $f^\ast \to
\gamma f$ splitting functions with polarized fermions, and
in~\cite{dittmaier2008polarized} for \gls{is} $\gamma \to f \bar f^\ast$
splittings with polarized photons.
These terms have, however, not been explicitly presented in the
literature so far.\footnotemark\
\footnotetext{%
  The results from~\cite{Cuomo:2019siu} do contain corresponding
  contributions, as discussed in Section~\ref{sect:wulzer-vecchi-cuomo}.
  These results are, however, derived and presented in terms of~$1 \to
  2$ splitting amplitudes.
  The relation to the
  structures~\eqref{eq:ktslash-terms} that appear in
  our factorization of squared amplitudes is not obvious,
  see~\eqref{eq:ktslash-terms-amplitudes-antiferm}
  and~\eqref{eq:ktslash-terms-amplitudes-ferm}.
}%
This is due to the fact that the structures~\eqref{eq:ktslash-terms}
always appear with prefactors that vanish in the limit of massless
fermions, and that taking their azimuthal average around the collinear
axis provides further
suppression.\footnotemark%
\footnotetext{%
  With the specific definition of the azimuthal integral from
  Section~\ref{sect:F-az-avg}, these terms even vanish. 
}%
\textsuperscript{,}\footnotemark\
\footnotetext{%
  In~\cite{dittmaier2008polarized}, the \gls{is} $\gamma \to f \bar
  f^\ast$ splitting functions are considered in the context of the
  application to the dipole subtraction formalism.
  In principle, even subtraction terms that vanish for $m_f\to0$
  may contribute to the massless limit of the integrated counterterms.
  Due to the suppression from the azimuthal integral within the emission phase space,
  this is not the case for the terms~\eqref{eq:ktslash-terms}.
  Thus, they are irrelevant for the construction of the dipole
  subtraction formalism for light fermions.
}%


Actually, it turns out that~\eqref{eq:ktslash-terms-amplitudes-antiferm}
and~\eqref{eq:ktslash-terms-amplitudes-ferm} can be subject to
additional mass suppression, on top of the naive quasi-collinear scaling
used to obtain our results.
In that case, terms in~\eqref{eq:ktslash-terms} can be dropped even
before taking the azimuthal average.
As we discuss in the following, such a suppression mechanism indeed
exists in the \gls{ewsm} and is due to the chirality structure of the
underlying amplitudes.


%
We first show that for any given hard process with an external
fermion, either the left-chiral amplitude
$\mathcal{M}^{(n)\,\mathrm{L}}_{f X}$
is suppressed with respect to the right-chiral counterpart
$\mathcal{M}^{(n)\,\mathrm{R}}_{f X}$,
or vice versa.
This can be understood with the help of the following
argument from~\cite{Cuomo:2019siu}.\footnotemark\
\footnotetext{%
  Note that the authors of~\cite{Cuomo:2019siu} use their
  ``mass--parity'' argument in an entirely different context.
  In fact, they do not consider the potential suppression of terms that
  correspond to~\eqref{eq:ktslash-terms}
  (via~\eqref{eq:id-ktslash-terms}).
}%
The \gls{ewsm} before SSB and gauge fixing is invariant under the
``mass--parity''~$\mathbb{Z}_2$-symmetry
\begin{align}
  \Phi \to - \Phi
  , \qquad
  f^\mathrm{L} \to - f^\mathrm{L}
  ,
  \label{eq:Z2-symmetry}
\end{align}
which is part of $\mathrm{SU}(2)_\mathrm{w}$.
In the spontaneously broken and 
gauge-fixed theory, this transformation
corresponds to
\begin{align}
  h &\to - h,
  &
  G_a &\to - G_a,
  &
  m &\to - m,
  \nonumber\\
  V_a &\to + V_a,
  &
  f^\mathrm{L} &\to - f^\mathrm{L},
  &
  f^\mathrm{R} &\to + f^\mathrm{R}
  ,
\end{align}
where~$m$ collectively denotes all masses and the \gls{vev}~$v$.
Invariance under this transformation implies that an amplitude
must scale with an even/odd power of~$m$, respectively,
if it involves an even/odd number of external left-handed fermions, Higgs
bosons, and would-be Goldstone bosons.
Hence, for some $l \in \mathbb{Z}$, we have
\begin{align}
  \left.
  \mathcal{M}^{(n)\, \mathrm{L}}_{f X}
  \middle/
  \mathcal{M}^{(n)\, \mathrm{R}}_{f X}
  \right.
  \sim
  m^{2l+1}
  ,
  \label{eq:amp-suppression-flipped-chir}
\end{align}
such that one of the amplitudes is suppressed with respect to the other.
As a consequence, the products
$\mathcal{M}^\mathrm{L} \mathcal{M}^{\mathrm{R}\,\ast}$
and
$\mathcal{M}^\mathrm{R} \mathcal{M}^{\mathrm{L}\,\ast}$
are mass suppressed w.r.t.\
$\left|\mathcal{M}^\mathrm{L}\right|^2$
or
$\left|\mathcal{M}^\mathrm{R}\right|^2$.
If either of the amplitudes vanishes, these statements become trivial.
Our factorization procedure implicitly assumes that one of them is non-zero.
Finally, this implies that the
contributions~\eqref{eq:ktslash-terms-amplitudes-ferm} are suppressed
w.r.t.~\eqref{eq:non-ktslash-terms-amplitudes-ferm}, because the
amplitudes of definite polarization can, at leading order in~$m$, be
substituted by amplitudes of definite chiralities,
\begin{align}
  \mathcal{M}^{(n)\, \mathrm{L}/\mathrm{R}}_{f X}
  &= 
  \big( \bar u_{\mp}(\bar p) + \orderof{m} \big)
  T^{(n)}_{f X}
  =
  \mathcal{M}^{(n)\, \mp}_{f X}
  +
  \orderof{m}
  .
\end{align}
The same arguments apply for amplitudes with an external antifermion and
generalize to more general gauge theories, as long as  they exhibit a
symmetry like~\eqref{eq:Z2-symmetry}.
This concludes our claim that the Dirac structures
involving~$\slashed{k}_\perp$ are subleading.


Note, however, that the above arguments require all particle masses~$m$ to be
small compared to the relevant energy scales.
This is generally assumed in the computation of our splitting functions.
In~\cite{dittmaier2008polarized} and~\cite{Kleiss:1986ct}, however, only
the mass of the fermion that participates in the considered splittings
needs to be small.
This limits the applicability of our
statements in the corresponding situations.


\acknowledgments

We acknowledge support by the Research Training Group RTG2044 of the
German Research Foundation (DFG).


\renewcommand{\bibname}{References}
\bibliographystyle{JHEP}
\bibliography{./literature}

\end{document}